\documentstyle[12pt]{article}
\newfont{\twelvemsb}{msbm10 scaled\magstep1}
\newfont{\eightmsb}{msbm8}
\newfam\msbfam
\textfont\msbfam=\twelvemsb
\scriptfont\msbfam=\eightmsb
\catcode`\@=11
\def\Bbb{\ifmmode\let\next\Bbb@\else
  \def\next{\errmessage{Use \string\Bbb\space only in math mode}}\fi\next}
\def\Bbb@#1{{\fam\msbfam{{#1}}}}
\newfont{\twelvegoth}{eufm10 scaled\magstep1}
\newfont{\eightgoth}{eufm8}
\newfam\gothfam
\textfont\gothfam=\twelvegoth
\scriptfont\gothfam=\eightgoth
\def\frak{\ifmmode\let\next\frak@\else
  \def\next{\errmessage{Use \string\frak\space only in math mode}}\fi\next}
\def\frak@#1{{\fam\gothfam{{#1}}}}
\catcode`@=12

\newcommand{\beo}{\begin{eqnarray*}}
\newcommand{\eno}{\end{eqnarray*}}

\newcommand{\bc}{{\bar c}}

\newcommand{\bi}{{\bar\imath}}
\newcommand{\bj}{{\bar\jmath}}
\newcommand{\bk}{{\bar k}}
\newcommand{\bl}{{\bar l}}

\newcommand{\bD}{{\bar D}}

\newcommand{\bF}{{\bar F}}

\newcommand{\bQ}{{\bar Q}}

\newcommand{\balpha}{{\bar\alpha}}
\newcommand{\bbeta}{{\bar\beta}}
\newcommand{\bchi}{{\bar\chi}}
\newcommand{\blambda}{{\bar\lambda}}

\newcommand{\bpsi}{{\bar\psi}}
\newcommand{\bsigma}{{\bar\sigma}}
\newcommand{\btheta}{{\bar\theta}}
\newcommand{\bzeta}{{\bar\zeta}}

\newcommand{\cA}{{\cal A}}
\newcommand{\cB}{{\cal B}}
\newcommand{\cC}{{\cal C}}
\newcommand{\cD}{{\cal D}}
\newcommand{\cF}{{\cal F}}
\newcommand{\cG}{{\cal G}}
\newcommand{\cH}{{\cal H}}
\newcommand{\cI}{{\cal I}}
\newcommand{\cK}{{\cal K}}
\newcommand{\cM}{{\cal M}}
\newcommand{\cN}{{\cal N}}

\newcommand{\cR}{{\cal R}}
\newcommand{\cS}{{\cal S}}

\newcommand{\cU}{{\cal U}}
\newcommand{\cV}{{\cal V}}
\newcommand{\cW}{{\cal W}}

\newcommand{\cZ}{{\cal Z}}

\newcommand{\dalpha}{{\dot\alpha}}
\newcommand{\dbeta}{{\dot\beta}}

\newcommand{\vv}{{\vec v}}

\newcommand{\CC}{{\Bbb C}}
\newcommand{\HH}{{\Bbb H}}
\newcommand{\II}{{\Bbb I}}
\newcommand{\JJ}{{\Bbb J}}
\newcommand{\KK}{{\Bbb K}}
\newcommand{\NN}{{\Bbb N}}

\newcommand{\RR}{{\Bbb R}}
\newcommand{\ZZ}{{\Bbb Z}}

\newcommand{\cle}{\Big[}
\newcommand{\cri}{\Big]}
\newcommand{\ale}{\Big\{}
\newcommand{\ari}{\Big\}}
\newcommand{\sle}{\Big[\hspace{-3pt}\Big[}
\newcommand{\sri}{\Big]\hspace{-3pt}\Big]}
\newcommand{\zle}{[\hspace{-2pt}[}
\newcommand{\zri}{]\hspace{-2pt}]}

\newcommand{\del}{\delta}
\newcommand{\eps}{\varepsilon}
\newcommand{\prt}{\partial}
\newcommand{\ad}{\mbox{ad}\,}
\newcommand{\aut}{\mbox{Aut}}
\newcommand{\autint}{\mbox{Int}}
\newcommand{\autout}{\mbox{Out}}
\newcommand{\ch}{\mbox{ch}\,}
\newcommand{\der}{\mbox{Der}\,}
\newcommand{\diag}{\mbox{diag}}
\newcommand{\fund}{\mbox{fund}\,}
\newcommand{\ind}{\mbox{Ind}\,}
\newcommand{\inder}{\mbox{Inder}\,}
\newcommand{\rank}{\mbox{rank}\,}
\newcommand{\sdet}{\mbox{sdet}}
\newcommand{\sdim}{\mbox{sdim}\,}
\newcommand{\sch}{\mbox{sch}\,}
\newcommand{\str}{\mbox{str}}
\newcommand{\tr}{\mbox{tr}}
\newcommand{\evn}{{\overline{0}}}
\newcommand{\odd}{{\overline{1}}}
\newcommand{\invdots}{{\mathinner{\mkern1mu\raise1pt
   \vbox{\kern7pt\hbox{.}}\mkern2mu\raise4pt\hbox{.}
   \mkern2mu\raise7pt\hbox{.}\mkern1mu}}}

\newcommand{\lsemisum}{{\mathinner{\ni\mkern-12.3mu\rule{0.15mm}
   {2.0mm}\mkern10mu}}} 
\newcommand{\rsemisum}{{\mathinner{\in\mkern-11mu\rule{0.15mm}
   {2.0mm}\mkern10mu}}}   
\newcommand{\ket}[1]{{\vert #1 \rangle}} 
 
\newcommand{\braket}[2]{{\langle #1 \vert #2 \rangle}}

\newcommand{\sfrac}[2]{\textstyle{\frac{#1}{#2}}}
\newcommand{\half}{\textstyle{\frac{1}{2}}}
\newcommand{\third}{\textstyle{\frac{1}{3}}}
\newcommand{\smbox}[1]{\ \mbox{#1}\ }
\newcommand{\medbox}[1]{\quad\mbox{#1}\quad}
\newcommand{\bigbox}[1]{\qquad\mbox{#1}\qquad}

\newcommand{\udef}{{\underline{Definition}}}
\newcommand{\uth}{{\underline{Theorem}}}
\newcommand{\uppt}{{\underline{Property}}}
\newcommand{\uppts}{{\underline{Properties}}}
\newcommand{\uex}{{\underline{Example}}}
\newcommand{\see}{{\rightarrow}}

\input epsf
\begin{document}

\pagestyle{empty}
\setcounter{page}{0}
\begin{minipage}{4.9cm}
\begin{center}
{\bf Groupe d'Annecy\\ \ \\
Laboratoire d'Annecy-le-Vieux de Physique des Particules}
\end{center}
\end{minipage}
\hfill
\hfill
\begin{minipage}{4.2cm}
\begin{center}
{\bf Groupe de Lyon\\ \ \\
Ecole Normale Sup\'erieure de Lyon}
\end{center}
\end{minipage}

\begin{center}
\rule{14cm}{.4mm}
\vfill
\vfill
{\LARGE {\bf {\sf Dictionary on Lie Superalgebras}}} 
\vfill
{\Large L. Frappat, P. Sorba}

\vspace{3mm}

{\it
Laboratoire de Physique Th\'eorique ENSLAPP
\footnote{URA 1436 du CNRS, associ\'ee \`a l'Ecole Normale 
Sup\'erieure de Lyon et \`a l'Universit\'e de Savoie.} \\
Chemin de Bellevue, BP 110, F-74941 Annecy-le-Vieux cedex, France \\
and \\
46, avenue d'Italie, F-69364 Lyon cedex 07, France \\
e-mail: 
{\tt frappat@lapp.in2p3.fr} \it and {\tt sorba@lapp.in2p3.fr} \\ 
}

\vspace{3mm}

and

\vspace{3mm}

{\Large A. Sciarrino}

\vspace{3mm}

{\it
Universit\`a di Napoli "Federico II" and I.N.F.N. Sezione di Napoli \\
Mostra d'Oltremare, Pad. 19, I-80125 Napoli, Italy \\
e-mail: {\tt sciarrino@.napoli.infn.it} \\
}
\end{center}
\vfill
\vfill
\vfill
\rightline{ENSLAPP-AL-600/96}
\rightline{DSF-T-30/96}
\rightline{hep-th/9607161}
\rightline{July 1996}

\newpage
\phantom{emptypage}
\newpage
\begin{center}
\section*{Foreword}
\end{center}

\vspace{10mm}

The main definitions and properties of Lie superalgebras are proposed
{\em \`a la fa\c con de} a short dictionary, the different items
following the alphabetical order. The main topics deal with the
structure of simple Lie superalgebras and their finite dimensional
representations; rather naturally, a few pages are devoted to
supersymmetry.

This modest booklet has two ambitious goals: to be elementary and easy to
use. The beginner is supposed to find out here the main concepts on
superalgebras, while a more experimented theorist should recognize the
necessary tools and informations for a specific use.

It has not been our intention to provide an exhaustive set of references
but, in the quoted papers, the reader must get the proofs and
developments of the items which are presented hereafter, as well as a
more detailed bibliography.

Actually, this work can be considered as the continuation of a first
section, entitled "Lie algebras for physicists" written fifteen years
ago (see ref. \cite{ScS}). The success of this publication as well as
the encouragements of many of our collegues convinced us to repeat the
same exercise for superalgebras. During the preparation of the
following pages, it has appeared to us necessary to update the Lie
algebra part. In this respect we are writing a new version of this
first section, by adding or developing some properties which are of
some interest these recent years in the domains of theoretical physics
where continuous symmetries are intensively used (elementary particle
physics, integrable systems, statistical mechanics, e.g.). Finally, a
third section is in preparation and deals with infinite dimensional
symmetries -- Kac-Moody algebras and superalgebras, two-dimensional
conformal symmetry and its extensions. When completed, we have in mind
to gather the three parts in an unique volume. However, we have
preferred to display right now the second part on superalgebras since
we do not see any reason to keep in a drawer this document which might
be of some help for physicists. Moreover, we hope to receive from the
interested readers suggestions and corrections for a better final
version.

\newpage
\phantom{emptypage}
\newpage
\section*{Contents}

\contentsline {section}{\numberline {1}\sf Automorphisms}{1}
\contentsline {section}{\numberline {2}\sf Cartan matrices}{1}
\contentsline {section}{\numberline {3}\sf Cartan subalgebras}{3}
\contentsline {section}{\numberline {4}\sf Cartan type superalgebras}{3}
\contentsline {section}{\numberline {5}\sf Casimir invariants}{7}
\contentsline {section}{\numberline {6}\sf Centralizer, Center, Normalizer of a Lie superalgebra}{10}
\contentsline {section}{\numberline {7}\sf Characters and supercharacters}{10}
\contentsline {section}{\numberline {8}\sf Classical Lie superalgebras}{12}
\contentsline {section}{\numberline {9}\sf Classification of simple Lie superalgebras}{14}
\contentsline {section}{\numberline {10}\sf Clifford algebras}{14}
\contentsline {section}{\numberline {11}\sf Decompositions w.r.t. $osp(1|2)$ subalgebras}{16}
\contentsline {section}{\numberline {12}\sf Decompositions w.r.t. $sl(1|2)$ subalgebras}{17}
\contentsline {section}{\numberline {13}\sf Derivation of a Lie superalgebra}{20}
\contentsline {section}{\numberline {14}\sf Dirac matrices}{20}
\contentsline {section}{\numberline {15}\sf Dynkin diagrams}{20}
\contentsline {section}{\numberline {16}\sf Embeddings of $osp(1|2)$}{21}
\contentsline {section}{\numberline {17}\sf Embeddings of $sl(2|1)$}{22}
\contentsline {section}{\numberline {18}\sf Exceptional Lie superalgebra $F(4)$}{24}
\contentsline {section}{\numberline {19}\sf Exceptional Lie superalgebra $G(3)$}{26}
\contentsline {section}{\numberline {20}\sf Exceptional Lie superalgebras $D(2,1;\alpha )$}{28}
\contentsline {section}{\numberline {21}\sf Gelfand-Zetlin basis}{30}
\contentsline {section}{\numberline {22}\sf Grassmann algebras}{32}
\contentsline {section}{\numberline {23}\sf Killing form}{32}
\contentsline {section}{\numberline {24}\sf Lie superalgebra, subalgebra, ideal}{34}
\contentsline {section}{\numberline {25}\sf Matrix realizations of the classical Lie superalgebras}{35}
\contentsline {section}{\numberline {26}\sf Nilpotent and solvable Lie superalgebras}{36}
\contentsline {section}{\numberline {27}\sf Orthosymplectic superalgebras}{37}
\contentsline {section}{\numberline {28}\sf Oscillator realizations: Cartan type superalgebras}{40}
\contentsline {section}{\numberline {29}\sf Oscillator realizations: orthosymplectic and unitary series}{41}
\contentsline {section}{\numberline {30}\sf Oscillator realizations: strange series}{44}
\contentsline {section}{\numberline {31}\sf Real forms}{45}
\contentsline {section}{\numberline {32}\sf Representations: basic definitions}{45}
\contentsline {section}{\numberline {33}\sf Representations: exceptional superalgebras}{46}
\contentsline {section}{\numberline {34}\sf Representations: highest weight representations}{49}
\contentsline {section}{\numberline {35}\sf Representations: induced modules}{50}
\contentsline {section}{\numberline {36}\sf Representations: orthosymplectic superalgebras}{52}
\contentsline {section}{\numberline {37}\sf Representations: reducibility}{55}
\contentsline {section}{\numberline {38}\sf Representations: star and superstar representations}{56}
\contentsline {section}{\numberline {39}\sf Representations: typicality and atypicality}{58}
\contentsline {section}{\numberline {40}\sf Representations: unitary superalgebras}{59}
\contentsline {section}{\numberline {41}\sf Roots, root systems}{60}
\contentsline {section}{\numberline {42}\sf Schur's lemma}{62}
\contentsline {section}{\numberline {43}\sf Serre-Chevalley basis}{62}
\contentsline {section}{\numberline {44}\sf Simple root systems}{63}
\contentsline {section}{\numberline {45}\sf Simple and semi-simple Lie superalgebras}{66}
\contentsline {section}{\numberline {46}\sf Spinors (in the Lorentz group)}{66}
\contentsline {section}{\numberline {47}\sf Strange superalgebras $P(n)$}{69}
\contentsline {section}{\numberline {48}\sf Strange superalgebras $Q(n)$}{71}
\contentsline {section}{\numberline {49}\sf Subsuperalgebras (regular)}{72}
\contentsline {section}{\numberline {50}\sf Subsuperalgebras (singular)}{73}
\contentsline {section}{\numberline {51}\sf Superalgebra, subsuperalgebra}{75}
\contentsline {section}{\numberline {52}\sf Superalgebra $osp(1|2)$}{75}
\contentsline {section}{\numberline {53}\sf Superalgebra $sl(1|2)$}{77}
\contentsline {section}{\numberline {54}\sf Superconformal algebra}{80}
\contentsline {section}{\numberline {55}\sf Supergroups}{81}
\contentsline {section}{\numberline {56}\sf Supergroups of linear transformations}{82}
\contentsline {section}{\numberline {57}\sf Supermatrices}{82}
\contentsline {section}{\numberline {58}\sf Superspace and superfields}{85}
\contentsline {section}{\numberline {59}\sf Supersymmetry algebra: definition}{87}
\contentsline {section}{\numberline {60}\sf Supersymmetry algebra: representations}{89}
\contentsline {section}{\numberline {61}\sf Unitary superalgebras}{91}
\contentsline {section}{\numberline {62}\sf Universal enveloping algebra}{93}
\contentsline {section}{\numberline {63}\sf Weyl group}{94}
\contentsline {section}{\numberline {64}\sf Z-graded Lie superalgebras}{95}
\contentsline {section}{\numberline {}}{}
\contentsline {section}{\numberline {}\sf List of Tables}{96}
\contentsline {section}{\numberline {}\sf Tables}{98}
\contentsline {section}{\numberline {}\sf Bibliography}{135}

\newpage
\section*{Main Notations}

\vspace{10mm}

\[
\begin{array}{ll}
[\,.\,,\,.\,] & \qquad \mbox{commutator} \cr
\{\,.\,,\,.\,\} & \qquad \mbox{anticommutator} \cr
\zle \,.\,,\,.\, \zri & \qquad \mbox{super or $\ZZ_2$-graded 
   commutator (Lie superbracket)} \cr 
(\,.\,,\,.\,) & \qquad \mbox{inner product, Killing form} \cr
\NN, \ZZ, \RR, \CC, \HH
   & \qquad \mbox{sets of positive integers, of integers, of real
   numbers,} \cr    
   & \qquad \mbox{~~~ of complex numbers, of quaternions} \cr 
\KK & \qquad \mbox{commutative field of characteristic zero} \cr    
\cA, \cA_\evn, \cA_\odd & \qquad \mbox{(super)algebra, even/odd 
   part of a superalgebra} \cr
\cB & \qquad \mbox{Borel subalgebra} \cr
\cG, \cG_\evn, \cG_\odd & \qquad \mbox{Lie superalgebra, even/odd
   part of a Lie superalgebra} \cr
\cH & \qquad \mbox{Cartan subalgebra} \cr
\cN & \qquad \mbox{nilpotent subalgebra} \cr
\cV & \qquad \mbox{module, representation space} \cr
A(m,n) \simeq sl(m+1|n+1) & \qquad \mbox{unitary basic superalgebras} \cr
B(m,n) \simeq osp(2m+1|2n) & \qquad \mbox{orthosymplectic basic
superalgebras} \cr
C(n+1) \simeq osp(2|2n) & \qquad
\mbox{~~~~~~~~~~~"~~~~~~~~~~~~"~~~~~~~~~~~"} \cr
D(m,n) \simeq osp(2m|2n) & \qquad 
\mbox{~~~~~~~~~~~"~~~~~~~~~~~~"~~~~~~~~~~~"} \cr
\evn, \odd & \qquad \mbox{$\ZZ_2$-gradation of a superalgebra} \cr
\aut & \qquad \mbox{automorphism group} \cr
\der & \qquad \mbox{derivation algebra} \cr
\autint & \qquad \mbox{inner automorphism group} \cr
\autout & \qquad \mbox{set of outer automorphisms} \cr
\Delta, \Delta_\evn, \Delta_\odd & \qquad \mbox{root system, even root
   system, odd root system} \cr
\Delta^+, \Delta_\evn^+, \Delta_\odd^+ & \qquad \mbox{positive roots,
   positive even roots, positive odd roots} \cr
\Delta^0 & \qquad \mbox{simple root system} \cr
\end{array}
\]

\newpage
\pagestyle{plain}
\baselineskip=18pt
\setcounter{page}{1}
\renewcommand{\thetable}{\Roman{table}}
\section{\sf Automorphisms}
\indent

Let $\cG = \cG_\evn \oplus \cG_\odd$ be a simple Lie superalgebra. An
automorphism $\Phi$ of $\cG$ is a bijective homomorphism from $\cG$
into itself which respects the $\ZZ_2$-gradation, that is
$\Phi(\cG_\evn) \subset \cG_\evn$ and $\Phi(\cG_\odd) \subset
\cG_\odd$. The automorphisms of $\cG$ form a group denoted by
$\aut(\cG)$. The group $\autint(\cG)$ of inner automorphisms of $\cG$
is the group generated by the automorphisms of the form $X \mapsto
gXg^{-1}$ with $g = \exp Y$ where $X \in \cG$ and $Y \in \cG_\evn$.
Every inner automorphism of $\cG_\evn$ can be extended to an inner
automorphism of $\cG$. The automorphisms of $\cG$ which are not inner
are called outer automorphisms.

\medskip

In the case of a simple Lie algebra $\cA$, the quotient of the
automorphism group by the inner automorphism group $\aut(\cA) /
\autint(\cA)$ -- called the factor group $F(\cA)$ -- is isomorphic to
the group of symmetries of the Dynkin diagram of $\cA$.

\medskip

In the same way, the outer automorphisms of a basic Lie superalgebra
$\cG$ can also be connected with some Dynkin diagram ($\see$) of $\cG$. It
is possible to write $\autout(\cG) = \aut(\cG) / \autint(\cG)$, where
$\autint(\cG) \simeq \cG_\evn$, and $\autout(\cG)$ can be reconstructed
in general by looking at the symmetries of the Dynkin diagrams of
$\cG$. More precisely, when $\autout(\cG)$ is not trivial, there
exists at least one Dynkin diagram of $\cG$ which exhibits a symmetry
associated to $\autout(\cG)$ -- except in the case of $sl(2m+1|2n+1)$. 
The table \ref{table1} lists the outer automorphisms of the basic Lie 
superalgebras. For more details, see ref. \cite{Ser85}.

\begin{table}[htbp]
\centering
\begin{tabular}{|c|c||c|c|} \hline
superalgebra $\cG$ & $\autout(\cG)$ & superalgebra $\cG$ 
& $\autout(\cG)$ \\ 
\hline &&& \\
$A(m,n) ~ (m \ne n \ne 0)$ & $\ZZ_2$ & $D(m,n)$ & $\ZZ_2$ \\ 
$A(1,1)$ & $\ZZ_2$ & $D(2,1;-2)$ & $\ZZ_2$ \\ 
$A(0,2n-1)$ & $\ZZ_2$ & $D(2,1;-1/2)$ & $\ZZ_2$ \\
$A(n,n) ~ (n \ne 0,1)$ & $\ZZ_2 \times \ZZ_2$ 
& $D(2,1;e^{2i\pi/3})$ & $\ZZ_3$ \\
$A(0,2n)$ & $\ZZ_4$ & $D(2,1;e^{4i\pi/3})$ & $\ZZ_3$ \\ 
$B(m,n)$ & $\II$ & $D(2,1;\alpha)$ for generic $\alpha$ & $\II$ \\
$C(n+1)$ & $\ZZ_2$ & $F(4),G(3)$ & $\II$ \\
&&& \\
\hline
\end{tabular}
\caption{Outer automorphisms of the basic Lie superalgebras.
\label{table1}}
\end{table}

\medskip

$\see$ Dynkin diagram, Roots, Weyl group.

\section{\sf Cartan matrices}
\indent

Let $\cG$ be a basic Lie superalgebra with Cartan subalgebra $\cH$. To a
simple root system $\Delta^0 = (\alpha_1, \dots, \alpha_r)$ of $\cG$,
it is always possible to associate a matrix $A = (a_{ij})$, called the
Cartan matrix, with the following conditions:
\beo
&& \cle H_i, H_j \cri = 0 \,, \\
&& \cle H_i, E_{\pm \alpha_j} \cri = \pm a_{ij} E_{\pm \alpha_j} \,, \\
&& \sle E_{\alpha_i}, E_{-\alpha_j} \sri = \delta_{ij} H_i \,,
\eno
the set $(H_1, \dots, H_r)$ generating the Cartan subalgebra $\cH$.

\medskip

For all basic Lie superalgebras, there exists a non-degenerate inner
product $(\,.\,,\,.\,)$ such that
\beo
&& (E_{\alpha_i},E_{-\alpha_j}) = (E_{\alpha_j},E_{-\alpha_j}) 
\delta_{ij} \\
&& (H_i,H_j) = (E_{\alpha_j},E_{-\alpha_j}) a_{ij}
\eno
Notice that this inner product coincides with the Killing form ($\see$)
except for $A(n,n)$, $D(n+1,n)$ and $D(2,1;\alpha)$ for which the
Killing form vanishes.
\\
A non-degenerate bilinear form on $\cH^*$ ($\see$ Simple root systems)
is then defined by $(\alpha_i,\alpha_j) \equiv (H_i,H_j)$, where
$\alpha_i,\alpha_j$ are simple roots, such a form being invariant under
the Weyl group of $\cG$, generated by the reflections of the even
roots.

\medskip

\udef: For each basic Lie superalgebra, there exists a simple root
system for which the number of odd roots is the smallest one. Such a
simple root system is called the distinguished simple root
system ($\see$). The associated Cartan matrix is called the {\em
distinguished Cartan matrix}.

The distinguished Cartan matrices can be found in Tables \ref{table11}
to \ref{table32}.

\medskip

One can also use symmetric Cartan matrices. A symmetric Cartan matrix
$A^s = (a'_{ij})$ can be obtained by rescaling the Cartan generators
$H_i \rightarrow H'_i = H_i/(E_{\alpha_i},E_{-\alpha_i})$. The
commutation relations then become $\cle H'_i, E_{\pm \alpha_j} \cri =
\pm a'_{ij} E_{\pm \alpha_j}$ and $\sle E_{\alpha_i}, E_{-\alpha_j}
\sri = (E_{\alpha_i},E_{-\alpha_i}) H'_i \delta_{ij}$, from which it
follows that $a'_{ij} = (H'_i,H'_j)$.
\\
If one defines the matrix $D_{ij} = d_i \delta_{ij}$ where the rational
coefficients $d_i$ satisfy $d_i a_{ij} = d_j a_{ji}$, the (distinguished)
symmetric Cartan matrix is given by $A^s = DA$. One has:
\begin{verse}
$d_i = (\underbrace{1,\dots,1}_{m+1},\underbrace{-1,\dots,-1}_{n})$
for $A(m,n)$, \\
$d_i = (\underbrace{1,\dots,1}_{n},\underbrace{-1,\dots,-1}_{m-1},-1/2)$
for $B(m,n)$, \\ 
$d_i = (\underbrace{1,\dots,1}_{n-1},1/2)$ for $B(0,n)$, \\
$d_i = (1,\underbrace{-1,\dots,-1}_{n-1},-2)$ for $C(n+1)$, \\
$d_i = (\underbrace{1,\dots,1}_{n},\underbrace{-1,\dots,-1}_{m})$ 
for $D(m,n)$.
\end{verse}

\medskip

$\see$ Killing form, Simple root systems.

\section{\sf Cartan subalgebras}
\indent

Let $\cG = \cG_\evn \oplus \cG_\odd$ be a classical Lie superalgebra. A
Cartan subalgebra $\cH$ of $\cG$ is defined as the maximal nilpotent
($\see$) subalgebra of $\cG$ coinciding with its own normalizer, that is
\[
\cH \smbox{nilpotent and}
\ale X \in \cG ~ \Big\vert ~ \cle X, \cH \cri \subseteq \cH \ari = \cH
\]
In most cases (for basic Lie superalgebras e.g.), a Cartan subalgebra
$\cH$ reduces to the Cartan subalgebra of the even part $\cG_\evn$
(then the Cartan subalgebras of a Lie superalgebra are conjugate since
the Cartan subalgebras of a Lie algebra are conjugate and any inner
automorphism of the even part $\cG_\evn$ can be extended to an inner
automorphism of $\cG$).

\medskip

In the case of the strange superalgebra $Q(n)$, the Cartan subalgebra
$\cH$ does not coincide with the Cartan subalgebra of the even part
$sl(n)$, but admits also an odd part: $\cH \cap \cG_\odd \ne
\emptyset$. Since the odd generators of $\cH$ change the gradation of
the generators on which they act, it is rather convenient to give the
root decomposition of $Q(n)$ with respect to $\cH_\evn = \cH \cap
\cG_\evn$ instead of $\cH$.

\medskip

{}From what precedes, all Cartan subalgebras of a classical superalgebra
$\cG$ have the same dimension. By definition, the dimension of a Cartan
subalgebra $\cH$ is the rank of $\cG$:
\[
\rank \cG = \dim \cH
\]

\section{\sf Cartan type superalgebras}
\indent

The Cartan type Lie superalgebras are superalgebras in which the
representation of the even subalgebra on the odd part is not completely
reducible ($\see$ Classification of simple Lie algebras). 
The Cartan type simple Lie superalgebras are classified into
four infinite families called $W(n)$ with $n \ge 2$, $S(n)$ with $n \ge
3$, $\tilde S(n)$ and $H(n)$ with $n \ge 4$. $S(n)$ and $\tilde S(n)$
are called special Cartan type Lie superalgebras and $H(n)$ Hamiltonian
Cartan type Lie superalgebras. Strictly speaking, $W(2)$, $S(3)$ and
$H(4)$ are not Cartan type superalgebras since they are isomorphic to
classical ones (see below).

\subsection{\sf Cartan type superalgebras $W(n)$}

Consider $\Gamma(n)$ the Grassmann algebra ($\see$) of order $n$ with
generators $\theta_1, \dots, \theta_n$ and relations $\theta_i \theta_j = 
-\theta_j \theta_i$. The $\ZZ_2$-gradation is induced by setting $\deg
\theta_i = \odd$. Let $W(n)$ be the derivation superalgebra of
$\Gamma(n)$: $W(n) = \der\Gamma(n)$. Any derivation $D \in W(n)$ is
written as
\[
D = \sum_{i = 1}^n P_i \frac{\prt}{\prt \theta_i}
\]
where $P_i \in \Gamma(n)$ and the action of the $\theta$-derivative is 
defined by
\[
\frac{\prt \theta_j}{\prt \theta_i} = \delta_{ij}
\]
The $\ZZ_2$-gradation of $\Gamma(n)$ induces a consistent $\ZZ$-gradation
of  $W(n)$ ($\see$ $\ZZ$-graded superalgebras) by
\[
W(n)_k = \ale \sum_{i = 1}^n P_i \frac{\prt}{\prt \theta_i}, \quad P_i \in 
\Gamma(n), \quad \deg P_i = k+1 \ari \medbox{where} -1 \le k \le n-1
\]
One has
\[
W(n) = \bigoplus_{k = -1}^{n-1} W(n)_k
\]
where
\[
\sle W(n)_i,W(n)_j \sri \subset W(n)_{i+j} 
\]
The superalgebra $W(n)$ has the following properties: 
\begin{itemize}
\item
$W(n)$ has dimension $n2^n$, the number of even generators being equal 
to the number of odd generators.
\item
The superalgebra $W(n)$ is simple for $n \ge 2$.
\item
The semi-simple part of $W(n)_0$ is isomorphic to $gl(n)$. 
\item
The superalgebra $W(2)$ is isomorphic to $A(1,0)$.
\item
Every automorphism of $W(n)$ with $n \ge 3$ is induced by an 
automorphism of $\Gamma(n)$.
\item
The superalgebra $W(n)$ is transitive ($\see$ $\ZZ$-graded superalgebras).
\item
$W(n)$ is universal as a $\ZZ$-graded Lie superalgebra. More precisely,
if $\cG = \oplus_{i \ge -1} \cG_i$ is a transitive $\ZZ$-graded
superalgebra with $\dim \cG_{-1} = n$, then there is an embedding of
$\cG$ in $W(n)$ preserving the $\ZZ$-gradation.
\item
The representations of $sl(n)$ in the subspace $W(n)_i$
$(i=-1,0,..,n-1)$ are in Young tableaux notation $[2^{i+1}1^{n-2-i}] ~
\oplus ~ [1^{i}]$ where the second representation appears only for $i
\ge 0$ and $[1^{0}]$ has to be read as the singlet. For example we
have (the subscripts stand for the $\ZZ$-gradation indices $i$):

for $W(3)$ ~~~ $(\overline{3})_{-1} \oplus (8 \oplus 1)_{0} \oplus
(\overline{6} \oplus 3)_{1} \oplus (\overline{3})_{2} $

for $W(4)$ ~~~ $(\overline{4})_{-1} \oplus (15 \oplus 1)_{0} \oplus
(\overline{20} \oplus 4)_{1} \oplus (\overline{10} \oplus 6)_{2} \oplus
(\overline{4})_{3} $

for $W(5)$ ~~~ $(\overline{5})_{-1} \oplus (24 \oplus 1)_{0} \oplus
(\overline{45} \oplus 5)_{1} \oplus (\overline{40} \oplus 10)_{2}
\oplus (\overline{15} \oplus \overline{10})_{3} \oplus
(\overline{5})_{4} $
\end{itemize}

\subsection{\sf Cartan type superalgebras $S(n)$ and $\tilde S(n)$}

The Cartan type Lie superalgebras $S(n)$ and $\tilde S(n)$, called
special Lie superalgebras, are constructed as follows. Consider
$\Theta(n)$ the associative superalgebra over $\Gamma(n)$ with 
generators denoted by $\xi\theta_1,\dots,\xi\theta_n$ and relations 
$\xi\theta_i \wedge \xi\theta_j = -\xi\theta_j \wedge \xi\theta_i$. 
A $\ZZ_2$-gradation is induced by setting $\deg \xi\theta_i = \odd$.
Any element of $\Theta(n)$ is written as
\[
\omega_k = \sum_{i_1 < \dots < i_k} a_{i_1 \dots i_k} ~ 
\xi\theta_{i_1} \wedge \dots \wedge \xi\theta_{i_k} 
\]
where $a_{i_1 \dots i_k} \in \Gamma(n)$.
\\
One defines then the volume form superalgebra $S(\omega)$ as a $W(n)$ 
subsuperalgebra by 
\[
S(\omega) = \ale D \in W(n) ~ \Big\vert ~ D(\omega) = 0 \ari 
\]
where $\omega = a(\theta_1,\dots,\theta_n) ~ \xi\theta_1 \wedge \dots
\wedge \xi\theta_n$ and $a \in \Gamma(n)_\evn$, $a(0) \ne 0$.
\\
Any element of $S(\omega)$ has the form
\[
\sum_{i = 1}^n P_i \frac{\prt}{\prt\theta_i} \bigbox{with}
\sum_{i = 1}^n \frac{\prt (a P_i)}{\prt\theta_i} = 0
\]
One sets also
\[
S(n) = S \Big( \omega = \xi\theta_1 \wedge \dots \wedge \xi\theta_n 
\Big) = \ale D \in W(n) ~ \Big\vert ~ D 
\Big( \xi\theta_1 \wedge \dots \wedge \xi\theta_n \Big) = 0 \ari 
\]
and
\beo
\tilde S(n) & = & S \Big( \omega = (1 + \theta_1 \dots \theta_n) ~ 
\xi\theta_1 \wedge \dots \wedge \xi\theta_n \Big) \\
& = & \ale D \in W(n) ~ \Big\vert ~ D \Big((1 + \theta_1 \dots \theta_n) 
 ~ \xi\theta_1 \wedge \dots \wedge \xi\theta_n \Big) = 0
\ari 
\medbox{where $n$ is even}
\eno
Elements of $S(n)$ are thus divergenceless derivations of $W(n)$:
\[
S(n) = \ale \sum_{i = 1}^n P_i \frac{\prt}{\prt\theta_i} \in W(n) ~ 
\Big\vert ~ \sum_{i = 1}^n \frac{\prt P_i}{\prt\theta_i} = 0 \ari
\]
The Lie superalgebras $S(n)$ and $\tilde S(n)$ have the following
properties: 
\begin{itemize}
\item 
$S(n)$ and $\tilde S(n)$ have dimension $(n-1)2^n+1$, the number
of even generators being less (resp. greater) by 1 than the number of
odd generators for $n$ even (resp. odd).
\item
The superalgebra $S(n)$ is simple for $n \ge 3$ and $\tilde S(n)$ is 
simple for $n \ge 4$. 
\item
The semi-simple part of $S(n)_0$ and $\tilde S(n)_0$ is isomorphic to 
$sl(n)$. 
\item
The superalgebra $S(3)$ is isomorphic to $P(3)$.
\item
the $\ZZ$-graded Lie superalgebra $S(n)$ is transitive ($\see$  
$\ZZ$-graded superalgebras).
\item
Every automorphism of $S(n)$ with $n \ge 3$ and $\tilde S(n)$ with
$n \ge 4 $ is induced by an automorphism of $\Gamma(n)$.
\item
Every superalgebra $S(\omega)$ is isomorphic either to $S(n)$ or 
$\tilde S(n)$.
\item
The representation of $sl(n)$ in the subspace $S(n)_i$ 
$(i=-1,0,..,n-2)$ is in Young tableaux notation $[2^{i+1}1^{n-2-i}]$. 
For example we have (the subscripts stand for the $\ZZ$-gradation 
indices $i$):

for $S(4)$ ~~~ $(\overline{4})_{-1} \oplus (15)_{0} \oplus
(\overline{20})_{1} \oplus (\overline{10})_{2} $

for $S(5)$ ~~~ $(\overline{5})_{-1} \oplus (24)_{0} \oplus
(\overline{45})_{1} \oplus (\overline{40})_{2} \oplus
(\overline{15})_{3} $

for $S(6)$ ~~~ $(\overline{6})_{-1} \oplus (35)_{0} \oplus
(\overline{84})_{1} \oplus (\overline{105})_{2} \oplus
(\overline{70})_{3} \oplus (\overline{21})_{4} $
\end{itemize}

\subsection{\sf Cartan type superalgebras $H(n)$}

The Cartan type Lie superalgebras $H(n)$ and $\tilde H(n)$, called
Hamiltonian Lie superalgebras, are constructed as follows. Consider
$\Omega(n)$ the associative superalgebra over $\Gamma(n)$ with
generators denoted by $d\theta_1,\dots,d\theta_n$ and relations
$d\theta_i \circ d\theta_j = d\theta_j \circ d\theta_i$. The
$\ZZ_2$-gradation is induced by setting $\deg d\theta_i = \evn$. Any
element of $\Omega(n)$ is written as
\[
\omega_k = \sum_{i_1 \le \dots \le i_k} a_{i_1 \dots i_k} ~ 
d\theta_{i_1} \circ \dots \circ d\theta_{i_k} 
\]
where $a_{i_1 \dots i_k} \in \Gamma(n)$.
\\
Among them are the Hamiltonian forms defined by
\[
\omega = \sum_{i,j = 1}^n a_{ij} ~ d\theta_i \circ d\theta_j 
\]
where $a_{ij} \in \Gamma(n)$, $a_{ij} = a_{ji}$ and $\det(a_{ij}(0))
\ne 0$. 
One defines then for each Hamiltonian form $\omega$ the Hamiltonian
form superalgebra $\tilde H(\omega)$ as a $W(n)$ subsuperalgebra by 
\[
\tilde H(\omega) = \ale D \in W(n) ~ \Big\vert ~ D(\omega) = 0 \ari 
\]
and
\[
H(\omega) = \sle \tilde H(\omega) , \tilde H(\omega) \sri
\]
Any element of $\tilde H(\omega)$ has the form
\[
\sum_{i = 1}^n P_i \frac{\prt}{\prt\theta_i} \bigbox{with}
\frac{\prt}{\prt\theta_j} \sum_{t = 1}^n a_{it} P_t +
\frac{\prt}{\prt\theta_i} \sum_{t = 1}^n a_{jt} P_t = 0
\]
One sets also
\beo
&& \tilde H(n) = \tilde H \Big( (d\theta_1)^2 + \dots + (d\theta_n)^2
\Big) \\
&& H(n) = \sle \tilde H(n) , \tilde H(n) \sri
\eno
The Lie superalgebra $H(n)$ has the following properties:
\begin{itemize}
\item 
$H(n)$ has dimension $2^n-2$, the number of even generators being
equal (resp. less by 2) to (than) the number of odd generators for $n$
odd (resp. even).
\item
The superalgebra $H(n)$ is simple for $n \ge 4$.
\item
The semi-simple part of $\tilde H(n)_0$ is isomorphic to $so(n)$. 
\item
The superalgebra $H(4)$ is isomorphic to $A(1,1)$.
\item
The $\ZZ$-graded Lie superalgebras $H(n)$ and $\tilde H(n)$ are
transitive ($\see$ $\ZZ$-graded superalgebras). 
\item
Every automorphism of $H(n)$ with $n \ge 4$ and of $\tilde H(n)$ with 
$n \ge 3$ is induced by an automorphism of $\Gamma(n)$. 
\item
The representation of $so(n)$ in the subspace $H(n)_i$ $(i =
-1,0,..,n-3)$ is given by the antisymmetric tensor of rank $i+2$.
For example we have (the subscripts stand for the $\ZZ$-gradation 
indices $i$):

for $H(4)$ ~~~ $(4)_{-1} \oplus (6)_{0} \oplus (4)_{1} $

for $H(5)$ ~~~ $(5)_{-1} \oplus (10)_{0} \oplus (10)_{1} \oplus
(5)_{2} $

for $H(10)$ ~~~ $(10)_{-1} \oplus (45)_{0} \oplus (120)_{1}
\oplus (210)_{2} \oplus (252)_{3} \oplus (210)_{4} \oplus (120)_{5}
\oplus (45)_{6} \oplus (10)_{7} $
\end{itemize}

\medskip

For more details, see ref. \cite{Kac77a}.

\section{\sf Casimir invariants}
\indent

The study of Casimir invariants plays a great role in the representation
theory of simple Lie algebras since their eigenvalues on a finite
dimensional highest weight irrreducible representation completely
characterize this representation. In the case of Lie superalgebras, the
situation is different. In fact, the eigenvalues of the Casimir
invariants {\em do not} always characterize the finite dimensional
highest weight irrreducible representations of a Lie superalgebra. More
precisely, their eigenvalues on a {\em typical} representation
completely characterize this representation while they are identically
vanishing on an {\em atypical} representation ($\see$ Representations:
typicality and atypicality).

\medskip

\udef: Let $\cG = \cG_\evn \oplus \cG_\odd$ be a classical Lie
superalgebra and $\cU(\cG)$ its universal enveloping superalgebra
($\see$). An element $C \in \cU(\cG)$ such that $\zle C,X \zri = 0$ for
all $X \in \cU(\cG)$ is called a {\em Casimir element} of $\cG$ ($\zle
~,~ \zri$ denotes the $\ZZ_2$-graded commutator). The algebra of the
Casimir elements of $\cG$ is the $\ZZ_2$-center of $\cU(\cG)$, denoted
by $\cZ(\cG)$. It is a ($\ZZ_2$-graded) subalgebra of $\cU(\cG)$.

\medskip

Standard sequences of Casimir elements of the basic Lie superalgebras
can be constructed as follows.
Let $\cG = sl(m|n)$ with $m \ne n$ or $osp(m|n)$ be a basic Lie
superalgebra with non-degenerate bilinear form. Let $\{E_{IJ}\}$ be a
matrix basis of generators of $\cG$ where $I,J = 1,\dots,m+n$ with
$\deg I = 0$ for $1 \le I \le m$ and $\deg I = 1$ for $m+1 \le I \le
m+n$. Then defining $(\bar E^0)_{IJ} = \delta_{IJ}$ and $(\bar
E^{p+1})_{IJ} = (-1)^{\deg K} E_{IK} (\bar E^p)_{KJ}$, a standard
sequence of Casimir operators is given by
\[
C_p = \str(\bar E^p) = (-1)^{\deg I} (\bar E^p)_{II} = E_{II_1} 
(-1)^{\deg I_1} \dots E_{I_kI_{k+1}} (-1)^{\deg I_{k+1}} 
\dots E_{I_{p-1}I}
\]
Consider the $(m+n)^2$ elementary matrices $e_{IJ}$ of order $m+n$
satisfying $(e_{IJ})_{KL} = \delta_{IL} \delta_{JK}$.
\\
In the case of $sl(m|n)$ with $m \ne n$, a basis $\{E_{IJ}\}$ is given
by the matrices $E_{ij} = e_{ij} - \frac{1}{m} \delta_{ij}
\sum_{q=1}^{q=m} e_{qq}$, $E_{kl} = e_{kl} - \frac{1}{n} \delta_{kl}
\sum_{q=m+1}^{q=m+n} e_{qq}$ and $Y = \frac{1}{m-n}(n \sum_{q=1}^{q=m}
e_{qq} + m \sum_{q=m+1}^{q=m+n} e_{qq})$, for the even part and $E_{ik} =
e_{ik}$, $E_{kj} = e_{kj}$ for the odd part, where $1 \le i,j \le m$ and
$m+1 \le k,l \le m+n$. One finds for example
\beo
&& C_1 = 0 \\
&& C_2 = E_{ij} E_{ji} - E_{kl} E_{lk} + E_{ki} E_{ik} - E_{ik} E_{ki} 
- \frac{m-n}{mn} Y^2
\eno
In the case of $osp(m|n)$, a basis $\{E_{IJ}\}$ is given $E_{IJ} =
G_{IK} e_{KJ} + (-1)^{(1+\deg I)(1+\deg J)}G_{JK} e_{KI}$ where the
matrix $G_{IJ}$ is defined in "Orthosymplectic superalgebras" ($\see$). 
One finds for example
\beo
&& C_1 = 0 \\
&& C_2 = E_{ij} E_{ji} - E_{kl} E_{lk} + E_{ki} E_{ik} - E_{ik} E_{ki} 
\eno
where $1 \le i,j \le m$ and $m+1 \le k,l \le m+n$.

\medskip

One has to stress that unlike the algebraic case, the center $\cZ(\cG)$
for the classical Lie superalgebras is in general {\em not finitely
generated}. More precisely, the only classical Lie superalgebras for
which the center $\cZ(\cG)$ is finitely generated are $osp(1|2n)$.
In that case, $\cZ(\cG)$ is generated by $n$ Casimirs invariants of
degree $2,4,\dots, 2n$.

\uex\ 1: 
Consider the superalgebra $sl(1|2)$ with generators $H,Z$, $E^+,E^-$,
$F^+,F^-$, $\bF^+,\bF^-$ ($\see$ Superalgebra $sl(1|2)$). Then one can
prove that a generating system of the center $\cZ(\cG)$ is given by,
for $p \in \NN$ and $H_\pm \equiv H \pm Z$:
\beo
&& C_{p+2} = H_+ H_- Z^p + E^-E^+(Z-\half)^p 
+ \bF^-F^+ \Big( H_+ Z^p - (H_+ + 1)(Z+\half)^p \Big) \\
&& ~~~ + F^-\bF^+ \Big( (H_- + 1)(Z-\half)^p - H_- Z^p \Big) 
+ (E^-\bF^+F^+ + \bF^-F^-E^+) \Big( Z^p - (Z-\half)^p \Big) \\
&& ~~~ + \bF^-F^-\bF^+F^+ \Big( (Z+\half)^p+(Z-\half)^p-2Z^p \Big)
\eno
In that case, the Casimir elements $C_p$ satisfy the polynomial relations
 $C_p C_q = C_r C_s$ for $p+q = r+s$ where $p,q,r,s \ge 2$.

\uex\ 2: 
Consider the superalgebra $osp(1|2)$ with generators $H$, $E^+,E^-$,
$F^+,F^-$ ($\see$ Superalgebra $osp(1|2)$). In that case, the center
$\cZ(\cG)$ is finitely generated by
\[
C_2 = H^2 + \half (E^-E^+ + E^+E^-) - (F^+F^- - F^-F^+) 
\]
Moreover, there exists in the universal enveloping superalgebra $\cU$ of 
$osp(1|2)$ an {\em even} operator $S$ which is a square root of the
Casimir operator $C_2$ such that it commutes with the even generators and
anticommutes with the odd ones, given by
\[
S = 2(F^+F^- - F^-F^+) + \sfrac{1}{4}
\]
More precisely, it satisfies $S^2 = C_2 + \sfrac{1}{16}$.
\\
Such an operator exists for any superalgebra of the type $osp(1|2n)$ 
\cite{ABF96}.

\medskip

\underline{Harish--Chandra homomorphism}:
\\
Consider a Borel decomposition $\cG = \cN^+ \oplus \cH \oplus \cN^-$ of
$\cG$ ($\see$ Simple root systems) where $\cH$ is a Cartan subalgebra
of $\cG$ and set $\rho = \rho_0 - \rho_1$ where $\rho_0$ is the
half-sum of positive even roots and $\rho_1$ the half-sum of positive
odd roots. The universal enveloping superalgebra $\cU(\cG)$ can be
decomposed as follows: 
\[
\cU(\cG) = \cU(\cH) \oplus (\cN^- \cU(\cG) + \cU(\cG) \cN^+)
\]
Then any element of the center $\cZ(\cG)$ can be written as $z = z_0 +
z'$ where $z_0 \in \cU(\cH)$ and $z' \in \cN^- \cU(\cG) + \cU(\cG)
\cN^+$. Let $S(\cH) \subset \cU(\cH)$ be the symmetric algebra over
$\cH$. Consider the projection $\bar h: ~ \cZ(\cG) \rightarrow S(\cH)$,
$z \mapsto z_0$ and $\gamma$ the automorphism of $S(\cH)$ such that for
all $H \in \cH$ and $\lambda \in \cH^*$, $\gamma(H(\lambda)) =
H(\lambda-\rho)$. The mapping
\[
h = \gamma \circ \bar h: ~ \cZ(\cG) \rightarrow S(\cH), ~ z \mapsto
\gamma(z_0)
\]
is called the Harish--Chandra homomorphism \cite{Kac77c,Kac78}. 

\uppt: Let $S(\cH)^W$ be the subset of elements of $S(\cH)$ invariant
under the Weyl group of $\cG$ ($\see$). Then the  image of $\cZ(\cG)$
by the Harish--Chandra homomorphism is a subset of $S(\cH)^W$.

\uex:
Consider the Casimir elements $C_p$ of $sl(1|2)$ given above. In the
fermionic basis of $sl(1|2)$ ($\see$ Simple root systems), the positive
(resp. negative) root generators are $E^+,F^+,\bF^+$ (resp.
$E^-,F^-,\bF^-$) and $\rho = 0$. It follows that the image of $C_p$ by
the Harish--Chandra homomorphism is given by 
\[
h(C_{p+2}) = H_+ H_- Z^p = 2^{-p} H_+ H_- (H_+ - H_-)^p 
\]
which is obviously invariant under the action of the Weyl group 
$H_+ \leftrightarrow -H_-$.

\medskip

For more details, see refs. \cite{GrJ79,JaM83,RiS82,Sch83}.

\section{\sf Centralizer, Center, Normalizer of a Lie superalgebra}
\indent

The definitions of the centralizer, the center, the normalizer of a Lie
superalgebra follow those of a Lie algebra.

\medskip

\udef: Let $\cG$ be a Lie superalgebra and $\cS$ a subset of elements in
$\cG$.
\\
- The centralizer $\cC_\cG(\cS)$ is the subset of $\cG$ given by
\[
\cC_\cG(\cS) = \ale X \in \cG ~ \Big\vert ~ \sle X,Y \sri = 0, ~ \forall
~ Y \in  \cS \ari
\]
- The center $\cZ(\cG)$ of $\cG$ is the set of elements of $\cG$ which
commute with any element of $\cG$ (in other words, it is the centralizer
of $\cG$ in $\cG$): 
\[
\cZ(\cG) = \ale X \in \cG ~ \Big\vert ~ \sle X,Y \sri = 0, ~ \forall ~ Y
\in \cG \ari 
\]
- The normalizer $\cN_\cG(\cS)$ is the subset of $\cG$ given by
\[
\cN_\cG(\cS) = \ale X \in \cG ~ \Big\vert ~ \sle X,Y \sri \in \cS, ~
\forall ~ Y \in  \cS \ari
\]

\section{\sf Characters and supercharacters}
\indent

Let $\cG$ be a basic Lie superalgebra with Cartan subalgebra $\cH$.
Consider $\cV(\Lambda)$ a highest weight representation ($\see$) of $\cG$
with highest weight $\Lambda$, the weight decomposition of $\cV$ with
respect to $\cH$ is
\[
\cV(\Lambda) = \bigoplus_{\lambda} \cV_{\lambda}
\medbox{where}
\cV_{\lambda} = \ale \vv \in \cV ~ \Big\vert ~ h(\vv) = \lambda(h)\vv,\,
h \in \cH \ari 
\]
Let $e^{\lambda}$ be the formal exponential, function on $\cH^*$ (dual
of $\cH$) such that $e^{\lambda}(\mu) = \delta_{\lambda,\mu}$ for two
elements $\lambda,\mu \in \cH^*$, which satisfies $e^{\lambda}e^{\mu} =
e^{\lambda+\mu}$.
\\
The {\em character} and {\em supercharacter} of $\cV(\Lambda)$ are
defined by 
\beo
&& \ch \cV(\Lambda) = \sum_{\lambda} (\dim \cV_{\lambda}) e^{\lambda} \\
&& \sch \cV(\Lambda) = \sum_{\lambda} (-1)^{\deg \lambda}
(\dim \cV_{\lambda}) e^{\lambda}
\eno

Let $W(\cG)$ be the Weyl group ($\see$) of $\cG$, $\Delta$ the root
system of $\cG$, $\Delta_\evn^+$ the set of positive even roots,
$\Delta_\odd^+$ the set of positive odd roots,
$\overline{\Delta}_\evn^+$ the subset of roots $\alpha \in
\Delta_\evn^+$ such that $\alpha/2 \notin \Delta_\odd^+$. We set for an
element $w \in W(\cG)$, $\eps(w) = (-1)^{\ell(w)}$ and $\eps'(w) =
(-1)^{\ell'(w)}$ where $\ell(w)$ is the number of reflections in the
expression of $w \in W(\cG)$ and $\ell'(w)$ is the number of
reflections with respect to the roots of $\overline{\Delta}_\evn^+$ in
the expression of $w \in W(\cG)$. We denote by $\rho_0$ and $\rho_1$
the half-sums of positive even roots and positive odd roots, and $\rho =
\rho_0 - \rho_1$. The characters and supercharacters of the {\em
typical} finite dimensional representations $\cV(\Lambda)$ ($\see$) of
the basic Lie superalgebras are given by
\beo
&&\ch \cV(\Lambda) = L^{-1} \sum_w \eps(w) e^{w(\Lambda+\rho)} \\
&&\sch \cV(\Lambda) = L'^{-1} \sum_w \eps'(w) e^{w(\Lambda+\rho)}
\eno
where
\[
L = \frac
{\prod_{\alpha \in \Delta_\evn^+} (e^{\alpha/2} - e^{-\alpha/2})}
{\prod_{\alpha \in \Delta_\odd^+} (e^{\alpha/2} + e^{-\alpha/2})}
\bigbox{and}
L' = \frac
{\prod_{\alpha \in \Delta_\evn^+} (e^{\alpha/2} - e^{-\alpha/2})}
{\prod_{\alpha \in \Delta_\odd^+} (e^{\alpha/2} - e^{-\alpha/2})}
\]

\medskip

In the case of the superalgebra $B(0,n)$ all the representations are
typical. One finds then explicitly
\beo
&&\ch \cV(\Lambda) = \frac{\sum_w \eps(w) e^{w(\Lambda+\rho)}}
{\sum_w \eps(w) e^{w(\rho)}} \\
&&\sch \cV(\Lambda) = \frac{\sum_w \eps'(w) e^{w(\Lambda+\rho)}}
{\sum_w \eps'(w) e^{w(\rho)}}
\eno

\medskip

In the case of the superalgebra $A(m,n)$, the character of the typical
representation $\cV(\Lambda)$ is given by
\[
\ch \cV(\Lambda) = \frac{1}{L_0}
\sum_w \eps(w) w \left( e^{\Lambda+\rho_0} \prod_{\beta\in\Delta_\odd^+}
(1+e^{-\beta}) \right) 
\]
and the character of the singly atypical representation by (see ref.
\cite{JHK89})
\[
\ch \cV(\Lambda) = \frac{1}{L_0}
\sum_w \eps(w) w \left( e^{\Lambda+\rho_0}
\prod_{\beta\in\Delta_\odd^+,\braket{\Lambda+\rho}{\beta}\ne 0} 
(1+e^{-\beta}) \right) 
\]
where $L_0$ is defined as 
\[
\prod_{\alpha\in\Delta_\evn^+} (e^{\alpha/2} - e^{-\alpha/2})
\]

\medskip

In the case of the superalgebra $C(n+1)$, the highest weight irreducible
representations are either typical or singly atypical. It follows that
the character formulae of the typical and atypical representations of
$C(n+1)$ are the same as for $A(m,n)$ above (with the symbols being those
of $C(n+1)$).

\medskip

$\see$ Representations: highest weight, induced modules, typical and
atypical.

For more details, see refs. \cite{Kac77c,JHK89}.

\section{\sf Classical Lie superalgebras}
\indent

\udef: A simple Lie superalgebra $\cG = \cG_\evn \oplus \cG_\odd$
is called {\em classical} if the representation of the even subalgebra
$\cG_\evn$ on the odd part $\cG_\odd$ is completely reducible.

\medskip

\uth: A simple Lie superalgebra $\cG$ is classical if and only if 
its even part $\cG_\evn$ is a reductive Lie algebra.

\medskip

Let $\cG = \cG_\evn \oplus \cG_\odd$ be a classical Lie superalgebra. 
Then the representation of $\cG_\evn$ on $\cG_\odd$ is either $(i)$
irreducible or $(ii)$ the direct sum of two irreducible representations
of $\cG_\evn$. In that case, one has (see below)
\[
\cG_\odd = \cG_{-1} \oplus \cG_1
\]
with
\[
\ale \cG_{-1},\cG_1 \ari = \cG_\evn
\bigbox{and}
\ale \cG_1,\cG_1 \ari = \ale \cG_{-1},\cG_{-1} \ari = 0
\]
In the case $(i)$, the superalgebra is said of the type I and in the case
$(ii)$ of the type II.

\medskip

\uth: Let $\cG = \cG_\evn \oplus \cG_\odd$ be a classical Lie
superalgebra. Then there exists a consistent $\ZZ$-gradation $\cG =
\oplus_{i\in\ZZ} ~ \cG_i$ of $\cG$ (called the distinguished
$\ZZ$-gradation) such that
\\
- for the superalgebras of type I, $\cG_i = 0$ for $|i| > 1$ and 
$\cG_\evn = \cG_0$, $\cG_\odd = \cG_{-1} \oplus \cG_1$.
\\
- for the superalgebras of type II, $\cG_i = 0$ for $|i| > 2$ and
$\cG_\evn = \cG_{-2} \oplus \cG_0 \oplus \cG_2$, $\cG_\odd = \cG_{-1}
\oplus \cG_1$.

\medskip

\udef: A classical Lie superalgebra $\cG$ is called {\em basic} if there
exists a non-degenerate invariant bilinear form on $\cG$ ($\see$ Killing 
form). The classical Lie superalgebras which are not basic are called
{\em strange}. 

\medskip

The Table \ref{table3} resumes the classification and the Table
\ref{table4} gives the $\cG_\evn$ and $\cG_\odd$ structure of the
classical Lie superalgebras.

\begin{table}[htbp]
\centering
\begin{tabular}{|c|cl|cl|} \hline
&& type I ~~~~~~~~ && type II ~~~~~~~~~~~~~ \cr 
\hline
BASIC & $A(m,n)$ & $m > n \ge 0$ & $B(m,n)$ & 
$m \ge 0, n \ge 1$ \cr 
(non-degenerate & $C(n+1)$ & $n \ge 1$ & $D(m,n)$ & 
$\left\{\begin{array}{l} m \ge 2, n\ge 1 \cr m \ne n+1 \cr 
\end{array} \right.$ \cr
Killing form) & $F(4)$ &&& \cr
& $G(3)$ &&& \cr
\hline
BASIC & $A(n,n)$ & $n \ge 1$ & $D(n+1,n)$ & $n \ge 1$ \cr
(zero Killing form) &&& $D(2,1;\alpha)$ & $\alpha \in
\CC\setminus \{0,-1\}$ \cr 
\hline
STRANGE & $P(n)$ & $n \ge 2$ & $Q(n)$ & $n \ge 2$ \cr
\hline
\end{tabular}
\caption{Classical Lie superalgebras.\label{table3}}
\end{table}

\begin{table}[htbp]
\centering
\begin{tabular}{|c|c|c|} \hline
superalgebra $\cG$ & $\cG_\evn$ & $\cG_\odd$ \\ 
\hline
$A(m,n)$ & $A_m \oplus A_n \oplus U(1)$ & $(\overline{m},n) \oplus
(m,\overline{n})$ \\ 
$A(n,n)$ & $A_n \oplus A_n$ & $(\overline{n},n) \oplus
(n,\overline{n})$ \\
$C(n+1)$ & $C_n \oplus U(1)$ & $(2n) \oplus (2n)$ \\
\hline
$B(m,n)$ & $B_m \oplus C_n$ & $(2m+1,2n)$ \\
$D(m,n)$ & $D_m \oplus C_n$ & $(2m,2n)$ \\
$F(4)$ & $A_1 \oplus B_3$ & $(2,8)$ \\
$G(3)$ & $A_1 \oplus G_2$ & $(2,7)$ \\
$D(2,1;\alpha)$ & $A_1 \oplus A_1 \oplus A_1$ & $(2,2,2)$ \\
\hline
$P(n)$ & $A_n$ & $[2] \oplus [1^{n-1}]$ \\
$Q(n)$ & $A_n$ & $\ad(A_n)$ \\
\hline
\end{tabular}
\caption{$\cG_\evn$ and $\cG_\odd$ structure of the
classical Lie superalgebras.\label{table4}}
\end{table}

\medskip

$\see$ Exceptional Lie superalgebras, Orthosymplectic superalgebras,
Strange superalgebras, Unitary superalgebras.

\section{\sf Classification of simple Lie superalgebras}
\indent

Among Lie superalgebras appearing in the classification of simple Lie
superalgebras, one distinguishes two general families: the classical
Lie superalgebras in which the representation of the even subalgebra on
the odd part is completely reducible and the Cartan type superalgebras
in which such a property is no more valid. Among the classical
superalgebras ($\see$), one naturally separates the basic series from
the strange ones.

The basic (or contragredient) Lie superalgebras split into four infinite
families denoted by $A(m,n)$ or $sl(m+1|n+1)$ for $m \ne n$ and
$A(n,n)$ or $sl(n+1|n+1)/\cZ$ where $\cZ$ is a one-dimensional center
for $m=n$ (unitary series), $B(m,n)$ or $osp(2m+1|2n)$, $C(n)$ or
$osp(2|2n)$, $D(m,n)$ or $osp(2m|2n)$ (orthosymplectic series) and
three exceptional superalgebras $F(4)$, $G(3)$ and $D(2,1;\alpha)$, the
last one being actually a one-parameter family of superalgebras.
Two infinite families denoted by $P(n)$ and $Q(n)$ constitute the
strange (or non-contragredient) superalgebras.

The Cartan type superalgebras ($\see$) are classified into four infinite
families, $W(n)$, $S(n)$, ${\tilde S}(n)$ and $H(n)$.

The following scheme resumes this classification:
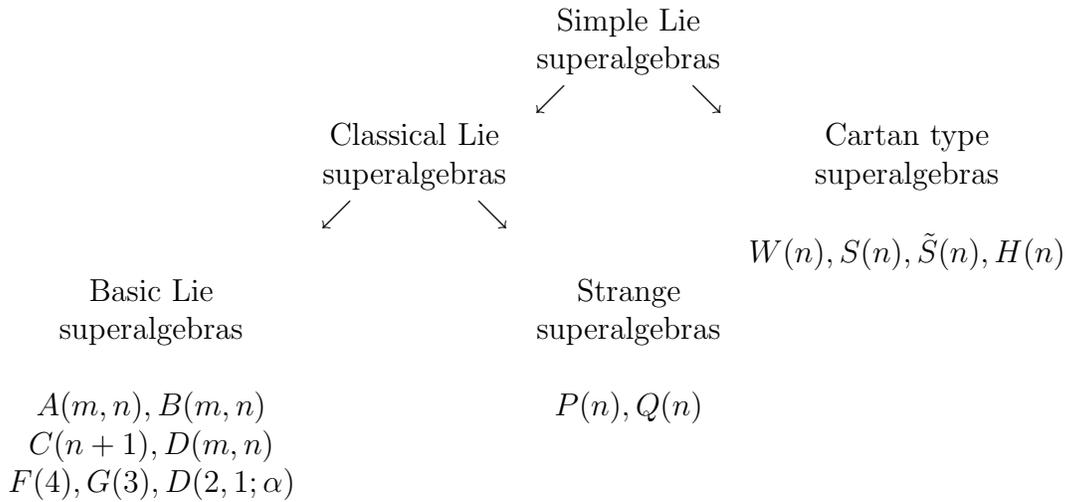
\[
\begin{array}{cccc}
&& \mbox{Simple Lie} & \cr
&& \mbox{superalgebras} & \cr
&& \swarrow \qquad \qquad \searrow & \cr
&\mbox{Classical Lie} && \mbox{Cartan type} \cr
&\mbox{superalgebras} && \mbox{superalgebras} \cr
& \swarrow \qquad \qquad \searrow && \cr
&&& W(n),S(n),{\tilde S}(n),H(n) \cr
\mbox{Basic Lie} && \mbox{Strange} & \cr
\mbox{superalgebras} && \mbox{superalgebras} & \cr
&&& \cr
A(m,n),B(m,n) && P(n),Q(n) & \cr
C(n+1),D(m,n) &&& \cr 
F(4),G(3),D(2,1;\alpha) &&& \cr
\end{array}
\]

\medskip

$\see$ Cartan type superalgebras, Classical Lie superalgebras. 

For more details, see refs. \cite{Kac77a,Kac77b,NRS76}.

\section{\sf Clifford algebras}
\indent

Let $\{\gamma_i\}$ ($i=1,\dots,n$) be a set of square matrices such that
\[
\ale \gamma_i,\gamma_j \ari = \gamma_i\gamma_j + \gamma_j\gamma_i = 2
\delta_{ij} \II 
\]
where $\II$ is the unit matrix. The algebra spanned by the $n$ matrices
$\gamma_i$ is called the Clifford algebra. These relations can be
satisfied by matrices of order $2^p$ when $n=2p$ or $n=2p+1$.

\medskip

Consider the $2 \times 2$ Pauli matrices $\sigma_1,\sigma_2,\sigma_3$:
\[
\sigma_1 = \left( \begin{array}{rr} 0 & 1 \cr 1 & 0 
\end{array} \right) \qquad
\sigma_2 = \left( \begin{array}{rr} 
0 & -i \cr i & 0 \end{array} \right) \qquad
\sigma_3 = \left( \begin{array}{rr} 
1 & 0 \cr 0 & -1 \end{array} \right)
\]
Then the matrices $\gamma_i$ can be expressed in terms of a $p$-fold 
tensor product of the Pauli matrices.

\medskip

\uppt: There exists a representation such that
\\
i) if $n$ is even, the matrices $\gamma_i$ 
are hermitian, half of them being symmetric, half of them being
antisymmetric.
\\
ii) if $n$ is odd, the matrices $\gamma_i$ with $i=1,\dots,2p$ are
hermitian, half of them being symmetric, half of them being
antisymmetric and the matrix $\gamma_{2p+1}$ is diagonal.

\medskip

In this representation, the matrices $\gamma$ can be written as 
($i=1,\dots,p-1$)
\beo
&& \gamma_1 = \sigma_1^{(1)} \otimes \dots \otimes \sigma_1^{(p)} \\
&& \gamma_{2i} = \sigma_1^{(1)} \otimes \dots \otimes \sigma_1^{(p-i)}
\otimes \sigma_2^{(p-i+1)} \otimes \II^{(p-i+2)} \otimes \dots \otimes
\II^{(p)} 
\\ && \gamma_{2i+1} = \sigma_1^{(1)} \otimes \dots \otimes \sigma_1^{(p-i)}
\otimes \sigma_3^{(p-i+1)} \otimes \II^{(p-i+2)} \otimes \dots \otimes
\II^{(p)} \\
&& \gamma_{2p} = \sigma_2^{(1)} \otimes \II^{(2)} \otimes \dots \otimes
\II^{(p)} \\
&& \gamma_{2p+1} = \sigma_3^{(1)} \otimes \II^{(2)} \otimes \dots
\otimes \II^{(p)} 
\eno
One can check that with this representation, one has ($i=1,\dots,p$)
\[
\gamma_{2i}^t = -\gamma_{2i} \bigbox{and} \gamma_{2i+1}^t = \gamma_{2i+1}
\,, \gamma_{2p+1}^t = \gamma_{2p+1}
\]

\medskip

\udef: The matrix $C = \prod_{i=1}^p \gamma_{2i-1}$ for $n=2p$ and
$C = \prod_{i=1}^{p+1} \gamma_{2i-1}$ for $n=2p+1$ is called the charge
conjugation matrix.

\medskip

\uppt: The charge conjugation matrix satisfies
\beo
&& \bullet ~ C^t C = 1 \\
&& \bullet ~ \mbox{for $n=2p$} \\ 
&& ~~~~~ C^t = (-1)^{p(p-1)/2} C = \left\{
\begin{array}{ll}  C & \quad \smbox{for} p=0,1 ~~~ (mod ~ 4) \cr
-C & \quad \smbox{for} p=2,3 ~~~ (mod ~ 4) \cr \end{array} \right. \\
&& ~~~~~ C \gamma_i = (-1)^{p+1} \gamma_i^t C \quad (i=1,\dots,2p) \\
&& \bullet ~ \mbox{for $n=2p+1$} \\ 
&& ~~~~~ C^t = (-1)^{p(p+1)/2} C = \left\{
\begin{array}{ll}  C & \quad \smbox{for} p=0,3 ~~~ (mod ~ 4) \cr
-C & \quad \smbox{for} p=1,2 ~~~ (mod ~ 4) \cr \end{array} \right. \\
&& ~~~~~ C \gamma_i = (-1)^p \gamma_i^t C \quad (i=1,\dots,2p+1) \\
\eno

\section{\sf Decompositions w.r.t. $osp(1|2)$ subalgebras}
\indent

The method for finding the decompositions of the fundamental and the
adjoint representations of the basic Lie superalgebras with respect to
their different $osp(1|2)$ subsuperalgebras is the following:
\\
- one considers an $osp(1|2)$ embedding in a basic Lie superalgebra $\cG
= \cG_\evn \oplus \cG_\odd$, determined by a certain subsuperalgebra
$\cK$ in $\cG$ ($\see$ Embeddings of $osp(1|2)$), which is expressed as
a direct sum of simple components: $\cK = \oplus_i ~ \cK_i$.
\\
- to each couple $(\cG,\cK_i)$ one associates $osp(1|2)$ representations
given in Tables \ref{table7} (regular embeddings) and \ref{table8} 
(singular embeddings); the notations $\cR$ and $\cR''$ are explained
below. 
\\
- the decomposition of the fundamental representation of $\cG$ with
respect to the $osp(1|2)$ subalgebra under consideration is then given
by a direct sum of $osp(1|2)$ representations. 
\\
- starting from a decomposition of the fundamental representation of 
$\cG$ of the form
\[
\fund_\cK ~ \cG = \Big( \oplus_i \cR_{j_i} \Big) \oplus 
\Big( \oplus_k \cR''_{j_k} \Big)
\]
the decomposition of the adjoint representation $\ad_\cK ~ \cG$ is
given in the unitary series by
\beo
&& \ad_\cK ~ \cG = \Big( \oplus_i \cR_{j_i} \oplus_k \cR''_{j_k}
\Big) \otimes \Big( \oplus_i \cR_{j_i} \oplus_k \cR''_{j_k} \Big) 
- \cR_0 \hspace{10mm} \mbox{for } sl(m|n) \,, m \ne n \\
&& \ad_{\cK} ~ \cG = \Big( \oplus_i \cR_{j_i} \oplus_k \cR''_{j_k}
\Big) \otimes \Big( \oplus_i \cR_{j_i} \oplus_k \cR''_{j_k} \Big) 
- 2\cR_0 \hspace{8mm} \mbox{for } sl(n|n)
\eno
and in the orthosymplectic series by
\[
\ad_\cK ~ \cG = \Big( \oplus_i \cR_{j_i} \Big) \otimes
\Big( \oplus_i \cR_{j_i} \Big) \Big\vert_A \oplus
\Big( \oplus_k \cR''_{j_k} \Big) \otimes \Big( \oplus_k \cR''_{j_k} 
\Big) \Big\vert_S \oplus \Big( \oplus_i \cR_{j_i} \Big) \otimes
\Big( \oplus_k \cR''_{j_k} \Big)
\]
The symmetrized and antisymmetrized products of $osp(1|2)$
representations $\cR_j$ are expressed, with analogy with the Lie
algebra case, by (in the following formulae $j$ and $q$ are integer)
\beo
&& \cR_j \otimes \cR_j \Big\vert_A = \bigoplus_{q=1}^{j}
\bigg( \cR_{2q-1} \oplus \cR_{2q-1/2} \bigg) \\
&& \cR_j \otimes \cR_j \Big\vert_S = \bigoplus_{q=0}^{j-1}
\bigg( \cR_{2q} \oplus \cR_{2q+1/2} \bigg) \oplus \cR_{2j} \\
&& \cR_{j-1/2} \otimes \cR_{j-1/2} \Big\vert_A = \bigoplus_{q=0}^{j-1}
\bigg( \cR_{2q} \oplus \cR_{2q+1/2} \bigg) \\
&& \cR_{j-1/2} \otimes \cR_{j-1/2} \Big\vert_S = \bigoplus_{q=1}^{j-1}
\bigg( \cR_{2q-1} \oplus \cR_{2q-1/2} \bigg) \oplus \cR_{2j-1}
\eno
together with (for $j,k$ integer or half-integer)
\beo
\Big( (\cR_j \oplus \cR_k) \otimes (\cR_j \oplus
\cR_k) \Big) \Big\vert_A &=& \Big(\cR_j \otimes \cR_j \Big)
\Big\vert_A ~ \oplus ~  \Big(\cR_k \otimes \cR_k \Big)
\Big\vert_A ~ \oplus ~  (\cR_j \oplus \cR_k) \\
\Big( (\cR_j \oplus \cR_k) \otimes (\cR_j \oplus
\cR_k) \Big) \Big\vert_S &=& \Big(\cR_j \otimes \cR_j \Big)
\Big\vert_S ~ \oplus ~  \Big(\cR_k \otimes \cR_k \Big)
\Big\vert_S ~ \oplus ~  (\cR_j \oplus \cR_k) 
\eno
and ($n$ integer)
\beo
(n\cR_j \otimes n\cR_j) \Big\vert_A &=& 
\frac{n(n+1)}{2} (\cR_j \otimes \cR_j) \Big\vert_A ~ \oplus ~ 
\frac{n(n-1)}{2} (\cR_j \otimes \cR_j) \Big\vert_S \\
(n\cR_j \otimes n\cR_j) \Big\vert_S &=& 
\frac{n(n+1)}{2} (\cR_j \otimes \cR_j) \Big\vert_S ~ \oplus ~ 
\frac{n(n-1)}{2} (\cR_j \otimes \cR_j) \Big\vert_A 
\eno
The same formulae also hold for the $\cR''$ representations.

Let us stress that one has to introduce here two different notations for
the $osp(1|2)$ representations which enter in the decomposition of the
fundamental representation of $\cG$, depending on the origin of the two
factors $\cD_j$ and $\cD_{j-1/2}$ of a representation $\cR_j$
(let us recall that an $osp(1|2)$ representation $\cR_j$ decomposes
under the $sl(2)$ part as $\cR_j = \cD_j \oplus \cD_{j-1/2}$). 
For $\cG = sl(m|n)$ (resp. $\cG = osp(m|n)$), an $osp(1|2)$
representation is denoted $\cR_j$ if the representation $\cD_j$ comes
from the decomposition of the fundamental of $sl(m)$ (resp. $so(m)$),
and $\cR''_j$ if the representation $\cD_j$ comes from the decomposition
of the fundamental of $sl(n)$ (resp. $sp(n)$).

In the same way, considering the tensor products of $\cR$ and $\cR''$
representations given above, one has to distinguish the $osp(1|2)$
representations in the decomposition of the adjoint representations:
the $\cR_j$ representations are such that the $\cD_j$ comes from the
decomposition of the even part $\cG_\evn$ for $j$ integer or of the odd
part $\cG_\odd$ for $j$ half-integer and the $\cR'_j$ representations
are such that $\cD_j$ comes from the decomposition of the even part
$\cG_\evn$ for $j$ half-integer or of the odd part $\cG_\odd$ for $j$
integer.

Finally, the products between unprimed and primed representations obey 
the following rules
\beo
&& \cR_{j_1} \otimes \cR_{j_2} = \left\{ \begin{array}{ll}
\oplus \cR_{j_3} & \quad \mbox{if $j_1+j_2$ is integer} \cr
\oplus \cR'_{j_3} & \quad \mbox{if $j_1+j_2$ is half-integer} \cr
\end{array} \right. \\
&& \cR''_{j_1} \otimes \cR''_{j_2} = \left\{ \begin{array}{ll}
\oplus \cR_{j_3} & \quad \mbox{if $j_1+j_2$ is integer} \cr
\oplus \cR'_{j_3} & \quad \mbox{if $j_1+j_2$ is half-integer} \cr
\end{array} \right. \\
&& \cR_{j_1} \otimes \cR''_{j_2} = \left\{ \begin{array}{ll}
\oplus \cR'_{j_3} & \quad \mbox{if $j_1+j_2$ is integer} \cr
\oplus \cR_{j_3} & \quad \mbox{if $j_1+j_2$ is half-integer} \cr
\end{array} \right.
\eno

The tables \ref{table7} to \ref{table24} give the different
decompositions of the fundamental and adjoint representations of the
basic Lie superalgebras with respect to the different $osp(1|2)$
embeddings. For more details, see ref. \cite{FRS93}.

\section{\sf Decompositions w.r.t. $sl(1|2)$ subalgebras}
\indent

The method for finding the decompositions of the fundamental and the
adjoint representations of the basic Lie superalgebras with respect to
their different $sl(1|2)$ subsuperalgebras is the following:
\\
- one considers a $sl(1|2)$ embedding in a basic Lie superalgebra $\cG =
\cG_\evn \oplus \cG_\odd$, determined by a certain subsuperalgebra
$\cK$ in $\cG$ ($\see$ Embeddings of $sl(1|2)$), which is expressed as
a direct sum of simple components: $\cK = \oplus_i ~ \cK_i$.
\\
- to each couple $(\cG,\cK_i)$ one associates (atypical) $sl(1|2)$
representations $\pi(\pm j_i,j_i) \equiv \pi_\pm(j_i)$ or $osp(2|2)$
representations $\pi(0,\half)$ ($\see$ Superalgebra $sl(1|2)$)
given in the following table:
\[
\begin{array}{c|c|c}
\cG & \cK & \fund_\cK ~ \cG \cr 
\hline && \cr
sl(m|n)
 & sl(p+1|p) & \pi_+(\sfrac{p}{2}) \cr
 & sl(p|p+1) & \pi''_+(\sfrac{p}{2}) \cr && \cr
osp(m|2n)
 & sl(p+1|p)
 & \pi_+(\sfrac{p}{2}) \oplus \pi_-(\sfrac{p}{2}) \cr
 & sl(p|p+1)
 & \pi''_+(\sfrac{p}{2}) \oplus \pi''_-(\sfrac{p}{2}) \cr 
 & osp(2|2) & \pi''(0,\half) \cr
\end{array}
\]
(The notation $\pi$ or $\pi''$ is just to distinguish the superalgebras
$sl(p+1|p)$ or $sl(p|p+1)$ they come from. This will be used below).
\\
In the case of $sl(m|n)$, one could also use $\pi_-$ and $\pi''_-$
representations as well, leading to different but equivalent
decompositions of the adjoint representation of $\cG$. This fact is
related to the existence of non-trivial outer automorphims for $sl(1|2)$.
\\
- the decomposition of the fundamental representation of $\cG$ with
respect to the $sl(1|2)$ subalgebra under consideration is then given
by a direct sum of $sl(1|2)$ representations of the above type,
eventually completed by trivial representations.
\\
- starting from a decomposition of the fundamental representation of 
$\cG$ of the form
\[
\fund_\cK ~ \cG = \Big( \oplus_i \pi_\pm(j_i) \Big) \oplus 
\Big( \oplus_k \pi''_\pm(j_k) \Big)
\]
the decomposition of the adjoint representation $\ad_\cK ~ \cG$ is
given in the unitary series by
\beo
&& \ad_\cK ~ \cG = \Big( \oplus_i \pi_\pm(j_i) \oplus_k 
\pi''_\pm(j_k) \Big)^2 - \pi(0,0) 
\hspace{12mm} \mbox{for } A(m|n) \,, m \ne n \\
&& \ad_{\cK} ~ \cG = \Big( \oplus_i \pi_\pm(j_i) \oplus_k 
\pi''_\pm(j_k) \Big)^2 - 2\pi(0,0) 
\hspace{10mm} \mbox{for } A(n|n)
\eno
and in the orthosymplectic series by
\beo
&& \ad_\cK ~ \cG = \Big( \oplus_i \pi_\pm(j_i) \Big)^2 \Big\vert_A 
\oplus \Big( \oplus_k \pi''_\pm(j_k) \Big)^2 \Big\vert_S 
\oplus \Big( \oplus_i \pi_\pm(j_i) \oplus_i \pi''_\pm(j_k) \Big)
\eno
The symmetrized and antisymmetrized products of atypical $sl(1|2)$
representations are given by 
\beo
\Big( \pi_\pm(j) \oplus \pi_\pm(k) \Big)^2 \Big\vert_A &=&
\Big( \pi_\pm(j) \otimes \pi_\pm(j) \Big)^2 \Big\vert_A \oplus
\Big( \pi_\pm(k) \otimes \pi_\pm(k) \Big)^2 \Big\vert_A 
\oplus \Big( \pi_\pm(j) \otimes \pi_\pm(k) \Big) \\
\Big( \pi_\pm(j) \oplus \pi_\pm(k) \Big)^2 \Big\vert_S &=&
\Big( \pi_\pm(j) \otimes \pi_\pm(j) \Big)^2 \Big\vert_S \oplus 
\Big( \pi_\pm(k) \otimes \pi_\pm(k) \Big)^2 \Big\vert_S 
\oplus \Big( \pi_\pm(j) \otimes \pi_\pm(k) \Big)
\eno
and ($n$ integer)
\beo
\Big( n\pi_\pm(j) \otimes n\pi_\pm(j) \Big) \Big\vert_A &=& 
\frac{n(n+1)}{2} \Big( \pi_\pm(j) \otimes \pi_\pm(j) \Big) \Big\vert_A 
~ \oplus ~ \frac{n(n-1)}{2} \Big( \pi_\pm(j) \otimes 
\pi_\pm(j) \Big) \Big\vert_S \\
\Big( n\pi_\pm(j) \otimes n\pi_\pm(j) \Big) \Big\vert_S &=& 
\frac{n(n+1)}{2} \Big( \pi_\pm(j) \oplus \pi_\pm(j) \Big) \Big\vert_S 
~ \oplus ~ \frac{n(n-1)}{2} \Big( \pi_\pm(j) \otimes 
\pi_\pm(j) \Big) \Big\vert_A 
\eno
where (in the following formulae $j$ and $q$ are integer)
\beo
&& \Big( \pi_+(j) \oplus \pi_-(j) \Big)^2 \Big\vert_A = 
\oplus_{q=0}^{2j} ~ \pi(0,q) ~ 
\oplus_{q=1}^j ~ \pi(2j+\half,2q-\half) ~ 
\oplus_{q=1}^j ~ \pi(-2j-\half,2q-\half) \\
&& \Big( \pi_+(j) \oplus \pi_-(j) \Big)^2 \Big\vert_S = 
\oplus_{q=0}^{2j} ~ \pi(0,q) ~ 
\oplus_{q=0}^{j-1} ~ \pi(2j+\half,2q+\half) ~ 
\oplus_{q=0}^j ~ \pi(-2j-\half,2q+\half) \\ 
&& \hspace{40mm} \oplus \pi_+(2j) \oplus \pi_-(2j) \\
&& \Big( \pi_+(j+\half) \oplus \pi_-(j+\half) \Big)^2 \Big\vert_A = 
\oplus_{q=0}^{2j+1} ~ \pi(0,q) ~ 
\oplus_{q=0}^j ~ \pi(2j+\sfrac{3}{2},2q+\half) \\ 
&& \hspace{55mm} \oplus_{q=0}^j ~ \pi(-2j-\sfrac{3}{2},2q+\half) \\
&& \Big( \pi_+(j+\half) \oplus \pi_-(j+\half) \Big)^2 \Big\vert_S = 
\oplus_{q=0}^{2j+1} ~ \pi(0,q) ~ 
\oplus_{q=1}^j ~ \pi(2j+\sfrac{3}{2},2q-\half) \\ 
&& \hspace{55mm} \oplus_{q=1}^j ~ \pi(-2j-\sfrac{3}{2},2q-\half) ~ 
\oplus \pi_+(2j+1) \oplus \pi_-(2j+1) 
\eno
Finally, in the case of $osp(2|2)$ embeddings, the product of the
$\pi(0,\half)$ representation by itself is not fully reducible but
gives rise to the indecomposable $sl(1|2)$ representation of the type
$\pi(0;-\half,\half;0)$ ($\see$ Superalgebra $sl(1|2)$).

Considering the tensor products of $\pi$ and $\pi''$ representations
given above, one has to distinguish the $sl(1|2)$ representations in
the decomposition of the adjoint representations. Let us recall that a
$sl(1|2)$ representation $\pi(b,j)$ decomposes under the $sl(2) \oplus
U(1)$ part as $\pi(b,j) = D_j(b) \oplus D_{j-1/2}(b-1/2) \oplus
D_{j-1/2}(b+1/2) \oplus D_{j-1}(b)$ and $\pi_\pm(j) = D_j(\pm j) \oplus
D_{j-1/2}(\pm j \pm 1/2)$.
The $\pi(b,j)$ representations are such that the $\cD_j$ comes from the
decomposition of the even part $\cG_\evn$ for $j$ integer or of the odd
part $\cG_\odd$ for $j$ half-integer and the $\pi'(b,j)$ representations
are such that $\cD_j$ comes from the decomposition of the even part
$\cG_\evn$ for $j$ half-integer or of the odd part $\cG_\odd$ for $j$
integer.
Finally, the products between unprimed and primed representations obey 
the following rules
\beo
&& \pi(b_1,j_1) \otimes \pi(b_2,j_2) = \left\{ \begin{array}{ll}
\oplus \pi(b_3,j_3) & \quad \mbox{if $j_1+j_2$ is integer} \cr
\oplus \pi'(b_3,j_3) & \quad \mbox{if $j_1+j_2$ is half-integer} \cr
\end{array} \right. \\
&& \pi''(b_1,j_1) \otimes \pi''(b_2,j_2) = \left\{ \begin{array}{ll}
\oplus \pi(b_3,j_3) & \quad \mbox{if $j_1+j_2$ is integer} \cr
\oplus \pi'(b_3,j_3) & \quad \mbox{if $j_1+j_2$ is half-integer} \cr
\end{array} \right. \\
&& \pi(b_1,j_1) \otimes \pi''(b_2,j_2) = \left\{ \begin{array}{ll}
\oplus \pi'(b_3,j_3) & \quad \mbox{if $j_1+j_2$ is integer} \cr
\oplus \pi(b_3,j_3) & \quad \mbox{if $j_1+j_2$ is half-integer} \cr
\end{array} \right.
\eno

The tables \ref{table33} to \ref{table38} give the different
decompositions of the fundamental and adjoint representations of the
basic Lie superalgebras with respect to the different $sl(1|2)$
embeddings. For more details, see ref. \cite{RSS96}.

\section{\sf Derivation of a Lie superalgebra}
\indent

\udef: Let $\cG = \cG_\evn \oplus \cG_\odd$ be a Lie superalgebra. A
{\em derivation} $D$ of degree $\deg D \in \ZZ_2$ of the superalgebra
$\cG$ is an endomorphism of $\cG$ such that
\[
D \sle X,Y \sri = \sle D(X),Y \sri + (-1)^{\deg D.\deg X} 
\sle X,D(Y) \sri
\]
If $\deg D = \evn$, the derivation is even, otherwise $\deg D = \odd$ and
the derivation is odd.
\\
The space of all the derivations of $\cG$ is denoted by $\der\cG = 
\mbox{Der}_\evn\,\cG \oplus \mbox{Der}_\odd\,\cG$.
\\
If $D$ and $D'$ are two derivations of $\cG$, then $\sle D,D' \sri \in 
\der\cG$, that is the space $\der\cG$ closes under the Lie superbracket.
\\
The space $\der\cG$ is called the superalgebra of derivations of $\cG$.
In particular, 
\[
\ad_X : Y \mapsto \ad_X(Y) = \sle X,Y \sri
\]
is a derivation of $\cG$. These derivations are called inner derivations
of $\cG$. They form an ideal $\inder\cG$ of $\der\cG$. Every derivation
of a simple Lie superalgebra with non-degenerate Killing form is inner.

\section{\sf Dirac matrices}
\indent

$\see$ Clifford algebra, Spinors (in the Lorentz group), Supersymmetry
algebra, Superconformal algebra.

\section{\sf Dynkin diagrams}
\indent

Let $\cG$ be a basic Lie superalgebra of rank $r$ and dimension $n$ with
Cartan subalgebra $\cH$. Let $\Delta^0 = (\alpha_1,\dots,\alpha_r)$ be
a simple root system ($\see$) of $\cG$ and $A^s = (a'_{ij})$ be the
corresponding {\em symmetric} Cartan matrix ($\see$), defined by
$a'_{ij} = (\alpha_i,\alpha_j)$. One can associate to $\Delta^0$ a
Dynkin diagram according to the following rules:
\begin{enumerate}
\item 
one associates to each simple even root a white dot, to each simple odd
root of non-zero length ($a'_{ii} \ne 0$) a black dot and to each simple
odd root of zero length ($a'_{ii} = 0$) a grey dot.
\item 
the $i$th and $j$th dots will be joined by $\eta_{ij}$ lines where 
\[
\begin{array}{ll}
\bigg. \displaystyle{
\eta_{ij} = \frac{2|a'_{ij}|}{\min (|a'_{ii}|,|a'_{jj}|)} }
& \smbox{if} a'_{ii}.a'_{jj} \ne 0 \cr
\bigg. \displaystyle{
\eta_{ij} = \frac{2|a'_{ij}|}{\min_{a'_{kk} \ne 0} |a'_{kk}|} }
& \smbox{if} a'_{ii} \ne 0 \smbox{and} a'_{jj} = 0 \cr
\bigg. \displaystyle{
\eta_{ij} = |a'_{ij}| }
& \smbox{if} a'_{ii} = a'_{jj} = 0
\end{array}
\]
\item 
we add an arrow on the lines connecting the $i$th and $j$th dots when
$\eta_{ij} > 1$, pointing from $i$ to $j$ if $a'_{ii}.a'_{jj} \ne 0$ and
$|a'_{ii}| > |a'_{jj}|$ or if $a'_{ii} = 0$, $a'_{jj} \ne 0$, $|a'_{jj}| <
2$, and pointing from $j$ to $i$ if $a'_{ii} = 0$, $a'_{jj} \ne 0$,
$|a'_{jj}| > 2$.
\end{enumerate}

Since a basic Lie superalgebra possesses many inequivalent simple root
systems, there will be for a basic Lie superalgebra many inequivalent
Dynkin diagrams. For each basic Lie superalgebra, there is a particular
Dynkin diagram which can be considered as canonical. Its characteristic
is that it contains the smallest number of odd roots. Such a Dynkin
diagram is called distinguished.

\udef: The {\em distinguished Dynkin diagram} is the Dynkin diagram
associated to the distinguished simple root system ($\see$) to which
corresponds the distinguished Cartan matrix ($\see$). It is constructed
as follows: the even dots are given by the Dynkin diagram of the even
part $\cG_\evn$ (it may be not connected) and the odd dot corresponds to
the lowest weight of the representation $\cG_\odd$ of $\cG_\evn$.

The list of the distinguished Dynkin diagrams of the basic Lie 
superalgebras are given in Table \ref{table5} (see also Table
\ref{table11} to \ref{table32}).

\medskip

$\see$ Cartan matrices, Simple root systems.

For more details, see refs. \cite{FSS89,Kac77a,Kac77b}.

\section{\sf Embeddings of $osp(1|2)$}
\indent

The determination of the possible $osp(1|2)$ subsuperalgebras of a basic
Lie superalgebra $\cG$ can be seen as the supersymmetric version of the
Dynkin classification of $sl(2)$ subalgebras in a simple Lie algebra.
Interest for this problem appeared recently in the framework of
supersymmetric integrable models (in particular super-Toda theories)
and super-$W$ algebras \cite{FRS93,LSS86}. As in the algebraic case, it
uses the notion of principal embedding (here superprincipal).

\medskip

\udef: Let $\cG$ be a basic Lie superalgebra of rank $r$ with simple
root system $\Delta^0 = \{\alpha_1,\dots,\alpha_r\}$ and corresponding
simple root generators $E_{\pm\alpha_i}$ in the Serre--Chevalley basis
($\see$). The generators of the  $osp(1|2)$ {\em superprincipal
embedding} in $\cG$ are defined by
\[
F^+ = \sum_{i=1}^r E_{\alpha_i} \,, \qquad 
F^- = \sum_{i=1}^r \sum_{j=1}^r a^{ji} E_{-\alpha_i} 
\]
$a_{ij}$ being the Cartan matrix of $\cG$ and $a^{ij}=(a^{-1})_{ij}$.
The even generators of the superprincipal $osp(1|2)$ are given by 
anticommutation of the odd generators $F^+$ and $F^-$:
\[
H = 2 \ale F^+,F^- \ari \,, \qquad E^+ = 2 \ale F^+,F^+\ari \,, \qquad
E^- = -2 \ale F^-,F^-\ari
\]

Not all the basic Lie superalgebras admit an $osp(1|2)$ superprincipal
embedding. It is clear from the expression of the $osp(1|2)$ generators
that a superprincipal embedding can be defined only if the superalgebra
under consideration admits a completely odd simple root system (which
corresponds to a Dynkin diagram with no white dot). This condition is
however necessary but not sufficient (the superalgebra $A(n|n)$ does
not admit a superprincipal embedding although it has a completely odd
simple root system). The basic Lie superalgebras admitting a
superprincipal $osp(1|2)$ are thus the following:
\[
sl(n \pm 1|n), osp(2n \pm 1|2n), osp(2n|2n), osp(2n+2|2n), 
D(2,1;\alpha) \smbox{with} \alpha \ne 0, \pm 1
\]
The classification of the $osp(1|2)$ embeddings of a basic Lie
superalgebra $\cG$ is given by the following theorem.

\medskip

\uth:
\begin{enumerate}
\item
Any $osp(1|2)$ embedding in a basic Lie superalgebra $\cG$ can be 
considered as the superprincipal $osp(1|2)$ subsuperalgebra of a regular
subsuperalgebra $\cK$ of $\cG$.
\item
For $\cG = osp(2n\pm 2|2n)$ with $n \ge 2$, besides the $osp(1|2)$
superprincipal embeddings of item 1, there exist $osp(1|2)$ embeddings
associated to the singular embeddings $osp(2k \pm 1|2k) \oplus
osp(2n-2k \pm 1|2n-2k) \subset osp(2n\pm 2|2n)$ with $1 \le k \le
\left[\frac{n-1}{2}\right]$.
\item
For $\cG = osp(2n|2n)$ with $n \ge 2$, besides the $osp(1|2)$
superprincipal embeddings of item 1, there exist $osp(1|2)$ embeddings
associated to the singular embeddings $osp(2k \pm 1|2k) \oplus
osp(2n-2k \mp 1|2n-2k) \subset osp(2n|2n)$ with $1 \le k \le
\left[\frac{n-2}{2}\right]$.
\end{enumerate}

\section{\sf Embeddings of $sl(2|1)$}
\indent

In the same way one can consider $osp(1|2)$ embeddings of a basic Lie
superalgebra, it is possible to determine the $sl(2|1)$
subsuperalgebras of a basic Lie superalgebra $\cG$. This problem was
recently considered for an exhaustive classification and
characterization of all extended $N=2$ superconformal algebras and all
string theories obtained by gauging $N=2$ Wess--Zumino--Witten models
\cite{RSS96}. Let us consider the basic Lie superalgebra $sl(n+1|n)$
with completely odd simple root system $\Delta^0$:
\[
\Delta^0 = \ale ~ \eps_1-\del_1, ~ \del_1-\eps_2, ~ \eps_2-\del_2, ~ 
\dots, ~ \del_{n-1}-\eps_n, ~ \eps_n-\del_n, ~ \del_n-\eps_{n+1} ~ \ari
\]
Denote by $E_{\pm(\eps_i-\del_i)}$, $E_{\pm(\del_i-\eps_{i+1})}$ ($1 \le
i \le n$) the corresponding simple root generators in the
Serre--Chevalley basis ($\see$). The $sl(2|1)$ {\em superprincipal
embedding} in $sl(n+1|n)$ is defined as follows. The odd part of the
superprincipal $sl(2|1)$ is generated by $F_{\pm\alpha}$ and
$F_{\pm\beta}$ where
\[
\begin{array}{lll}
\bigg. \displaystyle{ 
F_{+\alpha} = \sum_{i=1}^n E_{\del_i-\eps_{i+1}} }
&\qquad& \displaystyle{ 
F_{-\alpha} = \sum_{i=1}^n \sum_{j=1}^{2n} a^{j,2i} 
E_{-\del_i+\eps_{i+1}} } \cr
\bigg. \displaystyle{ 
F_{+\beta} = \sum_{i=1}^n E_{\eps_i-\del_i} }
&\qquad& \displaystyle{ 
F_{-\beta} = \sum_{i=1}^n \sum_{j=1}^{2n} a^{j,2i-1} 
E_{-\eps_i+\del_i} } \cr
\end{array}
\]
$a_{ij}$ being the Cartan matrix of $sl(n+1|n)$, $a^{ij}=(a^{-1})_{ij}$
and $a_{ij}$ is chosen as
\[
a = \left( \begin{array}{ccccccc}
0 & 1 \cr
1 & 0 & -1 \cr
& -1 & 0 & 1 \cr
&& 1 & 0 \cr
&&&& \ddots \cr
&&&&& 0 & 1 \cr
&&&&& 1 & 0  \cr
\end{array} \right)
\]
The even generators of the superprincipal $sl(2|1)$ are given by 
anticommutation of the odd generators $F_{\pm\alpha}$ and $F_{\pm\beta}$:
\[
\begin{array}{lll}
\bigg. \ale F_{+\alpha},F_{-\alpha} \ari = H_+ + H_- 
&\qquad& \ale F_{+\beta},F_{-\beta} \ari = H_+ - H_- \cr
\bigg. \ale F_{\pm\alpha},F_{\pm\alpha} \ari = 
\ale F_{\pm\beta},F_{\pm\beta} \ari = 0 
&\qquad& \ale F_{\pm\alpha},F_{\pm\beta} \ari = E_\pm \cr
\end{array}
\]
One obtains finally
\[
\begin{array}{lll}
\bigg. \cle E_\pm,F_{\pm\alpha} \cri = 0 
&\qquad& \cle E_\pm,F_{\pm\beta} \cri = 0 \cr
\bigg. \cle E_\pm,F_{\mp\alpha} \cri = \mp F_{\pm\beta} 
&\qquad& \cle E_\pm,F_{\mp\beta} \cri = \mp F_{\pm\alpha} \cr
\bigg. \cle H_\pm,F_{+\alpha} \cri = \pm \half F_{+\alpha} 
&\qquad& \cle H_\pm,F_{-\alpha} \cri = \mp \half F_{-\alpha} \cr
\bigg. \cle H_\pm,F_{+\beta} \cri = \half F_{+\beta} 
&\qquad& \cle H_\pm,F_{-\beta} \cri = - \half F_{-\beta} \cr
\bigg. \cle H_+,E_\pm \cri = \pm E_\pm 
&\qquad& \cle H_-,E_\pm \cri = 0 \cr
\end{array}
\]
where
\beo
H_\pm = {\half} \sum_{i=1}^n \sum_{j=1}^{2n} (a^{j,2i} H_{2i} \pm
a^{j,2i-1} H_{2i-1})
\eno
Notice that this $sl(2|1)$ superprincipal embedding contains as maximal
subsuperalgebra the superprincipal $osp(1|2)$ with generators $H_+,E_\pm$
and $F_\pm = F_{\pm\alpha} + F_{\pm\beta}$.

\medskip

The classification of the $sl(2|1)$ embeddings of a basic Lie
superalgebra $\cG$ is given by the following theorem.

\medskip

\uth: Let $\cG$ be a basic Lie superalgebra. Any $sl(2|1)$ embedding
into $\cG$ can be seen as the principal embedding of a (sum of) regular
$sl(n|n \pm 1)$ subsuperalgebra of $\cG$, except in the case of
$osp(m|n)$ ($m>1)$, $F(4)$ and $D(2,1;\alpha)$ where the (sum of)
regular $osp(2|2)$ has also to be considered.

\section{\sf Exceptional Lie superalgebra $F(4)$}
\indent

The Lie superalgebra $F(4)$ of rank 4 has dimension 40. The even part
(of dimension 24) is a non-compact form of $sl(2) \oplus o(7)$ and the
odd part (of dimension 16) is the spinorial representation $(2,8)$ of
$sl(2) \oplus o(7)$. In terms of the vectors $\eps_1,\eps_2,\eps_3$ and
$\del$ such that $\eps_i.\eps_j=\delta_{ij}, \del^2=-3, \eps_i.\del=0$,
the root system $\Delta = \Delta_\evn \cup \Delta_\odd$ is given by
\[
\Delta_\evn = \ale ~ \pm\del, ~ \pm\eps_i\pm\eps_j, ~ \pm\eps_i ~ \ari 
\medbox{and}
\Delta_\odd = \ale ~ \half (\pm\del\pm\eps_1\pm\eps_2\pm\eps_3) ~ \ari 
\]
The different simple root systems of $F(4)$ with the corresponding 
Dynkin diagrams and Cartan matrices are the following:
\\
Simple root system
$\Delta^0 = \ale 
\alpha_1 = {\half} (\del-\eps_1-\eps_2-\eps_3), \alpha_2 = \eps_3,
\alpha_3 = \eps_2-\eps_3, \alpha_4 = \eps_1-\eps_2 \ari$ 
\begin{center}
\begin{picture}(140,20)
\thicklines
\put(0,0){\circle{14}}
\put(42,0){\circle{14}}
\put(84,0){\circle{14}}
\put(126,0){\circle{14}}
\put(0,15){\makebox(0.4,0.6){$\alpha_1$}}
\put(42,15){\makebox(0.4,0.6){$\alpha_2$}}
\put(84,15){\makebox(0.4,0.6){$\alpha_3$}}
\put(126,15){\makebox(0.4,0.6){$\alpha_4$}}
\put(-5,-5){\line(1,1){10}}\put(-5,5){\line(1,-1){10}}
\put(7,0){\line(1,0){28}}
\put(48,-3){\line(1,0){30}}
\put(48,3){\line(1,0){30}}
\put(59,0){\line(1,1){10}}\put(59,0){\line(1,-1){10}}
\put(91,0){\line(1,0){28}}
\end{picture}
\qquad \qquad Cartan matrix = 
$\left(\begin{array}{rrrr} 
0 & 1 & 0 & 0 \cr -1 & 2 & -2 & 0 \cr 
0 & -1 & 2 & -1 \cr 0 & 0 & -1 & 2 \cr
\end{array}\right)$
\end{center}
Simple root system
$\Delta^0 = \ale 
\alpha_1 = {\half} (-\del+\eps_1+\eps_2+\eps_3), \alpha_2 = {\half} 
(\del-\eps_1-\eps_2+\eps_3), \alpha_3 = \eps_2-\eps_3, 
\alpha_4 = \eps_1-\eps_2 \ari$ 
\begin{center}
\begin{picture}(140,20)
\thicklines
\put(0,0){\circle{14}}
\put(42,0){\circle{14}}
\put(84,0){\circle{14}}
\put(126,0){\circle{14}}
\put(0,15){\makebox(0.4,0.6){$\alpha_1$}}
\put(42,15){\makebox(0.4,0.6){$\alpha_2$}}
\put(84,15){\makebox(0.4,0.6){$\alpha_3$}}
\put(126,15){\makebox(0.4,0.6){$\alpha_4$}}
\put(-5,-5){\line(1,1){10}}\put(-5,5){\line(1,-1){10}}
\put(37,-5){\line(1,1){10}}\put(37,5){\line(1,-1){10}}
\put(7,0){\line(1,0){28}}
\put(48,-3){\line(1,0){30}}
\put(48,3){\line(1,0){30}}
\put(59,0){\line(1,1){10}}\put(59,0){\line(1,-1){10}}
\put(91,0){\line(1,0){28}}
\end{picture}
\qquad \qquad Cartan matrix = 
$\left(\begin{array}{rrrr} 
0 & 1 & 0 & 0 \cr -1 & 0 & 2 & 0 \cr 
0 & -1 & 2 & -1 \cr 0 & 0 & -1 & 2 \cr
\end{array}\right)$
\end{center}
Simple root system
$\Delta^0 = \ale 
\alpha_1 = \eps_1-\eps_2, 
\alpha_2 = {\half} (\del-\eps_1+\eps_2-\eps_3), 
\alpha_3 = {\half} (-\del+\eps_1+\eps_2-\eps_3), 
\alpha_4 = \eps_3 \ari$ 
\begin{center}
\begin{picture}(100,40)
\thicklines
\put(0,20){\circle{14}}
\put(42,20){\circle{14}}
\put(84,20){\circle{14}}
\put(0,35){\makebox(0.4,0.6){$\alpha_1$}}
\put(42,35){\makebox(0.4,0.6){$\alpha_2$}}
\put(84,35){\makebox(0.4,0.6){$\alpha_3$}}
\put(84,-20){\makebox(0.4,0.6){$\alpha_4$}}
\put(27,20){\line(-1,1){10}}\put(27,20){\line(-1,-1){10}}
\put(37,15){\line(1,1){10}}\put(37,25){\line(1,-1){10}}
\put(79,15){\line(1,1){10}}\put(79,25){\line(1,-1){10}}
\put(6,17){\line(1,0){30}}
\put(6,23){\line(1,0){30}}
\put(48,17){\line(1,0){30}}
\put(48,23){\line(1,0){30}}
\put(61,-20){\circle{14}}
\put(56,-15){\line(-1,2){14}}
\put(66,-15){\line(1,2){14}}
\end{picture}
\qquad \qquad Cartan matrix = 
$\left(\begin{array}{rrrr} 
2 & -1 & 0 & 0 \cr -2 & 0 & 2 & 1 \cr 
0 & -2 & 0 & 1 \cr 0 & -1 & -1 & 2 \cr
\end{array}\right)$
\end{center}
Simple root system
$\Delta^0 = \ale 
\alpha_1 = {\half} (\del+\eps_1-\eps_2-\eps_3), 
\alpha_2 = {\half} (\del-\eps_1+\eps_2+\eps_3),
\alpha_3 = {\half} (-\del+\eps_1-\eps_2+\eps_3), 
\alpha_4 = \eps_2-\eps_3 \ari$
\begin{center}
\begin{picture}(120,40)
\thicklines
\put(11,20){\circle{14}}
\put(11,-20){\circle{14}}
\put(-9,20){\makebox(0.4,0.6){$\alpha_1$}}
\put(-9,-20){\makebox(0.4,0.6){$\alpha_2$}}
\put(42,0){\circle{14}}
\put(42,15){\makebox(0.4,0.6){$\alpha_3$}}
\put(6,15){\line(1,1){10}}\put(6,25){\line(1,-1){10}}
\put(6,-25){\line(1,1){10}}\put(6,-15){\line(1,-1){10}}
\put(37,-5){\line(1,1){10}}\put(37,5){\line(1,-1){10}}
\put(7,-14){\line(0,1){28}}
\put(11,-13){\line(0,1){26}}
\put(15,-14){\line(0,1){28}}
\put(17,-17){\line(2,1){21}}
\put(17,15){\line(2,-1){20}}\put(18,19){\line(2,-1){23}}
\put(84,0){\circle{14}}
\put(84,15){\makebox(0.4,0.6){$\alpha_4$}}
\put(48,-3){\line(1,0){30}}
\put(48,3){\line(1,0){30}}
\put(69,0){\line(-1,1){10}}\put(69,0){\line(-1,-1){10}}
\end{picture}
\qquad \qquad Cartan matrix = 
$ \left(\begin{array}{rrrr}
0 & 3 & 2 & 0 \cr -3 & 0 & 1 & 0 \cr 
-2 & -1 & 0 & 1 \cr 0 & 0 & -2 & 2 \cr
\end{array}\right)$
\end{center}
Simple root system
$\Delta^0 = \ale 
\alpha_1 = \del, \alpha_2 = {\half} (-\del+\eps_1-\eps_2-\eps_3), 
\alpha_3 = \eps_3, \alpha_4 = \eps_2-\eps_3 \ari$ 
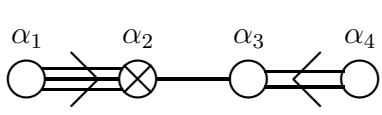
\begin{center}
\begin{picture}(140,20)
\thicklines
\put(0,0){\circle{14}}
\put(42,0){\circle{14}}
\put(84,0){\circle{14}}
\put(126,0){\circle{14}}
\put(0,15){\makebox(0.4,0.6){$\alpha_1$}}
\put(42,15){\makebox(0.4,0.6){$\alpha_2$}}
\put(84,15){\makebox(0.4,0.6){$\alpha_3$}}
\put(126,15){\makebox(0.4,0.6){$\alpha_4$}}
\put(27,0){\line(-1,1){10}}\put(27,0){\line(-1,-1){10}}
\put(37,-5){\line(1,1){10}}\put(37,5){\line(1,-1){10}}
\put(6,-4){\line(1,0){30}}
\put(7,0){\line(1,0){28}}
\put(6,4){\line(1,0){30}}
\put(49,0){\line(1,0){28}}
\put(90,-3){\line(1,0){30}}
\put(90,3){\line(1,0){30}}
\put(101,0){\line(1,1){10}}\put(101,0){\line(1,-1){10}}
\end{picture}
\qquad \qquad Cartan matrix = 
$\left(\begin{array}{rrrr} 
2 & -1 & 0 & 0 \cr -3 & 0 & 1 & 0 \cr 
0 & -1 & 2 & -2 \cr 0 & 0 & -1 & 2 \cr
\end{array}\right)$
\end{center}
Simple root system
$\Delta^0 = \ale 
\alpha_1 = \del, \alpha_2 = {\half} (-\del-\eps_1+\eps_2+\eps_3), 
\alpha_3 = \eps_1-\eps_2, \alpha_4 = \eps_2-\eps_3 \ari$ 
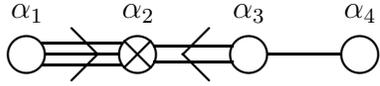
\begin{center}
\begin{picture}(140,20)
\thicklines
\put(0,0){\circle{14}}
\put(42,0){\circle{14}}
\put(84,0){\circle{14}}
\put(126,0){\circle{14}}
\put(0,15){\makebox(0.4,0.6){$\alpha_1$}}
\put(42,15){\makebox(0.4,0.6){$\alpha_2$}}
\put(84,15){\makebox(0.4,0.6){$\alpha_3$}}
\put(126,15){\makebox(0.4,0.6){$\alpha_4$}}
\put(37,-5){\line(1,1){10}}\put(37,5){\line(1,-1){10}}
\put(27,0){\line(-1,1){10}}\put(27,0){\line(-1,-1){10}}
\put(6,-4){\line(1,0){30}}
\put(7,0){\line(1,0){28}}
\put(6,4){\line(1,0){30}}
\put(48,-3){\line(1,0){30}}
\put(48,3){\line(1,0){30}}
\put(59,0){\line(1,1){10}}\put(59,0){\line(1,-1){10}}
\put(91,0){\line(1,0){28}}
\end{picture}
\qquad \qquad Cartan matrix = 
$\left(\begin{array}{rrrr} 
2 & -1 & 0 & 0 \cr -3 & 0 & 2 & 0 \cr 
0 & -1 & 2 & -1 \cr 0 & 0 & -1 & 2 \cr
\end{array}\right)$
\end{center}

Denoting by $T_i$ where $i = 1,2,3$ the generators of $sl(2)$, by
$M_{pq} = -M_{qp}$ where $1 \le p \ne q \le 7$ the generators of
$so(7)$ and by $F_{\alpha\mu}$ where $\alpha = +,-$ and $1 \le \mu \le
8$ the generators of the odd part, the commutation relations of $F(4)$
read as:
\beo 
&& \cle T_i , T_j \cri = i \eps_{ijk} T_k \hspace{25mm}
\cle T_i , M_{pq} \cri = 0 \\
&& \cle M_{pq} , M_{rs} \cri = \delta_{qr} M_{ps} + \delta_{ps}
M_{qr} - \delta_{pr} M_{qs} - \delta_{qs} M_{pr} \\
&& \cle T_i , F_{\alpha\mu} \cri = \half \sigma^i_{\beta\alpha} 
F_{\beta\mu} \hspace{20mm} 
\cle M_{pq} , F_{\alpha\mu} \cri = \half 
(\gamma_p\gamma_q)_{\nu\mu} F_{\alpha\nu} \\
&& \ale F_{\alpha\mu} , F_{\beta\nu} \ari = 2 C^{(8)}_{\mu\nu} 
(C^{(2)}\sigma^i)_{\alpha\beta} T_i + \third C^{(2)}_{\alpha\beta} 
(C^{(8)}\gamma_p\gamma_q)_{\mu\nu} M_{pq}
\eno
where $\sigma^1,\sigma^2,\sigma^3$ are the Pauli matrices and $C^{(2)} ~
(=i\sigma^2)$ is the $2 \times 2$ charge conjugation matrix. The
8-dimensional matrices $\gamma_p$ form a Clifford algebra $\{\gamma_p,
\gamma_q\} = 2\delta_{pq}$ and $C^{(8)}$ is the $8 \times 8$ charge
conjugation matrix. They can be chosen, $\II$ being the $2 \times 2$
unit matrix, as ($\see$ Clifford algebra):
\[
\begin{array}{lll}
\Big. \gamma_1 = \sigma^1 \otimes \sigma^3 \otimes \II \,,
&\quad \gamma_2 = \sigma^1 \otimes \sigma^1 \otimes \sigma^3 \,,
&\quad \gamma_3 = \sigma^1 \otimes \sigma^1 \otimes \sigma^1 \cr
\Big. \gamma_4 = \sigma^2 \otimes \II \otimes \II \,,
&\quad \gamma_5 = \sigma^1 \otimes \sigma^2 \otimes \II \,,
&\quad \gamma_6 = \sigma^1 \otimes \sigma^1 \otimes \sigma^2 \cr
\Big. \gamma_7 = \sigma^3 \otimes \II \otimes \II &&
\end{array}
\]
The generators in the Cartan-Weyl basis are given by (with obvious
notations): 
\beo
&& \hspace{-10mm} H_1 = T_3 ~~~~~~~~~~~~~~~~~~~~~~~~~~ E_{\pm\del} 
= T_1 \pm i T_2 \\ 
&& \hspace{-10mm} H_2 = i M_{41} ~~~~~~~~~~~~~~~~~~~~~~~ H_3 = i M_{52} 
~~~~~~~~~~~~~~~~~~~~~~~ H_4 = i M_{63} \\
&& \hspace{-10mm} E_{\pm\eps_1} = \sfrac{i}{\sqrt{2}} (M_{17} \pm i
M_{47}) ~~~~~~ E_{\pm\eps_2} = \sfrac{i}{\sqrt{2}} (M_{27} \pm i M_{57}) 
~~~~~~ E_{\pm\eps_3} = \sfrac{i}{\sqrt{2}} (M_{37} \pm i M_{67}) \\
&& \hspace{-10mm} E_{\pm(\eps_1+\eps_2)} = \sfrac{i}{2} (M_{12} \pm i
M_{42} + M_{54} \pm i M_{15}) ~~~~~ E_{\pm(\eps_1-\eps_2)} =
\sfrac{i}{2} (M_{12} \pm i M_{42} - M_{54} \mp i M_{15}) \\
&& \hspace{-10mm} E_{\pm(\eps_2+\eps_3)} = \sfrac{i}{2} (M_{23} \pm i
M_{53} +  M_{65} \pm i M_{26}) ~~~~~ E_{\pm(\eps_2-\eps_3)} =
\sfrac{i}{2} (M_{23} \pm i M_{53} - M_{65} \mp i M_{26}) \\
&& \hspace{-10mm} E_{\pm(\eps_1+\eps_3)} = \sfrac{i}{2} (M_{13} \pm i
M_{43} +  M_{64} \pm i M_{16}) ~~~~~ E_{\pm(\eps_1-\eps_3)} =
\sfrac{i}{2} (M_{13} \pm i M_{43} - M_{64} \mp i M_{16}) \\
&& \hspace{-10mm} E_{\frac{1}{2}(\pm\del\pm\eps_1\pm\eps_2\pm\eps_3)} 
= F_{\alpha,\mu}
\eno
where in the last equation the index $\alpha$ and the sign in $\pm\del$
are in one-to-one correspondence and the correspondence between the
index $\mu$ and the signs in $\pm\eps_1\pm\eps_2\pm\eps_3$ is given by 
$(1,2,3,4,5,6,7,8)=(+++,+--,--+,-+-,-++,---,+-+,++-)$.

\section{\sf Exceptional Lie superalgebra $G(3)$}
\indent

The Lie superalgebra $G(3)$ of rank 3 has dimension 31. The even part
(of dimension 17) is a non-compact form of $sl(2) \oplus G_2$ and the
odd part (of dimension 14) is the representation $(2,7)$ of $sl(2)
\oplus G_2$. In terms of the vectors $\eps_1,\eps_2,\eps_3$ with
$\eps_1+\eps_2+\eps_3 = 0$ and $\del$ such that
$\eps_i.\eps_j=1-3\delta_{ij}, \del^2=2, \eps_i.\del=0$, the root
system $\Delta = \Delta_\evn \cup \Delta_\odd$ is given by
\[
\Delta_\evn = \ale ~ \pm 2\del, ~ \eps_i-\eps_j, ~ \pm\eps_i ~ \ari 
\medbox{and}
\Delta_\odd = \ale ~ \pm\eps_i\pm\del, ~ \pm\del ~ \ari 
\]
The different simple root systems of $G(3)$ with the corresponding
Dynkin diagrams and Cartan matrices are the following:
\\
Simple root system
$\Delta^0 = \ale 
\alpha_1 = \del+\eps_3, \alpha_2 = \eps_1, \alpha_3 = \eps_2-\eps_1 \ari$ 
\begin{center}
\begin{picture}(100,20)
\thicklines
\put(0,0){\circle{14}}
\put(42,0){\circle{14}}
\put(84,0){\circle{14}}
\put(0,15){\makebox(0.4,0.6){$\alpha_1$}}
\put(42,15){\makebox(0.4,0.6){$\alpha_2$}}
\put(84,15){\makebox(0.4,0.6){$\alpha_3$}}
\put(-5,-5){\line(1,1){10}}\put(-5,5){\line(1,-1){10}}
\put(7,0){\line(1,0){28}}
\put(48,-4){\line(1,0){30}}
\put(49,0){\line(1,0){28}}
\put(48,4){\line(1,0){30}}
\put(59,0){\line(1,1){10}}\put(59,0){\line(1,-1){10}}
\end{picture}
\qquad \qquad Cartan matrix = 
$\left(\begin{array}{rrr} 
0 & 1 & 0 \cr -1 & 2 & -3 \cr 0 & -1 & 2 \cr 
\end{array}\right)$
\end{center}
Simple root system
$\Delta^0 = \ale 
\alpha_1 = -\del-\eps_3, \alpha_2 = \del-\eps_2, 
\alpha_3 = \eps_2-\eps_1 \ari$ 
\begin{center}
\begin{picture}(100,20)
\thicklines
\put(0,0){\circle{14}}
\put(42,0){\circle{14}}
\put(84,0){\circle{14}}
\put(0,15){\makebox(0.4,0.6){$\alpha_1$}}
\put(42,15){\makebox(0.4,0.6){$\alpha_2$}}
\put(84,15){\makebox(0.4,0.6){$\alpha_3$}}
\put(-5,-5){\line(1,1){10}}\put(-5,5){\line(1,-1){10}}
\put(37,-5){\line(1,1){10}}\put(37,5){\line(1,-1){10}}
\put(7,0){\line(1,0){28}}
\put(48,-4){\line(1,0){30}}
\put(49,0){\line(1,0){28}}
\put(48,4){\line(1,0){30}}
\put(59,0){\line(1,1){10}}\put(59,0){\line(1,-1){10}}
\end{picture}
\qquad \qquad Cartan matrix = 
$\left(\begin{array}{rrr}
0 & 1 & 0 \cr -1 & 0 & 3 \cr 0 & -1 & 2 \cr
\end{array}\right)$
\end{center}
Simple root system
$\Delta^0 = \ale 
\alpha_1 = \del, \alpha_2 = -\del+\eps_1, \alpha_3 = \eps_2-\eps_1 \ari$ 
\begin{center}
\begin{picture}(100,20)
\thicklines
\put(0,0){\circle*{14}}
\put(42,0){\circle{14}}
\put(84,0){\circle{14}}
\put(0,15){\makebox(0.4,0.6){$\alpha_1$}}
\put(42,15){\makebox(0.4,0.6){$\alpha_2$}}
\put(84,15){\makebox(0.4,0.6){$\alpha_3$}}
\put(37,-5){\line(1,1){10}}\put(37,5){\line(1,-1){10}}
\put(6,-3){\line(1,0){30}}
\put(6,3){\line(1,0){30}}
\put(48,-4){\line(1,0){30}}
\put(49,0){\line(1,0){28}}
\put(48,4){\line(1,0){30}}
\put(59,0){\line(1,1){10}}\put(59,0){\line(1,-1){10}}
\end{picture}
\qquad \qquad Cartan matrix = 
$\left(\begin{array}{rrr}
2 & -2 & 0 \cr -2 & 0 & 3 \cr 0 & -1 & 2 \cr
\end{array}\right)$
\end{center}
Simple root system
$\Delta^0 = \ale 
\alpha_1 = \del-\eps_1, \alpha_2 = -\del+\eps_2, \alpha_3 = \eps_1 \ari$ 
\begin{center}
\begin{picture}(120,30)
\thicklines
\put(11,20){\circle{14}}
\put(11,-20){\circle{14}}
\put(-9,20){\makebox(0.4,0.6){$\alpha_1$}}
\put(-9,-20){\makebox(0.4,0.6){$\alpha_2$}}
\put(42,0){\circle{14}}
\put(42,15){\makebox(0.4,0.6){$\alpha_3$}}
\put(6,15){\line(1,1){10}}\put(6,25){\line(1,-1){10}}
\put(6,-25){\line(1,1){10}}\put(6,-15){\line(1,-1){10}}
\put(7,-14){\line(0,1){28}}
\put(11,-13){\line(0,1){26}}
\put(15,-14){\line(0,1){28}}
\put(17,-17){\line(2,1){21}}
\put(17,15){\line(2,-1){20}}\put(18,19){\line(2,-1){23}}
\end{picture}
\qquad \qquad Cartan matrix = 
$ \left(\begin{array}{rrr}
0 & 3 & 2 \cr -3 & 0 & 1 \cr -2 & -1 & 2 \cr
\end{array}\right)$
\end{center}

In order to write the commutation relations of $G(3)$, it is convenient
to use a $so(7)$ basis. Consider the $so(7)$ generators $M_{pq} =
-M_{qp}$ where $1 \le p \ne q \le 7$. The singular embedding $G_2
\subset so(7)$ is obtained by imposing to the generators $M_{pq}$ the
constraints
\[ 
\xi_{ijk} M_{ij} = 0
\]
where the tensor $\xi_{ijk}$ is completely antisymmetric and whose 
non-vanishing components are
\[
\xi_{123} = \xi_{145} = \xi_{176} = \xi_{246} = \xi_{257} = 
\xi_{347} = \xi_{365} = 1
\]
Denoting by $T_i$ where $i = 1,2,3$ the generators of $sl(2)$, by 
$F_{\alpha p}$ where $\alpha = +,-$ and $1 \le p \le 7$ the generators 
of the odd part, the commutation relations of $G(3)$ read as: 
\beo
&& \cle T_i , T_j \cri = i \eps_{ijk} T_k \hspace{25mm} 
\cle T_i , M_{pq} \cri = 0 \\
&& \cle M_{pq} , M_{rs} \cri = \delta_{qr} M_{ps} + \delta_{ps}
M_{qr} - \delta_{pr} M_{qs} - \delta_{qs} M_{pr} +
\sfrac{1}{3} \xi_{pqu}\xi_{rsv} M_{uv} \\ 
&& \cle T_i , F_{\alpha p} \cri = \half \sigma^i_{\alpha\beta} 
F_{\beta p} \hspace{20mm} 
\cle M_{pq} , F_{\alpha r} \cri = \sfrac{2}{3} \delta_{qr} F_{\alpha p}
- \sfrac{2}{3} \delta_{pr} F_{\alpha q} 
+ \sfrac{1}{3} \zeta_{pqrs} F_{\alpha s} \\
&& \ale F_{\alpha p} , F_{\beta q} \ari = 2 \delta_{pq} 
(C\sigma^i)_{\alpha\beta} T_i + \sfrac{3}{2} C_{\alpha\beta} M_{pq}
\eno
where the tensor $\zeta_{pqrs}$ is completely antisymmetric and whose 
non-vanishing components are
\[
\zeta_{1247} = \zeta_{1265} = \zeta_{1364} = \zeta_{1375} = 
\zeta_{2345} = \zeta_{2376} = \zeta_{4576} = 1
\]
It can be written as
\[
\zeta_{pqrs} = \delta_{ps} \delta_{qr} - \delta_{pr} \delta_{qs} 
+ \sum_{u=1}^7 \xi_{pqu} \xi_{rsu}
\]
The $\sigma^i$'s are the Pauli matrices and $C ~ (=i\sigma^2)$ is the 
$2 \times 2$ charge conjugation matrix.
\\
In terms of the $M_{pq}$, the generators of $G_2$ are given by
\beo
&& E_1 = i (M_{17} - M_{24}) \qquad 
E'_1 = i\sqrt{3} (M_{17} + M_{24}) \\ 
&& E_2 = i (M_{21} - M_{74}) \qquad 
E'_2 = -i\sqrt{3} (M_{21} + M_{74}) \\ 
&& E_3 = i (M_{72} - M_{14}) \qquad 
E'_3 = i\sqrt{3} (M_{72} + M_{14}) = -E_8 \\ 
&& E_4 = i (M_{43} - M_{16}) \qquad 
E'_4 = i\sqrt{3} (M_{43} + M_{16}) \\ 
&& E_5 = i (M_{31} - M_{46}) \qquad 
E'_5 = i\sqrt{3} (M_{31} + M_{46}) \\ 
&& E_6 = i (M_{62} - M_{73}) \qquad 
E'_6 = i\sqrt{3} (M_{62} + M_{73}) \\ 
&& E_7 = i (M_{32} - M_{67}) \qquad 
E'_7 = i\sqrt{3} (M_{32} + M_{67}) 
\eno
The $E_a$'s with $a = 1,\dots,8$ generate $sl(3)$ and satisfy the 
commutation relations
\[
\cle E_a, E_b \cri = 2i f_{abc} E_c
\]
where $f_{abc}$ are the usual totally antisymmetric Gell-Mann
structure constants. The commutation relations between the $G_2$ 
generators $E_a$ and $E'_i$ ($i = 1,2,4,5,6,7$) are
\beo
&& \cle E_a , E'_i \cri = 2i c_{aij} E'_j \\
&& \cle E'_i , E'_j \cri = 2i (c_{aij} E_a + c'_{ijk} E'_k)
\eno
where the structure constants $c_{aij}$ (antisymmetric in the indices
$i,j$) and  $c'_{ijk}$ (totally antisymmetric) are
\beo
&& c_{147} = c_{156} = c_{257} = c_{345} = c_{367} = c_{417} = 
c_{725} = 1/2 \\
&& c_{246} = c_{426} = c_{516} = c_{527} = c_{615} = 
c_{624} = c_{714} = -1/2 \\
&& c_{845} = c_{876} = -1/2\sqrt{3} \qquad c_{812} = -1/\sqrt{3} \\
&& c'_{147} = c'_{165} = c'_{246} = c'_{257} = -1/\sqrt{3}
\eno
The generators $E_3$ and $E_8$ constitute a Cartan basis of the $G_2$
algebra. One can also take a basis $H_1,H_2,H_3$ such that $H_1+H_2+H_3
= 0$ given by $H_1 = \half(E_3+\frac{\sqrt{3}}{3} E_8)$, $H_2 =
\half(-E_3+\frac{\sqrt{3}}{3} E_8)$,  $H_3 = -\frac{\sqrt{3}}{3} E_8$.
The generators in the Cartan-Weyl basis are given by (with obvious
notations):
\[
\begin{array}{lll}
\bigg. H_1 = \half(E_3+\frac{\sqrt{3}}{3} E_8) &\quad
H_2 = \half(-E_3+\frac{\sqrt{3}}{3} E_8) &\quad
H_3 = -\frac{\sqrt{3}}{3} E_8 \\
\bigg. E_{\pm(\eps_1-\eps_2)} = E_1 \pm i E_2 &\quad
E_{\pm(\eps_2-\eps_3)} = E_6 \pm i E_7 &\quad
E_{\pm(\eps_1-\eps_3)} = E_4 \pm i E_5 \\
\bigg. E_{\pm\eps_1} = E'_7 \mp i E'_6 &\quad
E_{\pm\eps_2} = E'_4 \mp i E'_5 &\quad
E_{\pm\eps_3} = E'_1 \mp i E'_2 \\
\bigg. E_{\pm\del+\eps_1} = F_{\pm 1} + i F_{\pm 4} &\quad
E_{\pm\del+\eps_2} = F_{\pm 7} + i F_{\pm 2} &\quad
E_{\pm\del+\eps_3} = F_{\pm 3} + i F_{\pm 6} \\
\bigg. E_{\pm\del-\eps_1} = F_{\pm 1} - i F_{\pm 4} &\quad
E_{\pm\del-\eps_2} = F_{\pm 7} - i F_{\pm 2} &\quad
E_{\pm\del-\eps_3} = F_{\pm 3} - i F_{\pm 6} \\
\bigg. H_4 = T_3 &\quad
E_{\pm 2\del} = T_1 \pm i T_2 &\quad
E_{\pm\del} = F_{\pm 5} \\
\end{array}
\]

\section{\sf Exceptional Lie superalgebras $D(2,1;\alpha)$}
\indent

The Lie superalgebras $D(2,1;\alpha)$ with $\alpha \ne 0,-1,\infty$ form
a one-parameter family of superalgebras of rank 3 and dimension 17. The
even part (of dimension 9) is a non-compact form of $sl(2) \oplus sl(2)
\oplus sl(2)$ and the odd part (of dimension 8) is the spinorial
representation $(2,2,2)$ of the even part. In terms of the vectors
$\eps_1, \eps_2, \eps_3$ such that $\eps_1^2 = -(1+\alpha)/2,
\eps_2^2=1/2, \eps_3^2=\alpha/2$ and $\eps_i.\eps_j=0$ if $i \ne j$,
the root system $\Delta = \Delta_\evn \cup \Delta_\odd$ is given by
\[
\Delta_\evn = \ale \pm 2\eps_i \ari \medbox{and}
\Delta_\odd = \ale \pm\eps_1\pm\eps_2\pm\eps_3 \ari 
\]
$D(2,1;\alpha)$ is actually a deformation of the superalgebra $D(2,1)$ 
which corresponds to the case $\alpha = 1$.
\\
The different simple root systems of $D(2,1;\alpha)$ with the 
corresponding Dynkin diagrams and Cartan matrices are the following: 
\\
Simple root system
$\Delta^0 = \ale 
\alpha_1 = \eps_1-\eps_2-\eps_3, \alpha_2 = 2\eps_2, \alpha_3 = 2\eps_3
\ari$ 
\begin{center}
\begin{picture}(80,20)
\thicklines
\put(20,0){\circle{14}}
\put(15,-5){\line(1,1){10}}\put(15,5){\line(1,-1){10}}
\put(51,20){\circle{14}}
\put(51,-20){\circle{14}}
\put(0,0){\makebox(0.4,0.6){$\alpha_1$}}
\put(71,20){\makebox(0.4,0.6){$\alpha_2$}}
\put(71,-20){\makebox(0.4,0.6){$\alpha_3$}}
\put(32,-17){\makebox(0.4,0.6){{\footnotesize $\alpha$}}}
\put(25,5){\line(2,1){20}}\put(25,-5){\line(2,-1){20}}
\end{picture}
\qquad \qquad Cartan matrix = 
$\left(\begin{array}{rrr} 
0 & 1 & \alpha \cr -1 & 2 & 0 \cr -1 & 0 & 2 \cr 
\end{array}\right)$
\end{center}
\vspace{2mm}
Simple root system
$\Delta^0 = \ale 
\alpha_1 = 2\eps_2, \alpha_2 = -\eps_1-\eps_2+\eps_3, 
\alpha_3 = 2\eps_1 \ari$ 
\begin{center}
\begin{picture}(80,20)
\thicklines
\put(20,0){\circle{14}}
\put(15,-5){\line(1,1){10}}\put(15,5){\line(1,-1){10}}
\put(51,20){\circle{14}}
\put(51,-20){\circle{14}}
\put(0,0){\makebox(0.4,0.6){$\alpha_2$}}
\put(71,20){\makebox(0.4,0.6){$\alpha_3$}}
\put(71,-20){\makebox(0.4,0.6){$\alpha_1$}}
\put(32,17){\makebox(0.4,0.6){{\footnotesize 1+$\alpha$}}}
\put(25,5){\line(2,1){20}}\put(25,-5){\line(2,-1){20}}
\end{picture}
\qquad \qquad Cartan matrix = 
$\left(\begin{array}{rrc} 
2 & -1 & 0 \cr 1 & 0 & 1+\alpha \cr 0 & -1 & 2 \cr 
\end{array}\right)$
\end{center}
\vspace{2mm}
Simple root system
$\Delta^0 = \ale 
\alpha_1 = 2\eps_3, \alpha_2 = 2\eps_1, 
\alpha_3 = -\eps_1+\eps_2-\eps_3 \ari$ 
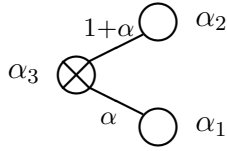
\begin{center}
\begin{picture}(80,20)
\thicklines
\put(20,0){\circle{14}}
\put(15,-5){\line(1,1){10}}\put(15,5){\line(1,-1){10}}
\put(51,20){\circle{14}}
\put(51,-20){\circle{14}}
\put(0,0){\makebox(0.4,0.6){$\alpha_3$}}
\put(71,20){\makebox(0.4,0.6){$\alpha_2$}}
\put(71,-20){\makebox(0.4,0.6){$\alpha_1$}}
\put(32,17){\makebox(0.4,0.6){{\footnotesize 1+$\alpha$}}}
\put(32,-17){\makebox(0.4,0.6){{\footnotesize $\alpha$}}}
\put(25,5){\line(2,1){20}}\put(25,-5){\line(2,-1){20}}
\end{picture}
\qquad \qquad Cartan matrix = 
$\left(\begin{array}{rrc} 
2 & 0 & \alpha \cr 0 & 2 & 1+\alpha \cr -1 & -1 & 0 \cr 
\end{array}\right)$
\end{center}
\vspace{2mm}
Simple root system
$\Delta^0 = \ale 
\alpha_1 = -\eps_1+\eps_2+\eps_3, \alpha_2 = \eps_1+\eps_2-\eps_3,
\alpha_3 = \eps_1-\eps_2+\eps_3 \ari$ 
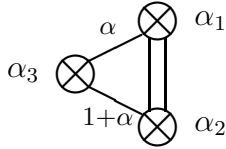
\begin{center}
\begin{picture}(80,20)
\thicklines
\put(20,0){\circle{14}}
\put(15,-5){\line(1,1){10}}\put(15,5){\line(1,-1){10}}
\put(51,20){\circle{14}}
\put(51,-20){\circle{14}}
\put(0,0){\makebox(0.4,0.6){$\alpha_3$}}
\put(71,20){\makebox(0.4,0.6){$\alpha_1$}}
\put(71,-20){\makebox(0.4,0.6){$\alpha_2$}}
\put(32,17){\makebox(0.4,0.6){{\footnotesize $\alpha$}}}
\put(32,-17){\makebox(0.4,0.6){{\footnotesize 1+$\alpha$}}}
\put(25,5){\line(2,1){20}}\put(25,-5){\line(2,-1){20}}
\put(46,15){\line(1,1){10}}\put(46,25){\line(1,-1){10}}
\put(46,-25){\line(1,1){10}}\put(46,-15){\line(1,-1){10}}
\put(48,-14){\line(0,1){28}}
\put(54,-14){\line(0,1){28}}
\end{picture}
\qquad \qquad Cartan matrix = 
$\left(\begin{array}{rcc} 
0 & 1 & \alpha \cr 1 & 0 & -1-\alpha \cr \alpha & -1-\alpha & 0 \cr 
\end{array}\right)$
\end{center}
\vspace{2mm}
\noindent
(the labels on the links are equal to the absolute values of the scalar
products of the simple roots which are linked.)

Denoting by $T_i^{(a)}$ where $i = 1,2,3$ and $a = 1,2,3$ the generators
of the three $sl(2)$ and by $F_{\beta\beta'\beta''}$ where $\beta,
\beta', \beta'' = +,-$, the generators of the odd part, the commutation
relations of $D(2,1;\alpha)$ read as:
\beo
&& \cle T_i^{(a)} , T_j^{(b)} \cri = i \delta_{ab} \eps_{ijk} T_k^{(a)} \\
&& \cle T_i^{(1)} , F_{\beta\beta'\beta''} \cri = \half
\sigma^i_{\gamma\beta} F_{\gamma\beta'\beta''} \\ 
&& \cle T_i^{(2)} , F_{\beta\beta'\beta''} \cri = \half
\sigma^i_{\gamma'\beta'} F_{\beta\gamma'\beta''} \\ 
&& \cle T_i^{(3)} , F_{\beta\beta'\beta''} \cri = \half
\sigma^i_{\gamma''\beta''} F_{\beta\beta'\gamma''} \\
&& \ale F_{\beta\beta'\beta''} , F_{\gamma\gamma'\gamma''} \ari = 
s_1 C_{\beta'\gamma'} C_{\beta''\gamma''} (C\sigma^i)_{\beta\gamma} 
T_i^{(1)} + s_2 C_{\beta''\gamma''} C_{\beta\gamma} 
(C\sigma^i)_{\beta'\gamma'} T_i^{(2)} \\
&& \hspace{78mm} + s_3 C_{\beta\gamma} C_{\beta'\gamma'} 
(C\sigma^i)_{\beta''\gamma''} T_i^{(3)} 
\eno
where $s_1+s_2+s_3 = 0$ is imposed by the generalized Jacobi identity.
The $\sigma^i$'s are the Pauli matrices and $C ~ (=i\sigma^2)$ is the $2
\times 2$ charge conjugation matrix. Since the superalgebras defined by
the triplets $\lambda s_1, \lambda s_2, \lambda s_3$ ($\lambda \in \CC$)
are isomorphic, one can set $s_2/s_1 = \alpha$ and $s_3/s_1 = -1-\alpha$
(the normalization of the roots given above corresponds to the choice
$s_1=1$, $s_2=\alpha$ and $s_3=-1-\alpha$). One can deduce after some
simple calculation that:

\uppt: The superalgebras defined by the parameters $\alpha$,
$\alpha^{-1}$, $-1-\alpha$ and $\displaystyle{\frac{-\alpha}{1+\alpha}}$ 
are isomorphic. Moreover, for the values 1, $-2$ and $-1/2$ of the 
parameter $\alpha$, the superalgebra $D(2,1;\alpha)$ is isomorphic to 
$D(2,1)$.

\medskip

In the Cartan-Weyl basis, the generators are given by:
\beo
&& H_1 = T_3^{(1)} \hspace{30mm} H_2 = T_3^{(2)} 
\hspace{30mm} H_3 = T_3^{(3)} \\ 
&& E_{\pm 2\eps_1} = T_1^{(1)} \pm i T_2^{(1)} \hspace{10mm} 
E_{\pm 2\eps_2} = T_1^{(2)} \pm i T_2^{(2)} \hspace{10mm} 
E_{\pm 2\eps_3} = T_1^{(3)} \pm i T_2^{(3)} \hspace{10mm} \\ 
&& E_{\pm\eps_1\pm\eps_2\pm\eps_3} = F_{\beta\beta'\beta''}
\eno
where in the last equation the signs in the indices
$\pm\eps_1\pm\eps_2\pm\eps_3$ and the indices $\beta\beta'\beta''$
are in one-to-one correspondence.

\section{\sf Gelfand-Zetlin basis}
\indent

Consider a finite dimensional irreducible representation $\pi$ of
$gl(m|n)$ with highest weight $\Lambda=(\lambda_1, \dots,
\lambda_{m+n})$. The coefficients $\lambda_i$ are complex numbers such
that the differences $\lambda_i-\lambda_{i+1} \in \NN$ for $i \ne m$.
The construction of the Gelfand-Zetlin basis stands on the reduction
of $\pi$ with respect to the chain of subalgebras
\[
gl(m|n) \supset gl(m|n-1) \supset \dots \supset gl(m) \supset gl(m-1)
\supset \dots \supset gl(1) 
\]
It is sufficient of course to achieve the first reduction.

This construction has been done up to now only in the case of the star
representations of $gl(m|n)$.
Let us recall that the superalgebra $gl(m|n)$ has two classes of star
representations and two classes of superstar representations ($\see$
Star and superstar representations): if $e_{IJ}$ is the standard basis
of $gl(m|n)$ ($\see$ Unitary superalgebras) where $1 \le I,J \le m+n$,
the two star representations $\pi$ are defined  by $\pi^\dagger(e_{IJ})
= \pi(e_{IJ}^\dagger)$ where
\beo
&& e_{IJ}^\dagger = e_{JI} ~~~~~~~~~~\medbox{adjoint 1} \\
&& e_{IJ}^\dagger = (-1)^{\deg(e_{IJ})} e_{JI} \medbox{adjoint 2}
\eno
and the two superstar representations $\pi$ by $\pi^\ddagger(e_{IJ}) =
\pi(e_{IJ}^\ddagger)$ where
\beo
&& e_{IJ}^\ddagger = \left\{
\begin{array}{l}
e_{IJ} \smbox{for} \deg(e_{IJ}) = 0 \cr
e_{IJ} ~ sign(J-I) \smbox{for} \deg(e_{IJ}) = 1 \cr
\end{array} \right. \medbox{superadjoint 1} \\
&& e_{IJ}^\ddagger = \left\{
\begin{array}{l}
e_{IJ} \smbox{for} \deg(e_{IJ}) = 0 \cr
e_{IJ} ~ sign(I-J) \smbox{for} \deg(e_{IJ}) = 1 \cr
\end{array} \right. \medbox{superadjoint 2}
\eno
In the following, we will concentrate on the star representations.

\medskip

\uth: The irreducible representation of $gl(m|n)$ with highest weight
$\Lambda=(\lambda_1, \dots,$ $\lambda_{m+n})$ is a star representation if
and only if all the $\lambda_i$ are real and the following conditions are
satisfied:
\begin{enumerate}
\item
$\lambda_m + \lambda_{m+n} - n + 1 \ge 0$.
\item
there is some $k \in \{1,\dots,n-1\}$ such that 
$0 \le \lambda_m + \lambda_{m+k} - k + 1 \le 1$ and
$\lambda_{m+k} = \lambda_{m+k+1} = \dots = \lambda_{m+n}$.
\item
$\lambda_1 + \lambda_{m+1} + m - 1 \le 0$.
\end{enumerate}

\medskip

\uth: Let $\pi$ be a star representation of $gl(m|n)$ with highest
weight $\Lambda=(\lambda_1,\dots,$ $\lambda_{m+n})$. The reduction of
the representation $\pi$ of $gl(m|n)$ with respect to of $gl(m|n-1)$
gives exactly once the irreducible representations of $gl(m|n-1)$ with
highest weights $\Lambda'=(\lambda'_1,\dots,\lambda'_{m+n-1})$ where
the differences $\lambda_i-\lambda'_i \in \NN$ for $1 \le i \le m+n-1$
and the coefficients satisfy the following inequalities:
\beo
&& \cdot ~ \lambda_1 \ge \lambda_2 \ge \dots \ge \lambda_{m-1} \ge
\lambda_m \\ 
&& \cdot ~ \lambda'_1 \ge \lambda'_2 \ge \dots \ge \lambda'_{m-1} \ge 
\lambda'_m \\
&& \cdot ~ \lambda_i \ge \lambda'_i \ge \lambda_i - \eps_i
\qquad (1 \le i \le m) \\
&& \cdot ~ \lambda_{m+1} \ge \lambda'_{m+1} \ge \lambda_{m+2} \ge 
\lambda'_{m+2} \dots \ge \lambda_{m+n-1} \ge \lambda'_{m+n-1} \ge 
\lambda_{m+n} 
\eno
where
\beo
&&\eps_1 = \left\{ \begin{array}{l}
0 \medbox{if} \lambda_1+\lambda_{m+1}+m-1=0 \cr 
1 \medbox{otherwise} \cr
\end{array} \right. \\
&&\eps_2 = \dots = \eps_{m-1} = 1 \cr
&&\eps_m = \left\{ \begin{array}{l}
0 \medbox{if} \lambda_1+\lambda_{m+1}+m-1=0 
\smbox{or} \lambda_m+\lambda_{m+k}+k-1=0 \cr
~~~ \smbox{with} \lambda'_{m+k-1} \ne \lambda_{m+k} ~
(2 \le k \le n) \cr
1 \medbox{otherwise} \cr
\end{array} \right.
\eno
$\eps_m=0$ in the case of the star adjoint 1 representations
while $\eps_1=0$ for the star adjoint 2 representations.

\medskip

It follows that the Gelfand-Zetlin basis for the star representations of
$gl(m|n)$ is given by the following theorem:

\uth: Let $\pi$ be a star representation of $gl(m|n)$ with highest
weight $\Lambda=(\lambda_1,\dots,$ $\lambda_{N})
\equiv (\lambda_{1N},\dots,$ $\lambda_{NN})$ where $N=m+n$. 
The Gelfand-Zetlin basis in the representation space $\cV(\Lambda)$ is
given by
\[
e_\Lambda = \left\vert \begin{array}{c}
\lambda_{1N} ~~~~~ \lambda_{2N} ~~~~~ \lambda_{3N} ~~~~~ \dots 
~~~~~ \lambda_{NN} \cr
\lambda_{1,N-1} ~ \lambda_{2,N-1} ~ \dots ~ \lambda_{N-1,N-1} \cr
\ddots ~~~~~~~~~~~~ \invdots \cr
\lambda_{12} ~~~ \lambda_{22} \cr
\lambda_{11} \cr
\end{array} \right.
\begin{picture}(20,40)
\put(5,4){\line(-1,3){12}}\put(5,4){\line(-1,-3){12}}
\end{picture}
\]
where the real numbers $\lambda_{ij}$, with $\lambda_{i,j+1} - 
\lambda_{ij} \in \NN$, satisfy the following inequalities: 
\beo
&& \cdot ~ \lambda_{1,j+1} \ge \lambda_{2,j+1} \ge \dots \ge 
\lambda_{m-1,j+1} \ge \lambda_{m,j+1} \\
&& \cdot ~ \lambda_{1j} \ge \lambda_{2j} \ge \dots \ge \lambda_{m-1,j} 
\ge \lambda_{mj} \\
&&  \cdot ~ \lambda_{i,j+1} \ge \lambda_{ij} \ge \lambda_{i,j+1} - 
\eps_{i,j+1} \qquad (1 \le i \le m) \\
&& \cdot ~ \lambda_{i,j+1} \ge \lambda_{ij} \ge \lambda_{i+1,j+1}
\medbox{for} j \ge i \ge m+1 \smbox{or} i \le j \le m-1 
\eno
where 
\beo
&&\eps_{1,j+1} = \left\{ \begin{array}{l}
0 \medbox{if} \lambda_{1,j+1}+\lambda_{m+1,j+1}+m-1=0 \cr 
1 \medbox{otherwise} \cr
\end{array} \right. \\
&&\eps_{m,j+1} = \left\{ \begin{array}{l}
0 \medbox{if} \lambda_{m,j+1}+\lambda_{m+1,j+1}=0 
\smbox{or} \lambda_{m,j+1}+\lambda_{m+k,j+1}-k+1=0 \cr
~~~ \smbox{with} \lambda_{m+k-1,j} \ne \lambda_{m+k,j+1} ~ 
(1 \le k \le j+1) \cr
1 \medbox{otherwise} \cr
\end{array} \right.
\eno
and $\eps_{2,j+1} = \dots = \eps_{m-1,j+1} = 1$.
Notice that $\eps_{m,j+1}=0$ in the case of the star adjoint 1 
representations while $\eps_{1,j+1}=0$ for the star adjoint 2
representations.

\medskip

For more details on the Gelfand-Zetlin basis for $gl(m|n)$, in
particular the action of the $gl(m|n)$ generators on the basis vectors,
see ref. \cite{IST85}.

\section{\sf Grassmann algebras}
\indent

\udef: The real (resp. complex) Grassmann algebra $\Gamma(n)$ of order
$n$ is the algebra over $\RR$ (resp. $\CC$) generated from the unit
element $1$ and the $n$ quantities $\theta_i$ (called Grassmann
variables) which satisfy the anticommutation relations
\[
\ale \theta_i,\theta_j \ari = 0
\]
This algebra has $2^n$ generators $1,\, \theta_i,\, \theta_i \theta_j,\, 
\theta_i \theta_j \theta_k,\, \dots,\, \theta_1 \dots \theta_n$.

\medskip

Putting $\deg \theta_i = \odd$, the algebra $\Gamma(n)$ acquires the
structure of a superalgebra: $\Gamma(n) = \Gamma(n)_\evn \oplus
\Gamma(n)_\odd$, where $\Gamma(n)_\evn$ is generated by the monomials
in $\theta_i$ with an even number of $\theta_i$ (even generators) and
$\Gamma(n)_\odd$ by the monomials in $\theta_i$ with an odd number of
$\theta_i$ (odd generators). Since $\dim \Gamma(n)_\evn =
\dim \Gamma(n)_\odd = 2^{n-1}$, the superalgebra $\Gamma(n)$ is
supersymmetric. The Grassmann superalgebra is associative and
commutative (in the sense of the superbracket).

\medskip

It is possible to define the complex conjugation on the Grassmann
variables. However, there are two possibilities to do so. If $c$ is a
complex number and $\bc$ its complex conjugate, $\theta_i,\theta_j$ being
Grassmann variables, the star operation, denoted by $^*$, is defined by
\[
(c \theta_i)^* = \bc \theta_i^* \,, \qquad
\theta_i^{**} = \theta_i \,, \qquad
(\theta_i \theta_j)^* = \theta_j^* \theta_i^*
\]
and the superstar operation, denoted by $^\#$, is defined by
\[
(c \theta_i)^{\#} = \bc \theta_i^{\#} \,, \qquad
\theta_i^{\#\#} = -\theta_i \,, \qquad
(\theta_i \theta_j)^{\#} = \theta_i^{\#} \theta_j^{\#}
\]

\medskip

Let us mention that the derivation superalgebra ($\see$) $\der\Gamma(n)$
of $\Gamma(n)$ is the Cartan type ($\see$) simple Lie superalgebra $W(n)$.

\section{\sf Killing form}
\indent

\udef: Let $\cG$ be a Lie superalgebra. One defines the bilinear form
$B_{\pi}$ associated to a representation $\pi$ of $\cG$ as a bilinear
form from $\cG \times \cG$ into the field of real numbers $\RR$ such
that
\[
B_{\pi}(X,Y) = \str(\pi(X),\pi(Y)) \,, \qquad \forall ~ X,Y \in \cG
\]
where $\pi(X)$ are the matrices of the generators $X \in \cG$ in the 
representation $\pi$.

\medskip

If $\{X_i\}$ is the basis of generators of $\cG$ ($i=1,\dots,\dim\cG$),
one has therefore 
\[
B_{\pi}(X_i,Y_j) = \str(\pi(X_i),\pi(Y_j)) = g_{ij}^{\pi}
\]

\medskip

\udef: A bilinear form $B$ on $\cG = \cG_\evn\oplus\cG_\odd$ is called 
\\
- consistent if $B(X,Y) = 0$ for all $X \in \cG_\evn$ and all $Y \in
\cG_\odd$. 
\\
- supersymmetric if $B(X,Y) = (-1)^{\deg X.\deg Y} B(Y,X)$, for all $X,Y 
\in \cG$.
\\
- invariant if $B( \zle X,Y \zri,Z) = B(X, \zle Y,Z \zri)$, 
for all $X,Y,Z \in \cG$.

\medskip

\uppt: An invariant form on a simple Lie superalgebra $\cG$ is either
non-degenerate (that is its kernel is zero) or identically zero, and two
invariant forms on $\cG$ are proportional.

\medskip

\udef: A bilinear form on $\cG$ is called an {\em inner product} on $\cG$
if it is consistent, supersymmetric and invariant.

\medskip

\udef: The bilinear form associated to the adjoint representation of
$\cG$ is called the {\em Killing form} on $\cG$ and is denoted
$K(X,Y)$:
\[
K(X,Y) = \str(\ad(X),\ad(Y)) \,, \qquad \forall ~ X,Y \in \cG
\]
We recall that $\ad(X) Z = \zle X,Z \zri$ and $\bigg(\ad({X_i})\bigg)_j^k
= -C_{ij}^k$ where $C_{ij}^k$ are the structure constants for the basis
$\{X_i\}$ of generators of $\cG$. We can therefore write
\[
K(X_i,X_j) = (-1)^{\deg X_j} C_{mi}^n C_{nj}^m = g_{ij}
\]

\medskip

\uppt: The Killing form $K$ of a Lie superalgebra $\cG$ is consistent,
supersymmetric and invariant (in other words, it is an inner product).

\medskip

\uppt: The Killing form $K$ of a Lie superalgebra $\cG$ satisfies
\[
K(\phi(X),\phi(Y)) = K(X,Y)
\]
for all $\phi \in \aut(\cG)$ and $X,Y \in \cG$.

\medskip

The following theorems give the fundamental results concerning the
Killing form of the (simple) Lie superalgebras:

\uth:
\begin{enumerate}
\item 
A Lie superalgebra $\cG$ with a non-degenerate Killing form is a direct 
sum of simple Lie superalgebras each having a non-degenerate Killing form.
\item 
A simple finite dimensional Lie superalgebra $\cG$ with a non-degenerate 
Killing form is of the type $A(m,n)$ where $m \ne n$, $B(m,n)$, 
$C(n+1)$, $D(m,n)$ where $m \ne n+1$, $F(4)$ or $G(3)$.
\item 
A simple finite dimensional Lie superalgebra $\cG$ with a zero Killing 
form is of the type $A(n,n)$, $D(n+1,n)$, $D(2,1;\alpha)$, $P(n)$ or 
$Q(n)$. 
\end{enumerate}

\medskip

$\see$ Cartan matrices.

For more details, see refs. \cite{Kac77a,Kac77b}.

\section{\sf Lie superalgebra, subalgebra, ideal}
\indent

\udef: A {\em Lie superalgebra} $\cG$ over a field $\KK$ of
characteristic zero (usually $\KK=\RR$ or $\CC$) is a $\ZZ_2$-graded
algebra, that is a vector space, direct sum of two vector spaces
$\cG_\evn$ and $\cG_\odd$, in which a product $\zle ~,~ \zri$, is
defined as follows:
\begin{itemize}
\item 
$\ZZ_2$-gradation:
\[
\sle \cG_i,\cG_j \sri \subset \cG_{i+j ~ (mod ~ 2)}
\]
\item 
graded-antisymmetry:
\[
\sle X_i,X_j \sri = -(-1)^{\deg X_i.\deg X_j} \sle X_j,X_i \sri
\]
where $\deg X_i$ is the degree of the vector space. $\cG_\evn$ is
called the even space and $\cG_\odd$ the odd space. If $\deg X_i.\deg
X_j = 0$, the bracket $\zle ~,~ \zri$ defines the usual commutator,
otherwise  it is an anticommutator.
\item 
generalized Jacobi identity:
\beo
&&(-1)^{\deg X_i.\deg X_k} \sle X_i,\sle X_j,X_k \sri \sri +
(-1)^{\deg X_j.\deg X_i} \sle X_j,\sle X_k,X_i \sri \sri \\
&&+ (-1)^{\deg X_k.\deg X_j} \sle X_k,\sle X_i,X_j \sri \sri = 0
\eno
\end{itemize}

\medskip

Notice that $\cG_\evn$ is a Lie algebra -- called the even or bosonic
part of $\cG$ -- while $\cG_\odd$ -- called the odd or fermionic part
of $\cG$ -- is not an algebra. 

\medskip

An associative superalgebra $\cG = \cG_\evn\oplus\cG_\odd$ over the field
$\KK$ acquires the structure of a Lie superalgebra by taking for the 
product $\zle ~,~ \zri$ the the {\em Lie superbracket} or {\em
supercommutator} (also called generalized or graded commutator)
\[
\sle X,Y \sri = XY - (-1)^{\deg X.\deg Y} YX
\]
for two elements $X,Y \in \cG$.

\medskip

\udef: Let $\cG = \cG_\evn \oplus \cG_\evn$ be a Lie superalgebra.
\\
- A {\em subalgebra} $\cK = \cK_\evn \oplus \cK_\odd$ of $\cG$ is a
subset of elements of $\cG$ which forms a vector subspace of $\cG$ that
is closed with respect to the Lie product of $\cG$ such that $\cK_\evn
\subset \cG_\evn$ and $\cK_\odd \subset \cG_\odd$. A subalgebra $\cK$
of $\cG$ such that $\cK \ne \cG$ is called a proper subalgebra of
$\cG$.
\\
- An {\em ideal} $\cI$ of $\cG$ is a subalgebra of $\cG$ such
that $\sle \cG,\cI \sri \subset \cI$, that is
\[
X \in \cG, Y \in \cI ~ \Rightarrow ~ \sle X,Y \sri \in \cI
\]
An ideal $\cI$ of $\cG$ such that $\cI \ne \cG$ is called a proper ideal
of $\cG$.

\medskip

\uppt: Let $\cG$ be a Lie superalgebra and $\cI$, $\cI'$ two ideals
of $\cG$. Then $\sle \cI,\cI' \sri$ is an ideal of $\cG$.

\section{\sf Matrix realizations of the classical Lie superalgebras}
\indent

The classical Lie superalgebras can be described as matrix superalgebras
as follows. Consider the $\ZZ_2$-graded vector space $\cV = \cV_\evn
\oplus \cV_\odd$ with $\dim\cV_\evn = m$ and $\dim\cV_\odd = n$. Then
the algebra $End\,\cV$ acquires naturally a superalgebra structure by
\[
End\,\cV = End_\evn \cV \oplus End_\odd \cV \medbox{where}
End_i \cV = \ale \phi \in End\,\cV ~ \Big\vert ~ \phi (\cV_j) \subset
\cV_{i+j} \ari 
\]
The Lie superalgebra $\ell(m,n)$ is defined as the superalgebra
$End\,\cV$ supplied with the Lie superbracket ($\see$ Lie superalgebras).
$\ell(m,n)$ is spanned by matrices of the form
\[
M = \left(\begin{array}{cc} A & B \cr C & D \end{array}\right) 
\]
where $A$ and $D$ are $gl(m)$ and $gl(n)$ matrices, $B$ and $C$ are 
$m \times n$ and $n \times m$ rectangular matrices.
\\
One defines on $\ell(m,n)$ the supertrace function denoted by $\str$:
\[
\str(M) = \tr(A) - \tr(D)
\]

The unitary superalgebra $A(m-1,n-1) = sl(m|n)$ is defined as the
superalgebra of matrices $M \in \ell(m,n)$ satisfying the supertrace 
condition $\str(M) = 0$.
\\
In the case $m = n$, $sl(n|n)$ contains a one-dimensional ideal $\cI$ 
generated by $\II_{2n}$ and one sets $A(n-1,n-1) = sl(n|n)/\cI$.

\medskip

The orthosymplectic superalgebra $osp(m,2n)$ is defined as the 
superalgebra of matrices $M \in \ell(m,n)$ satisfying the conditions 
\[
A^t = -A \,, \qquad D^t G = -GD \,, \qquad B = C^t G
\]
where $^t$ denotes the usual sign of transposition and the matrix $G$ 
is given by
\[
G = \left( \begin{array}{cc} 0 & \II_n \cr -\II_n & 0 \end{array}
\right) 
\]

\medskip

The strange superalgebra $P(n)$ is defined as the superalgebra of
matrices $M \in \ell(n,n)$ satisfying the conditions
\[
A^t = -D \,, \qquad B^t = B \,, \qquad C^t = -C \,, \qquad \tr(A) = 0
\]

\medskip

The strange superalgebra $\tilde Q(n)$ is defined as the superalgebra 
of matrices $M \in \ell(n,n)$ satisfying the conditions
\[
A = D \,, \qquad B = C \,, \qquad \tr(B) = 0
\]
The superalgebra $\tilde Q(n)$ has a one-dimensional center $\cZ$.
The simple superalgebra $Q(n)$ is given by $Q(n) = \tilde Q(n)/\cZ$.

\medskip

$\see$ Orthosymplectic superalgebras, Strange superalgebras, Unitary 
superalgebras.

For more details, see refs. \cite{Kac77b,Rit77}.

\section{\sf Nilpotent and solvable Lie superalgebras}
\indent

\udef: Let $\cG = \cG_\evn \oplus \cG_\odd$ be a Lie superalgebra. 
$\cG$ is said {\em nilpotent} if, considering the series
\beo
&& \sle \cG,\cG \sri = \cG^{[1]} \\
&& \sle \cG,\cG^{[1]} \sri = \cG^{[2]} \\
&& \dots \\
&& \sle \cG,\cG^{[i-1]} \sri = \cG^{[i]} 
\eno
then it exists an integer $n$ such that $\cG^{[n]}=\{0\}$.

\medskip

\udef: $\cG$ is said {\em solvable} if, considering the series
\beo
&& \sle \cG,\cG \sri = \cG^{(1)} \\
&& \sle \cG^{(1)},\cG^{(1)} \sri = \cG^{(2)} \\
&& \dots \\
&& \sle \cG^{(i-1)},\cG^{(i-1)} \sri = \cG^{(i)} 
\eno
then it exists an integer $n$ such that $\cG^{(n)}=\{0\}$.

\medskip

\uth: The Lie superalgebra $\cG$ is solvable if and only if $\cG_\evn$
is  solvable.

\medskip

\uppt: Let $\cG = \cG_\evn \oplus \cG_\odd$ be a solvable Lie
superalgebra. Then the irreducible representations of $\cG$ are
one-dimensional if and only if $\ale \cG_\odd,\cG_\odd \ari \subset
\cle \cG_\evn,\cG_\evn \cri$ -- let us recall that in the case of a
solvable Lie {\em algebra}, the irreducible finite dimensional
representation are one-dimensional.

\medskip

\uppt: Let $\cG = \cG_\evn \oplus \cG_\odd$ be a solvable Lie
superalgebra and let $\cV = \cV_\evn \oplus \cV_\odd$ be the space of
irreducible finite dimensional representations. Then either $\dim
\cV_\evn = \dim \cV_\odd$ and $\dim \cV = 2^s$ with $1 \le s \le \dim
\cG_\odd$, or $\dim \cV = 1$.

\section{\sf Orthosymplectic superalgebras}
\indent

The orthosymplectic superalgebras form three infinite families of basic
Lie superalgebras. The superalgebra $B(m,n)$ or $osp(2m+1|2n)$ defined
for $m \ge 0,n \ge 1$ has as even part the Lie algebra $so(2m+1) \oplus
sp(2n)$ and as odd part the $(2m+1,2n)$ representation of the even
part; it has rank $m+n$ and dimension $2(m+n)^2+m+3n$. The superalgebra
$C(n+1)$ or $osp(2|2n)$ where $n \ge 1$ has as even part the Lie
algebra $so(2) \oplus sp(2n)$ and the odd part is twice the fundamental
representation $(2n)$ of $sp(2n)$; it has rank $n+1$ and dimension
$2n^2+5n+1$. The superalgebra $D(m,n)$ or $osp(2m|2n)$ defined for
$m \ge 2,n \ge 1$ has as even part the Lie algebra $so(2m) \oplus
sp(2n)$ and its odd part is the $(2m,2n)$ representation of the even
part; it has rank $m+n$ and dimension $2(m+n)^2-m+n$.

The root systems can be expressed in terms of the orthogonal vectors
$\eps_1, \dots, \eps_m$ and $\del_1, \dots, \del_n$ as follows.
\\
- for $B(m,n)$ with $m \ne 0$:
\[
\Delta_\evn = \ale ~ \pm\eps_i\pm\eps_j, ~ \pm\eps_i,
 ~ \pm\del_i\pm\del_j, ~ \pm 2 \del_i ~ \ari 
\medbox{and} 
\Delta_\odd = \ale ~ \pm\eps_i\pm\del_j, ~ \pm\del_j ~ \ari \,,
\]
- for $B(0,n)$:
\[
\Delta_\evn = \ale ~ \pm\del_i\pm\del_j, ~ \pm 2 \del_i ~ \ari 
\medbox{and} 
\Delta_\odd = \ale ~ \pm\del_j ~ \ari \,,
\]
- for $C(n+1)$:
\[
\Delta_\evn = \ale ~ \pm\del_i\pm\del_j, ~ \pm 2 \del_i ~ \ari 
\medbox{and} 
\Delta_\odd = \ale ~ \pm\eps\pm\del_j ~ \ari \,,
\]
- for $D(m,n)$:
\[
\Delta_\evn = \ale ~ \pm\eps_i\pm\eps_j, ~ \pm\del_i\pm\del_j,
~ \pm 2\del_i ~ \ari 
\medbox{and} 
\Delta_\odd = \ale ~ \pm\eps_i\pm\del_j ~ \ari \,. 
\]
The Dynkin diagrams of the orthosymplectic superalgebras are of the
following types:
\\
- for the superalgebra $B(m,n)$
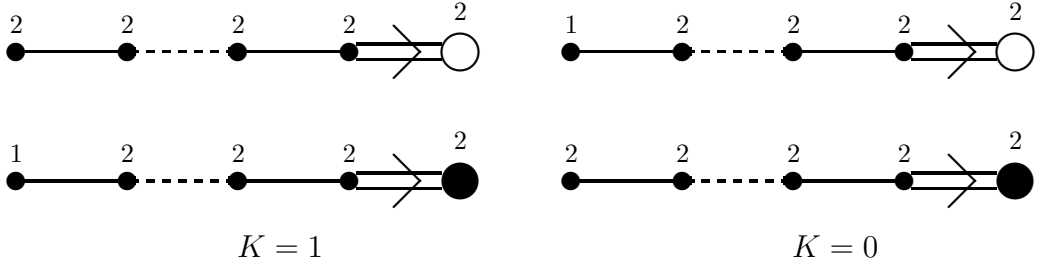
\[
\begin{array}{cc}
\begin{picture}(200,30)
\thicklines
\multiput(0,0)(42,0){4}{\circle*{7}}
\put(0,10){\makebox(0.4,0.6){{\footnotesize 2}}}
\put(42,10){\makebox(0.4,0.6){{\footnotesize 2}}}
\put(84,10){\makebox(0.4,0.6){{\footnotesize 2}}}
\put(126,10){\makebox(0.4,0.6){{\footnotesize 2}}}
\put(3,0){\line(1,0){36}}
\put(45,0){\dashbox{3}(36,0)}
\put(87,0){\line(1,0){36}}
\put(168,0){\circle{14}}
\put(168,15){\makebox(0.4,0.6){{\footnotesize 2}}}
\put(129,3){\line(1,0){32}}
\put(129,-3){\line(1,0){32}}
\put(153,0){\line(-1,1){10}}\put(153,0){\line(-1,-1){10}}
\end{picture}
&
\begin{picture}(200,30)
\thicklines
\multiput(0,0)(42,0){4}{\circle*{7}}
\put(0,10){\makebox(0.4,0.6){{\footnotesize 1}}}
\put(42,10){\makebox(0.4,0.6){{\footnotesize 2}}}
\put(84,10){\makebox(0.4,0.6){{\footnotesize 2}}}
\put(126,10){\makebox(0.4,0.6){{\footnotesize 2}}}
\put(3,0){\line(1,0){36}}
\put(45,0){\dashbox{3}(36,0)}
\put(87,0){\line(1,0){36}}
\put(168,0){\circle{14}}
\put(168,15){\makebox(0.4,0.6){{\footnotesize 2}}}
\put(129,3){\line(1,0){32}}
\put(129,-3){\line(1,0){32}}
\put(153,0){\line(-1,1){10}}\put(153,0){\line(-1,-1){10}}
\end{picture}
\\
&\\
\begin{picture}(200,30)
\thicklines
\multiput(0,0)(42,0){4}{\circle*{7}}
\put(0,10){\makebox(0.4,0.6){{\footnotesize 1}}}
\put(42,10){\makebox(0.4,0.6){{\footnotesize 2}}}
\put(84,10){\makebox(0.4,0.6){{\footnotesize 2}}}
\put(126,10){\makebox(0.4,0.6){{\footnotesize 2}}}
\put(3,0){\line(1,0){36}}
\put(45,0){\dashbox{3}(36,0)}
\put(87,0){\line(1,0){36}}
\put(168,0){\circle*{14}}
\put(168,15){\makebox(0.4,0.6){{\footnotesize 2}}}
\put(129,3){\line(1,0){32}}
\put(129,-3){\line(1,0){32}}
\put(153,0){\line(-1,1){10}}\put(153,0){\line(-1,-1){10}}
\end{picture}
&
\begin{picture}(200,30)
\thicklines
\multiput(0,0)(42,0){4}{\circle*{7}}
\put(0,10){\makebox(0.4,0.6){{\footnotesize 2}}}
\put(42,10){\makebox(0.4,0.6){{\footnotesize 2}}}
\put(84,10){\makebox(0.4,0.6){{\footnotesize 2}}}
\put(126,10){\makebox(0.4,0.6){{\footnotesize 2}}}
\put(3,0){\line(1,0){36}}
\put(45,0){\dashbox{3}(36,0)}
\put(87,0){\line(1,0){36}}
\put(168,0){\circle*{14}}
\put(168,15){\makebox(0.4,0.6){{\footnotesize 2}}}
\put(129,3){\line(1,0){32}}
\put(129,-3){\line(1,0){32}}
\put(153,0){\line(-1,1){10}}\put(153,0){\line(-1,-1){10}}
\end{picture}
\\
&\\
K = 1 & K = 0
\end{array}
\]
- for the superalgebra $C(n+1)$
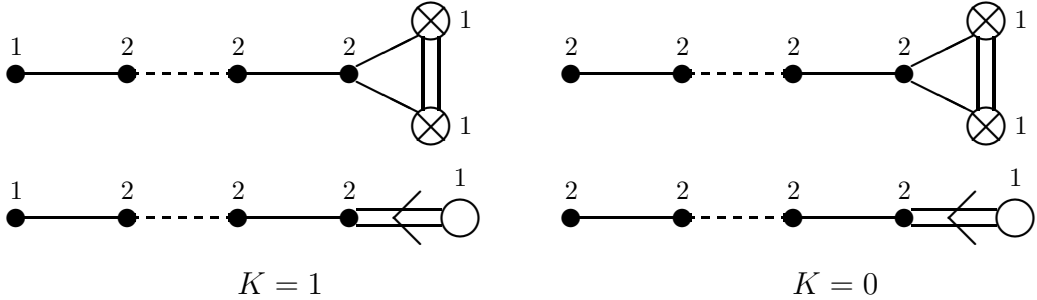
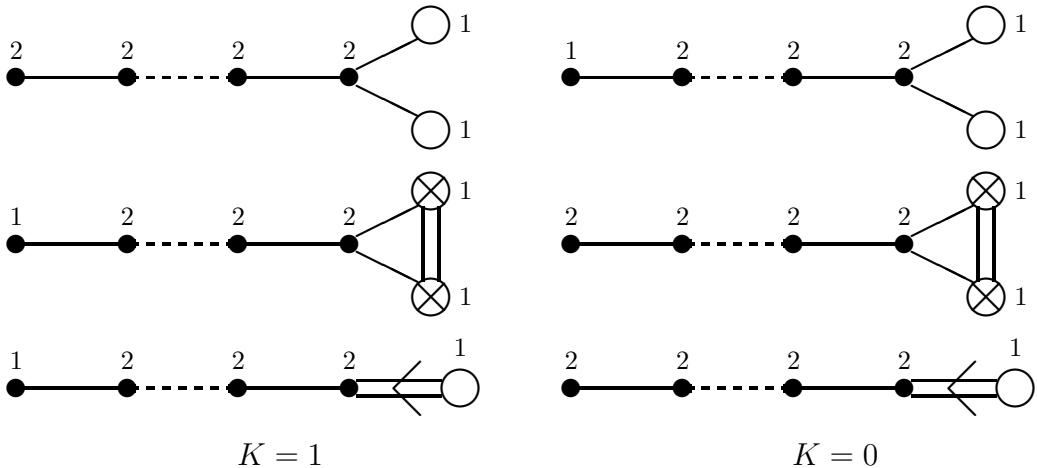
\[
\begin{array}{cc}
\begin{picture}(200,30)
\thicklines
\multiput(0,0)(42,0){4}{\circle*{7}}
\put(0,10){\makebox(0.4,0.6){{\footnotesize 1}}}
\put(42,10){\makebox(0.4,0.6){{\footnotesize 2}}}
\put(84,10){\makebox(0.4,0.6){{\footnotesize 2}}}
\put(126,10){\makebox(0.4,0.6){{\footnotesize 2}}}
\put(3,0){\line(1,0){36}}
\put(45,0){\dashbox{3}(36,0)}
\put(87,0){\line(1,0){36}}
\put(129,3){\line(2,1){23}}\put(129,-3){\line(2,-1){23}}
\put(157,20){\circle{14}}
\put(170,20){\makebox(0.4,0.6){{\footnotesize 1}}}
\put(157,-20){\circle{14}}
\put(170,-20){\makebox(0.4,0.6){{\footnotesize 1}}}
\put(152,15){\line(1,1){10}}\put(152,25){\line(1,-1){10}}
\put(152,-25){\line(1,1){10}}\put(152,-15){\line(1,-1){10}}
\put(154,-14){\line(0,1){28}}
\put(160,-14){\line(0,1){28}}
\end{picture}
&
\begin{picture}(200,30)
\thicklines
\multiput(0,0)(42,0){4}{\circle*{7}}
\put(0,10){\makebox(0.4,0.6){{\footnotesize 2}}}
\put(42,10){\makebox(0.4,0.6){{\footnotesize 2}}}
\put(84,10){\makebox(0.4,0.6){{\footnotesize 2}}}
\put(126,10){\makebox(0.4,0.6){{\footnotesize 2}}}
\put(3,0){\line(1,0){36}}
\put(45,0){\dashbox{3}(36,0)}
\put(87,0){\line(1,0){36}}
\put(129,3){\line(2,1){23}}\put(129,-3){\line(2,-1){23}}
\put(157,20){\circle{14}}
\put(170,20){\makebox(0.4,0.6){{\footnotesize 1}}}
\put(157,-20){\circle{14}}
\put(170,-20){\makebox(0.4,0.6){{\footnotesize 1}}}
\put(152,15){\line(1,1){10}}\put(152,25){\line(1,-1){10}}
\put(152,-25){\line(1,1){10}}\put(152,-15){\line(1,-1){10}}
\put(154,-14){\line(0,1){28}}
\put(160,-14){\line(0,1){28}}
\end{picture}
\\
\vspace{2mm} &\\
\begin{picture}(200,30)
\thicklines
\multiput(0,0)(42,0){4}{\circle*{7}}
\put(0,10){\makebox(0.4,0.6){{\footnotesize 1}}}
\put(42,10){\makebox(0.4,0.6){{\footnotesize 2}}}
\put(84,10){\makebox(0.4,0.6){{\footnotesize 2}}}
\put(126,10){\makebox(0.4,0.6){{\footnotesize 2}}}
\put(3,0){\line(1,0){36}}
\put(45,0){\dashbox{3}(36,0)}
\put(87,0){\line(1,0){36}}
\put(168,0){\circle{14}}
\put(168,15){\makebox(0.4,0.6){{\footnotesize 1}}}
\put(129,3){\line(1,0){32}}
\put(129,-3){\line(1,0){32}}
\put(143,0){\line(1,1){10}}\put(143,0){\line(1,-1){10}}
\end{picture}
&
\begin{picture}(200,30)
\thicklines
\multiput(0,0)(42,0){4}{\circle*{7}}
\put(0,10){\makebox(0.4,0.6){{\footnotesize 2}}}
\put(42,10){\makebox(0.4,0.6){{\footnotesize 2}}}
\put(84,10){\makebox(0.4,0.6){{\footnotesize 2}}}
\put(126,10){\makebox(0.4,0.6){{\footnotesize 2}}}
\put(3,0){\line(1,0){36}}
\put(45,0){\dashbox{3}(36,0)}
\put(87,0){\line(1,0){36}}
\put(168,0){\circle{14}}
\put(168,15){\makebox(0.4,0.6){{\footnotesize 1}}}
\put(129,3){\line(1,0){32}}
\put(129,-3){\line(1,0){32}}
\put(143,0){\line(1,1){10}}\put(143,0){\line(1,-1){10}}
\end{picture}
\\
&\\
K = 1 & K = 0
\end{array}
\]
-for the superalgebra $D(m,n)$
\[
\begin{array}{cc}
\begin{picture}(200,30)
\thicklines
\multiput(0,0)(42,0){4}{\circle*{7}}
\put(0,10){\makebox(0.4,0.6){{\footnotesize 2}}}
\put(42,10){\makebox(0.4,0.6){{\footnotesize 2}}}
\put(84,10){\makebox(0.4,0.6){{\footnotesize 2}}}
\put(126,10){\makebox(0.4,0.6){{\footnotesize 2}}}
\put(3,0){\line(1,0){36}}
\put(45,0){\dashbox{3}(36,0)}
\put(87,0){\line(1,0){36}}
\put(129,3){\line(2,1){23}}\put(129,-3){\line(2,-1){23}}
\put(157,20){\circle{14}}
\put(170,20){\makebox(0.4,0.6){\footnotesize 1}}
\put(157,-20){\circle{14}}
\put(170,-20){\makebox(0.4,0.6){{\footnotesize 1}}}
\end{picture}
&
\begin{picture}(200,30)
\thicklines
\multiput(0,0)(42,0){4}{\circle*{7}}
\put(0,10){\makebox(0.4,0.6){{\footnotesize 1}}}
\put(42,10){\makebox(0.4,0.6){{\footnotesize 2}}}
\put(84,10){\makebox(0.4,0.6){{\footnotesize 2}}}
\put(126,10){\makebox(0.4,0.6){{\footnotesize 2}}}
\put(3,0){\line(1,0){36}}
\put(45,0){\dashbox{3}(36,0)}
\put(87,0){\line(1,0){36}}
\put(129,3){\line(2,1){23}}\put(129,-3){\line(2,-1){23}}
\put(157,20){\circle{14}}
\put(170,20){\makebox(0.4,0.6){\footnotesize 1}}
\put(157,-20){\circle{14}}
\put(170,-20){\makebox(0.4,0.6){{\footnotesize 1}}}
\end{picture}
\\
\vspace{5mm} &\\
\begin{picture}(200,30)
\thicklines
\multiput(0,0)(42,0){4}{\circle*{7}}
\put(0,10){\makebox(0.4,0.6){{\footnotesize 1}}}
\put(42,10){\makebox(0.4,0.6){{\footnotesize 2}}}
\put(84,10){\makebox(0.4,0.6){{\footnotesize 2}}}
\put(126,10){\makebox(0.4,0.6){{\footnotesize 2}}}
\put(3,0){\line(1,0){36}}
\put(45,0){\dashbox{3}(36,0)}
\put(87,0){\line(1,0){36}}
\put(129,3){\line(2,1){23}}\put(129,-3){\line(2,-1){23}}
\put(157,20){\circle{14}}
\put(170,20){\makebox(0.4,0.6){{\footnotesize 1}}}
\put(157,-20){\circle{14}}
\put(170,-20){\makebox(0.4,0.6){{\footnotesize 1}}}
\put(152,15){\line(1,1){10}}\put(152,25){\line(1,-1){10}}
\put(152,-25){\line(1,1){10}}\put(152,-15){\line(1,-1){10}}
\put(154,-14){\line(0,1){28}}
\put(160,-14){\line(0,1){28}}
\end{picture}
&
\begin{picture}(200,30)
\thicklines
\multiput(0,0)(42,0){4}{\circle*{7}}
\put(0,10){\makebox(0.4,0.6){{\footnotesize 2}}}
\put(42,10){\makebox(0.4,0.6){{\footnotesize 2}}}
\put(84,10){\makebox(0.4,0.6){{\footnotesize 2}}}
\put(126,10){\makebox(0.4,0.6){{\footnotesize 2}}}
\put(3,0){\line(1,0){36}}
\put(45,0){\dashbox{3}(36,0)}
\put(87,0){\line(1,0){36}}
\put(129,3){\line(2,1){23}}\put(129,-3){\line(2,-1){23}}
\put(157,20){\circle{14}}
\put(170,20){\makebox(0.4,0.6){{\footnotesize 1}}}
\put(157,-20){\circle{14}}
\put(170,-20){\makebox(0.4,0.6){{\footnotesize 1}}}
\put(152,15){\line(1,1){10}}\put(152,25){\line(1,-1){10}}
\put(152,-25){\line(1,1){10}}\put(152,-15){\line(1,-1){10}}
\put(154,-14){\line(0,1){28}}
\put(160,-14){\line(0,1){28}}
\end{picture}
\\
\vspace{2mm} &\\
\begin{picture}(200,30)
\thicklines
\multiput(0,0)(42,0){4}{\circle*{7}}
\put(0,10){\makebox(0.4,0.6){{\footnotesize 1}}}
\put(42,10){\makebox(0.4,0.6){{\footnotesize 2}}}
\put(84,10){\makebox(0.4,0.6){{\footnotesize 2}}}
\put(126,10){\makebox(0.4,0.6){{\footnotesize 2}}}
\put(3,0){\line(1,0){36}}
\put(45,0){\dashbox{3}(36,0)}
\put(87,0){\line(1,0){36}}
\put(168,0){\circle{14}}
\put(168,15){\makebox(0.4,0.6){{\footnotesize 1}}}
\put(129,3){\line(1,0){32}}
\put(129,-3){\line(1,0){32}}
\put(143,0){\line(1,1){10}}\put(143,0){\line(1,-1){10}}
\end{picture}
&
\begin{picture}(200,30)
\thicklines
\multiput(0,0)(42,0){4}{\circle*{7}}
\put(0,10){\makebox(0.4,0.6){{\footnotesize 2}}}
\put(42,10){\makebox(0.4,0.6){{\footnotesize 2}}}
\put(84,10){\makebox(0.4,0.6){{\footnotesize 2}}}
\put(126,10){\makebox(0.4,0.6){{\footnotesize 2}}}
\put(3,0){\line(1,0){36}}
\put(45,0){\dashbox{3}(36,0)}
\put(87,0){\line(1,0){36}}
\put(168,0){\circle{14}}
\put(168,15){\makebox(0.4,0.6){{\footnotesize 1}}}
\put(129,3){\line(1,0){32}}
\put(129,-3){\line(1,0){32}}
\put(143,0){\line(1,1){10}}\put(143,0){\line(1,-1){10}}
\end{picture}
\\
&\\
K = 1 & K = 0
\end{array}
\]
In these diagrams, the labels are the Dynkin labels which give the
decomposition of the highest root in terms of the simple roots. The
small black dots represent either white dots (associated to even roots)
or grey dots (associated to odd roots of zero length), $K$ is the
parity of the number of grey dots. The Dynkin diagrams of the
orthosymplectic Lie superalgebras up to rank 4 are given in Table
\ref{table6}.

The orthosymplectic superalgebras $osp(M|N)$ (with $M = 2m$ or $2m+1$
and $N = 2n$) can be generated as matrix superalgebras by taking a
basis of $(M+N)^2$ elementary matrices $e_{IJ}$ of order $M+N$
satisfying $(e_{IJ})_{KL} = \delta_{IL} \delta_{JK}$ ($I,J,K,L =
1,\dots,M+N$). One defines the following graded matrices
\beo
&& G_{IJ} = \left(\begin{array}{c|c}
\begin{array}{cc} 0 & \II_m \cr \II_m & 0 \end{array} & 0 \cr
\hline
0 & \begin{array}{cc} 0 & \II_n \cr -\II_n & 0 \end{array}
\end{array}\right) \medbox{if $M = 2m$} \\
&& \\
&& G_{IJ} = \left(\begin{array}{c|c}
\begin{array}{ccc} 0 & \II_m & 0 \cr \II_m & 0 & 0 \cr 0 & 0 & 1 
\end{array} & 0 \cr 
\hline
0 & \begin{array}{cc} 0 & \II_n \cr -\II_n & 0 \end{array}
\end{array}\right) \medbox{if $M = 2m+1$}
\eno
where $\II_m$ and $\II_n$ are the $m \times m$ and $n \times n$
identity matrices respectively.
\\
Dividing the capital indices $I,J,\dots$ into small unbared indices
$i,j,\dots$ running from 1 to $M$ and small bared indices
$\bi,\bj,\dots$ running from $M+1$ to $M+N$, the generators of
$osp(M|N)$ are given by 
\beo
&&E_{ij} = G_{ik} e_{kj} - G_{jk} e_{ki} \\
&&E_{\bi\bj} = G_{\bi\bk} e_{\bk\bj} + G_{\bj\bk} e_{\bk\bi} \\ 
&&E_{i\bj} = E_{\bj i} = G_{ik} e_{k\bj}
\eno
Then the $E_{ij}$ (antisymmetric in the indices $i,j$) generate the
$so(M)$ part, the $E_{\bi\bj}$ (symmetric in the indices $\bi,\bj$)
generate the $sp(N)$ part and $E_{i\bj}$ transform as the $(M,N)$ 
representation of $osp(M|N)$. 
They satisfy the following (super)commutation relations: 
\beo
&&\cle E_{ij},E_{kl} \cri = 
G_{jk} E_{il} + G_{il} E_{jk} - G_{ik} E_{jl} - G_{jl} E_{ik} \\
&&\cle E_{\bi\bj},E_{\bk\bl} \cri = 
- G_{\bj\bk} E_{\bi\bl} - G_{\bi\bl} E_{\bj\bk} 
- G_{\bj\bl} E_{\bi\bk} - G_{\bi\bk} E_{\bj\bl} \\
&&\cle E_{ij},E_{\bk\bl} \cri = 0 \\
&&\cle E_{ij},E_{k\bl} \cri = G_{jk} E_{i\bl} - G_{ik} E_{j\bl} \\
&&\cle E_{i\bj},E_{\bk\bl} \cri = -G_{\bj\bk} E_{i\bl} - G_{\bj\bl}
E_{i\bk} \\ 
&&\ale E_{i\bj},E_{k\bl}\ari = G_{ik} E_{\bj\bl} - G_{\bj\bl} E_{ik} 
\eno
In the case of the superalgebra $osp(1|N)$, the commutation relations
greatly simplify. One obtains
\beo
&&\cle E_{\bi\bj},E_{\bk\bl} \cri = 
- G_{\bj\bk} E_{\bi\bl} - G_{\bi\bl} E_{\bj\bk} 
- G_{\bj\bl} E_{\bi\bk} - G_{\bi\bk} E_{\bj\bl} \\
&&\cle E_{\bi},E_{\bj\bk} \cri = -G_{\bi\bj} E_{\bk} - G_{\bi\bk} E_{\bj} \\ 
&&\ale E_{\bi},E_{\bj}\ari = E_{\bi\bj} 
\eno
where $E_{\bi}$ denote the odd generators.

\section{\sf Oscillator realizations: Cartan type superalgebras} 
\indent

Oscillator realizations of the Cartan type superalgebras can be 
obtained as follows. Take a set of $2n$ fermionic oscillators $a_i^-$ 
and $a_i^+$ with standard anticommutation relations
\[
\ale a_i^- , a_j^- \ari = \ale a_i^+ , a_j^+ \ari = 0
\medbox{and}
\ale a_i^+ , a_j^- \ari = \del_{ij} 
\]
In the case of the $W(n)$ superalgebra, one defines the following 
subspaces:
\beo
&& \cG_{-1} = \ale a_{i_0}^- \ari \\
&& \cG_0 = \ale a_{i_0}^+ a_{i_1}^- \ari \\
&& \cG_1 = \ale a_{i_0}^+ a_{i_1}^+ a_{i_2}^- \ari \quad i_0 \ne i_1 \\
&& \dots \\
&& \cG_{n-1} = \ale a_{i_0}^+ a_{i_1}^+ \dots a_{i_{n-1}}^+ a_{i_n}^- 
\ari \quad i_0 \ne i_1 \ne \dots \ne i_{n-1}
\eno
the superalgebra $W(n)$ is given by
\[
W(n) = \bigoplus_{i = -1}^{n-1} \cG_i
\]
with $\ZZ$-gradation $\cle \cG_i,\cG_j \cri \subset \cG_{i+j}$.
\\
In the case of $S(n)$ and $\tilde S(n)$, defining the following 
subspaces:
\beo
&& \cG_{-1} = \ale a_{i_0}^- \ari \bigbox{and}
\cG'_{-1} = \ale (1+a_1^+ \dots a_n^+) a_{i_0}^- \ari \\
&& \cG_0 = \ale a_1^+a_1^- - a_{i_0}^+a_{i_0}^- ~~~ (i_0 \ne 1), 
~~~ a_{i_0}^+a_{i_1}^- ~~~ (i_1 \ne i_0) \ari \\
&& \cG_1 = \ale a_{i_1}^+ (a_1^+a_1^- - a_{i_0}^+a_{i_0}^-) ~~~ 
(i_1 \ne i_0 \ne 1), \\
&& \quad \quad \quad a_1^+ (a_2^+a_2^- - a_{i_0}^+a_{i_0}^-) ~~~ 
(i_0 \ne 1,2), \\
&& \quad \quad \quad a_{i_2}^+ a_{i_1}^+ a_{i_0}^- 
~~~ (i_2 \ne i_1 \ne i_0) \ari \\
&& \cG_{2} = \ale a_{i_2}^+ a_{i_1}^+ (a_1^+a_1^- - a_{i_0}^+a_{i_0}^-)
 ~~~ (i_2 \ne i_1 \ne i_0 \ne 1), \\
&& \quad \quad \quad a_{i_1}^+ a_1^+ (a_2^+a_2^- - a_{i_0}^+a_{i_0}^-)
 ~~~ (i_1 \ne i_0 \ne 1,2), \\
&& \quad \quad \quad a_1^+ a_2^+ (a_3^+a_3^- - a_{i_0}^+a_{i_0}^-) ~~~ 
(i_0 \ne 1,2,3), \\
&& \quad \quad \quad a_{i_3}^+ a_{i_2}^+ a_{i_1}^+ a_{i_0}^- 
~~~ (i_3 \ne i_2 \ne i_1 \ne i_0) \ari \\
&& \dots 
\eno
the superalgebra $S(n)$ is given by
\[
S(n) = \bigoplus_{i = 0}^{n-2} \cG_i \oplus \cG_{-1}
\]
and the superalgebra $\tilde S(n)$ by
\[
\tilde S(n) = \bigoplus_{i = 0}^{n-2} \cG_i \oplus \cG'_{-1}
\]

Finally, in the case of $H(n)$ one defines the following subspaces:
\beo
&&\cG_{-1} = \ale a_{i_0}^- \ari \\
&&\cG_0 = \ale a_{i_0}^+ a_{i_1}^- - a_{i_1}^+ a_{i_0}^- \ari \\
&&\cG_1 = \ale a_{i_0}^+ a_{i_1}^+ a_{i_2}^- 
- a_{i_0}^+ a_{i_2}^+ a_{i_1}^- - a_{i_2}^+ a_{i_0}^+ a_{i_1}^- 
+ a_{i_1}^+ a_{i_2}^+ a_{i_0}^- + a_{i_2}^+ a_{i_0}^+ a_{i_1}^- 
- a_{i_2}^+ a_{i_1}^+ a_{i_0}^- 
\ari \\
&&\dots
\eno
The superalgebra $H(n)$ is given by
\[
H(n) = \bigoplus_{i = -1}^{n-3} \cG_i
\]

\medskip

For more details, see ref. \cite{Rit77}.

\section{\sf Oscillator realizations: orthosymplectic and unitary
series} 
\indent

Let us consider a set of $2n$ bosonic oscillators $b_i^-$ and $b_i^+$ 
with commutation relations: 
\[
\cle b_i^-,b_j^- \cri = \cle b_i^+,b_j^+ \cri = 0 
\medbox{and}
\cle b_i^-,b_j^+ \cri = \delta_{ij} 
\]
and a set of $2m$ fermionic oscillators $a_i^-$ and $a_i^+$ with
anticommutation relations: 
\[
\ale a_i^-,a_j^- \ari = \ale a_i^+,a_j^+ \ari = 0 
\medbox{and}
\ale a_i^-,a_j^+ \ari = \delta_{ij}
\]
the two sets commuting each other: 
\[
\cle b_i^-,a_j^- \cri = \cle b_i^-,a_j^+ \cri = \cle b_i^+,a_j^- \cri 
 = \cle b_i^+,a_j^+ \cri = 0 
\]

\underline{Oscillator realization of $A(m-1,n-1)$}

Let
\[
\Delta = \ale ~ \eps_i-\eps_j, ~ \del_i-\del_j, ~ \eps_i-\del_j,
~ -\eps_i+\del_j ~ \ari 
\] 
be the root system of $A(m-1,n-1)$ expressed in terms of the orthogonal 
vectors $\eps_1, \dots, \eps_m$ and $\del_1, \dots, \del_n$. An oscillator 
realization of the simple generators in the distinguished basis is given
by  
\[
\begin{array}{llll}
\bigg. H_i = b_i^+ b_i^- - b_{i+1}^+ b_{i+1}^- 
&E_{\del_i-\del_{i+1}} = b_i^+ b_{i+1}^- 
&E_{\del_{i+1}-\del_i} = b_{i+1}^+ b_i^-
&(1 \le i \le n-1) \cr
\bigg. H_n = b_{n+1}^+ b_{n+1}^- + a_1^+ a_1^- 
&E_{\del_n-\eps_1} = b_n^+ a_1^- 
&E_{\eps_1-\del_n} = a_1^+ b_n^- & \cr
\bigg. H_{n+i} = a_i^+ a_i^- - a_{i+1}^+ a_{i+1}^- 
&E_{\eps_i-\eps_{i+1}} = a_i^+ a_{i+1}^- 
&E_{\eps_{i+1}-\eps_i} = a_{i+1}^+ a_i^-
&(1 \le i \le m-1)
\end{array}
\]
By commutation relation, one finds the realization of the whole set of 
root generators: 
\[
\begin{array}{lll}
\bigg. E_{\eps_i-\eps_j} = a_i^+ a_j^- 
& \qquad & E_{\del_i-\del_j} = b_i^+ b_j^- \cr
\bigg. E_{\eps_i-\del_j} = a_i^+ b_j^- 
& \qquad & E_{-\eps_i+\del_j} = a_i^- b_j^+ 
\end{array}
\] 

\medskip

\underline{Oscillator realization of $B(m,n)$}

Let
\[
\Delta = \ale ~ \pm\eps_i\pm\eps_j, ~ \pm\eps_i, ~ \pm\del_i\pm\del_j,
 ~ \pm 2\del_i, ~ \pm\eps_i\pm\del_j, ~ \pm\del_i ~ \ari 
\]
be the root system of $B(m,n)$ expressed in terms of the orthogonal 
vectors $\eps_1, \dots, \eps_m$ and $\del_1, \dots, \del_n$.
An oscillator realization of the simple generators in the distinguished 
basis is given by
\[
\begin{array}{llll}
\bigg. H_i = b_i^+ b_i^- - b_{i+1}^+ b_{i+1}^- 
&E_{\del_i-\del_{i+1}} = b_i^+ b_{i+1}^- 
&E_{\del_{i+1}-\del_i} = b_{i+1}^+ b_i^-
&(1 \le i \le n-1) \cr
\bigg. H_n = b_n^+ b_n^- + a_1^+ a_1^- 
&E_{\del_n-\eps_1} = b_n^+ a_1^- 
&E_{\eps_1-\del_n} = a_1^+ b_n^- & \cr
\bigg. H_{n+i} = a_i^+ a_i^- - a_{i+1}^+ a_{i+1}^- 
&E_{\eps_i-\eps_{i+1}} = a_i^+ a_{i+1}^- 
&E_{\eps_{i+1}-\eps_i} = a_{i+1}^+ a_i^-
&(1 \le i \le m-1) \cr
\bigg. H_{n+m} = 2 a_m^+ a_m^- - 1
&E^+_{\eps_m} = (-1)^N a_m^+ 
&E^-_{\eps_m} = a_m^- (-1)^N &
\end{array}
\]
where $N = \sum_{k = 1}^m a_k^+ a_k^-$.
\\
By commutation relation, one finds the realization of the whole set of 
root generators: 
\[
\begin{array}{lll}
\bigg. E_{\pm\eps_i\pm\eps_j} = a_i^{\pm} a_j^{\pm} 
& \qquad & E_{\pm\eps_i\pm\del_j} = a_i^{\pm} b_j^{\pm} \cr
\bigg. E_{\pm\del_i\pm\del_j} = b_i^{\pm} b_j^{\pm} 
& \qquad & E_{\pm 2\del_i} = (b_i^{\pm})^2 \cr
\bigg. E_{\eps_i} = (-1)^N a_i^+ 
& \qquad & E_{-\eps_i} = a_i^- (-1)^N \cr
\bigg. E_{\del_i} = (-1)^N b_i^+
& \qquad & E_{-\del_i} = b_i^- (-1)^N \cr
 \end{array}
\] 

\medskip

\underline{Oscillator realization of $B(0,n)$}

The case $B(0,n)$ requires special attention. The root system of
$B(0,n)$ can be expressed in terms of the orthogonal vectors $\del_1,
\dots, \del_n$ and reduces to
\[
\Delta = \ale ~ \pm\del_i\pm\del_j, ~ \pm 2\del_i, ~ \pm\del_i ~ \ari 
\]
An oscillator realization of the generators of $B(0,n)$ can be obtained
only with the help of bosonic oscillators. It is given for the simple
generators by 
\[
\begin{array}{llll}
\bigg. H_i = b_i^+ b_i^- - b_{i+1}^+ b_{i+1}^- 
&E_{\del_i-\del_{i+1}} = b_i^+ b_{i+1}^- 
&E_{\del_{i+1}-\del_i} = b_{i+1}^+ b_i^- 
&(1 \le i \le n-1) \cr
\bigg. H_n = b_n^+ b_n^- + \half 
&E_{\del_n} = b_n^+ 
&E_{-\del_n} = b_n^- 
& \cr
\end{array}
\]
By commutation relation, one finds the realization of the whole set of 
root generators:
\[
E_{\pm\del_i\pm\del_j} = b_i^{\pm} b_j^{\pm} 
\qquad 
E_{\pm 2\del_i} = (b_i^{\pm})^2 
\qquad 
E_{\pm\del_i} = \sfrac{1}{\sqrt{2}} b_i^{\pm}
\] 

\medskip

\underline{Oscillator realization of $C(n+1)$}

Let
\[
\Delta = \ale ~ \pm\del_i\pm\del_j, ~ \pm 2 \del_i,
 ~ \pm\eps\pm\del_j ~ \ari 
\]
be the root system of $C(n+1)$ expressed in terms of the orthogonal 
vectors $\eps, \del_1, \dots, \del_n$. An oscillator realization of the
simple generators in the distinguished basis is given by
\[
\begin{array}{llll}
\bigg. H_1 = a_1^+ a_1^- + b_1^+ b_1^- 
&E_{\eps-\del_1} = a_1^+ b_1^- 
&E_{\del_1-\eps} = b_1^+ a_1^- & \cr
\bigg. H_i = b_i^+ b_i^- - b_{i+1}^+ b_{i+1}^- 
&E_{\del_i-\del_{i+1}} = b_i^+ b_{i+1}^- 
&E_{\del_{i+1}-\del_i} = b_{i+1}^+ b_i^- 
&(2 \le i \le n) \cr
\bigg. H_{n+1} = b_n^+ b_n^- + 1/2
&E_{2\del_n} = {\half} (b_n^+)^2 
&E_{-2\del_n} = {\half} (b_n^-)^2 &
\end{array}
\]
By commutation relation, one finds the realization of the whole set of 
root generators: 
\[
\begin{array}{lll}
\bigg. E_{\pm\del_i\pm\del_j} = b_i^{\pm} b_j^{\pm} 
& \qquad & E_{\pm 2\del_i} = (b_i^{\pm})^2/2 \cr
\bigg. E_{\eps\pm\del_j} = a_1^+ b_j^{\pm}
& \qquad & E_{-\eps\pm\del_j} = b_j^{\pm} a_1^-
\end{array}
\] 

\medskip

\underline{Oscillator realization of $D(m,n)$}

Let
\[
\Delta = \ale ~ \pm\eps_i\pm\eps_j, ~ \pm\del_i\pm\del_j,
 ~ \pm 2\del_i, ~ \pm\eps_i\pm\del_j ~ \ari 
\]
be the root system of $D(m,n)$ expressed in terms of the orthogonal 
vectors $\eps_1, \dots, \eps_m$ and $\del_1, \dots, \del_n$.
An oscillator realization of the simple generators in the distinguished 
basis is given by
\[
\begin{array}{llll}
\bigg. H_i = b_i^+ b_i^- - b_{i+1}^+ b_{i+1}^- 
&E_{\del_i-\del_{i+1}} = b_i^+ b_{i+1}^- 
&E_{\del_{i+1}-\del_i} = b_{i+1}^+ b_i^-
&(1 \le i \le n-1) \cr
\bigg. H_n = b_n^+ b_n^- + a_1^+ a_1^- 
&E_{\del_n-\eps_1} = b_n^+ a_1^- 
&E_{\eps_1-\del_n} = a_1^+ b_n^- & \cr
\bigg. H_{n+i} = a_i^+ a_i^- - a_{i+1}^+ a_{i+1}^-
&E_{\eps_i-\eps_{i+1}} = a_i^+ a_{i+1}^- 
&E_{\eps_{i+1}-\eps_i} = a_{i+1}^+ a_i^-
&(1 \le i \le m-1) \cr
\bigg. H_{n+m} = a_{m-1}^+ a_{m-1}^- + a_m^+ a_m^- - 1
&E_{\eps_{m-1}+\eps_m} = a_m^+ a_{m-1}^+ 
&E_{-\eps_m-\eps_{m-1}} = a_{m-1}^- a_m^- &
\end{array}
\]
By commutation relation, one finds the realization of the whole set of 
root generators: 
\[
\begin{array}{lll}
\bigg. E_{\pm\eps_i\pm\eps_j} = a_i^{\pm} a_j^{\pm} 
& \qquad & E_{\pm\eps_i\pm\del_j} = a_i^{\pm} b_j^{\pm} \cr
\bigg. E_{\pm\del_i\pm\del_j} = b_i^{\pm} b_j^{\pm} 
& \qquad & E_{\pm 2\del_i} = (b_i^{\pm})^2 \cr
\end{array}
\] 

\medskip

For more details, see refs. \cite{BBI81,Tan84}. Note that in ref.
\cite{BBI81}, oscillator realizations were used to analyse
supersymmetric structure in the spectra of complex nuclei; the first
reference of this interesting approach is \cite{Iac80}.

\section{\sf Oscillator realizations: strange series} 
\indent

Let us consider a set of $2n$ bosonic oscillators $b_i^-$ and $b_i^+$ 
with commutation relations: 
\[
\cle b_i^-,b_j^- \cri = \cle b_i^+,b_j^+ \cri = 0 
\medbox{and}
\cle b_i^-,b_j^+ \cri = \delta_{ij} 
\]
and a set of $2n$ fermionic oscillators $a_i^-$ and $a_i^+$ with
anticommutation relations: 
\[
\ale a_i^-,a_j^- \ari = \ale a_i^+,a_j^+ \ari = 0 
\medbox{and}
\ale a_i^-,a_j^+ \ari = \delta_{ij} 
\]
the two sets commuting each other: 
\[
\cle b_i^-,a_j^- \cri = \cle b_i^-,a_j^+ \cri = \cle b_i^+,a_j^- \cri 
 = \cle b_i^+,a_j^+ \cri = 0 
\]

\underline{Oscillator realization of $P(n)$}

An oscillator realization of the generators of $P(n)$ is obtained as 
follows:
\\
- the generators of the even $sl(n)$ part are given by
\beo
&&H_i = a_i^+ a_i^- - a_{i+1}^+ a_{i+1}^- + b_i^+ b_i^- -
b_{i+1}^+ b_{i+1}^- \medbox{with} 1 \le i \le n-1 \\ 
&&E^+_{ij} = a_i^+ a_j^- + b_i^+ b_j^- \medbox{with}
1 \le i < j \le n \\
&&E^-_{ij} = a_i^+ a_j^- + b_i^+ b_j^- \medbox{with}
1 \le j < i \le n
\eno
- the generators of the odd symmetric part $\cG_S$ of $P(n)$ by 
\beo
&&F^+_{ij} = b_i^+ a_j^+ + b_j^+ a_i^+ \medbox{with}
1 \le i \ne j \le n \\
&&F^+_{i} = b_i^+ b_j^+ \medbox{with} 1 \le i \le n
\eno
- the generators of the odd antisymmetric part $\cG_A$ of
$P(n)$ by 
\[
F^-_{ij} = b_i^- a_j^- + b_j^- a_i^- \medbox{with}
1 \le i \ne j \le n
\]

\medskip

\underline{Oscillator realization of $Q(n)$}

An oscillator realization of the generators of $Q(n)$ is obtained as 
follows:
\\
- the generators of the even $sl(n)$ part are given by
\beo
&&H_i = a_i^+ a_i^- - a_{i+1}^+ a_{i+1}^- + b_i^+ b_i^- -
b_{i+1}^+ b_{i+1}^- \\ 
&&E_{ij} = a_i^+ a_j^- + b_i^+ b_j^- 
\eno
- the generator of the $U(1)$ part by
\[
Z = \sum_{i = 1}^n a_i^+ a_i^- + b_i^+ b_i^-
\]
- the generators of the odd $sl(n)$ part by
\beo
&&K_i = a_i^+ b_i^- - a_{i+1}^+ b_{i+1}^- + b_i^+ a_i^- -
b_{i+1}^+ a_{i+1}^- \\ 
&&F_{ij} = a_i^+ b_j^- + b_i^+ a_j^- 
\eno

\medskip

For more details, see ref. \cite{FSS91}.

\section{\sf Real forms}
\indent

\udef: Let $\cG$ be a classical Lie superalgebra over $\CC$. A
semimorphism $\phi$ of $\cG$ is a semilinear transformation of $\cG$
which preserves the gradation, that is such that
\beo
&&\phi(X+Y) = \phi(X) + \phi(Y) \\
&&\phi(\alpha X) = \balpha \phi(X) \\
&&\zle \phi(X),\phi(Y) \zri = \phi(\zle X,Y \zri) 
\eno
for all $X,Y \in \cG$ and $\alpha \in \CC$.

\medskip

If $\phi$ is an involutive semimorphism of $\cG$, the superalgebra
$\cG^\phi = \{X+\phi(X) ~ \vert ~ X \in \cG \}$ is a real classical Lie
superalgebra. Moreover, two involutive semimorphisms $\phi$ and $\phi'$
of $\cG$ being given, the real forms $\cG^\phi$ and $\cG^{\phi'}$ are
isomorphic if and only if $\phi$ and $\phi'$ are conjugate by an
automorphism ($\see$) of $\cG$.

It follows that the real classical Lie superalgebras are either the
complex classical Lie superalgebras regarded as real superalgebras or 
the real forms obtained as subsuperalgebras of fixed points of the 
involutive semimorphisms of a complex classical Lie superalgebra. The 
real forms of a complex classical Lie superalgebra $\cG$ are thus 
classified by the involutive semimorphisms of $\cG$ in the automorphism
group of $\cG$. One can prove that the real forms of the complex 
classical Lie superalgebras are uniquely determined by the real forms 
$\cG_\evn^\phi$ of the even part $\cG_\evn$ of $\cG$. 
They are displayed in Table \ref{table12}. 

Notice that $m,n$ have to be even for $sl(m|n,\HH)$, $sl(n|n,\HH)$ and
$HQ(n)$. We recall that $su^*(2n)$ is the set of $2n \times 2n$
matrices of the form $\left( \begin{array}{cc} X_n & Y_n \cr -Y_n^* &
X_n^* \end{array} \right)$ such that $X_n,Y_n$ are matrices of order
$n$ and and $\tr(X_n)+\tr(X_n^*) = 0$ and $so^*(2n)$ is the set of $2n
\times 2n$ matrices of the form $\left( \begin{array}{cc} X_n & Y_n \cr
-Y_n^* & X_n^* \end{array} \right)$ such that $X_n$ and $Y_n$ are
antisymmetric and hermitian complex matrices of order $n$ respectively.

\medskip

For more details, see refs. \cite{Kac77a,Par80}.

\section{\sf Representations: basic definitions}
\indent

\udef: Let $\cG = \cG_\evn \oplus \cG_\odd$ be a classical Lie
superalgebra. Let $\cV = \cV_\evn \oplus \cV_\odd$ be a $\ZZ_2$-graded
vector space and consider the superalgebra $End\,\cV = End_\evn\cV
\oplus End_\odd\cV$ of endomorphisms of $\cV$. A linear representation
$\pi$ of $\cG$ is a homomorphism of $\cG$ into $End\,\cV$, that is, 
$\zle ~,~ \zri$ denoting the superbracket,
\beo
&& \pi(\alpha X) = \alpha \pi(X) \medbox{and} \pi(X+Y) = \pi(X)+\pi(Y) \\
&& \pi(\zle X,Y \zri) = \zle \pi(X),\pi(Y) \zri \\
&& \pi(\cG_\evn) \subset End_\evn\cV \medbox{and} 
\pi(\cG_\odd) \subset End_\odd\cV
\eno
for all $X,Y \in \cG$ and $\alpha \in \CC$.
\\
The vector space $\cV$ is the representation space. The vector space
$\cV$ has the structure of a $\cG$-module by $X(\vv) = \pi(X) \vv$ for
$X \in \cG$ and $\vv \in \cV$. 
\\
The dimension (resp. superdimension) of the representation $\pi$ is the
dimension (resp. graded dimension) of the vector  space $\cV$:
\beo
&& \dim \pi = \dim\cV_\evn + \dim\cV_\odd \\
&& \sdim \pi = \dim\cV_\evn - \dim\cV_\odd
\eno

\udef: The representation $\pi$ is said 
\\
- faithful if $\pi(X) \ne 0$ for all $X \in \cG$.
\\
- trivial if $\pi(X) = 0$ for all $X \in \cG$.

\medskip
\noindent
Every classical Lie superalgebra has a finite dimensional faithful 
representation.
\\
In particular, the representation $\ad : \cG \rightarrow End\,\cG$
($\cG$ being considered as a $\ZZ_2$-graded vector space) such that 
$\ad(X)Y = \zle X,Y \zri$ is called the {\em adjoint} representation of
$\cG$.

\section{\sf Representations: exceptional superalgebras}
\indent

\subsection{\sf Representations of $F(4)$}

A highest weight irreducible representation of $F(4)$ is characterized
by its Dynkin labels ($\see$ Highest weight representations) drawn on
the distinguished Dynkin diagram:
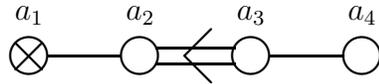
\begin{center}
\begin{picture}(140,45)
\thicklines
\multiput(0,20)(42,0){4}{\circle{14}}
\put(0,35){\makebox(0.4,0.6){$a_1$}}
\put(42,35){\makebox(0.4,0.6){$a_2$}}
\put(84,35){\makebox(0.4,0.6){$a_3$}}
\put(126,35){\makebox(0.4,0.6){$a_4$}}
\put(-5,15){\line(1,1){10}}\put(-5,25){\line(1,-1){10}}
\put(7,20){\line(1,0){28}}
\put(48,17){\line(1,0){30}}
\put(48,23){\line(1,0){30}}
\put(59,20){\line(1,1){10}}\put(59,20){\line(1,-1){10}}
\put(91,20){\line(1,0){28}}
\end{picture}
\end{center}
where $a_2, a_3, a_4$ are positive or null integers.
\\
For the $so(7)$ part, $a_2$ is the shorter root. The $sl(2)$
representation label is hidden by the odd root and its value is given
by $b = \frac{1}{3} (2a_1-3a_2-4a_3-2a_4)$. Since $b$ has to be a
non-negative integer, this relation implies $a_1$ to be a positive
integer or half-integer. Finally, a $F(4)$ representation with $b<4$
has to satisfy a consistency condition, that is
\beo
&&b = 0 \qquad \qquad a_1 = a_2 = a_3 = a_4 = 0 \\
&&b = 1 \qquad \qquad \mbox{not possible} \\
&&b = 2 \qquad \qquad a_2 = a_4 = 0 \\
&&b = 3 \qquad \qquad a_2 = 2a_4+1
\eno
The eight atypicality conditions for the $F(4)$ representations are 
the following:
\[
\begin{array}{lll}
\bigg. a_1 = 0 & \bigbox{or} 
 & b = 0 \cr
\bigg. a_1 = a_2+1 & \bigbox{or} 
 & b = \third (2-a_2-4a_3-2a_4) \cr
\bigg. a_1 = a_2+2a_3+3 & \bigbox{or} 
 & b = \third (6-a_2-2a_4) \cr
\bigg. a_1 = a_2+2a_3+2a_4+5 & \bigbox{or} 
 & b = \third (10-a_2+2a_4) \cr
\bigg. a_1 = 2a_2+2a_3+4 & \bigbox{or} 
 & b = \third (8+a_2-2a_4) \cr
\bigg. a_1 = 2a_2+2a_3+2a_4+6 & \bigbox{or} 
 & b = \third (12+a_2+2a_4) \cr
\bigg. a_1 = 2a_2+4a_3+2a_4+8 & \bigbox{or} 
 & b = \third (16+a_2+4a_3+2a_4) \cr
\bigg. a_1 = 3a_2+4a_3+2a_4+9 & \bigbox{or} 
 & b = \third (18+3a_2+4a_3+2a_4)
\end{array}
\]
Moreover, a necessary (but not sufficient) condition for a
representation to be typical is that $b \ge 4$. 
\\
The dimension of a typical representation with highest weight
$\Lambda = (a_1,a_2,a_3,a_4)$ is given by
\beo
&&\dim \cV(\Lambda) = \frac{32}{45} (a_2+1) (a_3+1) (a_4+1) (a_2+a_3+2) 
(a_3+a_4+2) (a_2+2a_3+3) \\
&&(a_2+a_3+a_4+3) (a_2+2a_3+2a_4+5) (a_2+2a_3+a_4+4) 
(2a_1-3a_2-4a_3-2a_4-9)
\eno

\medskip

For more details, see refs. \cite{Kac78,ScS86a}.

\subsection{\sf Representations of $G(3)$}

A highest weight irreducible representation of $G(3)$ is characterized
by its Dynkin labels ($\see$ Highest weight representations) drawn on
the distinguished Dynkin diagram:
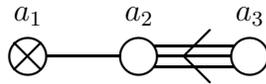
\begin{center}
\begin{picture}(100,45)
\thicklines
\multiput(0,20)(42,0){3}{\circle{14}}
\put(0,35){\makebox(0.4,0.6){$a_1$}}
\put(42,35){\makebox(0.4,0.6){$a_2$}}
\put(84,35){\makebox(0.4,0.6){$a_3$}}
\put(-5,15){\line(1,1){10}}\put(-5,25){\line(1,-1){10}}
\put(7,20){\line(1,0){28}}
\put(48,16){\line(1,0){30}}
\put(49,20){\line(1,0){28}}
\put(48,24){\line(1,0){30}}
\put(59,20){\line(1,1){10}}\put(59,20){\line(1,-1){10}}
\end{picture}
\end{center}
where $a_2,a_3$ are positive or null integers.
\\
For the $G(2)$ part, $a_2$ is the shorter root. The $sl(2)$
representation label is hidden by the odd root and its value is given
by $b = \frac{1}{2} (a_1-2a_2-3a_3)$. Since $b$ has to be a non-negative
integer, this relation implies $a_1$ to be a positive integer. Finally, a
$G(3)$ representation with $b<3$ has to satisfy a consistency condition,
that is 
\beo
&&b = 0 \qquad \qquad a_1 = a_2 = a_3 = 0 \\
&&b = 1 \qquad \qquad \mbox{not possible} \\
&&b = 2 \qquad \qquad a_2 = 0
\eno
The six atypicality conditions for the $G(3)$ representations are the 
following:
\[
\begin{array}{lll}
\bigg. a_1 = 0 & \bigbox{or} 
 & b = 0 \cr
\bigg. a_1 = a_2+1 & \bigbox{or} 
 & b = \half (1-a_2-3a_3) \cr
\bigg. a_1 = a_2+3a_3+4 & \bigbox{or} 
 & b = \half (4-a_2) \cr
\bigg. a_1 = 3a_2+3a_3+6 & \bigbox{or} 
 & b = \half (6+a_2) \cr
\bigg. a_1 = 3a_2+6a_3+9 & \bigbox{or} 
 & b = \half (4+a_2+3a_3) \cr
\bigg. a_1 = 4a_2+6a_3+10 & \bigbox{or} 
 & b = \half (10+2a_2+3a_3)
\end{array}
\]
Let us remark that the first condition corresponds to the trivial
representation and the second one is never satisfied.
\\
Moreover, a necessary (but not sufficient) condition for a
representation to be typical is that $b \ge 3$. 
\\
The dimension of a typical representation with highest weight
$\Lambda = (a_1,a_2,a_3)$ is given by
\beo
&&\dim \cV(\Lambda) = \frac{8}{15} (a_2+1) (a_3+1) (a_2+a_3+2) 
(a_2+3a_3+4) (a_2+2a_3+3) \\
&& ~~~~~~~~~~~~~~~~~~~~ (2a_2+3a_3+5) (a_1-2a_2-3a_3-5)
\eno

\medskip

For more details, see refs. \cite{Kac78,ScS86b}.

\subsection{\sf Representations of $D(2,1;\alpha)$}

A highest weight irreducible representation of $D(2,1;\alpha)$ is
characterized by its Dynkin labels ($\see$ Highest weight
representations) drawn on the distinguished Dynkin diagram:
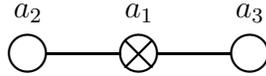
\begin{center}
\begin{picture}(100,45)
\thicklines
\multiput(0,20)(42,0){3}{\circle{14}}
\put(0,35){\makebox(0.4,0.6){$a_2$}}
\put(42,35){\makebox(0.4,0.6){$a_1$}}
\put(84,35){\makebox(0.4,0.6){$a_3$}}
\put(37,15){\line(1,1){10}}\put(37,25){\line(1,-1){10}}
\put(7,20){\line(1,0){28}}
\put(49,20){\line(1,0){28}}
\end{picture}
\end{center}
where $a_2, a_3$ are positive or null integers.
\\
The $sl(2)$ representation label is hidden by the odd root and its value
is given by $b = \frac{1}{1+\alpha} (2a_1-a_2-\alpha a_3)$, which has
to be a non-negative integer. Finally, a $D(2,1;\alpha)$ representation
with $b<2$ has to satisfy a consistency condition, that is
\beo
&&b = 0 \qquad \qquad a_1 = a_2 = a_3 = 0 \\
&&b = 1 \qquad \qquad \alpha (a_3+1) = \pm (a_2+1)
\eno
The four atypicality conditions for the $D(2,1;\alpha)$ representations 
are the following:
\[
\begin{array}{lll}
\bigg. a_1 = 0 & \bigbox{or}
 & b = 0 \cr
\bigg. a_1 = a_2+1 & \bigbox{or}
 & b = \frac{1}{1+\alpha} (2+2a_2-\alpha a_3) \cr
\bigg. a_1 = \alpha (a_3+1) & \bigbox{or}
 & b = \frac{1}{1+\alpha} (2\alpha-a_2-\alpha a_3) \cr
\bigg. a_1 = a_2+\alpha a_3+1+\alpha & \bigbox{or}
 & b = \frac{1}{1+\alpha} (2+2\alpha+a_2+\alpha a_3)
\end{array}
\]
The dimension of a typical representation with highest weight
$\Lambda = (a_1,a_2,a_3)$ is given by
\[
\dim \cV(\Lambda) = \frac{16}{1+\alpha} (a_2+1) (a_3+1)
(2a_1-a_2-\alpha a_3 - 1 - \alpha)
\]

\medskip

For more details, see refs. \cite{Kac78,VdJ85}.

\section{\sf Representations: highest weight representations}
\indent

Let $\cG = \cG_\evn \oplus \cG_\odd$ be a basic Lie superalgebra with
Cartan subalgebra $\cH$ and $\cH^*$ be the dual of $\cH$. We assume
that $\cG \ne A(n,n)$ but the following results still hold for
$sl(n+1|n+1)$. Let $\cG = \cN^+ \oplus \cH \oplus \cN^-$ be a Borel
decomposition of $\cG$ where $\cN^+$ (resp. $\cN^-$) is spanned by the
positive (resp. negative) root generators of $\cG$ ($\see$ Simple root
systems).

\medskip

\udef: A representation $\pi : \cG \rightarrow End\,\cV$ with
representation space $\cV$ is called a {\em highest weight}
representation with highest weight $\Lambda \in \cH^*$ if there exists
a non-zero vector $\vv_\Lambda \in \cV$ such that
\beo
&& \cN^+ \vv_\Lambda = 0 \\
&& h(\vv_\Lambda) = \Lambda(h) \vv_\Lambda \qquad (h \in \cH)
\eno
The $\cG$-module $\cV$ is called a highest weight module, denoted by
$\cV(\Lambda)$, and the vector $\vv_\Lambda \in \cV$ a highest weight 
vector.

\medskip

{}From now on, $\cH$ is the {\em distinguished} Cartan subalgebra ($\see$) 
of $\cG$ with basis of generators $(H_1, \dots, H_r)$ where $r = \rank\cG$ 
and $H_s$ denotes the Cartan generator associated to the odd simple root.
The Dynkin labels are defined by
\[
a_i = 2\frac{(\Lambda,\alpha_i)}{(\alpha_i,\alpha_i)} \smbox{for} i \ne s
\bigbox{and} a_s = (\Lambda,\alpha_s)
\]
A weight $\Lambda \in \cH^*$ is called a dominant weight if $a_i\ge 0$
for all $i \ne s$, integral if $a_i \in \ZZ$ for all $i \ne s$ and
integral dominant if $a_i \in \NN$ for all $i \ne s$.

\uppt: A necessary condition for the highest weight representation of
$\cG$ with highest weight $\Lambda$ to be finite dimensional is that
$\Lambda$ be an integral dominant weight.

\medskip

Following Kac (see ref. \cite{Kac78}), one defines the Kac module:

\udef: Let $\cG$ be a basic Lie superalgebra with the distinguished
$\ZZ$-gradation $\cG = \oplus_{i\in\ZZ} ~ \cG_i$ ($\see$ Classical Lie
superalgebras).
Let $\Lambda \in \cH^*$ be an integral dominant weight and $\cV_0
(\Lambda)$ be the $\cG_0$-module with highest weight $\Lambda \in
\cH^*$. Consider the $\cG$-subalgebra $\cK = \cG_0 \oplus \cN^+$ where
$\cN^+ = \oplus_{i>0} \cG_i$. The $\cG_0$-module $\cV_0(\Lambda)$ is
extended to a $\cK$-module by setting $\cN^+ \cV_0(\Lambda) = 0$. The
Kac module $\bar\cV (\Lambda)$ is defined as follows:
\\
{\em i)} if the superalgebra $\cG$ is of type I (the odd part is the
direct sum of two irreducible representations of the even part), the
Kac module is the induced module ($\see$ Representations: induced
modules)
\[
\bar\cV (\Lambda) = \ind_{\cK}^{\cG} ~ \cV_0 (\Lambda)
\]
{\em ii)} if the superalgebra $\cG$ is of type II (the odd part is an
irreducible representation of the even part), the induced module
$\ind_{\cK}^{\cG} ~ \cV_0 (\Lambda)$ contains a submodule $\cM(\Lambda)
= \cU(\cG) \cG_{-\psi}^{b+1} \cV_0(\Lambda)$ where $\psi$ is the
longest simple root of $\cG_\evn$ which is hidden behind the odd simple
root (that is the longest simple root of $sp(2n)$ in the case of
$osp(m|2n)$ and the simple root of $sl(2)$ in the case of $F(4)$,
$G(3)$ and $D(2,1;\alpha)$) and $b = 2(\Lambda,\psi)/(\psi,\psi)$ is
the component of $\Lambda$ with respect to $\psi$
($\see$ Representations: orthosymplectic superalgebras, exceptional
superalgebras for explicit expressions of $b$). The Kac module is
then defined as the quotient of the induced module $\ind_{\cK}^{\cG} ~
\cV_0 (\Lambda)$ by the submodule $\cM(\Lambda)$:
\[
\bar\cV (\Lambda) = \ind_{\cK}^{\cG} ~ \cV_0 (\Lambda) /
\cU(\cG) \cG_{-\psi}^{b+1} \cV_0(\Lambda)
\]

In case the Kac module is not simple, it contains a maximal submodule
$\cI(\Lambda)$ and the quotient module $\cV(\Lambda) = \bar\cV 
(\Lambda) / \cI(\Lambda)$ is a simple module.

The fundamental result concerning the representations of basic Lie 
superalgebras is the following:

\medskip

\uth:
\begin{itemize}
\item
Any finite dimensional irreducible representation of $\cG$ is of the
form $\cV(\Lambda) = \bar\cV (\Lambda) / \cI(\Lambda)$ where $\Lambda$
is an integral dominant weight. 
\item
Any finite dimensional simple $\cG$-module is uniquely characterized by
its integral dominant weight $\Lambda$: two $\cG$-modules $\cV(\Lambda)$ 
and $\cV(\Lambda')$ are isomorphic if and only if $\Lambda = \Lambda'$.
\item
The finite dimensional simple $\cG$-module $\cV(\Lambda) = \bar\cV
(\Lambda) / \cI(\Lambda)$ has the weight decomposition
\[
\cV(\Lambda) = \bigoplus_{\lambda\le\Lambda} \cV_\lambda 
\medbox{with}
\cV_\lambda = \ale \vv \in \cV ~ \Big\vert ~ h(\vv) = \lambda(h) \vv,\, h 
 \in \cH \ari 
\]
\end{itemize}

\section{\sf Representations: induced modules}
\indent

The method of induced representations is an elegant and powerful way to
construct the highest weight representations ($\see$) of the basic Lie
superalgebras. This section is quite formal compared to the rest of the
text but is fundamental for the representation theory of the Lie
superalgebras.

Let $\cG$ be a basic Lie superalgebra and $\cK$ be a subalgebra of
$\cG$. Denote by $\cU(\cG)$ and $\cU(\cK)$ the corresponding universal
enveloping superalgebras ($\see$). From a $\cK$-module $\cV$, it is
possible to construct a $\cG$-module in the following way. The vector
space $\cV$ is naturally extended to a $\cU(\cK)$-module. One considers
the factor space $\cU(\cG) \otimes_{\cU(\cK)} \cV$ consisting of
elements of $\cU(\cG) \otimes \cV$ such that the elements $h \otimes
\vv$ and $1 \otimes h(\vv)$ have been identified for $h \in \cK$ and
$\vv \in \cV$. This space acquires the structure of a $\cG$-module by
setting $g(u \otimes \vv)=gu \otimes \vv$ for $u \in \cU(\cG)$, $g \in
\cG$ and $\vv \in \cV$.

\udef: The $\cG$-module $\cU(\cG) \otimes_{\cU(\cK)} \cV$ is called {\em
induced module} from the $\cK$-module $\cV$ and denoted by
$\ind_{\cK}^{\cG} ~ \cV$.

\medskip

\uth: Let $\cK'$ and $\cK''$ be subalgebras of $\cG$ such that
$\cK'' \subset \cK' \subset \cG$. If $\cV$ is a $\cK''$-module, then
\[
\ind_{\cK'}^{\cG} ~ (\ind_{\cK''}^{\cK'} ~ \cV) = 
\ind_{\cK''}^{\cG} ~ \cV 
\]

\medskip

\uth: Let $\cG$ be a basic Lie superalgebra, $\cK$ be a subalgebra of 
$\cG$ such that $\cG_\evn \subset \cK$ and $\cV$ a $\cK$-module. If
$\{f_1,\dots,f_d\}$  denotes a basis of odd generators of $\cG/\cK$, then
$\ind_{\cK}^{\cG} ~ \cV = \bigoplus_{1 \le i_1 < \dots < i_k \le d} 
f_{i_1} \dots f_{i_k} \cV$ is a direct sum of subspaces and 
$\dim\ind_{\cK}^{\cG} ~ \cV = 2^d \dim\cV$.

\medskip

\uex:
Consider a basic Lie superalgebra $\cG$ of type I (the odd part is the
direct sum of two irreducible representations of the even part, that is
$\cG=sl(m|n)$ or $osp(2|2n)$) with $\ZZ$-gradation $\cG = \cG_{-1}
\oplus \cG_0 \oplus \cG_1$ ($\see$ Classical Lie superalgebras). Take
for $\cK$ the subalgebra $\cG_0 \oplus \cG_1$. Let
$\cV_0(\Lambda)$ be a $\cG_0$-module with highest weight $\Lambda$,
which is extended to a $\cK$-module by setting $\cG_1\cV_0(\Lambda)=0$.
Since $\ale \cG_{-1},\cG_{-1} \ari = 0$, only the completely
antisymmetric combinations of the generators of $\cG_{-1}$ can apply on
$\cV_0(\Lambda)$. In other words, the $\cG$-module $\cV(\Lambda)$ is
obtained by
\[
\cV = \bigwedge (\cG_{-1}) \otimes \cV_0 \simeq \cU(\cG_{-1}) \otimes
\cV_0 
\]
where 
\[
\bigwedge (\cG_{-1}) = \bigoplus_{k=0}^{\dim\cG_{-1}} \wedge^k (\cG_{-1})
\]
is the exterior algebra over $\cG_{-1}$ of dimension $2^d$ if 
$d=\dim\cG_{-1}$. 
\\
It follows that $\cV(\Lambda)$ is built from $\cV_0(\Lambda)$ by
induction of the generators of $\cG/\cK$: 
\[
\cV = \cU(\cG_{-1}) \otimes \cV_0 = \cU(\cG)
\otimes_{\cU(\cG_0\oplus\cG_1)} \cV_0 = 
\ind_{\cG_0\oplus\cG_1}^{\cG} ~ \cV_0
\]
Since $\dim \wedge^k (\cG_{-1}) = {d \choose k}$, the dimension of $\cV$ 
is given by
\[
\dim\cV(\Lambda) = \sum_{k=0}^d {d \choose k} \dim\cV_0(\Lambda) 
= 2^d \dim\cV_0(\Lambda)
\]
while its superdimension ($\see$ Representations: basic definitions) is
identically zero 
\[
\sdim\cV(\Lambda) = \sum_{k=0}^d (-1)^k {d \choose k} 
\dim\cV_0(\Lambda) = 0
\]
Let us stress that such a $\cG$-module is not always an irreducible one.

\medskip

For more details, see refs. \cite{Kac77c,Kac78,JHK89}.

\section{\sf Representations: orthosymplectic superalgebras} 
\indent

A highest weight irreducible representation of $osp(M|N)$ is
characterized by its Dynkin labels ($\see$ Highest weight
representations) drawn on the distinguished Dynkin diagram. The
different diagrams are the following:
\\
$\bullet$ $osp(2m+1|2n)$ with $\Lambda = (a_1,\dots,a_{m+n})$
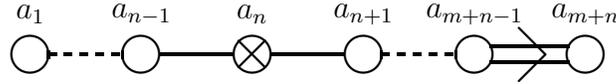
\begin{center}
\begin{picture}(220,45)
\thicklines
\multiput(0,20)(42,0){6}{\circle{14}}
\put(0,35){\makebox(0.4,0.6){$a_1$}}
\put(42,35){\makebox(0.4,0.6){$a_{n-1}$}}
\put(84,35){\makebox(0.4,0.6){$a_n$}}
\put(126,35){\makebox(0.4,0.6){$a_{n+1}$}}
\put(168,35){\makebox(0.4,0.6){$a_{m+n-1}$}}
\put(210,35){\makebox(0.4,0.6){$a_{m+n}$}}
\put(79,15){\line(1,1){10}}\put(79,25){\line(1,-1){10}}
\put(7,20){\dashbox{3}(28,0)}
\put(49,20){\line(1,0){28}}
\put(91,20){\line(1,0){28}}
\put(133,20){\dashbox{3}(28,0)}
\put(174,17){\line(1,0){30}}
\put(174,23){\line(1,0){30}}
\put(195,20){\line(-1,1){10}}\put(195,20){\line(-1,-1){10}}
\end{picture}
\end{center}
$\bullet$ $osp(2|2n)$ with $\Lambda = (a_1,\dots,a_{n+1})$
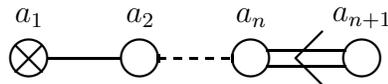
\begin{center}
\begin{picture}(140,45)
\thicklines
\multiput(0,20)(42,0){4}{\circle{14}}
\put(0,35){\makebox(0.4,0.6){$a_1$}}
\put(42,35){\makebox(0.4,0.6){$a_2$}}
\put(84,35){\makebox(0.4,0.6){$a_n$}}
\put(126,35){\makebox(0.4,0.6){$a_{n+1}$}}
\put(-5,15){\line(1,1){10}}\put(-5,25){\line(1,-1){10}}
\put(7,20){\line(1,0){28}}
\put(49,20){\dashbox{3}(28,0)}
\put(101,20){\line(1,1){10}}\put(101,20){\line(1,-1){10}}
\put(90,17){\line(1,0){30}}
\put(90,23){\line(1,0){30}}
\end{picture}
\end{center}
$\bullet$ $osp(2m|2n)$ with $\Lambda = (a_1,\dots,a_{m+n})$
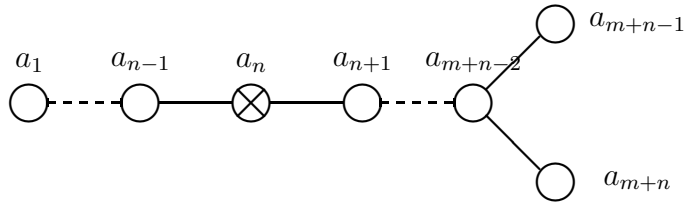
\begin{center}
\begin{picture}(260,80)
\thicklines
\multiput(0,40)(42,0){5}{\circle{14}}
\put(0,55){\makebox(0.4,0.6){$a_1$}}
\put(42,55){\makebox(0.4,0.6){$a_{n-1}$}}
\put(84,55){\makebox(0.4,0.6){$a_n$}}
\put(126,55){\makebox(0.4,0.6){$a_{n+1}$}}
\put(168,55){\makebox(0.4,0.6){$a_{m+n-2}$}}
\put(79,35){\line(1,1){10}}\put(79,45){\line(1,-1){10}}
\put(7,40){\dashbox{3}(28,0)}
\put(49,40){\line(1,0){28}}
\put(91,40){\line(1,0){28}}
\put(133,40){\dashbox{3}(28,0)}
\put(173,45){\line(1,1){20}}\put(173,35){\line(1,-1){20}}
\put(199,70){\circle{14}}
\put(230,70){\makebox(0.4,0.6){$a_{m+n-1}$}}
\put(199,10){\circle{14}}
\put(230,10){\makebox(0.4,0.6){$a_{m+n}$}}
\end{picture}
\end{center}
$\bullet$ $osp(1|2n)$ with $\Lambda = (a_1,\dots,a_n)$
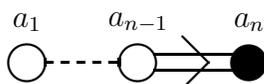
\begin{center}
\begin{picture}(100,45)
\thicklines
\put(0,20){\circle{14}}
\put(42,20){\circle{14}}
\put(84,20){\circle*{14}}
\put(0,35){\makebox(0.4,0.6){$a_1$}}
\put(42,35){\makebox(0.4,0.6){$a_{n-1}$}}
\put(84,35){\makebox(0.4,0.6){$a_n$}}
\put(7,20){\dashbox{3}(28,0)}
\put(48,17){\line(1,0){30}}
\put(48,23){\line(1,0){30}}
\put(69,20){\line(-1,1){10}}\put(69,20){\line(-1,-1){10}}
\end{picture}
\end{center}
Notice that the superalgebra $osp(2|2n)$ is of type I, while the
superalgebras $osp(2m+1|2n)$ and $osp(2m|2n)$ are of type II: in the
first case, the odd part is the direct sum of two irreducible
representations of the even part, in the second case it is an
irreducible representation of the even part. The numbers $a_i$ are
constrained to satisfy the following conditions:
\\
\phantom{~~~~~} $a_n$ is integer or half-integer for $osp(2m+1|2n)$ and
$osp(2m|2n)$, 
\\
\phantom{~~~~~} $a_1$ is an arbitrary complex number for $osp(2|2n)$.
\\
The coordinates of $\Lambda$ in the root space characterize a $so(M)
\oplus sp(2n)$ representation ($M = 2m$ or $M = 2m+1$). The $so(M)$
representation can be directly read on the Kac-Dynkin diagram, but the
longest simple root of $sp(2n)$ is hidden behind the odd simple roots.
{}From the knowledge of $(a_n,\dots,a_{m+n})$, it is possible to deduce
the component $b$ that $\Lambda$ would have with respect to the longest
simple root:
\\
\phantom{~~~~~} in the $osp(2m+1|2n)$ case, one has
$b = a_n - a_{n+1} - \dots - a_{m+n-1} - \frac{1}{2} a_{m+n}$
\\
\phantom{~~~~~} in the $osp(2m|2n)$ case, one has
$b = a_n - a_{n+1} - \dots - a_{m+n-2} - \frac{1}{2}
(a_{m+n-1} + a_{m+n})$
Notice that the number $b$ has to be a non-negative integer.
\\
The highest weight of a finite representation of $osp(M|2n)$ belongs 
therefore to a $so(M) \oplus sp(2n)$ representation and thus one must 
have the following consistency conditions: 
\beo
&& b \ge 0 \\
&& \smbox{for} osp(2m+1|2n), a_{n+b+1} = \dots = a_{n+m} = 0 
\smbox{if} b \le m-1 \\
&& \smbox{for} osp(2m|2n), a_{n+b+1} = \dots = a_{n+m} = 0 
\smbox{if} b \le m-2 \smbox{and} a_{n+m-1} = a_{n+m} \smbox{if} b = m-1
\eno
We give hereafter the atypicality conditions of the representations
for the superalgebras of the orthosymplectic series. If at least one
of these conditions is satisfied, the representation is an atypical
one. Otherwise, the representation is typical, the dimension of which 
is given by the number $\dim \cV(\Lambda)$.
\\
$\bullet$ superalgebras $osp(2m+1|2n)$
\\
The atypicality conditions are
\beo
&&\sum_{k = i}^n a_k - \sum_{k = n+1}^j a_k + 2n - i - j = 0 \\
&&\sum_{k = i}^n a_k - \sum_{k = n+1}^j a_k - 2 \sum_{k = j+1}^{m+n-1} 
a_k - a_{m+n} - 2m - i + j + 1 = 0 \\
&&\medbox{with} 1 \le i \le n \le j \le m+n-1
\eno
The dimensions of the typical representations are given by
\beo
\lefteqn{\dim\left(~~~ 
\begin{picture}(230,20)
\thicklines
\multiput(0,0)(42,0){6}{\circle{14}}
\put(0,20){\makebox(0.4,0.6){$a_1$}}
\put(42,20){\makebox(0.4,0.6){$a_{n-1}$}}
\put(84,20){\makebox(0.4,0.6){$a_n$}}
\put(126,20){\makebox(0.4,0.6){$a_{n+1}$}}
\put(168,20){\makebox(0.4,0.6){$a_{n+m-1}$}}
\put(210,20){\makebox(0.4,0.6){$a_{n+m}$}}
\put(79,-5){\line(1,1){10}}\put(79,5){\line(1,-1){10}}
\put(7,0){\dashbox{3}(28,0)}
\put(49,0){\line(1,0){28}}
\put(91,0){\line(1,0){28}}
\put(133,0){\dashbox{3}(28,0)}
\put(174,-3){\line(1,0){30}}
\put(174,3){\line(1,0){30}}
\put(195,0){\line(-1,1){10}}\put(195,0){\line(-1,-1){10}}
\end{picture}
\right) 
=} \\
&& \\
&& 2^{(2m+1)n} ~ \times ~ \dim\left(~~~ 
\begin{picture}(110,20)
\thicklines
\put(0,0){\circle{14}}
\put(42,0){\circle{14}}
\put(84,0){\circle{14}}
\put(0,20){\makebox(0.4,0.6){$a_1$}}
\put(42,20){\makebox(0.4,0.6){$a_{n-1}$}}
\put(84,20){\makebox(0.4,0.6){$b-m-\half$}}
\put(7,0){\dashbox{3}(28,0)}
\put(48,-3){\line(1,0){30}}
\put(48,3){\line(1,0){30}}
\put(69,0){\line(-1,1){10}}\put(69,0){\line(-1,-1){10}}
\end{picture}
\right)
~ \times ~ \dim\left(~~~ 
\begin{picture}(100,20)
\thicklines
\put(0,0){\circle{14}}
\put(42,0){\circle{14}}
\put(84,0){\circle{14}}
\put(0,20){\makebox(0.4,0.6){$a_1$}}
\put(42,20){\makebox(0.4,0.6){$a_{n+m-1}$}}
\put(84,20){\makebox(0.4,0.6){$a_{n+m}$}}
\put(7,0){\dashbox{3}(28,0)}
\put(48,-3){\line(1,0){30}}
\put(48,3){\line(1,0){30}}
\put(59,0){\line(1,1){10}}\put(59,0){\line(1,-1){10}}
\end{picture}
\right)
\eno
that is    
\beo
&&\dim \cV(\Lambda) = 2^{(2m+1)n} \prod_{1 \le i < j \le n-1}
\frac{j-i+1 + \sum_{k=i}^j a_k}{j-i+1}
\prod_{n+1 \le i \le j \le m+n-1}
\frac{j-i+1 + \sum_{k=i}^j a_k}{j-i+1} \\
&&\prod_{1 \le i \le j \le n}
\frac{\sum_{k=i}^{j-1} a_k + 2(\sum_{k=j}^{n} a_k - \sum_{k=n+1}^{m+n-1}
a_k) - a_{m+n} + 2n-2m-i-j+1}{2n-i-j+2} \\ 
&&\prod_{n+1 \le i \le j \le m+n-1} 
\frac{\sum_{k=i}^{j-1} a_k + 2 \sum_{k=j}^{m+n-1} a_k + a_{m+n}
+ 2m-i-j+1}{2m-i-j+1} 
\eno
$\bullet$ superalgebras $osp(2|2n)$
\\
The atypicality conditions are
\beo
&&a_1 - \sum_{k = 2}^i a_k - i + 1 = 0 \\
&&a_1 - \sum_{k = 2}^i a_k - 2 \sum_{k = i+1}^{n+1} 
a_k - 2n + i - 1 = 0 \\
&&\medbox{with} 1 \le i \le n
\eno
The dimensions of the typical representations are given by
\[
\dim\left(~~~ 
\begin{picture}(140,20)
\thicklines
\multiput(0,0)(42,0){4}{\circle{14}}
\put(0,20){\makebox(0.4,0.6){$a_1$}}
\put(42,20){\makebox(0.4,0.6){$a_2$}}
\put(84,20){\makebox(0.4,0.6){$a_n$}}
\put(126,20){\makebox(0.4,0.6){$a_{n+1}$}}
\put(-5,-5){\line(1,1){10}}\put(-5,5){\line(1,-1){10}}
\put(7,0){\line(1,0){28}}
\put(49,0){\dashbox{3}(28,0)}
\put(101,0){\line(1,1){10}}\put(101,0){\line(1,-1){10}}
\put(90,-3){\line(1,0){30}}
\put(90,3){\line(1,0){30}}
\end{picture}
\right)
=
2^n ~ \times ~ \dim\left(~~~ 
\begin{picture}(100,20)
\thicklines
\put(0,0){\circle{14}}
\put(42,0){\circle{14}}
\put(84,0){\circle{14}}
\put(0,20){\makebox(0.4,0.6){$a_2$}}
\put(42,20){\makebox(0.4,0.6){$a_n$}}
\put(84,20){\makebox(0.4,0.6){$a_{n+1}$}}
\put(7,0){\dashbox{3}(28,0)}
\put(48,-3){\line(1,0){30}}
\put(48,3){\line(1,0){30}}
\put(59,0){\line(1,1){10}}\put(59,0){\line(1,-1){10}}
\end{picture}
\right)
\]
that is
\beo
\dim \cV(\Lambda) & = & 2^{2n} \prod_{2 \le i \le j \le n}
\frac{a_i + \dots + a_j + j-i+1}{j-i+1} \\
&&\prod_{2 \le i \le j \le n+1}
\frac{a_i + \dots + a_{j-1} + 2a_j + \dots + 2a_{n+1}}
{2n-i-j+4}
\eno
$\bullet$ superalgebras $osp(2m|2n)$
\\
The atypicality conditions are
\beo
&&\sum_{k = i}^n a_k - \sum_{k = n+1}^j a_k + 2n - i - j = 0 \\
&&\medbox{with} 1 \le i \le n \le j \le m+n-1 \\
&&\sum_{k = i}^n a_k - \sum_{k = n+1}^{m+n-2} 
a_k - a_{m+n} + n - m - i + 1 = 0 \\
&&\medbox{with} 1 \le i \le n \\
&&\sum_{k = i}^n a_k - \sum_{k = n+1}^j a_k - 2 \sum_{k = j+1}^{m+n-2} 
a_k - a_{m+n-1} - a_{m+n} - 2m - i + j + 2 = 0 \\
&&\medbox{with} 1 \le i \le n \le j \le m+n-2
\eno
The dimensions of the typical representations are given by 
\beo
\lefteqn{\dim\left(~~~ 
\begin{picture}(260,20)
\thicklines
\multiput(0,0)(42,0){5}{\circle{14}}
\put(0,20){\makebox(0.4,0.6){$a_1$}}
\put(42,20){\makebox(0.4,0.6){$a_{n-1}$}}
\put(84,20){\makebox(0.4,0.6){$a_n$}}
\put(126,20){\makebox(0.4,0.6){$a_{n+1}$}}
\put(168,20){\makebox(0.4,0.6){$a_{m+n-2}$}}
\put(79,-5){\line(1,1){10}}\put(79,5){\line(1,-1){10}}
\put(7,0){\dashbox{3}(28,0)}
\put(49,0){\line(1,0){28}}
\put(91,0){\line(1,0){28}}
\put(133,0){\dashbox{3}(28,0)}
\put(173,5){\line(2,1){20}}\put(173,-5){\line(2,-1){20}}
\put(199,20){\circle{14}}
\put(230,20){\makebox(0.4,0.6){$a_{n+m-1}$}}
\put(199,-20){\circle{14}}
\put(230,-20){\makebox(0.4,0.6){$a_{n+m}$}}
\end{picture}
\right)
=} \\
&& \\
&& \\
&& 2^{2mn} ~ \times ~ \dim\left(~~~ 
\begin{picture}(105,20)
\thicklines
\put(0,0){\circle{14}}
\put(42,0){\circle{14}}
\put(84,0){\circle{14}}
\put(0,20){\makebox(0.4,0.6){$a_1$}}
\put(42,20){\makebox(0.4,0.6){$a_{n-1}$}}
\put(84,20){\makebox(0.4,0.6){$b-m$}}
\put(7,0){\dashbox{3}(28,0)}
\put(48,-3){\line(1,0){30}}
\put(48,3){\line(1,0){30}}
\put(59,0){\line(1,1){10}}\put(59,0){\line(1,-1){10}}
\end{picture}
\right) ~ \times ~ \dim\left(~~~ 
\begin{picture}(125,20)
\thicklines
\put(0,0){\circle{14}}
\put(42,0){\circle{14}}
\put(0,20){\makebox(0.4,0.6){$a_{n+1}$}}
\put(42,20){\makebox(0.4,0.6){$a_{n+m-2}$}}
\put(7,0){\dashbox{3}(28,0)}
\put(47,5){\line(2,1){20}}\put(47,-5){\line(2,-1){20}}
\put(73,20){\circle{14}}
\put(104,20){\makebox(0.4,0.6){$a_{n+m-1}$}}
\put(73,-20){\circle{14}}
\put(104,-20){\makebox(0.4,0.6){$a_{n+m}$}}
\end{picture}
\right)
\eno
and one obtains the same formula for $\dim \cV(\Lambda)$ as for
$osp(2m+1|2n)$. 
\\
$\bullet$ superalgebras $osp(1|2n)$
\\
The superalgebras $osp(1|2n)$ carry the property of having only typical 
representation (the Dynkin diagram of $osp(1|2n)$ does not contain any 
grey dot). One has 
\beo
\dim \cV(\Lambda) & = & \prod_{1 \le i < j \le n}
\frac{a_i + \dots + a_j + 2(a_{j+1} + \dots + a_{n-1}) + a_n 
+ 2n-i-j}{2n-i-j} \\
&&\prod_{1 \le i \le n} 
\frac{2(a_i + \dots + a_{n-1}) + a_n + 2n-2i+1}{2n-2i+1} 
\eno
Moreover, the representations of $osp(1|2n)$ can be put in a one-to-one
correspondence with those of $so(2n+1)$ \cite{RiS82}. More precisely,
one has
\[
\dim\left(~~~ 
\begin{picture}(100,20)
\thicklines
\put(0,0){\circle{14}}
\put(42,0){\circle{14}}
\put(84,0){\circle*{14}}
\put(0,15){\makebox(0.4,0.6){$a_1$}}
\put(42,15){\makebox(0.4,0.6){$a_{n-1}$}}
\put(84,15){\makebox(0.4,0.6){$a_n$}}
\put(7,0){\dashbox{3}(28,0)}
\put(48,-3){\line(1,0){30}}
\put(48,3){\line(1,0){30}}
\put(69,0){\line(-1,1){10}}\put(69,0){\line(-1,-1){10}}
\end{picture}
\right)
=
\dim\left(~~~ 
\begin{picture}(100,20)
\thicklines
\put(0,0){\circle{14}}
\put(42,0){\circle{14}}
\put(84,0){\circle{14}}
\put(0,15){\makebox(0.4,0.6){$a_1$}}
\put(42,15){\makebox(0.4,0.6){$a_{n-1}$}}
\put(84,15){\makebox(0.4,0.6){$a_n$}}
\put(7,0){\dashbox{3}(28,0)}
\put(48,-3){\line(1,0){30}}
\put(48,3){\line(1,0){30}}
\put(69,0){\line(-1,1){10}}\put(69,0){\line(-1,-1){10}}
\end{picture}
\right)
\]
as well as
\[
\sdim\left(~~~ 
\begin{picture}(100,20)
\thicklines
\put(0,0){\circle{14}}
\put(42,0){\circle{14}}
\put(84,0){\circle*{14}}
\put(0,15){\makebox(0.4,0.6){$a_1$}}
\put(42,15){\makebox(0.4,0.6){$a_{n-1}$}}
\put(84,15){\makebox(0.4,0.6){$a_n$}}
\put(7,0){\dashbox{3}(28,0)}
\put(48,-3){\line(1,0){30}}
\put(48,3){\line(1,0){30}}
\put(69,0){\line(-1,1){10}}\put(69,0){\line(-1,-1){10}}
\end{picture}
\right)
=
\frac{1}{2^{n-1}} ~ \dim\left(~~~ 
\begin{picture}(125,20)
\thicklines
\put(0,0){\circle{14}}
\put(42,0){\circle{14}}
\put(0,20){\makebox(0.4,0.6){$a_1$}}
\put(42,20){\makebox(0.4,0.6){$a_{n-2}$}}
\put(7,0){\dashbox{3}(28,0)}
\put(47,5){\line(2,1){20}}\put(47,-5){\line(2,-1){20}}
\put(73,20){\circle{14}}
\put(104,20){\makebox(0.4,0.6){$a_{n-1}$}}
\put(73,-20){\circle{14}}
\put(104,-15){\makebox(0.4,0.6){$a_{n-1}+$}}
\put(104,-25){\makebox(0.4,0.6){$a_n+1$}}
\end{picture}
\right)
\]

\vspace{5mm}
\noindent
(let us recall that $\dim \cV = \dim\cV_\evn + \dim\cV_\odd$ while 
$\sdim \cV = \dim\cV_\evn - \dim\cV_\odd$).

\medskip

For more details, see refs. \cite{Hur87,HuM82,Kac78,MSS85,RiS82}.

\section{\sf Representations: reducibility}
\indent

\udef: Let $\cG$ be a classical Lie superalgebra. A representation $\pi :
\cG \rightarrow End\,\cV$ is called {\em irreducible} if the $\cG$-module
$\cV$ has no $\cG$-submodules except trivial ones. The $\cG$-module
$\cV$ is then called a {\em simple} module. Otherwise the
representation $\pi$ is said {\em reducible}. In that case, one has
$\cV = \cV' \oplus \cV''$, $\cV''$ being a complementary subspace of
$\cV'$ in $\cV$ and the $\cG$-submodule $\cV'$ is an invariant subspace
under $\pi$. If the subspace $\cV''$ is also an invariant subspace
under $\pi$, the representation $\pi$ is said {\em completely
reducible}. The $\cG$-module $\cV$ is then called a {\em semi-simple}
module.

\medskip

\udef: Two representations $\pi$ and $\pi'$ of $\cG$ being given, with
representation spaces $\cV$ and $\cV'$, one defines the direct sum $\pi
\oplus \pi'$ with representation space $\cV \oplus \cV'$ and the direct
(or tensor) product $\pi \otimes \pi'$ with representation space $\cV
\otimes \cV'$ of the two representations. The action of the
representations $\pi \oplus \pi'$ and $\pi \otimes \pi'$ on the
corresponding representation spaces is given by, for $X \in \cG$, $\vv
\in \cV$ and $\vv' \in \cV'$:
\beo
&& (\pi\oplus\pi')(X) \vv \oplus \vv' = \pi(X) \vv ~ \oplus ~ \pi'(X)
\vv' \\ 
&& (\pi\otimes\pi')(X) \vv \otimes \vv' = \pi(X) \vv ~ \otimes ~ \vv' +
\vv ~ \otimes ~ \pi'(X) \vv' 
\eno

\medskip

The representations $\pi$ and $\pi'$ of $\cG$ being irreducible, the
tensor product $\pi \otimes \pi'$ is a representation which is in
general reducible. Notice however that, contrary to the Lie algebra
case, in the Lie superalgebra case, the tensor product of two
irreducible representations is not necessary completely reducible. In
fact, one has the following theorem: 

\medskip

\uth\, (Djokovic-Hochschild): The only Lie superalgebras for which all
finite dimensional representations are completely reducible are the
direct products of $osp(1|2n)$ superalgebras and semi-simple Lie
algebras.

\section{\sf Representations: star and superstar representations} 
\indent

The star and superstar representations of a classical Lie superalgebra 
are the generalization of the hermitian representations of a simple 
Lie algebra. The importance of the hermitian representations for 
simple Lie algebras comes from the fact that the finite dimensional
representations of a compact simple Lie algebra are equivalent to
hermitian representations.

Let $\cG = \cG_\evn \oplus \cG_\odd$ be a classical Lie superalgebra.
One can define two different adjoint operations as follows.

\udef: An adjoint operation in $\cG$, denoted by $\dagger$, is a mapping
from $\cG$ into $\cG$ such that: 

i) $X \in \cG_i ~ \Rightarrow ~ X^\dagger \in \cG_i$ for 
$i = \evn,\odd$,

ii) $(\alpha X + \beta Y)^\dagger = \balpha X^\dagger 
+ \bbeta Y^\dagger$,

iii) $\sle X,Y \sri ^\dagger = \sle Y^\dagger,X^\dagger \sri$,

iv) $(X^\dagger)^\dagger = X$,
\\
where $X,Y \in \cG$ and $\alpha,\beta \in \CC$.

\medskip

\udef: A superadjoint operation in $\cG$, denoted by $\ddagger$, is a
mapping from $\cG$ into $\cG$ such that:

i) $X \in \cG_i ~ \Rightarrow ~ X^\ddagger \in \cG_i$ for
$i = \evn,\odd$,

ii) $(\alpha X + \beta Y)^\ddagger = \balpha X^\ddagger 
+ \bbeta Y^\ddagger$,

iii) $\sle X,Y \sri ^\ddagger = (-1)^{\deg X.\deg Y}
\sle Y^\ddagger,X^\ddagger \sri$,

iv) $(X^\ddagger)^\ddagger = (-1)^{\deg X} X$,
\\
where $X,Y \in \cG$ and $\alpha,\beta \in \CC$.

\medskip

The definitions of the star and superstar representations follow
immediately. 

\medskip

\udef: Let $\cG$ be a classical Lie superalgebra and $\pi$ a
representation of $\cG$ acting in a $\ZZ_2$-graded vector space $\cV$. 
Then $\pi$ is a star representation of $\cG$ if $\pi(X^\dagger) = 
\pi(X)^\dagger$ and a superstar representation of $\cG$ if 
$\pi(X^\ddagger) = \pi(X)^\ddagger$ for all $X \in \cG$. 

\medskip

The following properties hold:
\medskip

\uppt:
\begin{enumerate}
\item
Any star representation $\pi$ of $\cG$ in a graded Hilbert space $\cV$ 
is completely reducible. 
\item
Any superstar representation $\pi$ of $\cG$ in a graded Hilbert space 
$\cV$ is completely reducible. 
\item
The tensor product $\pi\otimes\pi'$ of two star representations (resp. 
to superstar representations) $\pi$ and $\pi'$ is a star representation 
(resp. a superstar representation).
\item
The tensor product $\pi\otimes\pi'$ of two star representations $\pi$ 
and $\pi'$ is completely reducible.
\end{enumerate}
Let us emphasize that the last property does not hold for superstar 
representations (that is the tensor product of two superstar 
representations is in general not completely reducible).

\medskip

The classes of star and superstar representations of the classical Lie 
superalgebras are the following:
\\
- the superalgebra $A(m,n)$ has two classes of star representations
and two classes of superstar representations.
\\
- the superalgebras $B(m,n)$ and $D(m,n)$ have two classes of
superstar representations.
\\
- the superalgebra $C(n+1)$ has either two classes of star
representations and two classes of superstar representations, or one
class of superstar representations, depending on the definition of the 
adjoint operation in the Lie algebra part.
\\
- the superalgebras $F(4)$ and $G(3)$ have two classes of superstar 
representations.
\\
- the superalgebra $P(n)$ has neither star nor superstar representations.
\\
- the superalgebra $Q(n)$ has two classes of star representations.

\medskip

For more details, see ref. \cite{NRS77}.

\section{\sf Representations: typicality and atypicality} 
\indent

Any representation of a basic Lie superalgebra $\cG = \cG_\evn \oplus
\cG_\odd$ can be decomposed into a direct sum of irreducible
representations of the even subalgebra $\cG_\evn$. The generators
associated to the odd roots will transform a vector basis belonging to
a certain representation of $\cG_\evn$ into a vector in another
representation of $\cG_\evn$ (or into the null vector), while the
generators associated to the even roots will operate inside an
irreducible representation of $\cG_\evn$.

The presence of odd roots will have another important consequence in the
representation theory of superalgebras. Indeed, one might find in
certain representations weight vectors, different from the highest one
specifying the representation, annihilated by all the generators
corresponding to positive roots. Such vector have, of course, to be
decoupled from the representation. Representations of this kind are
called atypical, while the other irreducible representations not
suffering this pathology are called typical.

\medskip

More precisely, let $\cG = \cG_\evn \oplus \cG_\odd$ be a basic Lie
superalgebra with distinguished Cartan subalgebra $\cH$. Let $\Lambda 
\in \cH^*$ be an integral dominant weight.
Denote the root system of $\cG$ by $\Delta = \Delta_\evn \cup
\Delta_\odd$. One defines $\overline{\Delta}_\evn$ as the subset of
roots $\alpha \in \Delta_\evn$ such that $\alpha/2 \notin \Delta_\odd$
and $\overline{\Delta}_\odd$ as the subset of roots $\alpha \in
\Delta_\odd$ such that $2\alpha \notin \Delta_\evn$. Let $\rho_0$ be the
half-sum of the roots of $\Delta_\evn^+$, $\bar\rho_0$ the half-sum of
the roots of $\bar\Delta_\evn^+$, $\rho_1$ the half-sum of the roots of
$\Delta_\odd^+$, and $\rho = \rho_0-\rho_1$.

\medskip

\udef: The representation $\pi$ with highest weight $\Lambda$ is called
{\em typical} if 
\[
(\Lambda + \rho,\alpha) \ne 0 \qquad \smbox{for all} \alpha \in 
\overline{\Delta}_\odd^+
\]
The highest weight $\Lambda$ is then called typical.
\\
If there exists some $\alpha \in \overline{\Delta}_\odd^+$ such that
$(\Lambda + \rho,\alpha) = 0$, the representation $\pi$ and the highest
weight $\Lambda$ are called {\em atypical}. The number of distinct
elements $\alpha \in \overline{\Delta}_\odd^+$ for which $\Lambda$ is 
atypical is the degree of atypicality of the representation $\pi$.
If there exists one and only one $\alpha \in \overline{\Delta}_\odd^+$
such that $(\Lambda + \rho,\alpha) = 0$, the representation $\pi$ and
the highest weight $\Lambda$ are called {\em singly atypical}. 

\medskip

Denoting as before $\bar\cV (\Lambda)$ the Kac module ($\see$
Representations: highest weight representations) corresponding to the
integral dominant weight $\Lambda$, one has the following theorem:

\medskip

\uth: The Kac module $\bar\cV(\Lambda)$ is a simple $\cG$-module if and
only if the highest weight $\Lambda$ is typical.

\medskip

\uppts:
\\
1) All the finite dimensional representations of $B(0,n)$ are typical.
\\
2) All the finite dimensional representations of $C(n+1)$ are either 
typical or singly atypical.

\medskip

Let $\cV$ be a typical finite dimensional representation of $\cG$. 
Then the dimension of $\cV(\Lambda)$ is given by 
\[
\dim \cV(\Lambda) = 2^{\dim \Delta_\odd^+} \prod_{\alpha \in 
\Delta_\evn^+}\frac{(\Lambda+\rho,\alpha)}
{(\rho_0,\alpha)} 
\]
and
\beo
&&\dim \cV_\evn(\Lambda) - \dim \cV_\odd(\Lambda) = 0
\medbox{if} \cG \ne B(0,n) \\
&&\dim \cV_\evn(\Lambda) - \dim \cV_\odd(\Lambda) = \prod_{\alpha \in 
\bar\Delta_\evn^+}\frac{(\Lambda+\rho,\alpha)}{(\bar\rho_0,\alpha)}
= \prod_{\alpha \in \bar\Delta_\evn^+} 
\frac{(\Lambda+\rho,\alpha)}{(\rho,\alpha)} \medbox{if} \cG = B(0,n)
\eno

It follows that the fundamental representations of $sl(m|n)$ and
$osp(m|n)$ (of dimension $m+n$) and the adjoint representations of the
basic Lie superalgebras $\cG \ne sl(1|2), osp(1|2n)$ (of dimension
$\dim\cG$) are {\em atypical} ones (since $\dim\cV_\evn - \dim\cV_\odd
\ne 0$).

\medskip

For more details, see refs. \cite{Kac77c,Kac78}.

\section{\sf Representations: unitary superalgebras}
\indent

A highest weight irreducible representation of $sl(m|n)$ is
characterized by its Dynkin labels ($\see$ Highest weight
representations) drawn on the distinguished Dynkin diagram. The
different diagrams are the following:
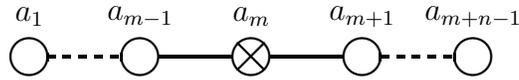
\begin{center}
\begin{picture}(180,45)
\thicklines
\multiput(0,20)(42,0){5}{\circle{14}}
\put(0,35){\makebox(0.4,0.6){$a_1$}}
\put(42,35){\makebox(0.4,0.6){$a_{m-1}$}}
\put(84,35){\makebox(0.4,0.6){$a_m$}}
\put(126,35){\makebox(0.4,0.6){$a_{m+1}$}}
\put(168,35){\makebox(0.4,0.6){$a_{m+n-1}$}}
\put(79,15){\line(1,1){10}}\put(79,25){\line(1,-1){10}}
\put(7,20){\dashbox{3}(28,0)}
\put(49,20){\line(1,0){28}}
\put(91,20){\line(1,0){28}}
\put(133,20){\dashbox{3}(28,0)}
\end{picture}
\end{center}
The numbers $a_i$ are constrained:
\\
~~~~~ $a_i$ are non-negative integer for 
$i = 1, \dots, m-1, m+1, \dots, m+n-1$,
\\
~~~~~ $a_m$ is an arbitrary real number.
\\
For the atypical representations, the numbers $a_i$ have to satisfy
one of the following atypicality conditions:
\[
\sum_{k = m+1}^j a_k - \sum_{k = 1}^{m-1} a_k - a_m - 2m + i - j = 0
\]
with $1 \le i \le m \le j \le m+n-1$.

Otherwise, the representation under consideration is a typical one.
Then its dimension is given by
\beo
\dim \cV(\Lambda) & = & 2^{(m+1)(n+1)} \prod_{1 \le i \le j \le m}
\frac{a_i + \dots + a_j + j-i+1}{j-i+1} \\
&&\prod_{m+2 \le i \le j \le m+n+1}
\frac{a_i + \dots + a_j + j-i+1}{j-i+1}
\eno

\medskip

For more details, see refs. \cite{Hur87,HuM83,Kac78}.

\section{\sf Roots, root systems}
\indent

Let $\cG = \cG_\evn \oplus \cG_\odd$ be a classical Lie superalgebra of
dimension $n$. Let $\cH$ be a Cartan subalgebra of $\cG$. The 
superalgebra $\cG$ can be decomposed as follows:
\[
\cG = \bigoplus_{\alpha} \cG_{\alpha}
\]
where
\[
\cG_{\alpha} = \ale x \in \cG ~ \Big\vert ~ \cle h,x \cri = 
\alpha (h) x,\, h \in \cH \ari 
\]
The set 
\[
\Delta = \ale \alpha \in \cH^* ~ \Big\vert ~ \cG_{\alpha} \ne 0 \ari 
\]
is by definition the {\em root system} of $\cG$. A root $\alpha$ is
called even (resp. odd) if $\cG_\alpha \cap \cG_\evn \ne \emptyset$
(resp. $\cG_\alpha \cap \cG_\odd \ne \emptyset$). The set of even roots
is denoted by $\Delta_\evn$ : it is the root system of the even part
$\cG_\evn$ of $\cG$. The set of odd roots is denoted by $\Delta_\odd$ :
it is the weight system of the representation of $\cG_\evn$ in
$\cG_\odd$. One has $\Delta = \Delta_\evn \cup \Delta_\odd$. Notice
that a root can be both even and odd (however this only occurs in the
case of the superalgebra $Q(n)$). The vector space spanned by all the
possible roots is called the root space. It is the dual $\cH^*$ of the
Cartan subalgebra $\cH$ as vector space. 
\\
Except for $A(1,1)$, $P(n)$ and $Q(n)$, using the invariant bilinear
form defined on the superalgebra $\cG$, one can define a bilinear form
on the root space $\cH^*$ by $(\alpha_i,\alpha_j) = (H_i,H_j)$ where
the $H_i$ form a basis of $\cH$ ($\see$ Cartan matrix, Killing form).

One has the following properties.

\medskip

\uppts:
\begin{enumerate}
\item
$\cG_{(\alpha = 0)} = \cH$ except for $Q(n)$.
\item
$\dim \cG_{\alpha} = 1$ when $\alpha \ne 0$ except for $A(1,1)$, 
$P(2)$, $P(3)$ and $Q(n)$.
\item
Except for $A(1,1)$, $P(n)$ and $Q(n)$, one has
\begin{itemize}
\item
$\sle \cG_{\alpha},\cG_{\beta} \sri \ne 0$ if and only if 
$\alpha,\beta,\alpha+\beta \in \Delta$
\item
$( \cG_{\alpha},\cG_{\beta} ) = 0$ for $\alpha+\beta \ne 0$
\item
if $\alpha \in \Delta$ (resp. $\Delta_\evn$, $\Delta_\odd$),
then $-\alpha \in \Delta$ (resp. $\Delta_\evn$, $\Delta_\odd$)
\item
$\alpha \in \Delta \Longrightarrow 2\alpha \in \Delta$ if and 
only if $\alpha \in \Delta_\odd$ and $(\alpha,\alpha) \ne 0$
\end{itemize}
\end{enumerate}

\medskip

The roots of a basic Lie superalgebra do not satisfy many properties of
the roots of a simple Lie algebra. In particular, the bilinear form on
$\cH^*$ has in general pseudo-euclidean signature (except in the case
of $B(0,n)$). The roots of a basic Lie superalgebra can be classified 
into three classes:
\\
- roots $\alpha$ such that $(\alpha,\alpha) \ne 0$ and $2\alpha$ is
not a root. Such roots will be called even or bosonic roots.
\\
- roots $\alpha$ such that $(\alpha,\alpha) \ne 0$ and $2\alpha$
is still a root (of bosonic type). Such roots will be called odd or
fermionic roots of non-zero length.
\\
- roots $\alpha$ such that $(\alpha,\alpha) = 0$. Such roots will
be called odd or fermionic roots of zero length (or also isotropic odd
roots).

The root systems of the basic Lie superalgebras are given in Table 
\ref{table15}.

\begin{table}[htbp]
\centering
\begin{tabular}{|c|c|c|}
\hline
superalgebra & even root system $\Delta_\evn$ & odd root system
$\Delta_\odd$ \\ \hline
$\bigg. A(m-1,n-1)$ 
&$\eps_i-\eps_j, ~ \del_i-\del_j$ &$\pm(\eps_i-\del_j)$ \\ 
$\bigg. B(m,n)$
&$\pm\eps_i\pm\eps_j, ~ \pm\eps_i, ~ \pm\del_i\pm\del_j, ~ \pm 2\del_i$
&$\pm\eps_i\pm\del_j, ~ \pm\del_i$  \\ 
$\bigg. B(0,n)$
&$\pm\del_i\pm\del_j, ~ \pm 2\del_i$ & $\pm\del_i$ \\ 
$\bigg. C(n+1)$
&$\pm\del_i\pm\del_j, ~ \pm 2\del_i$ & $\pm\eps\pm\del_i$ \\ 
$\bigg. D(m,n)$ 
&$\pm\eps_i\pm\eps_j, ~ \pm\del_i\pm\del_j, ~ \pm 2\del_i$ 
&$\pm\eps_i\pm\del_j$ \\ 
$\bigg. F(4)$ 
&$\pm\del, ~ \pm\eps_i\pm\eps_j, ~ \pm\eps_i$ 
&${\half} (\pm\eps_1\pm\eps_2\pm\eps_3\pm\del)$ \\ 
$\bigg. G(3)$ 
&$\pm 2\del, ~ \pm\eps_i, ~ \eps_i-\eps_j$ 
&$\pm\del, ~ \pm\eps_i\pm\del$ \\ 
$\bigg. D(2,1;\alpha)$ 
&$\pm 2\eps_i$ &$ \pm\eps_1\pm\eps_2\pm\eps_3$ \\ 
\hline
\end{tabular}
\caption{Root systems of the basic Lie superalgebras.\label{table15}}
\end{table}

\noindent
For the superalgebras $A(m-1,n-1), B(m,n), D(m,n)$, the indices $i \ne
j$ run from 1 to $m$ for the vectors $\eps$ and from 1 to $n$ for the
vectors $\del$. For the superalgebras $C(n+1)$, the indices $i \ne j$
run from 1 to $n$ for the vectors $\del$. For the superalgebras $F(4),
G(3), D(2,1;\alpha)$, the indices $i \ne j$ run from 1 to 3 for the
vectors $\eps$, with the condition $\eps_1+\eps_2+\eps_3=0$ in the case
of $G(3)$ (see Tables \ref{table11} to \ref{table32} for more details).

\medskip

$\see$ Cartan matrices, Killing form, Simple root systems.

For more details, see refs. \cite{Kac77a,Kac77b}.

\section{\sf Schur's lemma}
\indent

The Schur's lemma is of special importance. Let us stress however that
in the superalgebra case it takes a slightly different form than in the
algebra case \cite{Kac77b}.

\uth: Let $\cG = \cG_\evn \oplus \cG_\odd$ be a basic Lie superalgebra
and $\pi$ be an irreducible representation of $\cG$ in a complex linear 
vector space $\cV$. Let 
\[
\cC(\pi) = \ale \phi: \cV \rightarrow \cV ~ \Big\vert ~ 
\sle \pi(X),\phi \sri = 0 \,, ~ \forall X \in \cG \ari 
\]
where $\phi \in End\,\cV$. Then either

$\bullet$ $\cC(\pi)$ is a multiple of the identity operator $\II$.

$\bullet$  If $\dim \cG_\evn = \dim \cG_\odd$, $\cC(\pi) = \ale \II,
\sigma \ari$ where $\sigma$ is a non-singular operator in $\cG$
permuting $\cG_\evn$ and $\cG_\odd$.

\section{\sf Serre-Chevalley basis}
\indent

The Serre presentation of a Lie algebra consists to describe the algebra
in terms of simple generators and relations (called the Serre
relations), the only parameters being the entries of the Cartan matrix
of the algebra. For the basic Lie superalgebras, the presentation is
quite similar to the Lie algebra case but with some subtleties.

\medskip

Let $\cG$ be a basic Lie superalgebra of rank $r$ with Cartan subalgebra
$\cH$ and simple root system $\Delta^0$ and denote by $E_i^\pm$ ($1 \le
i \le r$) the raising/lowering generators associated to the simple root
system $\Delta^0$. If $\tau$ is a subset of $I = \ale 1,\dots,r \ari$,
the $\ZZ_2$-gradation is defined by $\deg E_i^\pm = \evn$ if $i \notin 
\tau$ and $\deg E_i^\pm = \odd$ if $i \in \tau$. The defining 
commutation relations are
\beo
&& \cle H_i,H_j \cri = 0 \qquad \qquad
\cle H_i,E_j^\pm \cri = \pm a_{ij} E_j^\pm \\ 
&& \cle E_i^+,E_j^- \cri = \delta_{ij} H_i \medbox{for} i \notin \tau \\
&& \ale E_i^+,E_j^- \ari = \delta_{ij} H_i \medbox{for} i \in \tau
\eno
and the Serre relations read as
\[
(\ad E_i^\pm)^{1-{\tilde a}_{ij}} E_j^\pm = 
\sum_{n = 0}^{1-{\tilde a}_{ij}} (-1)^n {1-{\tilde a}_{ij} \choose n}
(E_i^\pm)^{1-{\tilde a}_{ij}-n} E_j^\pm (E_i^\pm)^n = 0 
\]
where the matrix ${\tilde A} = ({\tilde a}_{ij})$ is deduced from the 
Cartan matrix $A = (a_{ij})$ of $\cG$ by replacing all its positive
off-diagonal entries by $-1$.

In the case of superalgebras however, the description given by these
Serre relations leads in general to a bigger superalgebra than the
superalgebra under consideration. It is necessary to write 
supplementary relations involving more than two generators, in order
to quotient the bigger superalgebra and recover the original one. 
As one can imagine, these supplementary conditions appear when one
deals with odd roots of zero length (that is $a_{ii} = 0$).

The supplementary conditions depend on the different kinds of vertices
which appear in the Dynkin diagrams. For the superalgebras $A(m,n)$ with
$m,n \ge 1$ and $B(m,n)$, $C(n+1)$, $D(m,n)$, the vertices can be of the
following type: 

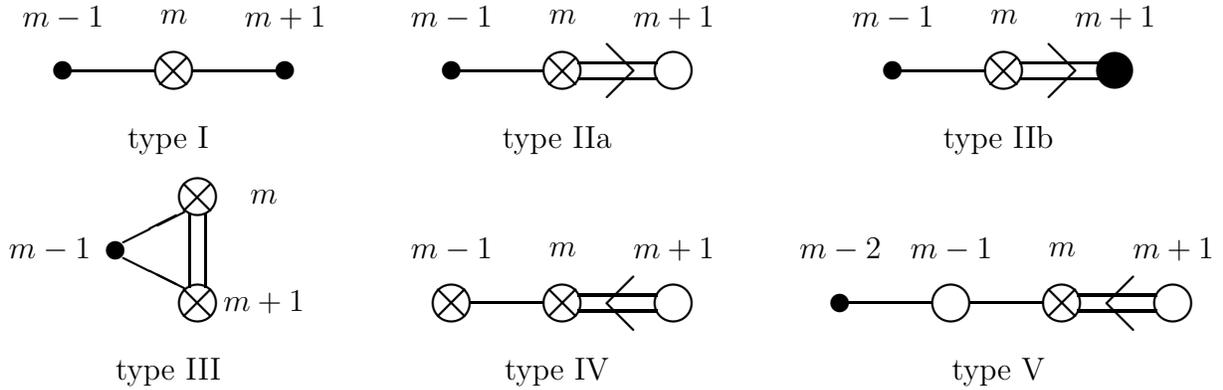
\[
\begin{array}{ccccc}
\begin{picture}(80,20)
\thicklines
\put(0,0){\circle*{7}}
\put(42,0){\circle{14}}
\put(84,0){\circle*{7}}
\put(37,-5){\line(1,1){10}}\put(37,5){\line(1,-1){10}}
\put(0,20){\makebox(0.4,0.6){$m-1$}}
\put(42,20){\makebox(0.4,0.6){$m$}}
\put(84,20){\makebox(0.4,0.6){$m+1$}}
\put(3,0){\line(1,0){32}}
\put(49,0){\line(1,0){32}}
\end{picture}
&\qquad\qquad&
\begin{picture}(80,20)
\thicklines
\put(0,0){\circle*{7}}
\put(42,0){\circle{14}}
\put(84,0){\circle{14}}
\put(37,-5){\line(1,1){10}}\put(37,5){\line(1,-1){10}}
\put(0,20){\makebox(0.4,0.6){$m-1$}}
\put(42,20){\makebox(0.4,0.6){$m$}}
\put(84,20){\makebox(0.4,0.6){$m+1$}}
\put(3,0){\line(1,0){32}}
\put(48,3){\line(1,0){30}}
\put(48,-3){\line(1,0){30}}
\put(69,0){\line(-1,-1){10}}\put(69,0){\line(-1,1){10}}
\end{picture}
&\qquad\qquad&
\begin{picture}(80,20)
\thicklines
\put(0,0){\circle*{7}}
\put(42,0){\circle{14}}
\put(84,0){\circle*{14}}
\put(37,-5){\line(1,1){10}}\put(37,5){\line(1,-1){10}}
\put(0,20){\makebox(0.4,0.6){$m-1$}}
\put(42,20){\makebox(0.4,0.6){$m$}}
\put(84,20){\makebox(0.4,0.6){$m+1$}}
\put(3,0){\line(1,0){32}}
\put(48,3){\line(1,0){30}}
\put(48,-3){\line(1,0){30}}
\put(69,0){\line(-1,-1){10}}\put(69,0){\line(-1,1){10}}
\end{picture}
\cr
&&&&\cr
\mbox{type I} && \mbox{type IIa} && \mbox{type IIb} \cr
&&&&\cr
\begin{picture}(40,40)
\thicklines
\put(0,20){\circle*{7}}
\put(-25,20){\makebox(0.4,0.6){$m-1$}}
\put(3,23){\line(2,1){23}}\put(3,17){\line(2,-1){23}}
\put(31,40){\circle{14}}
\put(56,40){\makebox(0.4,0.6){$m$}}
\put(31,0){\circle{14}}
\put(56,0){\makebox(0.4,0.6){$m+1$}}
\put(26,35){\line(1,1){10}}\put(26,45){\line(1,-1){10}}
\put(26,-5){\line(1,1){10}}\put(26,5){\line(1,-1){10}}
\put(28,6){\line(0,1){28}}
\put(34,6){\line(0,1){28}}
\end{picture}
&\qquad\qquad&
\begin{picture}(80,20)
\thicklines
\put(0,0){\circle{14}}
\put(42,0){\circle{14}}
\put(84,0){\circle{14}}
\put(-5,-5){\line(1,1){10}}\put(-5,5){\line(1,-1){10}}
\put(37,-5){\line(1,1){10}}\put(37,5){\line(1,-1){10}}
\put(0,20){\makebox(0.4,0.6){$m-1$}}
\put(42,20){\makebox(0.4,0.6){$m$}}
\put(84,20){\makebox(0.4,0.6){$m+1$}}
\put(7,0){\line(1,0){28}}
\put(48,3){\line(1,0){30}}
\put(48,-3){\line(1,0){30}}
\put(59,0){\line(1,1){10}}\put(59,0){\line(1,-1){10}}
\end{picture}
&\qquad\qquad&
\begin{picture}(120,20)
\thicklines
\put(0,0){\circle*{7}}
\put(42,0){\circle{14}}
\put(84,0){\circle{14}}
\put(126,0){\circle{14}}
\put(79,-5){\line(1,1){10}}\put(79,5){\line(1,-1){10}}
\put(0,20){\makebox(0.4,0.6){$m-2$}}
\put(42,20){\makebox(0.4,0.6){$m-1$}}
\put(84,20){\makebox(0.4,0.6){$m$}}
\put(126,20){\makebox(0.4,0.6){$m+1$}}
\put(3,0){\line(1,0){32}}
\put(49,0){\line(1,0){28}}
\put(90,3){\line(1,0){30}}
\put(90,-3){\line(1,0){30}}
\put(101,0){\line(1,1){10}}\put(101,0){\line(1,-1){10}}
\end{picture}
\cr
&&&&\cr
\mbox{type III} && \mbox{type IV} && \mbox{type V} \cr
\end{array}
\]
where the small black dots represent either white dots associated to even
roots or grey dots associated to isotropic odd roots.
\\
The supplementary conditions take the following form:
\beo
&& \mbox{- for the type I, IIa and IIb vertices:} \\
&& \qquad (\ad E_m^\pm)(\ad E_{m+1}^\pm)(\ad E_m^\pm) E_{m-1}^\pm = 
(\ad E_m^\pm)(\ad E_{m-1}^\pm)(\ad E_m^\pm) E_{m+1}^\pm = 0 \\
&& \mbox{- for the type III vertex:} \\
&& \qquad (\ad E_m^\pm)(\ad E_{m+1}^\pm) E_{m-1}^\pm - 
(\ad E_{m+1}^\pm)(\ad E_m^\pm) E_{m-1}^\pm = 0 \\
&& \mbox{- for the type IV vertex:} \\
&& \qquad (\ad E_m^\pm)
\left( \cle (\ad E_{m+1}^\pm)(\ad E_m^\pm) E_{m-1}^\pm , 
(\ad E_m^\pm) E_{m-1}^\pm \cri \right) = 0 \\
&& \mbox{- for the type V vertex:} \\
&& \qquad (\ad E_m^\pm)(\ad E_{m-1}^\pm)(\ad E_m^\pm)(\ad E_{m+1}^\pm)
(\ad E_m^\pm)(\ad E_{m-1}^\pm) E_{m-2}^\pm = 0
\eno
For $A(m,n)$ with $m=0$ or $n=0$, $F(4)$ and $G(3)$, it is not necessary
to impose supplementary conditions.

\medskip

For more details, see refs. \cite{FLV91,LSS90,Sch93,Yam94}.

\section{\sf Simple root systems}
\indent

Let $\cG = \cG_\evn \oplus \cG_\odd$ be a {\em basic} Lie superalgebra
with Cartan subalgebra $\cH$ and root system $\Delta = \Delta_\evn \cup
\Delta_\odd$. Then $\cG$ admits a Borel decomposition $\cG = \cN^+
\oplus \cH \oplus \cN^-$ where $\cN^+$ and $\cN^-$ are subalgebras such
that $\cle \cH,\cN^+ \cri \subset \cN^+$ and $\cle \cH,\cN^- \cri
\subset \cN^-$ with $\dim\cN^+ = \dim\cN^-$.

If $\cG = \cH \bigoplus_{\alpha} \cG_{\alpha}$ is the root decomposition
of $\cG$, a root $\alpha$ is called positive if $\cG_{\alpha} \cap \cN^+
\ne \emptyset$ and negative if $\cG_{\alpha} \cap \cN^- \ne \emptyset$.
A root is called simple if it cannot be decomposed into a sum of
positive roots. The set of all simple roots is called a {\em simple root
system} of $\cG$ and is denoted here by $\Delta^0$.

Let $\rho_0$ be the half-sum of the positive even roots, $\rho_1$ the
half-sum of the positive odd roots and $\rho = \rho_0-\rho_1$.
Then one has for a simple root $\alpha_i$,
$(\rho,\alpha_i) = \half(\alpha_i,\alpha_i)$. In particular, one has
$(\rho,\alpha_i) = 0$ if $\alpha_i \in \Delta^0_\odd$ with
$(\alpha_i,\alpha_i)=0$.

\medskip

We will call $\cB = \cH \oplus \cN^+$ a {\em Borel subalgebra} of $\cG$.
Notice that such a Borel subalgebra is solvable but not maximal
solvable. Indeed, adding to such a Borel subalgebra $\cB$ a negative
simple isotropic root generator (that is a generator associated to an
odd root of zero length, $\see$ Roots), the obtained subalgebra is
still solvable since the superalgebra $sl(1,1)$ is solvable. However,
$\cB$ contains a maximal solvable subalgebra $\cB_\evn$ of the even part
$\cG_\evn$.

In general, for a basic Lie superalgebra $\cG$, there are many
inequivalent classes of conjugacy of Borel subalgebras (while for the
simple Lie algebras, all Borel subalgebras are conjugate). To each
class of conjugacy of Borel subalgebras of $\cG$ is associated a simple
root system $\Delta^0$. Hence, contrary to the Lie algebra case, to a
given basic Lie superalgebra $\cG$ will be associated in general many
inequivalent simple root systems, up to a transformation of the Weyl
group $W(\cG)$ of $\cG$ (under a transformation of $W(\cG)$, a simple
root system will be transformed into an equivalent one with the same
Dynkin diagram). Let us recall that the Weyl group $W(\cG)$ of $\cG$ is
generated by the Weyl reflections $\omega$ with respect to the even
roots:
\[
\omega_{\alpha}(\beta) = \beta - 2 \frac{(\alpha,\beta)}
{(\alpha,\alpha)} \alpha 
\]
where $\alpha \in \Delta_\evn$ and $\beta \in \Delta$.

For the basic Lie superalgebras, one can extend the Weyl group $W(\cG)$
to a larger group by adding the following transformations (called
generalized Weyl transformations) associated to the odd roots of $\cG$ 
\cite{FSS89,LSS86}. For $\alpha \in \Delta_\odd$, one defines:
\[
\begin{array}{lll}
\bigg. \displaystyle{
\omega_{\alpha}(\beta) = \beta - 2 \frac{(\alpha,\beta)}
{(\alpha,\alpha)} \alpha }
 &\quad & \smbox{if} (\alpha,\alpha) \ne 0
 \cr
\bigg. \omega_{\alpha}(\beta) = \beta + \alpha 
 &\quad & \smbox{if} (\alpha,\alpha) = 0
 \smbox{and} (\alpha,\beta) \ne 0
 \cr
\bigg. \omega_{\alpha}(\beta) = \beta 
 &\quad & \smbox{if} (\alpha,\alpha) = 0 
 \smbox{and} (\alpha,\beta) = 0 
 \cr
\bigg. \omega_{\alpha}(\alpha) = - \alpha 
 &&
\end{array}
\]
Notice that the transformation associated to an odd root $\alpha$ of
zero length cannot be lifted to an automorphism of the superalgebra
since $\omega_{\alpha}$ transforms even roots into odd ones and
vice-versa, and the $\ZZ_2$-gradation would not be respected.

The generalization of the Weyl group for a basic Lie superalgebra $\cG$
gives a method for constructing all the simple root systems of $\cG$
and hence all the inequivalent Dynkin diagrams: a simple root system
$\Delta^0$ being given, from any root $\alpha \in \Delta^0$ such that
$(\alpha,\alpha) = 0$, one constructs the simple root system
$\omega_{\alpha} (\Delta^0)$ and repeats the procedure on the obtained
system until no  new basis arises.

In the set of all inequivalent simple root systems of a basic Lie
superalgebra, there is one simple root system that plays a particular
role: the distinguished simple root system.

\udef: For each basic Lie superalgebra, there exists a simple root
system for which the number of odd roots is the smallest one. It is
constructed as follows: the even simple roots are given by the simple
root system of the even part $\cG_\evn$ and the odd simple root is the
lowest weight of the representation $\cG_\odd$ of $\cG_\evn$. Such a
simple root system is called the {\em distinguished simple root
system}.

\medskip

\uex:
\\
Consider the basic Lie superalgebra $sl(2|1)$ with Cartan generators
$H,Z$ and root generators $E^\pm,F^\pm,\bF^\pm$. The root system is
given by $\Delta = \{ \pm(\eps_1-\eps_2), \pm(\eps_1-\del),
\pm(\eps_2-\del) \}$. One can find two inequivalent Borel subalgebras,
namely $\cB' = \{H,Z,E^+,\bF^+,\bF^-\}$ and $\cB'' =
\{H,Z,E^+,\bF^+,F^+\}$, with positive root systems ${\Delta'}^+ =
\{\eps_1-\eps_2, \eps_1-\del, \eps_2-\del\}$ and ${\Delta''}^+ =
\{\eps_1-\eps_2, \eps_1-\del, -\eps_2+\del\}$ respectively. The
corresponding simple root systems are ${\Delta'}^0 = \{\eps_1-\eps_2,
\eps_2-\del\}$ (called distinguished simple root system) and
${\Delta''}^0 = \{\eps_1-\del, -\eps_2+\del\}$ (called fermionic simple
root system). The fermionic simple root system ${\Delta''}^0$ is
obtained from the distinguished one ${\Delta'}^0$ by the Weyl
transformation associated to the odd root $\eps_2-\del$:
$\omega_{\eps_2-\del}(\eps_2-\del) = -\eps_2+\del$ and 
$\omega_{\eps_2-\del}(\eps_1-\eps_2) = \eps_1-\del$.

\medskip

We give in Table \ref{table13} the list of the distinguished simple root
systems of the basic Lie superalgebras in terms of the orthogonal vectors
$\eps_i$ and $\del_i$. For more details, see ref. \cite{Kac77a}.

\begin{table}[htbp]
\centering
\begin{tabular}{|c|c|} \hline
superalgebra $\cG$ & distinguished simple root system $\Delta^0$ \\ 
\hline
$\bigg. A(m-1,n-1)$
&$\del_1-\del_2, \dots, \del_{n-1}-\del_n, \del_n-\eps_1, 
\eps_1-\eps_2, \dots, \eps_{m-1}-\eps_{m}$ \\
$\bigg. B(m,n)$
&$\del_1-\del_2, \dots, \del_{n-1}-\del_n, \del_n-\eps_1, 
\eps_1-\eps_2, \dots, \eps_{m-1}-\eps_m, \eps_m$ \\
$\bigg. B(0,n)$
&$\del_1-\del_2, \dots, \del_{n-1}-\del_n, \del_n$ \\
$\bigg. C(n)$
&$\eps-\del_1, \del_1-\del_2, \dots, \del_{n-1}-\del_n, 2\del_n$ \\
$\bigg. D(m,n)$
&$\del_1-\del_2, \dots, \del_{n-1}-\del_n, \del_n-\eps_1, 
\eps_1-\eps_2, \dots, \eps_{m-1}-\eps_m, \eps_{m-1}+\eps_m$ \\ 
$\bigg. F(4)$
&${\half} (\del-\eps_1-\eps_2-\eps_3), \eps_3, \eps_2-\eps_3, 
\eps_1-\eps_2$ \\
$\bigg. G(3)$
&$\del+\eps_3, \eps_1, \eps_2-\eps_1$ \\
$\bigg. D(2,1;\alpha)$
&$\eps_1-\eps_2-\eps_3, 2\eps_2, 2\eps_3$ \\
\hline
\end{tabular}
\caption{Distinguished simple root systems of the basic Lie
superalgebras.\label{table13}}
\end{table}

\section{\sf Simple and semi-simple Lie superalgebras}
\indent

\udef: Let $\cG = \cG_\evn \oplus \cG_\odd$ be a Lie superalgebra.
\\
The Lie superalgebra $\cG$ is called {\em simple} if it does not contain
any non-trivial ideal. The Lie superalgebra $\cG$ is called {\em
semi-simple} if it does not contain any non-trivial solvable ideal.

\medskip

A necessary condition for a Lie superalgebra $\cG = \cG_\evn
\oplus  \cG_\odd$ to be simple is that the representation of $\cG_\evn$ on 
$\cG_\odd$ is faithful and $\ale \cG_\odd,\cG_\odd \ari = \cG_\evn$. 
If the representation of $\cG_\evn$ on $\cG_\odd$ is irreducible, then 
$\cG$ is simple.

\medskip

Recall that if $\cA$ is a {\em semi-simple Lie algebra}, then it can be
written as the direct sum of {\em simple Lie algebras} $\cA_i$: 
$\cA = \oplus_i \cA_i$. {\em It is not the case for superalgebras}.
However, the following properties hold.

\medskip

\uppts:
\begin{enumerate}
\item
If $\cG$ is a Lie superalgebra and $\cI$ is the maximal solvable ideal, 
then the quotient $\cG/\cI$ is a semi-simple Lie superalgebra.
However, opposed to the case of Lie algebras, one {\em cannot} write
here $\cG = {\bar\cG} \rsemisum \cI$ where $\bar\cG$ is a direct sum
of simple Lie superalgebras.
\item
If $\cG$ is a Lie superalgebra with a non-singular Killing form, then 
$\cG$ is a direct sum of simple Lie superalgebras with non-singular
Killing form.
\item
If $\cG$ is a Lie superalgebra whose all its finite dimensional
representations are completely reducible, then $\cG$ is a direct sum
of simple Lie algebras and $osp(1|n)$ simple superalgebras.
\item
Let $\cG = \cG_\evn \oplus \cG_\odd$ be a Lie superalgebra such that its
even part $\cG_\evn$ is a semi-simple Lie algebra. Then $\cG$ is an
elementary extension of a direct sum of Lie algebras or one of the Lie
superalgebras $A(n,n)$, $B(m,n)$, $D(m,n)$, $D(2,1;\alpha)$, $F(4)$,
$G(3)$, $P(n)$, $Q(n)$, $\der Q(n)$ or $G(S_1,\dots,S_r;L)$. (For the
definition of $G(S_1,\dots,S_r;L)$, see ref. \cite{Kac77a}).
\\
The elementary extension of a Lie superalgebra $\cG = \cG_\evn \oplus
\cG_\odd$ is defined as $\cG \rsemisum \cI$ where $\cI$ is an odd 
commutative ideal and $\ale \cG_\odd,\cI \ari = 0$.
\end{enumerate}

\medskip

For more details, see refs. \cite{Kac77a,Rit77}.

\section{\sf Spinors (in the Lorentz group)}
\indent

The algebra of the Lorentz group is $o(1,3)$ whose generators 
$M_{\mu\nu}=-M_{\nu\mu}$ satisfy the commutation relations 
($\mu,\nu=0,1,2,3$)
\[
\cle M_{\mu\nu},M_{\rho\sigma} \cri = i(
- g_{\nu\sigma}M_{\mu\rho} + g_{\nu\rho}M_{\mu\sigma}
+ g_{\mu\sigma}M_{\nu\rho} - g_{\mu\rho}M_{\nu\sigma}) 
\]
where the metric is 
$g^{\mu\nu} = 2\delta^{\mu 0} \delta^{\nu 0} - \delta^{\mu\nu} 
= \diag(1,-1,-1,-1)$ and $g^{\mu\sigma} g_{\sigma\nu} = \delta^\mu_\nu$.
\\
If we define $J_i = \sfrac{1}{2} \eps_{ijk} M^{jk}$ and $K_i = M^{0i}$, 
we have
\beo
&& \cle J_i,J_j \cri = i ~ \eps_{ijk} ~ J_k \\
&& \cle K_i,K_j \cri = -i ~ \eps_{ijk} ~ J_k \\
&& \cle J_i,K_i \cri = i ~ \eps_{ijk} ~ K_k 
\eno
where $i,j,k=1,2,3$ and $\eps_{ijk}$ is the completely antisymmetric rank
three tensor, $\eps_{123} = 1$.
\\
Defining $M_i = \half(J_i+iK_i)$ and $N_i = \half(J_i-iK_i)$, the
Lorentz algebra can be rewritten as: 
\beo
&& \cle M_i,M_j \cri = i ~ \eps_{ijk} ~ M_k \\
&& \cle N_i,N_j \cri = i ~ \eps_{ijk} ~ N_k \\
&& \cle M_i,N_j \cri = 0 
\eno

The finite dimensional irreducible representations of the Lorentz group
are labelled by a pair of integers or half-integers $(m,n)$. These
representations are non-unitary since the generators $M_i$ and $N_i$
can be represented by finite dimensional Hermitian matrices while $J_i$
and $K_i$ are not. Because of the relation $J_i=M_i+N_i$, the
combination $m+n$ is the spin of the representation. Representations
with half-integer spin (resp. integer spin) are called spinorial (resp.
tensorial) representations. The two representations $(1/2,0)$ and
$(0,1/2)$ are the fundamental spinorial representations: all the
spinorial and tensorial representations of the Lorentz group can be
obtained by tensorialization and symmetrization of these.
\\
The $\sigma^i$ being the Pauli matrices, one has in the representation
$(1/2,0)$
\[
M_i = \half \sigma^i \medbox{and} N_i = 0
\bigbox{that is}
J_i = \half \sigma^i \medbox{and} K_i = -\sfrac{i}{2} \sigma^i
\]
and in the representation $(0,1/2)$
\[
M_i = 0 \medbox{and} N_i = \half \sigma^i
\bigbox{that is}
J_i = \half \sigma^i \medbox{and} K_i = \sfrac{i}{2} \sigma^i
\]
The vectors of the representation spaces of the spinorial representations
are called (Weyl) {\em spinors} under the Lorentz group. 
Define $\sigma^\mu=(\II,\sigma^i)$ and $\bsigma^\mu=(\II,-\sigma^i)$. 
Under a Lorentz transformation $\Lambda^\mu_\nu$, a {\em covariant
undotted} spinor $\psi_\alpha$ (resp. {\em contravariant undotted}
spinor $\psi^\alpha$) transforms as
\[
\psi_\alpha ~~~ \mapsto ~~~ S_{\alpha}^{~\beta} ~ \psi_\beta 
\bigbox{and}
\psi^\alpha ~~~ \mapsto ~~~ \psi^\beta ~ (S^{-1})^{~\alpha}_\beta 
\]
where the matrix $S$ is related to the matrix $\Lambda^\mu_\nu$ by
\[
\Lambda^\mu_\nu = \half ~ \tr(S \sigma_\nu S^\dagger \bsigma^\mu)
\]
The spinors $\psi_\alpha$ (or $\psi^\alpha$) transform as the
$(1/2,0)$ representation of the Lorentz group.
Under a Lorentz transformation $\Lambda^\mu_\nu$, a {\em covariant
dotted} spinor $\bpsi_\dalpha$ (resp. {\em contravariant dotted} 
spinor $\bpsi^\dalpha$) transforms as
\[
\bpsi_\dalpha ~~~ \mapsto ~~~ \bpsi_\dbeta 
~ (S^\dagger)_{~\dalpha}^\dbeta \bigbox{and} \bpsi^\dalpha ~~~ 
\mapsto ~~~ ({S^\dagger}^{-1})^\dalpha_{~\dbeta} ~ \bpsi^\dbeta  
\]
where the matrix $S$ is related to the matrix $\Lambda^\mu_\nu$ by
\[
\Lambda^\mu_\nu = \half ~ \tr((S^\dagger)^{-1} \bsigma_\nu S^{-1} 
\sigma^\mu) 
\]
The spinors $\bpsi_\dalpha$ (or $\bpsi^\dalpha$)
transform as the $(0,1/2)$ representation of the Lorentz group.
\\
The relation between covariant and contravariant spinors is given by
means of the two-dimensional Levi-Civita undotted tensors
$\eps_{\alpha\beta},\eps^{\alpha\beta}$ and dotted tensors
$\eps_{\dalpha\dbeta},\eps^{\dalpha\dbeta}$ such that 
$\eps_{\alpha\beta} = \eps^{\alpha\beta} = -\eps_{\dalpha\dbeta} 
= -\eps^{\dalpha\dbeta}$ and $\eps_{12} = 1$: 
\[
\psi^\alpha = \eps^{\alpha\beta} \psi_\beta \,, \quad 
\psi_\alpha = \psi^\beta \eps_{\beta\alpha} \,, \quad 
\bpsi^\dalpha = \bpsi_\dbeta \eps^{\dbeta\dalpha} \,, \quad
\bpsi_\dalpha = \eps_{\dalpha\dbeta} \bpsi^\dbeta 
\]
Notice that $\bpsi^\dalpha = (\psi^\alpha)^*$ and $\bpsi_\dalpha =
(\psi_\alpha)^*$ where the star denotes the complex conjugation, and
also $\eps_{\dalpha\dbeta} = -(\eps_{\alpha\beta})^*$.
\\
Finally, the rule for contracting undotted and dotted spinor indices is
the following:
\[
\psi \zeta \equiv \psi^\alpha \zeta_\alpha = -\psi_\alpha \zeta^\alpha
\bigbox{and} \bpsi \bzeta \equiv \bpsi_\dalpha \bzeta^\dalpha 
= -\bpsi^\dalpha \bzeta_\dalpha
\]

The space inversion leaves the rotation generators $J_i$ invariant
but changes the sign of the boost generators $K_i$. It follows that
under the space inversion, the undotted Weyl spinors are transformed
into dotted ones and vice-versa. On the (reducible) representation
$(1/2,0) \oplus (0,1/2)$, the space inversion acts in a well-defined
way. The corresponding vectors in the representation space are
called {\em Dirac spinors}. In the Weyl representation, the Dirac
spinors are given by
\[
\Psi_D = \left(\begin{array}{c} \psi_\alpha \cr \bchi^\dalpha \cr
\end{array}\right)
\]
Under a Lorentz transformation $\Lambda^\mu_\nu$, a Dirac spinor
$\Psi_D$ transforms as 
\[
\Psi_D ~~~ \mapsto ~~~ L ~ \Psi_D = \left(\begin{array}{cc}
S(\Lambda^\mu_\nu) & 0 \cr 0 & S({\Lambda^\mu_\nu}^\dagger)^{-1} \cr
\end{array}\right) \left(\begin{array}{c} \psi_\alpha \cr \bchi^\dalpha
\cr \end{array}\right) = \left(\begin{array}{c} S(\Lambda^\mu_\nu) ~
\psi_\alpha \cr S({\Lambda^\mu_\nu}^\dagger)^{-1} ~ \bchi^\dalpha \cr
\end{array}\right)
\]
The generators of the Lorentz group in the $(1/2,0) \oplus (0,1/2)$
representation are given by $\Sigma^{\mu\nu} = \frac{i}{2}
\cle\gamma^\mu,\gamma^\nu\cri$ where the matrices $\gamma^\mu$ are
called the {\em Dirac matrices}. They satisfy the Clifford algebra in
four dimensions:
\[
\ale \gamma^\mu,\gamma^\nu \ari = 2 g^{\mu\nu}
\]
One defines also the $\gamma_5$ matrix by $\gamma_5 = Ði \gamma^0
\gamma^1 \gamma^2 \gamma^3$ such that $\ale \gamma_5,\gamma^\mu \ari =
0$ and $\gamma_5^2 = \II$.

\medskip

The adjoint spinor $\bar\Psi$ and the charge conjugated spinors $\Psi^c$
and $\bar\Psi^c$ of a Dirac spinor $\Psi = \left(\begin{array}{c}
\psi_\alpha \cr \bchi^\dalpha \end{array}\right)$ are defined by
$\bar\Psi = \left( \chi^\alpha ~ \bpsi_\dalpha \right)$, $\Psi^c =
\left( \begin{array}{c} \chi_\alpha \cr \bpsi^\dalpha \end{array}
\right)$ and $\bar\Psi^c = \left( \psi^\alpha ~ \bchi_\dalpha \right)$.
The spinors $\Psi$ and $\Psi^c$ are related through the charge
conjugation matrix $C$ by $\Psi^c = C \bar\Psi^t$.
Moreover, one has $C^{-1} \gamma^\mu C = -(\gamma^\mu)^t$. The six
matrices $C, \gamma^\mu\gamma_5C, \gamma_5 C$ are antisymmetric and the
ten matrices $\gamma^\mu C, \Sigma^{\mu\nu}C$ are symmetric. They form a
set of 16 linearly independent matrices.

\medskip

A Majorana spinor is a Dirac spinor such that $\Psi = \Psi^c$. For such
a spinor, there is a relation between the two Weyl components: a
Majorana spinor $\Psi$ has the form $\Psi = \left(\begin{array}{c}
\psi_\alpha \cr \bpsi^\dalpha \end{array}\right)$. In the Majorana
representation of the $\gamma$ matrices, the components of a Majorana
spinor are all real and the $\gamma$ matrices are all purely imaginary.
\\
The $\gamma$ matrices are given, in the Weyl representation, by 
\[
\gamma^0 = \left(\begin{array}{cc}
0 & \sigma^0 \cr \sigma^0 & 0 \cr \end{array}\right) \qquad
\gamma^i = \left(\begin{array}{cc}
0 & \sigma^i \cr -\sigma^i & 0 \cr \end{array}\right) \qquad
\gamma_5 = \left(\begin{array}{cc}
\II & 0 \cr 0 & -\II \cr \end{array}\right) \qquad
C = \left(\begin{array}{cc}
-i\sigma^2 & 0 \cr 0 & i\sigma^2 \cr \end{array}\right)
\]
Another often used representation of the $\gamma$ matrices is the Dirac
representation:
\[
\gamma_0 = \left(\begin{array}{cc}
\sigma_0 & 0 \cr 0 & -\sigma_0 \cr \end{array}\right) \qquad
\gamma^i = \left(\begin{array}{cc}
0 & \sigma^i \cr -\sigma^i & 0 \cr \end{array}\right) \qquad
\gamma^5 = \left(\begin{array}{cc}
0 & \II \cr \II & 0 \cr \end{array}\right) \qquad
C = \left(\begin{array}{cc}
0 & -i\sigma^2 \cr -i\sigma^2 & 0 \cr \end{array}\right)
\]
Finally, in the Majorana representation, one has:
\beo
&&\gamma^1 = \left(\begin{array}{cc}
i\sigma^3 & 0 \cr 0 & i\sigma^3 \cr \end{array}\right) \qquad
\gamma^2 = \left(\begin{array}{cc}
0 & -\sigma^2 \cr \sigma^2 & 0 \cr \end{array}\right) \qquad
\gamma^3 = \left(\begin{array}{cc}
-i\sigma^1 & 0 \cr 0 & -i\sigma^1 \cr \end{array}\right) \\
&&\\
&&\gamma^0 = \left(\begin{array}{cc}
0 & \sigma^2 \cr \sigma^2 & 0 \cr \end{array}\right) \qquad
\gamma_5 = \left(\begin{array}{cc}
\sigma^2 & 0 \cr 0 & -\sigma^2 \cr \end{array}\right) \qquad
C = \left(\begin{array}{cc}
0 & -i\sigma^2 \cr -i\sigma^2 & 0 \cr \end{array}\right)
\eno

\section{\sf Strange superalgebras $P(n)$}
\indent

We consider the superalgebra $A(n-1,n-1)$ and $P(n-1)$ the subalgebra 
of $A(n-1,n-1)$ generated by the $2n \times 2n$ matrices of the form 
\[
\left(\begin{array}{cc} \lambda & S \cr A & -\lambda^t \end{array}
\right) 
\]
where $\lambda$ are $sl(n)$ matrices, $S$ and $A$ are $n \times n$ 
symmetric and antisymmetric complex matrices which can be seen as
elements of the twofold symmetric representation ($[2]$ in Young
tableau notation) of dimension $n(n+1)/2$ and of the $(n-2)$-fold
antisymmetric representation ($[1^{n-2}]$ in Young tableau notation) of
dimension $n(n-1)/2$ respectively. The $\ZZ$-gradation of the
superalgebra $P(n-1)$ being $\cG = \cG_{-1} \oplus \cG_0 \oplus \cG_1$
where $\cG_0 = sl(n)$, $G_1 = [2]$ and $G_{-1} = [1^{n-2}]$, the
subspaces $\cG_i$ satisfy the following commutation relations
\[
\begin{array}{ll}
\cle \cG_0,\cG_0 \cri \subset \cG_0 & \qquad
\cle \cG_0,\cG_{\pm 1} \cri \subset \cG_{\pm 1} \cr
&\cr
\ale \cG_1,\cG_1 \ari = \ale \cG_{-1},\cG_{-1} \ari = 0 & \qquad
\ale \cG_1,\cG_{-1} \ari \subset \cG_0
\end{array}
\]
The $\ZZ$-gradation is consistent: $\cG_\evn = \cG_0$ and $\cG_\odd = 
\cG_{-1} \oplus \cG_1$.

Defining the Cartan subalgebra $\cH$ as the Cartan subalgebra of the
even part, the root system $\Delta = \Delta_\evn \cup \Delta_\odd$ of
$P(n-1)$ can be expressed in terms of the orthogonal vectors $\eps_1, 
\dots, \eps_n$ as 
\[
\Delta_\evn = \ale ~ \alpha_{ij} = \eps_i-\eps_j~ \ari 
\]
and
\[ 
\Delta_\odd = \ale ~ \pm\beta_{ij} = \pm \left( \eps_i+\eps_j - 
\frac{2}{n} \sum_{k = 1}^n \eps_k \right), ~ \gamma_i = 2\eps_i - 
\frac{2}{n} \sum_{k = 1}^n \eps_k ~ \ari 
\]

Denoting by $H_i$ the Cartan generators, by $E_\alpha$ the even root
generators and by $E_\beta,E_\gamma$ the odd root generators of $P(n-1)$,
the commutation relations in the Cartan-Weyl basis are the following: 
\beo
&&\cle H_k,E_{\alpha_{ij}} \cri = (\delta_{ik} - \delta_{jk} -
\delta_{i,k+1} + \delta_{j,k+1}) E_{\alpha_{ij}} \\
&&\cle H_k,E_{\beta_{ij}} \cri = (\delta_{ik} + \delta_{jk} -
\delta_{i,k+1} - \delta_{j,k+1}) E_{\beta_{ij}} \\
&&\cle H_k,E_{-\beta_{ij}} \cri = -(\delta_{ik} + \delta_{jk} -
\delta_{i,k+1} - \delta_{j,k+1}) E_{-\beta_{ij}} \\
&&\cle H_k,E_{\gamma_i} \cri = 2(\delta_{ik} - \delta_{i,k+1}) 
E_{\gamma_i} \\
&&\cle E_{\alpha_{ij}},E_{\alpha_{kl}} \cri = \delta_{jk} E_{\alpha_{il}}
- \delta_{il} E_{\alpha_{kj}} \\
&&\cle E_{\alpha_{ij}},E_{-\alpha_{ij}} \cri = \sum_{k = i}^{j-1} H_k \\
&&\cle E_{\alpha_{ij}},E_{\beta_{kl}} \cri = \left\{ \begin{array}{ll}
\delta_{jk} E_{\beta_{il}} + \delta_{jl} E_{\beta_{ik}}
& \smbox{if} (i,j) \ne (k,l) \cr
E_{\gamma_i} & \smbox{if} (i,j) = (k,l)
\end{array} \right. \\
&&\cle E_{\alpha_{ij}},E_{-\beta_{kl}} \cri = \left\{ \begin{array}{ll}
-\delta_{ik} E_{-\beta_{jl}} + \delta_{il} E_{-\beta_{jk}}
& \smbox{if} (i,j) \ne (k,l) \cr
0 & \smbox{if} (i,j) = (k,l)
\end{array} \right. \\
&&\cle E_{\alpha_{ij}},E_{\gamma_k} \cri = \delta_{jk} E_{\beta_{ik}} \\
&&\ale E_{-\beta_{ij}},E_{\gamma_k} \ari = -\delta_{ik} E_{\alpha_{kj}}
+ \delta_{jk} E_{\alpha_{ki}} \\
&&\ale E_{\beta_{ij}},E_{-\beta_{kl}} \ari = \left\{ \begin{array}{ll}
- \delta_{ik} E_{\alpha_{jl}} + \delta_{il} E_{\alpha_{jk}}
- \delta_{jk} E_{\alpha_{il}} + \delta_{jl} E_{\alpha_{ik}}
& \smbox{if} (i,j) \ne (k,l) \cr
\sum_{k = i}^{j-1} H_k & \smbox{if} (i,j) = (k,l)
\end{array} \right. \\
&&\ale E_{\beta_{ij}},E_{\beta_{kl}} \ari = 
\ale E_{-\beta_{ij}},E_{-\beta_{kl}} \ari = 
\ale E_{\beta_{ij}},E_{\gamma_k} \ari = 0 \\
\eno

Let us emphasize that $P(n)$ is a {\em non-contragredient} simple Lie
superalgebra, that is the number of positive roots and the number of
negative roots are not equal. Moreover, since every bilinear form is
identically vanishing in $P(n)$, it is impossible to define a
non-degenerate scalar product on the root space. It follows that the
notions of Cartan matrix and Dynkin diagram are not defined for $P(n)$.
However, using an extension of $P(n)$ by suitable diagonal matrices,
one can construct a non-vanishing bilinear form on the Cartan subalgebra
of this extension and therefore generalize in this case the notions 
of Cartan matrix and Dynkin diagram. 

\medskip

$\see$ Oscillator realization of the strange superalgebras.

For more details, see ref. \cite{FSS91}.

\section{\sf Strange superalgebras $Q(n)$}
\indent

We consider the superalgebra $sl(n|n)$ and $\widetilde{Q}(n-1)$ the
subalgebra of $sl(n|n)$ generated by the $2n \times 2n$ matrices 
of the form
\[
\left(\begin{array}{cc} A & B \cr B & A \end{array} \right)
\]
where $A$ and $B$ are $sl(n)$ matrices. The even part of the
superalgebra $\widetilde{Q}(n-1)$ is the Lie algebra $\cG_\evn = sl(n)
\oplus U(1)$ of dimension $n^2$ and the odd part is the adjoint
representation $\cG_\odd$ of $sl(n)$ of dimension $n^2-1$. The even
generators of $\cG_\evn$ are divided into the $sl(n)$ Cartan generators
$H_i$ with $1 \le i \le n-1$, the $U(1)$ generator $Z$ and the $n(n-1)$
root generators $E_{ij}$ with $1 \le i \ne j \le n$ of $sl(n)$. The odd
root generators of $\cG_\odd$ are also divided into two parts, $F_{ij}$
with $1 \le i \ne j \le n$ and $K_i$ with $1 \le i \le n-1$. This
superalgebra $\widetilde{Q}(n-1)$ is not a simple superalgebra: in
order to obtain a simple superalgebra, one should factor out the
one-dimensional center, as in the case of the $sl(n|n)$ superalgebra.
We will denote by $Q(n-1)$ the simple superalgebra
$\widetilde{Q}(n-1)/U(1)$.

Following the definition of the Cartan subalgebra ($\see$), the strange
superalgebra $Q(n-1)$ has the property that the Cartan subalgebra $\cH$
does not coincide with the Cartan subalgebra of the even part $sl(n)$,
but admits also an odd part: $\cH \cap \cG_\odd \ne \emptyset$. More
precisely, one has
\[
\cH = \cH_\evn \oplus \cH_\odd 
\]
where $\cH_\evn$ is spanned by the $H_i$ generators and $\cH_\odd$ by
the $K_i$ generators ($1 \le i \le n-1$). However, since the $K_i$
generators are odd, the root generators $E_{ij}$ and $F_{ij}$ are not
eigenvectors of $\cH_\odd$. It is convenient in this case to give the
root decomposition with respect to $\cH_\evn = \cH \cap \cG_\evn$
instead of $\cH$. The root system $\Delta$ of $Q(n-1)$ coincide then with
the root system of $sl(n)$. One has
\[
\cG = \cG_\evn \oplus \cG_\odd = \cH_\evn \oplus \Big( \bigoplus_{\alpha
\in \Delta} \cG_{\alpha} \Big) \bigbox{with} \dim \cG_{(\alpha \ne 0)} =
2 \smbox{and} \dim \cG_{(\alpha = 0)} = n
\]
Moreover, since $\dim \cG_{\alpha} \cap \cG_\evn \ne \emptyset$ and 
$\dim \cG_{\alpha} \cap \cG_\odd \ne \emptyset$ for any non-zero root
$\alpha$, the non-zero roots of $Q(n-1)$ are both even and odd.

Denoting by $H_i$ the Cartan generators, by $E_{ij}$ the even root
generators and by $F_{ij}$ the odd root generators of
$\widetilde{Q}(n)$, the commutation relations in the Cartan-Weyl basis
are the following:
\beo
&&\cle H_i,H_j \cri = \cle H_i,K_j \cri = 0 \\
&&\ale K_i,K_j\ari = \frac{2}{n} (2 \delta_{ij} - \delta_{i,j+1} -
\delta_{i,j-1}) \left( Z - \sum_{k = 1}^{n-1} k H_k \right) \\
&&\qquad \qquad + 2(\delta_{ij} - \delta_{i,j+1}) \sum_{k = i}^{n-1}
H_k + 2(\delta_{ij} - \delta_{i,j-1}) \sum_{k = i+1}^{n-1} H_k \\
&&\cle H_k,E_{ij} \cri = (\delta_{ik} - \delta_{jk} - \delta_{i,k+1} +
\delta_{j,k+1}) E_{ij} \\
&&\cle H_k,F_{ij} \cri = (\delta_{ik} - \delta_{jk} - \delta_{i,k+1} + 
\delta_{j,k+1}) F_{ij} \\
&&\cle K_k,E_{ij} \cri = (\delta_{ik} - \delta_{jk} - \delta_{i,k+1} +
\delta_{j,k+1}) F_{ij} \\
&&\ale K_k,F_{ij} \ari = (\delta_{ik} + \delta_{jk} - \delta_{i,k+1} -
\delta_{j,k+1}) E_{ij} \\
&&\cle E_{ij},E_{kl} \cri = \delta_{jk} E_{il} - \delta_{il} E_{kj} 
\quad (i,j) \ne (k,l) \\
&&\cle E_{ij},E_{ji} \cri = \sum_{k = i}^{j-1} H_k \\
&&\cle E_{ij},F_{kl} \cri = \delta_{jk} F_{il} - \delta_{il} F_{kj} 
\quad (i,j) \ne (k,l) \\
&&\cle E_{ij},F_{ji} \cri = \sum_{k = i}^{j-1} K_k\\
&&\ale F_{ij},F_{kl} \ari = \delta_{jk} E_{il} + \delta_{il} E_{kj} 
\quad (i,j) \ne (k,l) \\
&&\ale F_{ij},F_{ji} \ari = \frac{2}{n} Z + \frac{n-2}{n} \left(
2 \sum_{k = i}^{n-1} k H_k - n \sum_{k = i}^{n-1} H_k
- n \sum_{k = j}^{n-1} H_k \right) \\
\eno

\medskip

$\see$ Cartan subalgebras, Oscillator realization of the strange
superalgebras.

\section{\sf Subsuperalgebras (regular)}
\indent

\udef: Let $\cG$ be a basic Lie superalgebra and consider its canonical
root decomposition
\[
\cG = \cH \oplus \bigoplus_{\alpha \in \Delta} \cG_{\alpha}
\]
where $\cH$ is a Cartan subalgebra of $\cG$ and $\Delta$ its
corresponding root system ($\see$).
\\
A subsuperalgebra $\cG'$ of $\cG$ is called {\em regular} (by analogy
with the algebra case) if it has the root decomposition
\[
\cG' = \cH' \oplus \bigoplus_{\alpha' \in \Delta'} \cG'_{\alpha'}
\]
where $\cH' \subset \cH$ and $\Delta \subset \Delta'$. The
semi-simplicity of $\cG'$ will be insured if to each $\alpha' \in 
\Delta'$ then $-\alpha' \in \Delta'$ and $\cH'$ is the linear
closure of $\Delta'$.

\medskip

The method for finding the regular semi-simple sub(super)algebras of a
given basic Lie superalgebra $\cG$ is completely analogous to the usual
one for Lie algebras by means of extended Dynkin diagrams. However, one
has now to consider all the Dynkin diagrams associated to the
inequivalent simple root systems. For a given simple root system
$\Delta^0$ of $\cG$, one considers the associated Dynkin diagram. The
corresponding extended simple root system is ${\widehat\Delta^0} =
\Delta^0 \cup \{\Psi\}$ where $\Psi$ is the lowest root with respect to
$\Delta^0$, to which is associated the extended Dynkin diagram. Now,
deleting arbitrarily some dot(s) of the extended diagram, will yield to
some connected Dynkin diagram or a set of disjointed Dynkin diagrams
corresponding to a regular semi-simple sub(super)algebra of $\cG$.
Indeed, taking away one or more roots from ${\widehat\Delta^0}$, one is
left with a set of independent roots which constitute the simple root
system of a regular semi-simple subsuperalgebra of $\cG$. Then
repeating the same operation on the obtained Dynkin diagrams -- that is
adjunction of a dot associated to the lowest root of a simple part and
cancellation of one arbitrary dot (or two in the unitary case) -- as
many time as necessary, one obtains all the Dynkin diagrams associated
with regular semi-simple basic Lie sub(super)algebras. In order to get
the maximal regular semi-simple sub(super)algebras of the same rank $r$
of $\cG$, only the first step has to be achieved. The other possible
maximal regular subsuperalgebras of $\cG$ if they exist will be
obtained by deleting only one dot in the non-extended Dynkin diagram of
$\cG$ and will be therefore of rank $r-1$.

The table \ref{table14} presents the list of the maximal regular
semi-simple sub(super)algebras for the basic Lie superalgebras. 

\begin{table}[htbp]
\centering
\begin{tabular}{|c|c||c|c|} \hline
superalg. & subsuperalgebra & superalg. & subsuperalgebra \\ 
\hline
&&& \\
$A(m,n)$ & $A(i,j) \oplus A(m-i-1,n-j-1)$ 
 & $C(n+1)$ & $C_i \oplus C(n-i+1)$ \\
 & $A_m \oplus A_n$ && $C_n$ \\
&&& \\
$B(m,n)$ & $B(i,j) \oplus D(m-i,n-j)$
 & $D(m,n)$ & $D(i,j) \oplus D(m-i,n-j)$ \\
 & $B_m \oplus C_n$ && $D_m \oplus C_n$ \\
 & $D(m,n)$ && $A(m-1,n-1)$ \\
&&& \\
$G(3)$ & $A_1 \oplus G_2$ & $F(4)$ & $A_1 \oplus B_3$ \\
 & $A_1 \oplus B(1,1)$ && $A_2 \oplus A(0,1)$ \\
 & $A_2 \oplus B(0,1)$ && $A_1 \oplus D(2,1;2)$ \\
 & $A(0,2)$ && $A(0,3)$ \\
 & $D(2,1;3)$ && $C(3)$ \\
 & $G(3)$ && $ F(4)$ \\
&&& \\
$D(2,1;\alpha)$ & $A_1 \oplus A_1 \oplus A_1$ && \\
 & $A(0,1)$ && \\
 & $D(2,1;\alpha)$ && \\
&&& \\
\hline
\end{tabular}
\caption{Maximal regular sub(super)algebras of the basic Lie
superalgebras.\label{table14}} 
\end{table}

\medskip

$\see$ Cartan subalgebras, Dynkin diagrams, Roots, Simple and semi-simple
Lie superalgebras.

For more details, see ref. \cite{VdJ87}.

\section{\sf Subsuperalgebras (singular)}
\indent

\udef: Let $\cG$ be a basic Lie superalgebra and $\cG'$ a
subsuperalgebra of $\cG$. $\cG'$ is called a {\em singular}
subsuperalgebra of $\cG'$ if it is not regular ($\see$).

Some of the singular subsuperalgebras of the basic Lie superalgebras can
be found by the folding technique. Let $\cG$ be a basic Lie
superalgebra, with non-trivial outer automorphism ($\autout(\cG)$ does not
reduce to the identity). Then, there exists at least one Dynkin diagram
of $\cG$ which has the symmetry given by $\autout(\cG)$. One can notice
that each symmetry $\tau$ described on that Dynkin diagram induces a
direct construction of the subsuperalgebra $\cG'$ invariant under the
$\cG$ outer automorphisms associated to $\tau$. Indeed, if the simple
root $\alpha$ is transformed into $\tau(\alpha)$, then $\half
(\alpha+\tau(\alpha))$ is $\tau$-invariant since $\tau^2 = 1$, and
appears as a simple root of $\cG'$ associated to the generators
$E_{\alpha} + E_{\tau(\alpha)}$, the generator $E_{\alpha}$ (resp.
$E_{\tau(\alpha)}$ corresponding to the root $\alpha$ (resp.
$\tau(\alpha)$). A Dynkin diagram of $\cG'$ will therefore be obtained
by folding the $\ZZ_2$-symmetric Dynkin diagram of $\cG$, that is by
transforming each couple $(\alpha,\tau(\alpha))$ into the root $\half
(\alpha+\tau(\alpha))$ of $\cG'$. One obtains the following invariant
subsuperalgebras (which are singular):
\[
\begin{array}{c|c}
\mbox{superalgebra} \cG \quad &\quad \mbox{singular subsuperalgebra}  
\cG' \cr
\hline
sl(2m+1|2n) \quad &\quad osp(2m+1|2n) \cr
sl(2m|2n) \quad &\quad osp(2m|2n) \cr
osp(2m|2n) \quad &\quad osp(2m-1|2n) \cr
osp(2|2n) \quad &\quad osp(1|2n)
\end{array}
\]

\section{\sf Superalgebra, subsuperalgebra}
\indent

\udef: Let $\cA$ be an algebra over a field $\KK$ of characteristic zero
(usually $\KK=\RR$ or $\CC$) with internal laws $+$ and $\cdot$. One sets
$\ZZ_2 = \ZZ/2\ZZ = \{\evn,\odd\}$. 
$\cA$ is called a superalgebra or $\ZZ_2$-graded algebra if $\cA$
can be written into a direct sum of two spaces $\cA = \cA_\evn \oplus
\cA_\odd$, such that
\[
\cA_\evn \cdot \cA_\evn \subset \cA_\evn, \qquad
\cA_\evn \cdot \cA_\odd \subset \cA_\odd, \qquad
\cA_\odd \cdot \cA_\odd \subset \cA_\evn
\]
Elements $X \in \cA_\evn$ are called even or of degree $\deg X = 0$
while elements $X \in \cA_\odd$ are called odd or of degree $\deg X = 1$.

\medskip

One defines the {\em Lie superbracket} or {\em supercommutator}
of two elements $X$ and $Y$ by
\[
\sle X,Y \sri = X \cdot Y - (-1)^{\deg X.\deg Y} Y \cdot X
\]
A superalgebra $\cA$ is called associative if $(X \cdot Y) \cdot Z
=  X \cdot (Y \cdot Z)$ for all elements $X,Y,Z \in \cA$.
\\
A superalgebra $\cA$ is called commutative if $X \cdot Y = Y \cdot X$
for all elements $X,Y \in \cA$.

\medskip

\udef: A (graded) subalgebra $\cK = \cK_\evn \oplus \cK_\odd$ of a
superalgebra $\cA = \cA_\evn \oplus \cA_\odd$ is a non-empty set $\cK
\subset \cA$ which is a superalgebra with the two composition laws
induced by $\cA$ such that $\cK_\evn \subset \cA_\evn$ and $\cK_\odd
\subset \cA_\odd$.

\medskip

\udef: A homomorphism $\Phi$ from a superalgebra $\cA$ into a
superalgebra $\cA'$ is a linear application from $\cA$ into $\cA'$
which respects the $\ZZ_2$-gradation, that is $\Phi(\cA_\evn) \subset
\cA'_\evn$ and $\Phi(\cA_\odd) \subset \cA'_\odd$.

\medskip

Let $\cA$ and $\cA'$ be two superalgebras. One defines the tensor
product $\cA \otimes \cA'$ of the two superalgebras by
\[
(X_1 \otimes X'_1) (X_2 \otimes X'_2) = (-1)^{\deg X_2.\deg X'_1}
(X_1 X_2 \otimes X'_1 X'_2)
\]
if $X_1,X_2 \in \cA$ and $X'_1,X'_2 \in \cA'$.

\medskip

$\see$ Lie Superalgebras.

\section{\sf Superalgebra $osp(1|2)$}
\indent

The superalgebra $osp(1|2)$ is the simplest one and can be viewed as the
supersymmetric version of $sl(2)$. It contains three bosonic generators
$E^+,E^-,H$ which form the Lie algebra $sl(2)$ and two fermionic
generators $F^+,F^-$, whose non-vanishing commutation relations in the
Cartan-Weyl basis read as 
\[
\begin{array}{lll}
\cle H,E^\pm \cri = \pm E^\pm &\quad& \cle E^+,E^- \cri = 2H \cr &&\cr
\cle H,F^\pm \cri = \pm \half F^\pm &\quad& \ale F^+,F^-\ari = \half
H \cr &&\cr 
\cle E^\pm,F^\mp \cri = -F^\pm &\quad& \ale F^\pm,F^\pm\ari = \pm \half
E^\pm 
\end{array}
\]
The three-dimensional matrix representation (fundamental 
representation) is given by 
\beo
&& H = \left( \begin{array}{ccc} 
\half & 0 & 0 \cr 0 & -\half & 0 \cr 0 & 0 & 0
\end{array} \right) \qquad
E^+ = \left( \begin{array}{ccc} 
0 & 1 & 0 \cr 0 & 0 & 0 \cr 0 & 0 & 0
\end{array} \right) \qquad 
E^- = \left( \begin{array}{ccc} 
0 & 0 & 0 \cr 1 & 0 & 0 \cr 0 & 0 & 0
\end{array} \right) \\
&& F^+ = \left( \begin{array}{ccc} 
0 & 0 & \half \cr 0 & 0 & 0 \cr 0 & \half & 0
\end{array} \right) \qquad
F^- = \left( \begin{array}{ccc} 
0 & 0 & 0 \cr 0 & 0 & -\half \cr \half & 0 & 0
\end{array} \right)
\eno
The quadratic Casimir operator is
\[
C_2 = H^2 + \half (E^+E^- + E^-E^+) - (F^+F^- - F^-F^+)
\]

\medskip

The superalgebra $osp(1|2)$ reveals many features which make it very
close to the Lie algebras. In particular, one has the following results 
for the representation theory:
\begin{enumerate}
\item 
All finite dimensional representations of $osp(1|2)$ are completely
reducible.
\item 
Any irreducible representation of $osp(1|2)$ is typical.
\item 
An irreducible representation $\cR$ of $osp(1|2)$ is characterized by a
non-negative integer or half-integer $j = 0,1/2,1,3/2,\dots$ and
decomposes under the even part $sl(2)$ into two multiplets $\cR_j = D_j
\oplus D_{j-1/2}$ for $j \ne 0$, the case $j = 0$ reducing to the trivial
one-dimensional representation. The dimension of an irreducible
representation $\cR_j$ of $osp(1|2)$ is $4j+1$. The eigenvalue of the
quadratic Casimir $C_2$ in the representation $\cR_j$ is $j(j+\half)$.
\item 
The product of two irreducible $osp(1|2)$ representations decomposes 
as follows:
\[
\cR_{j_1} \otimes \cR_{j_2} = \bigoplus_{j=|j_1-j_2|}^{j=j_1+j_2} \cR_j
\]
$j$ taking integer and half-integer values.
\end{enumerate}

\medskip

$\see$ Casimir invariants, Decomposition w.r.t. $osp(1|2)$ subalgebras,
Embeddings of $osp(1|2)$.

For more details, see refs. \cite{BeT81,NRS77}.

\section{\sf Superalgebra $sl(1|2)$}
\indent

The superalgebra $sl(1|2) \simeq sl(2|1)$ is the (N=2) extended
supersymmetric version of $sl(2)$ and contains four bosonic generators
$E^+,E^-,H,Z$ which form the Lie algebra $sl(2) \oplus U(1)$ and four
fermionic generators $F^+,F^-,\bF^+,\bF^-$, whose commutation relations
in the Cartan-Weyl basis read as
\[
\begin{array}{lllll}
\cle H,E^\pm \cri = \pm E^\pm &\quad& \cle H,F^\pm \cri = \pm \half
F^\pm &\quad& \cle H,\bF^\pm \cri = \pm \half \bF^\pm \cr &&&&\cr
\cle Z,H \cri = \cle Z,E^\pm \cri = 0 &\quad& \cle Z,F^\pm \cri 
 = \half F^\pm &\quad& \cle Z,\bF^\pm \cri = -\half \bF^\pm \cr &&&&\cr
\cle E^\pm,F^\pm \cri = \cle E^\pm,\bF^\pm \cri = 0 &\quad&
\cle E^\pm,F^\mp \cri = -F^\pm &\quad& \cle E^\pm,\bF^\mp \cri 
 = \bF^\pm \cr &&&&\cr
\ale F^\pm,F^\pm \ari = \ale \bF^\pm,\bF^\pm \ari = 0 &\quad&
\ale F^\pm,F^\mp \ari = \ale \bF^\pm,\bF^\mp \ari = 0 &\quad&
\ale F^\pm,\bF^\pm \ari = E^\pm \cr &&&&\cr
\cle E^+,E^- \cri = 2H &\quad& \ale F^\pm,\bF^\mp \ari = Z \mp H &&
\end{array} 
\]
The three-dimensional matrix representation (fundamental 
representation) is given by
\beo
&& H = \left( \begin{array}{ccc} 
\half & 0 & 0 \cr 0 & -\half & 0 \cr 0 & 0 & 0
\end{array} \right) \quad
Z = \left( \begin{array}{ccc} 
\half & 0 & 0 \cr 0 & \half & 0 \cr 0 & 0 & 1
\end{array} \right) \quad
E^+ = \left( \begin{array}{ccc} 
0 & 1 & 0 \cr 0 & 0 & 0 \cr 0 & 0 & 0
\end{array} \right) \quad 
E^- = \left( \begin{array}{ccc} 
0 & 0 & 0 \cr 1 & 0 & 0 \cr 0 & 0 & 0
\end{array} \right) \\
&& F^+ = \left( \begin{array}{ccc} 
0 & 0 & 0 \cr 0 & 0 & 0 \cr 0 & 1 & 0
\end{array} \right) \quad
\bF^+ = \left( \begin{array}{ccc} 
0 & 0 & 1 \cr 0 & 0 & 0 \cr 0 & 0 & 0
\end{array} \right) \quad
F^- = \left( \begin{array}{ccc} 
0 & 0 & 0 \cr 0 & 0 & 0 \cr 1 & 0 & 0
\end{array} \right) \quad
\bF^- = \left( \begin{array}{ccc} 
0 & 0 & 0 \cr 0 & 0 & 1 \cr 0 & 0 & 0
\end{array} \right)
\eno
The quadratic and cubic Casimir operators are
\beo
&& C_2 = H^2 - Z^2 + E^-E^+ + F^-\bF^+ - \bF^-F^+ \\ 
&& C_3 = (H^2-Z^2)Z + E^-E^+(Z-\half) - \half F^-\bF^+(H-3Z+1) \\
&& ~~~~~~~~ - \half \bF^-F^+(H+3Z+1) + \half E^-\bF^+F^+ 
+ \half \bF^-F^-E^+ 
\eno

\medskip

The irreducible representations of $sl(1|2)$ are characterized by the
pair of labels $(b,j)$ where $j$ is a non-negative integer or
half-integer and $b$ an arbitrary complex number. The representations
$\pi(b,j)$ with $b \ne \pm j$ are {\em typical} and have dimension
$8j$. The representations $\pi(\pm j,j)$ are {\em atypical} and have
dimension $4j+1$. In the typical representation  $\pi(b,j)$, the
Casimir operators $C_2$ and $C_3$ have the eigenvalues $C_2 = j^2-b^2$ 
and $C_3 = b(j^2-b^2)$ while they are identically zero in the atypical
representations $\pi(\pm j,j)$.

The typical representation $\pi(b,j)$ of $sl(1|2)$ decomposes
under the even part $sl(2) \oplus U(1)$ for $j \ge 1$ as
\[
\pi(b,j) = D_j(b) \oplus D_{j-1/2}(b-1/2) \oplus D_{j-1/2}(b+1/2)
\oplus D_{j-1}(b) 
\]
the case $j=\half$ reducing to
\[
\pi(b,\half) = D_{1/2}(b) \oplus D_0(b-1/2) \oplus D_0(b+1/2)  
\]
where $D_j(b)$ denotes the representation of $sl(2) \oplus U(1)$ with
isospin $j$ and hypercharge $b$.

The irreducible atypical representations $\pi_\pm(j) \equiv \pi(\pm j,j)$ 
of $sl(1|2)$  decompose under the even part $sl(2) \oplus U(1)$ as
\beo
&& \pi_+(j) = D_j(j) \oplus D_{j-1/2}(j+1/2) \\
&& \pi_-(j) = D_j(-j) \oplus D_{j-1/2}(-j-1/2)
\eno

The not completely reducible atypical representations of $sl(1|2)$
decompose as {\em semi-direct} sums of $sl(1|2)$ irreducible (atypical)
representations. They are of the following types:
\beo
&&\pi_\pm(j;j-1/2) \equiv \pi_\pm(j) \lsemisum \pi_\pm(j-1/2) \\
&&\pi_\pm(j-1/2;j) \equiv \pi_\pm(j-1/2) \lsemisum \pi_\pm(j) \\
&&\pi_\pm(j-1/2,j+1/2;j) \equiv \pi_\pm(j-1/2) \lsemisum \pi_\pm(j)
\rsemisum \pi_\pm(j+1/2) \\ 
&&\pi_\pm(j;j-1/2,j+1/2) \equiv \pi_\pm(j-1/2) \rsemisum \pi_\pm(j)
\lsemisum \pi_\pm(j+1/2) \\ 
&&\pi_\pm(j,j \pm 1;j \pm 1/2;j \pm 3/2) \equiv \pi_\pm(j) \lsemisum
\pi_\pm(j \pm 1/2) \rsemisum \pi_\pm(j \pm 1) \lsemisum 
\pi_\pm(j \pm 3/2) \\
&&\pi_\pm(j;j-1/2,j+1/2;j) \equiv \pi_\pm(j) \begin{array}{c}
\lsemisum \pi_\pm(j-1/2) \lsemisum \cr
\lsemisum \pi_\pm(j+1/2) \lsemisum 
\end{array} \pi_\pm(j) 
\eno
where the symbol $\lsemisum$ (resp. $\rsemisum$) means that the
representation space on the left (resp. on the right) is an invariant
subspace of the whole representation space. 

\medskip

It is also possible to decompose the $sl(1|2)$ representations under the
superprincipal $osp(1|2)$ subsuperalgebra of $sl(1|2)$ ($\see$
Embeddings of $osp(1|2)$). One obtains for the typical representations
$\pi(b,j) = \cR_j \oplus \cR_{j-1/2}$ and for the irreducible atypical
representations $\pi_\pm(j) = \cR_j$ where $\cR_j$ denotes an
irreducible $osp(1|2)$-representation.

\medskip

We give now the formulae of the tensor products of two $sl(1|2)$
representations $\pi(b_1,j_1)$ and $\pi(b_2,j_2)$. In what follows, we
set $b = b_1+b_2$, $j =  j_1+j_2$ and $\bj = |j_1-j_2|$. Moreover, the
product of two irreducible representations will be called non-degenerate
if it decomposes into a direct sum of irreducible representations;
otherwise it is called degenerate.

$\bullet$ product of two typical representations
\\
The product of two typical representations $\pi(b_1,j_1)$ and 
$\pi(b_2,j_2)$ is non-degenerate when $b \ne \pm (j-n)$ for 
$n = 0,1,\dots,2\min(j_1,j_2)$. It is then given by
\beo
&&\pi(b_1,j_1) \otimes \pi(b_2,j_2) = \bigoplus_{n = 0}^{2\min(j_1,j_2)}
\pi(b,j-n) \bigoplus_{n = 1}^{2\min(j_1,j_2)-1} \pi(b,j-n) \\
&& ~~~~~~~~~~~~~~~~~~~~~~~
\bigoplus_{n = 0}^{2\min(j_1,j_2)-1} \pi(b+\half,j-\half-n) ~ \oplus ~ 
\pi(b-\half,j-\half-n) \\
&& \pi(b_1,j_1) \otimes \pi(b_2,\half) = \pi(b,j_1+\half) ~ \oplus ~ 
\pi(b,j_1-\half) ~ \oplus ~ \pi(b+\half,j_1) ~ \oplus ~ 
\pi(b-\half,j_1) \\
&& \pi(b_1,\half) \otimes \pi(b_2,\half) = \pi(b,1) ~ \oplus ~ 
\pi(b+\half,\half) ~ \oplus ~ \pi(b-\half,\half) 
\eno
When the product is degenerate, one has
\beo
&&\smbox{if} b = \pm j \\
&&~~~~~ \pi(b,j) \oplus \pi(b \mp 1/2,j-1/2) \smbox{is replaced by}
\pi_\pm(j-1/2;j-1,j;j-1/2) \\
&&\smbox{if} b = \pm\bj\ne 0 \\
&&~~~~~ \pi(b,\bj) \oplus \pi(b \pm 1/2,\bj+1/2) \smbox{is replaced by}
\pi_\pm(j;j-1/2,j+1/2;j) \\
&&\smbox{if} b = \bj = 0 \\
&&~~~~~ \pi(b+1/2,1/2) \oplus \pi(b-1/2,1/2) \smbox{is replaced by}
\pi(0;-1/2,1/2;0) \\
&&\smbox{if} b = \pm(j-n) \smbox{for} n=1,\dots,2\min(j_1,j_2) \\
&&~~~~~ \pi(b \pm 1/2,j+1/2-n) \oplus \pi(b,j-n) \oplus \pi(b,j-n) \oplus
\pi(b \mp 1/2,j-1/2-n) \\
&&~~~~~ \smbox{is replaced by} \pi_\pm(j-1/2-n;j-1-n,j-n;j-1/2-n) \\
&&~~~~~~~~ \phantom{is replaced by} \oplus \pi_\pm(j-n;j-1/2-n,j+1/2-n;j-n) 
\eno

$\bullet$ product of a typical with an atypical representation
\\
The non-degenerate product of a typical representation $\pi(b_1,j_1)$ with
an atypical one $\pi_\pm(j_2)$ ($b_2=\pm j_2$) is given by
\[
\pi(b_1,j_1) \otimes \pi_\pm(j_2) = \left\{ 
\begin{array}{l} 
{\displaystyle 
\bigoplus_{n = 0}^{2\min(j_1,j_2)-1} \pi(b,j-n) ~ \oplus ~ 
\pi(b\pm\half,j-\half-n)} \cr 
~~~~~~~~~~~~~~~~~~~~~~~~~~~~~~~~~~~~~~~~~~~~~~~~~~~~~~~~~~~~~~
\smbox{if} j_1 \le j_2 \cr
{\displaystyle 
\bigoplus_{n = 0}^{2\min(j_1,j_2)-1} \pi(b,j-n) ~ \oplus ~ 
\pi(b\pm\half,j-\half-n) ~ \oplus ~ \pi(b,|j_1-j_2|)} \cr
~~~~~~~~~~~~~~~~~~~~~~~~~~~~~~~~~~~~~~~~~~~~~~~~~~~~~~~~~~~~~~
\smbox{if} j_1 > j_2
\end{array} \right.
\]
When the product $\pi(b_1,j_1) \otimes \pi_+(j_2)$ is degenerate, one
has 
\beo
&&\smbox{if} b = -(j-n) \smbox{for} n=0,1,\dots,2\min(j_1,j_2)-1 \\
&&~~~~~ \pi(b,j-n) \oplus \pi(b+1/2,j-1/2-n) \smbox{is replaced by} \\
&&~~~~~ \pi_\pm(j-1/2-n;j-1-n,j-n;j-1/2-n) \\
&&\smbox{if} b = j-n \smbox{for} n=1,\dots,2\min(j_1,j_2) \\
&&~~~~~ \pi(b,j-n) \oplus \pi(b+1/2,j+1/2-n) \smbox{is replaced by} \\
&&~~~~~ \pi_\pm(j-n;j-1/2-n,j+1/2-n;j-n) 
\eno
The case of the degenerate product $\pi(b_1,j_1) \otimes \pi_-(j_2)$ is
similar.

$\bullet$ product of two atypical representations
\\
The product of two atypical representations $\pi_\pm(j_1)$ and 
$\pi_\pm(j_2)$ is always non-degenerate. It is given by
\beo
&& \pi_\pm(j_1) \otimes \pi_\pm(j_2) = \pi_\pm(j) \oplus
\bigoplus_{n = 0}^{2\min(j_1,j_2)-1} \pi(\pm (j+\half),j-\half-n) \\
&& \pi(j_1,j_1) \otimes \pi(-j_2,j_2) = 
\bigoplus_{n = 0}^{2\min(j_1,j_2)-1} 
\pi(j_1-j_2,j-n) ~ \oplus ~ \left\{ \begin{array}{ll} 
\pi(j_1-j_2,j_1-j_2) & \smbox{if} j_1 > j_2 \cr
\pi(j_1-j_2,j_2-j_1) & \smbox{if} j_1 < j_2 \cr
(0) & \smbox{if} j_1 = j_2 
\end{array} \right.
\eno

\medskip

$\see$ Casimir invariants, Decomposition w.r.t. $sl(1|2)$ subalgebras,
Embeddings of $sl(1|2)$.

For more details, see refs. \cite{Mar80,NRS77}.

\section{\sf Superconformal algebra}
\indent

For massless theory the concept of Fermi-Bose symmetry or supersymmetry
requires the extension of the conformal Lie algebra including the
generators of the supersymmetry transformations $Q_{\alpha}$,
$S_{\alpha}$ which transform bosonic fields into fermionic ones and
vice-versa. The conformal algebra in four space-time dimensions is
spanned by the 15 generators $M_{\mu\nu}$, $P_\mu$, $K_\mu$ and $D$
(with the greek labels running from 0 to 3). The generators
$M_{\mu\nu}$ and $P_\mu$ span the Poincar\'e algebra and their
commutation relations are given in "Supersymmetry algebra" ($\see$),
while $K_\mu$ and $D$ are respectively the generators of the
conformal transformations and of the dilatation. The commutation
relations of the $N=1$ superconformal algebra read as (the metric is
$g_{\mu\nu} = \diag(1,-1,-1,-1)$):
\[
\begin{array}{ll}
\bigg. \cle M_{\mu\nu},K_\rho \cri = i(g_{\nu\rho}K_{\mu} - 
g_{\mu\rho}K_{\nu}) 
& \cle P_{\mu},K_{\nu}\cri = 2i(g_{\mu\nu} D - M_{\mu\nu}) \cr
\bigg. \cle D,M_{\mu\nu} \cri = 0 
& \cle D,P_{\mu} \cri = -iP_{\mu} \cr
\bigg. \cle K_{\mu},K_{\nu} \cri = 0 
& \cle D,K_{\mu} \cri = iK_{\mu} \cr
\bigg. \cle M_{\mu\nu},Q_a \cri = -\half (\Sigma_{\mu\nu})_a^{~b} Q_b
& \cle M_{\mu\nu},S_a \cri = -\half (\Sigma_{\mu\nu})_a^{~b} S_b \cr
\bigg. \cle P_{\mu},Q_a \cri = 0 
& \cle P_{\mu},S_a \cri = -(\gamma_\mu)_a^{~b} Q_b \cr
\bigg. \cle K_{\mu},Q_a \cri = -(\gamma_\mu)_a^{~b} S_b 
& \cle K_{\mu},S_a \cri = 0 \cr
\bigg. \cle D,Q_a \cri = -\half i Q_a
& \cle D,S_a \cri = \half i S_a \cr
\bigg. \ale Q_a,Q_b \ari = 2 (\gamma_\mu C)_{ab} 
P^{\mu} 
& \ale S_a,S_b \ari = 2 (\gamma_\mu C)_{ab} 
K^{\mu} \cr
\bigg. \ale Q_a,S_b \ari = (\Sigma_{\mu\nu} C)_{ab} M^{\mu\nu} 
+ 2i C_{ab} D & \cr 
~~~~~~~~~~~~~~~~ + 3i(\gamma_5 C)_{ab} Y & \cr 
\bigg. \cle Y,Q_a \cri = i (\gamma_5)_a^{~b} Q_b
& \cle Y,S_a \cri = -i (\gamma_5)_a^{~b} S_b \cr 
\bigg. \cle Y,M_{\mu\nu} \cri = \cle Y,D \cri = 0
& \cle Y,P_\mu \cri = \cle Y,K_\mu \cri = 0 \cr
\end{array}
\]
where $\gamma$ are the Dirac matrices in Majorana representation, $C$ is
the charge conjugation matrix and $Y$ is the generator of the (chiral)
$U(1)$. The transformations of $Q_a$  and $S_a$ under
$M_{\mu\nu}$ show that the $Q_a$ and $S_a$ are spinors.
The superconformal algebra contains the super-Poincar\'e as
subsuperalgebra, however in the conformal case the are no central
charges for $N>1$.

\medskip

Let us emphasize that the superconformal algebra is isomorphic to
the simple Lie superalgebra $su(2,2|N)$, real form of $sl(4|N)$.

\medskip

$\see$ Spinors, Supersymmetry algebra.

For more details, see refs. \cite{Soh85,Wes86}.

\section{\sf Supergroups}
\indent

In order to construct the supergroup or group with Grassmann structure
associated to a (simple) superalgebra $\cA = \cA_\evn \oplus \cA_\odd$,
one starts from the complex Grassmann algebra ($\see$) $\Gamma(n)$ of
order $n$ with $n$ generators 1, $\theta_1, \dots, \theta_n$ satisfying
$\ale \theta_i , \theta_j \ari = 0$. The element
\[
\eta = \sum_{m \ge 0} ~ \sum_{i_1 < \dots < i_m} \eta_{i_1 \dots i_m}
\theta_{i_1} \dots \theta_{i_m}
\]
is called even (resp. odd) if each complex coefficient $\eta_{i_1 \dots
i_m}$ in the above expression of $\eta$ corresponds to an even (resp.
odd) value of $m$. As a vector space, one decomposes $\Gamma(n)$ as
$\Gamma(n) = \Gamma(n)_\evn \oplus \Gamma(n)_\odd$ with
$\Gamma(n)_\evn$ (resp. $\Gamma(n)_\odd$) made of homogeneous even
(resp. odd) elements.

The Grassmann envelope $A(\Gamma)$ of $\cA$ consists of formal linear
combinations $\sum_i \eta_i a_i$ where $\{a_i\}$ is a basis of $\cA$
and $\eta_i \in \Gamma(n)$ such that for a fixed index $i$, the
elements $a_i$ and $\eta_i$ are both even or odd. The commutator
between two arbitrary elements $X = \sum_i \eta_i a_i$ and $Y = \sum_j
\eta'_j a_j$ is naturally defined by $\cle X,Y \cri = \sum_{ij} \eta_i
\eta'_j \sle a_i,a_j \sri$ where $\sle a_i,a_j \sri$ means the
supercommutator in $\cA$. This commutator confers to the Grassmann
envelope $A(\Gamma)$ of $\cA$ a {\em Lie algebra} structure.

The relation between a supergroup and its superalgebra is analogous to
the Lie algebra case: the supergroup $A$ associated to the superalgebra
$\cA$ is the exponential mapping of the Grassmann envelope $A(\Gamma)$
of $\cA$, the even generators of the superalgebra $\cA$ corresponding
to even parameters (that is even elements of the Grassmann algebra) and
the odd generators of $\cA$ to odd parameters (that is odd elements of
the Grassmann algebra).

The above approach is due to Berezin. In particular, the case of
$osp(1|2)$ is worked out explicitly in ref. \cite{BeT81}. On classical
supergroups, see also refs. \cite{Kac77a,Kac77b}.

\medskip

$\see$ Grassmann algebras.

\section{\sf Supergroups of linear transformations}
\indent

Let $\Gamma = \Gamma_\evn \oplus \Gamma_\odd$ be a Grassmann algebra
($\see$) over a field $\KK = \RR$ or $\CC$ and consider the set of
$(m+n) \times (m+n)$ even supermatrices ($\see$) of the form
\[
M = \left( \begin{array}{cc} A & B \cr C & D \end{array} \right)
\]
where $A,B,C,D$ are $m \times m$, $m \times n$, $n \times m$ and $n
\times n$ submatrices respectively, with even entries in $\Gamma_\evn$
for $A,D$ and odd entries in $\Gamma_\odd$ for $B,C$.

The {\em general linear supergroup} $GL(m|n;\KK)$ is the supergroup of
even invertible supermatrices $M$, the product law being the usual
matrix multiplication.

The transposition and adjoint operations allow us to define the 
classical subsupergroups of $GL(m|n;\KK)$ corresponding to the classical
superalgebras.

\medskip

The special linear supergroup $SL(m|n;\KK)$ is the subsupergroup of
supermatrices $M \in GL(m|n;\KK)$ such that $\sdet(M) = 1$.

The unitary and superunitary supergroups $U(m|n)$ and $sU(m|n)$ are the
subsupergroups of supermatrices $M \in GL(m|n;\CC)$ such that
$MM^\dagger = 1$ and $MM^\ddagger = 1$ respectively (for the notations
$\dagger$ and $\ddagger$, $\see$ Supermatrices).

The orthosymplectic supergroup $OSP(m|n;\KK)$ is the subsupergroup of
supermatrices $M \in GL(m|n;\KK)$ such that $M^{st} H M = H$ where
($n=2p$) 
\[
H = \left( \begin{array}{cc} \II_m & 0 \cr 0 & \JJ_{2p} \end{array}
\right) \bigbox{and}
\JJ_{2p} = \left( \begin{array}{cc} 0 & \II_p \cr -\II_p & 0 \end{array} 
\right)
\]

The compact forms are $USL(m|n)$ and $sOSP(m|n)$, subsupergroups of
supermatrices $M \in GL(m|n;\CC)$ such that $\sdet(M) = 1, ~ MM^\dagger
= 1$ and $M^{st} H M = H, ~ MM^\ddagger = 1$ respectively.

Finally the strange supergroups are defined as follows. The supergroup
$P(n)$ is the subsupergroup of supermatrices $M \in GL(n|n;\KK)$ such
that $\sdet(M) = 1$ and $M \JJ_{2n} M^{st} = \JJ_{2n}$ with $\JJ_{2n}$
defined above. The supergroup $Q(n)$ is the subsupergroup of
supermatrices $M \in GL(n|n;\KK)$ with $A=D$ and $B=C$ such that $\tr
\ln ((A-B)^{-1}(A+B)) = 0$.

\medskip

For more details, see ref. \cite{Rit77}.

\section{\sf Supermatrices}
\indent

\udef: A matrix $M$ is called a complex (resp. real) {\em supermatrix} if
its entries have values in a complex (resp. real) Grassmann algebra
$\Gamma = \Gamma_\evn \oplus \Gamma_\odd$. More precisely, consider the
set of $(m+n) \times (p+q)$ supermatrices $M$ of the form
\[
M = \left( \begin{array}{cc} A & B \cr C & D \end{array} \right)
\]
where $A,B,C,D$ are $m \times p$, $m \times q$, $n \times p$ and $n
\times q$ submatrices respectively. The supermatrix $M$ is called {\em
even} (or of degree 0) if $A,D \in \Gamma_\evn$ and $B,C \in
\Gamma_\odd$, while it is called {\em odd} (or of degree 1) if $A,D \in
\Gamma_\odd$ and $B,C \in \Gamma_\evn$.

\medskip

The product of supermatrices is defined as the product of matrices: $M$
and $M'$ being two $(m+n) \times (p+q)$ and $(p+q) \times (r+s)$
supermatrices, the entries of the $(m+n) \times (r+s)$ supermatrix $MM'$
are given by 
\[
(MM')_{ij} = \sum_{k=1}^{p+q} M_{ik} M'_{kj}
\]
Since the Grassmann algebra $\Gamma$ is associative, the product of
supermatrices is also associative.

\medskip

{}From now on, we will consider only square supermatrices, that is such
that $m=p$ and $n=q$. The set of $(m+n) \times (m+n)$ complex (resp. real)
square supermatrices is denoted by $M(m|n;\CC)$ (resp. $M(m|n;\RR)$).

A square supermatrix $M$ is said to be invertible if there exists a
square supermatrix $M'$ such that $MM' = M'M = I$ where $I$ is the unit
supermatrix (even supermatrix with zero off-diagonal entries and diagonal
entries equal to the unit 1 of the Grassmann algebra $\Gamma$).

\udef: The {\em general linear supergroup} $GL(m|n;\CC)$ (resp.
$GL(m|n;\RR)$) is the super\-group of even invertible complex (resp.
real) supermatrices, the group law being the product of supermatrices.

\medskip

The usual operations of transposition, determinant, trace, adjoint
are defined as follows in the case of supermatrices.
\\
Let $M \in M(m|n;\CC)$ be a complex square supermatrix of the form 
\[
M = \left( \begin{array}{cc} A & B \cr C & D \end{array} \right)
\]

\medskip

The transpose and supertranspose of $M$ are defined by:
\[
\begin{array}{ll}
M^t = \left( \begin{array}{cc} A^t & C^t \cr B^t & D^t \end{array} 
\right) & \mbox{transpose} \cr &\cr
M^{st} = \left( \begin{array}{cc} A^t & (-1)^{\deg M} C^t \cr 
-(-1)^{\deg M}B^t & D^t \end{array} \right) 
& \mbox{supertranspose} \cr
\end{array}
\]
Explicitly, one finds
\[
\begin{array}{ll}
M^{st} = \left( \begin{array}{cc} A^t & C^t \cr 
-B^t & D^t \end{array} \right) 
& \mbox{if $M$ is even} \cr
 &\cr
M^{st} = \left( \begin{array}{cc} A^t & -C^t \cr 
B^t & D^t \end{array} \right) 
& \mbox{if $M$ is odd} \cr
\end{array}
\]
It follows that
\beo
&& ((M)^{st})^{st} = 
\left( \begin{array}{cc} A & -B \cr -C & D \end{array} \right) \\
&& ((((M)^{st})^{st})^{st})^{st} = M \\
&& (MN)^{st} = (-1)^{\deg M.\deg N} N^{st} M^{st} 
\eno
but $(MN)^t \ne N^t M^t$.

\medskip

The supertrace of $M$ is defined by 
\[
\str(M) = \tr(A) - (-1)^{\deg M} \tr(D) = 
\left\{ \begin{array}{l} 
\tr(A) - \tr(D) \medbox{if $M$ is even} \cr
\tr(A) + \tr(D) \medbox{if $M$ is odd}
\end{array} \right.
\]
One has the following properties for the supertrace:
\beo
&& \str(M+N) = \str(M) + \str(N) \medbox{if $\deg M = \deg N$} \\
&& \str(MN) = (-1)^{\deg M.\deg N} \str(M) \str(N) \\
&& \str(M^{st}) = \str(M) 
\eno

\medskip

If $M$ is even invertible, one defines the superdeterminant (or
Berezinian) of $M$ by 
\[
\sdet(M) = \frac{\det(A-BD^{-1}C)}{\det(D)}
= \frac{\det(A)}{\det(D-CA^{-1}B)}
\]
Notice that $M$ being an even invertible matrix, the inverse matrices
$A^{-1}$ and $D^{-1}$ exist.
\\
One has the following properties for the superdeterminant:
\beo
&& \sdet(MN) = \sdet(M)\,\sdet(N) \\
&& \sdet(M^{st}) = \sdet(M)  \\
&& \sdet(\exp(M)) = \exp(\str(M))
\eno
The adjoint operations on the supermatrix $M$ are defined by
\[
\begin{array}{ll}
M^\dagger = {(M^t)}^* & \mbox{adjoint} \cr
M^\ddagger = {(M^{st})}^{\#} & \mbox{superadjoint} 
\end{array}
\]
One has
\beo
&& (MN)^\dagger = N^\dagger M^\dagger \\
&& (MN)^\ddagger = N^\ddagger M^\ddagger \\
&& (M^\dagger)^\dagger = M \bigbox{and} (M^\ddagger)^\ddagger = M \\
&& \sdet(M^\dagger) = \overline{\sdet(M)} = (\sdet(M))^*
\eno
where the bar denotes the usual complex conjugation and the star the
Grassmann complex conjugation ($\see$ Grassmann algebra).

\medskip

$\see$ Supergroups of linear transformations.

For more details, see refs. \cite{Rit77,BaB81}.

\section{\sf Superspace and superfields}
\indent

It is fruitful to consider the supergroup associated to the
supersymmetry algebra, the super-Poincar\'e group. A group element $g$
is then given by the exponential of the supersymmetry algebra
generators. However, since $Q_\alpha$ and $\bQ_\dalpha$ are
fermionic, the corresponding parameters have to be anticommuting
($\see$ Grassmann algebra).
More precisely, a group element $g$ with parameters $x^\mu, 
\omega^{\mu\nu}, \theta^\alpha, \btheta^\dalpha$ is given by
\[
g(x^\mu,\omega^{\mu\nu},\theta^\alpha,\btheta^\dalpha) = 
\exp i( x^\mu P_\mu + \half \omega^{\mu\nu} M_{\mu\nu} + 
\theta^\alpha Q_\alpha + \bQ_\dalpha \btheta^\dalpha)
\]
One defines the superspace as the coset space of the super-Poincar\'e
group by the Lorentz group, parametrized by the coordinates 
$x^\mu, \theta^\alpha, \btheta^\dalpha$ subject to the 
condition $\btheta^\dalpha = (\theta^\alpha)^*$.
The multiplication of group elements is induced by the supersymmetry
algebra: 
\[
g(x^\mu,\theta^\alpha,\btheta^\dalpha) ~ 
g(y^\mu,\zeta^\alpha,\bzeta^\dalpha) =
g(x^\mu+y^\mu+i\theta\sigma^\mu\bzeta-i\zeta\sigma^\mu\btheta,
\theta+\zeta,\btheta+\bzeta) 
\]
If group element multiplication is considered as a {\em left} action,
one can write infinitesimally
\[
g(y^\mu,\zeta^\alpha,\bzeta^\dalpha) ~ 
g(x^\mu,\theta^\alpha,\btheta^\dalpha) =
\left[ 1 - i y^\mu P_\mu - i \zeta^\alpha Q_\alpha - i \bzeta_\dalpha 
\bQ^\dalpha \right] g(x^\mu,\theta^\alpha,\btheta^\dalpha)
\]
where the differential operators
\[
Q_\alpha = i\frac{\prt}{\prt\theta^\alpha} - (\sigma^\mu \btheta)_\alpha
\prt_\mu \bigbox{and} \bQ_\dalpha = -i\frac{\prt}{\prt\btheta^\dalpha}
+ (\theta \sigma^\mu)_\dalpha \prt_\mu
\]
are the supersymmetry generators of the supersymmetry algebra ($\see$).
\\
If group element multiplication is considered as a {\em right} action, 
one has infinitesimally
\[
g(x^\mu,\theta^\alpha,\btheta^\dalpha) ~ 
g(y^\mu,\zeta^\alpha,\bzeta^\dalpha) =
\left[ 1 - i y^\mu P_\mu - i \zeta^\alpha D_\alpha - i \bzeta_\dalpha 
\bD^\dalpha \right] g(x^\mu,\theta^\alpha,\btheta^\dalpha)
\]
where the differential operators
\[
D_\alpha = i\frac{\prt}{\prt\theta^\alpha} + (\sigma^\mu \btheta)_\alpha
\prt_\mu \bigbox{and} \bD_\dalpha = -i\frac{\prt}{\prt\btheta^\dalpha}
- (\theta \sigma^\mu)_\dalpha \prt_\mu
\]
satisfy the following algebra
\beo
&& \ale D_\alpha,\bD_\dbeta \ari = - 2i \sigma^\mu_{~\alpha\dbeta} ~ 
\prt_\mu \\
&& \ale D_\alpha,D_\beta \ari = \ale \bD_\dalpha,\bD_\dbeta \ari = 0
\eno
and anticommute with the $Q_\alpha$ and $\bQ_\dalpha$
generators.
\\
Unlike the $Q$ generators, the $D$ generators behave like 
{\em covariant derivatives} under the super-Poincar\'e group.

\medskip

One defines a {\em superfield} $\cF$ as a function of the superspace.
Since the parameters $\theta^\alpha,\btheta^\dalpha$ are
Grassmann variables, a Taylor expansion of $\cF$ in $\theta,\btheta$ 
has a finite number of terms:
\beo
\cF(x,\theta,\btheta) &=& f(x) + \theta \phi(x) +
\btheta\bchi(x) + \theta\theta m(x) + \btheta\btheta n(x) \\
&& + \theta\sigma^\mu\btheta A_\mu(x) + \theta\theta\btheta\blambda(x) 
+ \btheta\btheta\theta\lambda'(x) + \theta\theta\btheta\btheta d(x)
\eno
Notice the very important property that the product of two superfields is
again a superfield.

Under a superspace transformation, the variation of the superfield $\cF$
is given by the action of the supersymmetry generators $Q_\alpha$ and 
$\bQ_\dalpha$: 
\[
\delta\cF(x,\theta,\btheta) = -i(\zeta Q + \bQ\bzeta) \cF
\]
The superfield $\cF$ forms a representation of the supersymmetry algebra.
However, this representation is not irreducible. Irreducible
representations can be obtained by imposing constraints on the
superfields. The two main examples are the scalar (chiral or 
antichiral) and the vector superfields.
\\
- The {\em chiral} superfield $\cF$ is defined by the covariant
constraint $\bD_\dalpha \cF = 0$. It follows that the chiral
superfield $\cF$ can be expressed, in terms of $y^\mu = x^\mu - i\theta
\sigma^\mu\btheta$ and $\theta$, as
\[
\cF = A(y) + 2\theta\psi(y) + \theta\theta F(y)
\]
The transformation law for the chiral superfield is therefore
\beo
&& \delta A = 2\zeta \psi \\
&& \delta \psi = -i\sigma^\mu\bzeta\prt_\mu A + \zeta F \\
&& \delta F = 2i\prt_\mu\psi\sigma^\mu\bzeta
\eno
- In the same way, the {\em antichiral} superfield $\cF$ is defined by
the covariant constraint $D_\alpha \cF = 0$. The antichiral superfield 
$\cF$ can be expressed, in terms of $(y^\mu)^\dagger = x^\mu + i\theta
\sigma^\mu\btheta$ and $\btheta$, as
\[
\cF = A^*(y^\dagger) + 2\btheta\bpsi(y^\dagger) +
\btheta\btheta F^*(y^\dagger) 
\]
and the transformation law for the antichiral superfield is 
\beo
&& \delta A^\dagger = 2\bpsi \bzeta \\
&& \delta \bpsi = i\zeta\sigma^\mu\prt_\mu A^\dagger + F^\dagger\bzeta \\
&& \delta F^\dagger = -2i\zeta\sigma^\mu\prt_\mu\bpsi
\eno
- The {\em vector} superfield $\cF$ is defined by the reality constraint
$\cF^\dagger = \cF$. In terms of $x,\theta,\btheta$, it takes the form
(with standard notations)
\beo
\cF(x,\theta,\btheta) &=& C(x) + i \theta \chi(x) - i \btheta\bchi(x) +
\sfrac{i}{2} \theta\theta \Big(M(x)+iN(x)\Big)  - \sfrac{i}{2}
\btheta\btheta \Big(M(x)-iN(x)\Big) \\
&& - \theta\sigma^\mu\btheta A_\mu(x) + i \theta\theta\btheta
\Big(\blambda(x) + \sfrac{i}{2} \bsigma^\mu \prt_\mu\chi(x)\Big) - i
\btheta\btheta\theta \Big(\lambda(x) + \sfrac{i}{2}
\sigma^\mu\prt_\mu\bchi(x)\Big) \\
&& + \half\theta\theta\btheta\btheta \Big(D(x)+\half\Box C(x)\Big)
\eno
where $C,M,N,D$ are real scalar fields, $A_\mu$ is a real vector field
and $\chi,\lambda$ are spinor fields.

\medskip

$\see$ Grassmann algebras, Spinors, Supersymmetry algebra: definition,
representations.

For more details, see refs. \cite{BWe83,Soh85,Wes86}.

\section{\sf Supersymmetry algebra: definition}
\indent

The concept of Fermi-Bose symmetry or supersymmetry requires the
extension of the Poincar\'e Lie algebra including the generators of the
supersymmetry transformations $Q_{\alpha}$ and $\bQ_\dalpha$,
which are fermionic, that is transform bosonic fields into fermionic
ones and vice-versa. The supersymmetry generators $Q_{\alpha}$ and
$\bQ_\dalpha$ behave like (1/2,0) and (0,1/2) spinors under the
Lorentz group. 
\\
The metric being $g_{\mu\nu} = \diag(1,-1,-1,-1)$, the
$N=1$ supersymmetry algebra takes the following form in two-spinor
notation (the indices $\mu,\nu,\dots = 0,1,2,3$ are space-time indices
while the indices $\alpha, \beta = 1,2$ and $\dalpha, \dbeta =
\dot 1, \dot2$ are spinorial ones):
\beo 
&&~ \cle M_{\mu\nu},M_{\rho\sigma} \cri = i(
- g_{\nu\sigma}M_{\mu\rho} + g_{\nu\rho}M_{\mu\sigma}
+ g_{\mu\sigma}M_{\nu\rho} - g_{\mu\rho}M_{\nu\sigma}) \\
&&~ \cle M_{\mu\nu},P_\rho \cri = i(g_{\nu\rho}P_{\mu} -
g_{\mu\rho}P_{\nu}) \\ 
&&~ \cle P_\mu,P_\nu \cri = 0 \\
&&\begin{array}{ll}
\cle M_{\mu\nu},Q_\alpha \cri = -\half 
(\sigma_{\mu\nu})_{\alpha}^{~\beta} Q_{\beta} & \qquad
\cle M_{\mu\nu},\bQ_\dalpha \cri = -\half 
\bQ_\dbeta (\bsigma_{\mu\nu})_{~\dalpha}^\dbeta \cr
\bigg. \cle P_\mu,Q_\alpha \cri = \cle P_\mu,\bQ_\dalpha \cri = 0 \cr 
\bigg. \ale Q_\alpha,Q_\beta \ari = 
\ale \bQ_\dalpha,\bQ_\dbeta \ari = 0 & \qquad 
\ale Q_\alpha,\bQ_\dbeta \ari = 2\sigma^\mu_{~\alpha\dbeta} P_\mu \cr
\end{array}
\eno
where the $\sigma^i$ are the Pauli matrices, $\bsigma^i = -\sigma^i$ for
$i=1,2,3$ and $\sigma^0=\bsigma^0=\II$. The matrices
$\half\sigma^{\mu\nu}$ and $\half\bsigma^{\mu\nu}$ are the generators
of the Lorentz group in the two fundamental spinorial representations:
$\sigma^{\mu\nu} = \sfrac{i}{2}(\sigma^\mu\bsigma^\nu -
\sigma^\nu\bsigma^\mu)$ and $\bsigma^{\mu\nu} =
\sfrac{i}{2}(\bsigma^\mu\sigma^\nu - \bsigma^\nu\sigma^\mu)$.
\\
In four-spinor notation, the $N=1$ supersymmetry algebra reads as:
\beo
&&\cle M_{\mu\nu},Q_a \cri = -\half (\Sigma_{\mu\nu})_a^{~b} Q_b \\ 
&&\cle P_\mu,Q_a \cri = 0 \\
&&\ale Q_a,Q_b \ari = 2 (\gamma_\mu C)_{ab} P^\mu
\eno
where $Q_a = \left( \begin{array}{c} Q_\alpha \cr \bQ^\dalpha
\end{array} \right)$ is a Majorana spinor ($a=1,2,3,4$), $\gamma^\mu$
are the Dirac matrices in the Majorana representation, $C$ is the
charge conjugation matrix and the $\Sigma^{\mu\nu}$ are the generators
of the Lorentz group in the representation $(1/2,0) \oplus (0,1/2)$:
$\Sigma^{\mu\nu} = \sfrac{i}{2}(\gamma^\mu\gamma^\nu -
\gamma^\nu\gamma^\mu)$ ($\see$ Spinors).

\medskip

There is an extended version of this algebra if one considers many
supersymmetry generators $Q_\alpha^A$, $\bQ_\dalpha^A$ with 
$A = 1,\dots,N$ transforming under some symmetry group. The extended
$N$-supersymmetry algebra becomes then in two-spinor notation:
\[
\begin{array}{ll}
\bigg. \cle M_{\mu\nu},Q_\alpha^A \cri = -\half 
(\sigma_{\mu\nu})_{\alpha}^{~\beta} Q_{\beta}^A 
& \qquad \cle M_{\mu\nu},\bQ_\dalpha^A \cri = -\half 
\bQ_\dbeta^A (\bsigma_{\mu\nu})_{~\dalpha}^\dbeta \cr 
\bigg. \cle P_\mu,Q_\alpha^A \cri = 0
& \qquad \cle P_\mu,\bQ_\dalpha^A \cri = 0 \cr
\bigg. \ale Q_\alpha^A,Q_\beta^B \ari = 2 \eps_{\alpha\beta} Z^{AB} 
& \qquad \ale \bQ_\dalpha^A,\bQ_\dbeta^B \ari = 
-2 \eps_{\dalpha\dbeta} (Z^{AB})^\dagger \cr
\bigg. \ale Q_\alpha^A,\bQ_\dbeta^B \ari =
2\sigma^\mu_{~\alpha\dbeta} P_\mu ~ \delta_{AB} \cr
\bigg. \cle T_i,T_j \cri = i ~ f_{ij}^k T_k 
& \qquad \cle T_i,M_{\mu\nu} \cri = \cle T_i,P_\mu \cri = 0 \cr
\bigg. \cle T_i,Q_\alpha^A \cri = (\zeta_i)^A_{~B} Q_\alpha^B 
& \qquad \cle T_i,\bQ_\dalpha^A \cri = -\bQ_\dalpha^B 
(\zeta_i)^{~A}_B \cr
\bigg. \cle Z^{AB},\mbox{anything} \cri = 0
\end{array}
\]
while in four-spinor notation it takes the form (for the relations
involving the supersymmetry generators): 
\beo
&&\cle M_{\mu\nu},Q_a^A \cri = -\half (\Sigma_{\mu\nu})_a^{~b} Q_b^A \\ 
&&\cle P_\mu,Q_a^A \cri = 0 \\
&&\ale Q_a^A,Q_b^B \ari = 2 (\gamma^\mu C)_{ab} P_\mu
\delta^{AB} + C_{ab} U^{AB} + (\gamma_5 C)_{ab} V^{AB} \\
&&\cle T_i,Q_a^A \cri = (\xi_i)^A_B Q_a^B + (i\zeta_i)^A_B 
(\gamma_5)_a^b Q_b^B \\
&&\cle U^{AB},\mbox{anything} \cri = \cle V^{AB},\mbox{anything} \cri = 0
\eno
$U^{AB}$ and $V^{AB}$ being central charges and the matrices
$\xi_i,\zeta_i$ having to satisfy $(\xi_i + i\zeta_i) + 
(\xi_i + i\zeta_i)^\dagger = 0$. 
\\
Actually, the number of central charges $U^{AB}=-U^{BA}$ and
$V^{AB}=-V^{BA}$ present in the algebra imposes constraints on the
symmetry group of the matrices $\xi_i$ and $\zeta_i$. If there is no
central charge this symmetry group is $U(N)$, otherwise it is $USp(2N)$,
compact form of $Sp(2N)$.

\medskip

For more details, see refs. \cite{BWe83,Soh85,Wes86}.

\section{\sf Supersymmetry algebra: representations}
\indent

We will only consider the finite dimensional representations of the
$N$-supersymmetry algebra ($\see$ Supersymmetry algebra: definition).
Since the translation generators $P^\mu$ commute with the supersymmetry
generators $Q_\alpha^A$ and $\bQ_\dalpha^A$, the
representations of the $N$-supersymmetry algebra are labelled by the
mass $M$ if $M^2$ is the eigenvalue of the Casimir operator $P^2 =
P^\mu P_\mu$.

If $N_F$ denotes the fermion number operator, the states $\ket{B}$ such
that $(-1)^{N_F} \ket{B} = \ket{B}$ are bosonic states while the states
$\ket{F}$ such that $(-1)^{N_F} \ket{F} = -\ket{F}$ are fermionic ones.
In a finite dimensional representation, one has $\tr(-1)^{N_F} = 0$,
from which it follows that {\em the finite dimensional representations 
of the supersymmetry algebra contain an equal number of bosonic and
fermionic states}.

\medskip

For the massive representations ($M \ne 0$), the supersymmetry algebra
in the rest frame, where $P^\mu=(M,0,0,0)$, takes the form (with
vanishing central charges)
\beo
&& \ale Q_\alpha^A,\bQ_\dbeta^B \ari =
2M \delta_{\alpha\dbeta} \delta_{AB} \\
&& \ale Q_\alpha^A,Q_\beta^B \ari = 
\ale \bQ_\dalpha^A,\bQ_\dbeta^B \ari = 0
\eno
with $A,B=1,\dots,N$.
\\
The rescaled operators $a_\alpha^A = Q_\alpha^A/\sqrt{2M}$ and
$(a_\alpha^A)^\dagger = \bQ_\dalpha^A/\sqrt{2M}$ satisfy the
Clifford algebra ($\see$) in $2N$ dimensions.
The states of a representation can be arranged into spin multiplets of
some ground state -- or vacuum -- $\ket{\Omega}$ of given spin $s$,
annihilated by the $a_\alpha^A$ operators. The other states of the
representation are given by
\[
\ket{a_{\alpha_1}^{A_1} \dots a_{\alpha_n}^{A_n}}
= (a_{\alpha_1}^{A_1})^\dagger \dots (a_{\alpha_n}^{A_n})^\dagger
\ket{\Omega}
\]
When the ground state $\ket{\Omega}$ has spin $s$, the maximal spin state
has spin $s+\half N$ and the minimal spin state has spin 
$0$ if $s \le \half N$ or $s-\half N$ if $s \ge \half N$.
\\
When the ground state $\ket{\Omega}$ has spin zero, the total number of
states is equal to $2^{2N}$ with $2^{2N-1}$ fermionic states
(constructed with an odd number of $(a_\alpha^A)^\dagger$ operators)
and $2^{2N-1}$ bosonic states (constructed with an even number of
$(a_\alpha^A)^\dagger$ operators). The maximal spin is $\half N$ and
the minimal spin is $0$.
\\
In the case $N=1$, when the ground state $\ket{\Omega}$ has spin $j$,
the states of the multiplet have spins $(j,j+\half,j-\half,j)$. When
the ground state $\ket{\Omega}$ has spin $0$, the multiplet has two
states of spin 0 and one state of spin $\half$.
\\
The following table gives the dimensions of the massive 
representations with ground states $\Omega_s$ (of spin $s$) for
$N=1,2,3,4$.

\begin{table}[htbp]
\centering
$N=1$
\hspace{5mm}
\begin{tabular}{c|cccc}
spin & $\Omega_0$ & $\Omega_{1/2}$ & $\Omega_1$ & $\Omega_{3/2}$ \cr
\hline 
0 & 2 & 1 \cr
$\frac{1}{2}$ & 1 & 2 & 1 \cr
1 & & 1 & 2 & 1 \cr
$\frac{3}{2}$ & & & 1 & 2 \cr
2 & & & & 1
\end{tabular}
\hspace{10mm}
\begin{tabular}{c|c}
spin & $\Omega_0$ \cr
\hline 
0 & 42 \cr
$\frac{1}{2}$ & 48 \cr
1 & 27 \cr
$\frac{3}{2}$ & 8 \cr
2 & 1
\end{tabular}
\hspace{5mm}
$N=4$
\\
\vspace{5mm}
$N=2$
\hspace{5mm}
\begin{tabular}{c|ccc}
spin & $\Omega_0$ & $\Omega_{1/2}$ & $\Omega_1$ \cr
\hline 
0 & 5 & 4 & 1 \cr
$\frac{1}{2}$ & 4 & 6 & 4 \cr
1 & 1 & 4 & 6 \cr
$\frac{3}{2}$ & & 1 & 4 \cr
2 & & & 1
\end{tabular}
\hspace{10mm}
\begin{tabular}{c|cc}
spin & $\Omega_0$ & $\Omega_{1/2}$ \cr
\hline 
0 & 14 & 14 \cr
$\frac{1}{2}$ & 14 & 20 \cr
1 & 6 & 15 \cr
$\frac{3}{2}$ & 1 & 6 \cr
2 & & 1
\end{tabular}
\hspace{5mm}
$N=3$
\end{table}

\medskip

We consider now the massless representations corresponding to $P^2=0$. In
a reference frame where $P^\mu=(E,0,0,E)$, the supersymmetry algebra
become
\beo
&& \ale Q_\alpha^A,\bQ_\dbeta^B \ari = 
4E \delta^{AB} \delta_{\alpha\dbeta,1 \dot 1} \\
&& \ale Q_\alpha^A,Q_\beta^B \ari = 
\ale \bQ_\dalpha^A,\bQ_\dbeta^B \ari = 0 
\eno
The rescaled operators $a^A = Q_1^A/\sqrt{4E}$ and
$(a^A)^\dagger = \bQ_{\dot 1}^A/\sqrt{4E}$ satisfy the Clifford
algebra in $N$ dimensions while the operators $a'^A = Q_2^A/\sqrt{4E}$
and $(a'^A)^\dagger = \bQ_{\dot 2}^A/\sqrt{4E}$ mutually anticommute
and act as zero on the representation states.
A representation of the supersymmetry algebra is therefore characterized
by a Clifford ground state $\ket{\Omega}$ labelled by the energy $E$
and the helicity $\lambda$ and annihilated by the $a^A$ operators. The
other states of the representation are given by
\[
\ket{a^{A_1} \dots a^{A_n}} = (a^{A_1})^\dagger \dots
(a^{A_n})^\dagger \ket{\Omega}
\]
The number of states with helicity $\lambda+n$ with $0 \le n \le \half
N$ is $N \choose 2n$. The total number of states is therefore $2^N$
with $2^{N-1}$ bosonic states and $2^{N-1}$ fermionic states.

\medskip

For more details on the supersymmetry representations (in particular
when the central charges are not zero), see refs. \cite{BWe83, Soh85, 
Wes86}.

\section{\sf Unitary superalgebras}
\indent

{\bf The superalgebras $A(m-1,n-1)$ with $m \ne n$}

The unitary superalgebra $A(m-1,n-1)$ or $sl(m|n)$ with $m \ne n$
defined for $m > n \ge 0$ has as even part the Lie algebra $sl(m)
\oplus sl(n) \oplus U(1)$ and as odd part the $(\overline{m},n) +
(m,\overline{n})$ representation of the even part; it has rank $m+n-1$
and dimension $(m+n)^2-1$. One has $A(m-1,n-1) \simeq A(n-1,m-1)$.

\medskip

The root system $\Delta = \Delta_\evn\cup\Delta_\odd$ of $A(m-1,n-1)$
can be expressed in terms of the orthogonal vectors $\eps_1, \dots, 
\eps_m$ and $\del_1, \dots, \del_n$ such that $\eps_i^2 = 1$ and 
$\del_i^2 = -1$ as 
\[
\Delta_\evn = \ale ~ \eps_i-\eps_j, ~ \del_i-\del_j ~ \ari 
\medbox{and} 
\Delta_\odd = \ale ~ \eps_i-\del_j, ~ -\eps_i+\del_j ~ \ari 
\]
The Dynkin diagrams of the unitary superalgebras $A(m-1,n-1)$ are of the
following types:
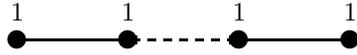
\begin{center}
\begin{picture}(140,15)
\thicklines
\multiput(0,0)(42,0){4}{\circle*{7}}
\put(0,10){\makebox(0.4,0.6){{\footnotesize 1}}}
\put(42,10){\makebox(0.4,0.6){{\footnotesize 1}}}
\put(84,10){\makebox(0.4,0.6){{\footnotesize 1}}}
\put(126,10){\makebox(0.4,0.6){{\footnotesize 1}}}
\put(3,0){\line(1,0){36}}
\put(45,0){\dashbox{3}(36,0)}
\put(87,0){\line(1,0){36}}
\end{picture}
\end{center}
where the small black dots represent either white dots (associated to
even roots) or grey dots (associated to odd roots of zero length). The
diagrams are drawn with their Dynkin labels which give the
decomposition of the highest root in terms of the simple ones. The
Dynkin diagrams of the unitary Lie superalgebras up to rank 4 are given
in Table \ref{table6}.

\medskip

The superalgebra $A(m-1,n-1)$ can be generated as a matrix superalgebra
by taking matrices of the form
\[
M = \left(\begin{array}{cc} X_{mm} & T_{mn} \cr T_{nm} & X_{nn}
\end{array}\right) 
\]
where $X_{mm}$ and $X_{nn}$ are $gl(m)$ and $gl(n)$ matrices, $T_{mn}$ 
and $T_{nm}$ are $m \times n$ and $n \times m$ matrices respectively, 
with the supertrace condition
\[
\str(X) = \tr(X_{mm}) - \tr(X_{nn}) = 0
\]

A basis of matrices can be constructed as follows. Consider $(m+n)^2$ 
elementary matrices $e_{IJ}$ of order $m+n$ such that $(e_{IJ})_{KL} = 
\delta_{IL} \delta_{JK}$ ($I,J,K,L = 1,\dots,m+n$) and define the 
$(m+n)^2-1$ generators
\beo
&&E_{ij} = e_{ij} - \frac{1}{m-n} \delta_{ij} (e_{kk}+e_{\bk\bk}) 
~~~~~~~~~~ E_{i \bj} = e_{i \bj} \\
&&E_{\bi\bj} = e_{\bi\bj} + \frac{1}{m-n} \delta_{\bi\bj}
(e_{kk}+e_{\bk\bk}) 
~~~~~~~~~~ E_{\bi j} = e_{\bi j}
\eno
where the indices $i,j,\dots$ run from 1 to $m$ and $\bi,\bj,\dots$ from
$m+1$ to $m+n$. Then the generator $Z = E_{kk} = -E_{\bk\bk} =
-\sfrac{1}{m-n} (n e_{kk} + m e_{\bk\bk})$ generate the $U(1)$ part,
the generators $E_{ij} - \sfrac{1}{m} \delta_{ij} Z$ generate the
$sl(m)$ part and the generators $E_{\bi\bj} + \sfrac{1}{n}
\delta_{\bi\bj} Z$ generate the $sl(n)$ part, while $E_{i\bj}$ and
$E_{\bi j}$ transform as the $(\overline{m},n)$ and $(m,\overline{n})$
representations of $sl(m) \oplus sl(n) \oplus U(1)$. In all these
expressions, summation over repeated indices is understood.
\\
The generators in the Cartan-Weyl basis are given by:
\begin{itemize}
\item
for the Cartan subalgebra
\beo
&&H_i = E_{ii} - E_{i+1,i+1} 
 \medbox{with} 1 \le i \le m-1 \\
&&H_{\bi} = E_{\bi\bi} - E_{\bi+1,\bi+1} 
 \medbox{with} m+1 \le \bi \le m+n-1 \\
&&H_m = E_{mm} + E_{m+1,m+1}
\eno
\item
for the raising operators
\[
E_{ij} \smbox{with} i < j \smbox{for} sl(m) \,, \quad
E_{\bi\bj} \smbox{with} \bi < \bj \smbox{for} sl(n) \,, \quad
E_{i \bj} \smbox{for the odd part}
\]
\item
for the lowering operators
\[
E_{ji} \smbox{with} i < j \smbox{for} sl(m) \,, \quad
E_{\bj\bi} \smbox{with} \bi < \bj \smbox{for} sl(n) \,, \quad
E_{\bi j} \smbox{for the odd part}
\]
\end{itemize}

The commutation relations in the Cartan-Weyl basis read as:
\beo
&&\cle H_I,H_J \cri = 0 \\
&&\cle H_K,E_{IJ} \cri = \delta_{IK} E_{KJ} - \delta_{I,K+1} E_{K+1,J} 
 - \delta_{KJ} E_{IK} + \delta_{K+1,J} E_{I,K+1} 
 \qquad (K \ne m) \\ 
&&\cle H_m,E_{IJ} \cri = \delta_{Im} E_{mJ} + \delta_{I,m+1} E_{m+1,J} 
 - \delta_{mJ} E_{Im} - \delta_{m+1,J} E_{I,m+1} \\
&&\cle E_{IJ},E_{KL} \cri = \delta_{JK} E_{IL} - \delta_{IL} E_{KJ} 
 \medbox{for $E_{IJ}$ and $E_{KL}$ even} \\
&&\cle E_{IJ},E_{KL} \cri = \delta_{JK} E_{IL} - \delta_{IL} E_{KJ}
 \medbox{for $E_{IJ}$ even and $E_{KL}$ odd} \\
&&\ale E_{IJ},E_{KL}\ari = \delta_{JK} E_{IL} + \delta_{IL} E_{KJ}
 \medbox{for $E_{IJ}$ and $E_{KL}$ odd} 
\eno

{\bf The superalgebras $A(n-1,n-1)$ with $n>1$}

The unitary superalgebra $A(n-1,n-1)$ defined for $n>1$ has as even
part the Lie algebra $sl(n) \oplus sl(n)$ and as odd part the
$(\overline{n},n) + (n,\overline{n})$ representation of the even part;
it has rank $2n-2$ and dimension $4n^2-2$. Note that the superalgebra
$A(0,0)$ is not simple.

\medskip

The root system $\Delta = \Delta_\evn\cup\Delta_\odd$ of $A(n-1,n-1)$
can be expressed in terms of the orthogonal vectors $\eps_1, \dots, 
\eps_n$ and $\del_1, \dots, \del_n$ such that $\eps_i^2 = 1$ and 
$\del_i^2 = -1$ as 
\[
\Delta_\evn = \ale ~ \eps_i-\eps_j, ~ \del_i-\del_j ~ \ari 
\medbox{and} 
\Delta_\odd = \ale ~ \eps_i-\del_j, ~ -\eps_i+\del_j ~ \ari 
\]
The Dynkin diagrams of the unitary superalgebras $A(n-1,n-1)$ are of the
same type as those of the $A(m-1,n-1)$ case.

\medskip

The superalgebra $A(n-1,n-1)$ can be generated as a matrix superalgebra
by taking matrices of $sl(n|n)$. However, $sl(n|n)$ contains a
one-dimensional ideal $\cI$ generated by $\II_{2n}$ and one sets
$A(n-1,n-1) \equiv sl(n|n)/\cI$, hence the rank and dimension of
$A(n-1,n-1)$. 

One has to stress that the rank of the superalgebra is $2n-2$ although
the Dynkin diagram has $2n-1$ dots: the $2n-1$ associated simple roots
are {\em not linearly independent} in that case.

Moreover, in the case of $A(1,1)$, one has the relations
$\eps_1+\eps_2=0$ and $\del_1+\del_2=0$ from which it follows that there
is only four distinct odd roots $\alpha$ such that $\dim\cG_\alpha = 2$
and each odd root is {\em both positive and negative}.

\section{\sf Universal enveloping algebra}
\indent

\udef: Let $\cG = \cG_\evn \oplus \cG_\odd$ be a Lie superalgebra over a
field $\KK = \RR$ or $\CC$. The definition of the universal enveloping
superalgebra $\cU(\cG)$ is similar to the definition in the algebraic
case. If $\cG^\otimes$ is the tensor algebra over $\cG$ with
$\ZZ_2$-graded tensor product ($\see$ Superalgebra) and $\cI$ the ideal
of $\cG$ generated by $\zle X,Y \zri - (X \otimes Y - (-1)^{\deg X.\deg
Y} Y \otimes X)$ where $X,Y \in \cG$, the universal enveloping
superalgebra $\cU(\cG)$ is the quotient $\cG^\otimes / \cI$.

\medskip

\underline{Poincar\'e--Birkhoff--Witt theorem}:
Let $b_1,\dots,b_B$ ($B = \dim\cG_\evn$) be a basis of the even part
$\cG_\evn$ and $f_1,\dots,f_F$ ($F = \dim\cG_\odd$) be a basis of the odd
part $\cG_\odd$. Then the elements
\[
b_1^{i_1} \dots b_B^{i_B} f_1^{j_1} \dots f_F^{j_F} \medbox{with}
i_1,\dots,i_B \ge 0 \smbox{and} j_1,\dots,j_F \in \{0,1\}
\]
form a basis of the universal enveloping superalgebra $\cU(\cG)$, called
the Poincar\'e--Birkhoff--Witt (PBW) basis.

\medskip

The universal enveloping superalgebra $\cU(\cG)$ contains in general
zero divisors (let us remind that $\cU(\cG_\evn)$ never contains zero
divisors). In fact, if $F \in \cG_\odd$ is a generator associated to an
isotropic root, one has $F^2 = \{F,F\} = 0$ in $\cU(\cG)$. More
precisely, one has the following property:

\uppt: The universal enveloping superalgebra $\cU(\cG)$ does not contain
any zero divisors if and only if $\cG = ops(1|2n)$. In that case,
$\cU(\cG)$ is said entire.

\medskip

\underline{Filtration of $\cG$}:
$\cU(\cG)$ can be naturally filtered as follows. Let $\cU_n$ be the
subspace of $\cU(\cG)$ generated by the PBW-basis monomials of degree
$\le n$ (e.g. $\cU_0 = \KK$ and $\cU_1 = \KK+\cG$). Then one has the
following filtration, with $\cU_i ~ \cU_j \subset \cU_{i+j}$:
\[
\cU_0 ~ \subset ~ \cU_1 ~ \subset ~ \dots ~ \subset ~ \cU_n ~ \subset
~ \dots ~ \subset ~ \cU(\cG) ~ = ~ \bigcup^\infty \cU_n
\]
Defining the quotient subspaces $\bar\cU_0 = \cU_0$ and
$\bar\cU_i = \cU_i/\cU_{i-1}$ for $i \ge 1$, one can associate to
$\cU(\cG)$ the following graded algebra $\mbox{Gr}(\cU(\cG))$:
\[
\mbox{Gr}(\cU(\cG)) = \bar\cU_0 ~ \oplus ~ \bar\cU_1 ~ \oplus ~ \dots
~ \oplus ~ \bar\cU_n ~ \oplus ~ \dots
\]
Then, one can show that
\[
\mbox{Gr}(\cU(\cG)) \simeq \KK[b_1,\dots,b_B] \otimes
\Lambda(f_1,\dots,f_F)
\]
where $\KK[b_1,\dots,b_B]$ is the ring of polynomials in the
indeterminates $b_1,\dots,b_B$ with coefficients in $\KK$ and
$\Lambda(f_1,\dots,f_F)$ is the exterior algebra over $\cG$.

\medskip

For more details, see ref. \cite{Kac77a}.

\section{\sf Weyl group}
\indent

Let $\cG = \cG_\evn \oplus \cG_\odd$ be a classical Lie superalgebra
with root system $\Delta  = \Delta_\evn \cup \Delta_\odd$.
$\Delta_\evn$ is the set of even roots and $\Delta_\odd$ the set of odd
roots. The Weyl group $W(\cG)$ of $\cG$ is generated by the Weyl
reflections $\omega$ with respect to the even roots:
\[
\omega_{\alpha}(\beta) = \beta - 2 \frac{(\alpha,\beta)}
{(\alpha,\alpha)} \alpha 
\]
where $\alpha \in \Delta_\evn$ and $\beta \in \Delta$.

\medskip

The properties of the Weyl group are the following.

\medskip

\uppts:
\begin{enumerate}
\item 
The Weyl group $W(\cG)$ leaves $\Delta$, $\Delta_\evn$, $\Delta_\odd$,
$\overline{\Delta}_\evn$, $\overline{\Delta}_\odd$ invariant, where
$\Delta$, $\Delta_\evn$, $\Delta_\odd$ are defined above,
$\overline{\Delta}_\evn$ is the subset of roots $\alpha \in
\Delta_\evn$ such that $\alpha/2 \notin \Delta_\odd$ and
$\overline{\Delta}_\odd$ is the subset of roots $\alpha \in
\Delta_\odd$ such that $2\alpha \notin \Delta_\evn$.
\item 
Let $e^{\lambda}$ be the formal exponential, function on $\cH^*$ such
that $e^{\lambda}(\mu) = \delta_{\lambda,\mu}$ for two elements
$\lambda,\mu \in \cH^*$, which satisfies $e^{\lambda}e^{\mu} =
e^{\lambda+\mu}$. One defines
\[
L = \frac
{\prod_{\alpha \in \Delta_\evn^+} (e^{\alpha/2} - e^{-\alpha/2})}
{\prod_{\alpha \in \Delta_\odd^+} (e^{\alpha/2} + e^{-\alpha/2})}
\bigbox{and}
L' = \frac
{\prod_{\alpha \in \Delta_\evn^+} (e^{\alpha/2} - e^{-\alpha/2})}
{\prod_{\alpha \in \Delta_\odd^+} (e^{\alpha/2} - e^{-\alpha/2})}
\]
where $\Delta_\evn^+$ and $\Delta_\odd^+$ are the sets of positive even
roots and positive odd roots respectively. Then one has
\[
w(L) = \eps(w) L \medbox{and} w(L') = \eps'(w) L' \medbox{where}
w \in W(\cG)
\]
with $\eps(w) = (-1)^{\ell(w)}$ and $\eps'(w) = (-1)^{\ell'(w)}$ where
$\ell(w)$ is the number of reflections in the expression of $w \in 
W(\cG)$ and $\ell'(w)$ is the number of reflections with respect to the
roots of $\overline{\Delta}_\evn^+$ in the expression of $w \in W(\cG)$.
\end{enumerate}

\medskip

For more details, see ref. \cite{Kac77a}.

\section{\sf Z-graded Lie superalgebras}
\indent

\udef: Let $\cG = \cG_\evn \oplus \cG_\odd$ be a Lie superalgebra. $\cG$
is a $\ZZ$-graded Lie superalgebra if it can be written as a direct 
sum of finite dimensional $\ZZ_2$-graded subspaces $\cG_i$ such that
\[
\cG = \bigoplus_{i \in \ZZ} \cG_i \bigbox{where}
\sle \cG_i , \cG_j \sri \subset \cG_{i+j}
\]
The $\ZZ$-gradation is said {\em consistent} with the $\ZZ_2$-gradation
if 
\[
\cG_\evn = \sum_{i \in \ZZ} \cG_{2i} \bigbox{and}
\cG_\odd = \sum_{i \in \ZZ} \cG_{2i+1}
\]

\medskip

\udef: Let $\cG$ be a $\ZZ$-graded Lie superalgebra. It is called
\\ 
- irreducible if the representation of $\cG_\evn$ in $\cG_{-1}$ is
irreducible,
\\
- transitive if 
$X \in \cG_{i \ge 0}, \sle X,\cG_{-1} \sri = 0 \Rightarrow X = 0$,
\\
- bitransitive if 
$X \in \cG_{i \ge 0}, \sle X,\cG_{-1} \sri = 0 \Rightarrow X = 0$
and $X \in \cG_{i \le 0}, \sle X,\cG_1 \sri = 0 \Rightarrow X = 0$.

\medskip

For more details, see ref. \cite{Kac77a}.

\newpage
\setcounter{table}{0}
\renewcommand{\thetable}{\arabic{table}}
\section*{List of Tables}

\contentsline {table}{\numberline {1}{\ignorespaces Classification of the simple Lie superalgebras.}}{98}
\contentsline {table}{\numberline {2}{\ignorespaces Classical Lie superalgebras.}}{98}
\contentsline {table}{\numberline {3}{\ignorespaces ${\cal G}_{\overline {0}}$ and ${\cal G}_{\overline {1}}$ structure of the classical Lie superalgebras.}}{98}
\contentsline {table}{\numberline {4}{\ignorespaces The basic Lie superalgebra $A(m-1,n-1) = sl(m|n)$. }}{99} 
\contentsline {table}{\numberline {5}{\ignorespaces The basic Lie superalgebra $A(n-1,n-1) = sl(n|n)/{\cal Z}$. }}{101} 
\contentsline {table}{\numberline {6}{\ignorespaces The basic Lie superalgebra $B(m,n) = osp(2m+1|2n)$. }}{103}
\contentsline {table}{\numberline {7}{\ignorespaces The basic Lie superalgebra $B(0,n) = osp(1|2n)$. }}{105}
\contentsline {table}{\numberline {8}{\ignorespaces The basic Lie superalgebra $C(n+1) = osp(2|2n)$. }}{107}
\contentsline {table}{\numberline {9}{\ignorespaces The basic Lie superalgebra $D(m,n) = osp(2m|2n)$. }}{109}
\contentsline {table}{\numberline {10}{\ignorespaces The basic Lie superalgebra $F(4)$.}}{111}
\contentsline {table}{\numberline {11}{\ignorespaces The basic Lie superalgebra $G(3)$.}}{113}
\contentsline {table}{\numberline {12}{\ignorespaces The basic Lie superalgebra $D(2,1;\alpha )$.}}{114}
\contentsline {table}{\numberline {13}{\ignorespaces Distinguished Dynkin diagrams of the basic Lie superalgebras.}}{116}
\contentsline {table}{\numberline {14}{\ignorespaces Dynkin diagrams of the basic Lie superalgebras of rank $\le $ 4 }}{117}
\contentsline {table}{\numberline {15}{\ignorespaces Distinguished extended Dynkin diagrams of the basic Lie superalgebras.}}{120}
\contentsline {table}{\numberline {16}{\ignorespaces Real forms of the classical Lie superalgebras. }}{121}
\contentsline {table}{\numberline {17}{\ignorespaces $osp(1|2)$ decompositions of the fundamental representations of the basic Lie superalgebras (regular cases).}}{122}
\contentsline {table}{\numberline {18}{\ignorespaces $osp(1|2)$ decompositions of the fundamental representations of the basic Lie superalgebras (singular cases).}}{123}
\contentsline {table}{\numberline {19}{\ignorespaces $osp(1|2)$ decompositions of the adjoint representations of the basic Lie superalgebras (regular cases).}}{124}
\contentsline {table}{\numberline {20}{\ignorespaces $osp(1|2)$ decompositions of the adjoint representations of the basic Lie superalgebras (singular cases).}}{126}
\contentsline {table}{\numberline {21}{\ignorespaces $osp(1|2)$ decompositions of the $A(m,n)$ superalgebras up to rank 4. }}{127}
\contentsline {table}{\numberline {22}{\ignorespaces $osp(1|2)$ decompositions of the $B(m,n)$ superalgebras of rank 2 and 3. }}{127}
\contentsline {table}{\numberline {23}{\ignorespaces $osp(1|2)$ decompositions of the $B(m,n)$ superalgebras of rank 4. }}{128}
\contentsline {table}{\numberline {24}{\ignorespaces $osp(1|2)$ decompositions of the $C(n+1)$ superalgebras up to rank 4. }}{129}
\contentsline {table}{\numberline {25}{\ignorespaces $osp(1|2)$ decompositions of the $D(m,n)$ superalgebras up to rank 4. }}{129}
\contentsline {table}{\numberline {26}{\ignorespaces $osp(1|2)$ decompositions of the superalgebra $F(4)$. }}{130}
\contentsline {table}{\numberline {27}{\ignorespaces $osp(1|2)$ decompositions of the superalgebra $G(3)$. }}{130}
\contentsline {table}{\numberline {28}{\ignorespaces $osp(1|2)$ decompositions of the superalgebra $D(2,1;\alpha )$. }}{130}
\contentsline {table}{\numberline {29}{\ignorespaces $sl(1|2)$ decompositions of the $A(m,n)$ superalgebras up to rank 4. }}{131}
\contentsline {table}{\numberline {30}{\ignorespaces $sl(1|2)$ decompositions of the $B(m,n)$ superalgebras of rank 2 and 3. }}{131}
\contentsline {table}{\numberline {31}{\ignorespaces $sl(1|2)$ decompositions of the $B(m,n)$ superalgebras of rank 4. }}{132}
\contentsline {table}{\numberline {32}{\ignorespaces $sl(1|2)$ decompositions of the $C(n+1)$ superalgebras up to rank 4. }}{133}
\contentsline {table}{\numberline {33}{\ignorespaces $sl(1|2)$ decompositions of the $D(m,n)$ superalgebras up to rank 4. }}{133}
\contentsline {table}{\numberline {34}{\ignorespaces $sl(1|2)$ decompositions of the exceptional superalgebras. }}{134}

\clearpage

\begin{table}[t]
\centering
\begin{tabular}{cccc}
&& Simple Lie & \cr
&& superalgebras & \cr
&& $\swarrow \qquad \qquad \searrow$ & \cr
&Classical Lie && Cartan type \cr
&superalgebras && superalgebras \cr
& $\swarrow \qquad \qquad \searrow$ && $W(n),S(n),{\tilde S}(n),H(n)$ \cr
Basic Lie && Strange & \cr
superalgebras && superalgebras & \cr
$A(m,n),B(m,n)$ && $P(n),Q(n)$ & \cr
$C(n+1),D(m,n)$ &&& \cr 
$F(4),G(3),D(2,1;\alpha)$ &&& \cr
\end{tabular}
\caption{Classification of the simple Lie superalgebras.\label{table39}}
\end{table}

\begin{table}[p]
\centering
\begin{tabular}{|c|cl|cl|} \hline
&& type I ~~~~~~~~ && type II ~~~~~~~~~~~~~ \cr 
\hline
BASIC & $A(m,n)$ & $m > n \ge 0$ & $B(m,n)$ & 
$m \ge 0, n \ge 1$ \cr 
(non-degenerate & $C(n+1)$ & $n \ge 1$ & $D(m,n)$ & 
$\left\{\begin{array}{l} m \ge 2, n\ge 1 \cr m \ne n+1 \cr 
\end{array} \right.$ \cr
Killing form) & $F(4)$ &&& \cr
& $G(3)$ &&& \cr
\hline
BASIC & $A(n,n)$ & $n \ge 1$ & $D(n+1,n)$ & $n \ge 1$ \cr
(zero Killing form) &&& $D(2,1;\alpha)$ & $\alpha \in
\CC\setminus \{0,-1\}$ \cr 
\hline
STRANGE & $P(n)$ & $n \ge 2$ & $Q(n)$ & $n \ge 2$ \cr
\hline
\end{tabular}
\caption{Classical Lie superalgebras.\label{table61}}
\end{table}

\begin{table}[p]
\centering
\begin{tabular}{|c|c|c|} \hline
superalgebra $\cG$ & even part $\cG_\evn$ & odd part $\cG_\odd$ \\ 
\hline
$A(m,n)$ & $A_m \oplus A_n \oplus U(1)$ & $(\overline{m},n) \oplus
(m,\overline{n})$ \\ 
$A(n,n)$ & $A_n \oplus A_n$ & $(\overline{n},n) \oplus
(n,\overline{n})$ \\
$C(n+1)$ & $C_n \oplus U(1)$ & $(2n) \oplus (2n)$ \\
\hline
$B(m,n)$ & $B_m \oplus C_n$ & $(2m+1,2n)$ \\
$D(m,n)$ & $D_m \oplus C_n$ & $(2m,2n)$ \\
$F(4)$ & $A_1 \oplus B_3$ & $(2,8)$ \\
$G(3)$ & $A_1 \oplus G_2$ & $(2,7)$ \\
$D(2,1;\alpha)$ & $A_1 \oplus A_1 \oplus A_1$ & $(2,2,2)$ \\
\hline
$P(n)$ & $A_n$ & $[2] \oplus [1^{n-1}]$ \\
$Q(n)$ & $A_n$ & $\ad(A_n)$ \\
\hline
\end{tabular}
\caption{$\cG_\evn$ and $\cG_\odd$ structure of the
classical Lie superalgebras.\label{table62}}
\end{table}

\clearpage

\begin{table}[htbp]
\caption{The basic Lie superalgebra $A(m-1,n-1)  = sl(m|n)$.
\label{table11}}
\end{table}

\noindent
Structure: $\cG_\evn = sl(m) \oplus sl(n) \oplus U(1)$ and
$\cG_\odd = (\overline{m},n) \oplus (m,\overline{n})$, type I.
\\
Rank: $m+n-1$, dimension: $(m+n)^2-1$.
\\
Root system:
\beo
&& \Delta = \{ \eps_i-\eps_j, ~ \del_k-\del_l, ~ \eps_i-\del_k, 
 ~ \del_k-\eps_i \} \\
&& \Delta_\evn = \{ \eps_i-\eps_j, ~ \del_k-\del_l \}, \quad 
\Delta_\odd = \{ \eps_i-\del_k, ~ \del_k-\eps_i \} \\
&& \overline{\Delta}_\evn = \Delta_\evn, \quad 
\overline{\Delta}_\odd = \Delta_\odd
\eno
where $1 \le i \ne j \le m$ and $1 \le k \ne l \le n$.
\\
$\dim\Delta_\evn = \dim\overline{\Delta}_\evn = m^2 + n^2 - m - n + 1$ 
and $\dim\Delta_\odd = \dim\overline{\Delta}_\odd = 2mn$.
\\
Distinguished simple root system:
\beo
&& \alpha_1=\del_1-\del_2, ~ \dots, ~ \alpha_{n-1}=\del_{n-1}-\del_n,
~ \alpha_n=\del_n-\eps_1, \\
&& \alpha_{n+1}=\eps_1-\eps_2, ~ \dots,
~ \alpha_{n+m-1}=\eps_{m-1}-\eps_{m}
\eno
Distinguished positive roots ($1 \le i < j \le m$ and 
$1 \le k < l \le n$): 
\beo
\del_k-\del_l &=& \alpha_k + \dots + \alpha_{l-1} \\
\eps_i-\eps_j &=& \alpha_{n+i} + \dots + \alpha_{n+j-1} \\
\del_k-\eps_i &=& \alpha_k + \dots + \alpha_{n+i-1} 
\eno
Sums of even/odd distinguished positive roots:
\beo
&& 2\rho_0 = (m-1)\eps_1 + (m-3)\eps_2 + (m-5)\eps_3 + \dots 
- (m-3)\eps_{m-1} - (m-1)\eps_m \\
&& \hspace{12mm} + (n-1)\del_1 + (n-3)\del_2 + (n-5)\del_3 + \dots  
- (n-3)\del_{n-1} - (n-1)\del_n \\
&& 2\rho_1 = m(\del_1 + \dots + \del_n) - n(\eps_1 + \dots + \eps_m)
\eno
Distinguished Dynkin diagram:
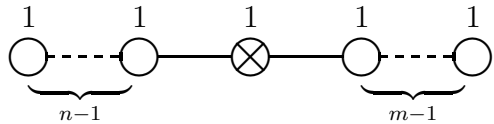
\begin{center}
\begin{picture}(160,20)
\thicklines
\multiput(0,0)(42,0){5}{\circle{14}}
\put(0,15){\makebox(0.4,0.6){1}}
\put(42,15){\makebox(0.4,0.6){1}}
\put(84,15){\makebox(0.4,0.6){1}}
\put(126,15){\makebox(0.4,0.6){1}}
\put(168,15){\makebox(0.4,0.6){1}}
\put(79,-5){\line(1,1){10}}\put(79,5){\line(1,-1){10}}
\put(7,0){\dashbox{3}(28,0)}
\put(49,0){\line(1,0){28}}
\put(91,0){\line(1,0){28}}
\put(133,0){\dashbox{3}(28,0)}
\put(0,-10){$\underbrace{~~~~~~~~~~}_{n-1}$}
\put(126,-10){$\underbrace{~~~~~~~~~~}_{m-1}$}
\end{picture}
\end{center}
\vspace{5mm}
Distinguished Cartan matrix:
\[
\left(\begin{array}{rrrrr|r|rrrrr}
2 & -1 & 0 & \cdots & 0 & \cdots & \cdots &&& \cdots & 0 \\
-1 & \ddots & \ddots & \ddots &&&&&&& \vdots \\
0 & \ddots && \ddots & 0 &&&&&& \\
\vdots & \ddots & \ddots & \ddots & -1 & \ddots &&&&& \\
0 && 0 & -1 & 2 & -1 & \ddots &&&& \vdots \\ \hline
\vdots &&& \ddots & -1 & 0 & 1 & \ddots &&& \vdots \\ \hline
\vdots &&&& \ddots & -1 & 2 & -1 & 0 && 0 \\
&&&&& \ddots & -1 & \ddots & \ddots & \ddots & \vdots \\
&&&&&& 0 & \ddots && \ddots & 0 \\
&&&&&&& \ddots & \ddots & \ddots & -1 \\
0 & \cdots &&& \cdots & \cdots & 0 & \cdots & 0 & -1 & 2 \\
\end{array}\right)
\]
Longest distinguished root:
\[
-\alpha_0 = \alpha_1 + \dots + \alpha_{n+m-1} = \delta_1 - \eps_m
\]
Distinguished extended Dynkin diagram:
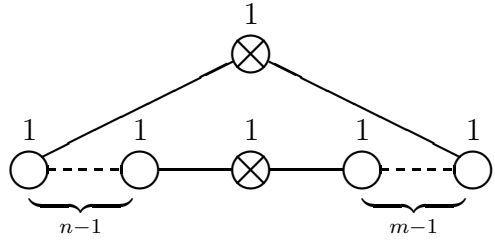
\vspace{5mm}
\begin{center}
\begin{picture}(180,40)
\thicklines
\multiput(0,0)(42,0){5}{\circle{14}}
\put(84,44){\circle{14}}
\put(0,15){\makebox(0.4,0.6){1}}
\put(42,15){\makebox(0.4,0.6){1}}
\put(84,15){\makebox(0.4,0.6){1}}
\put(126,15){\makebox(0.4,0.6){1}}
\put(168,15){\makebox(0.4,0.6){1}}
\put(84,59){\makebox(0.4,0.6){1}}
\put(79,-5){\line(1,1){10}}\put(79,5){\line(1,-1){10}}
\put(79,39){\line(1,1){10}}\put(79,49){\line(1,-1){10}}
\put(5,5){\line(2,1){72}}
\put(163,5){\line(-2,1){72}}
\put(7,0){\dashbox{3}(28,0)}
\put(49,0){\line(1,0){28}}
\put(91,0){\line(1,0){28}}
\put(133,0){\dashbox{3}(28,0)}
\put(0,-10){$\underbrace{~~~~~~~~~~}_{n-1}$}
\put(126,-10){$\underbrace{~~~~~~~~~~}_{m-1}$}
\end{picture}
\end{center}
\vspace{7mm}
Factor group $\autout(\cG) = \aut(\cG)/\autint(\cG)$: 
\beo
&& \autout(\cG) = \ZZ_2 \smbox{for} A(m,n) \smbox{with} 
m \ne n \ne 0 \smbox{and} A(0,2n-1) \\
&& \autout(\cG) = \ZZ_4 \smbox{for} A(0,2n)
\eno

\clearpage

\begin{table}[htbp]
\caption{The basic Lie superalgebra $A(n-1,n-1) = sl(n|n)/\cZ$.
\label{table25}} 
\end{table}

\noindent
Structure: $\cG_\evn = sl(n) \oplus sl(n)$ and
$\cG_\odd = (\overline{n},n) \oplus (n,\overline{n})$, type I.
\\
Rank: $2n-2$, dimension: $4n^2-2$.
\\
Root system:
\beo
&& \Delta = \{ \eps_i-\eps_j, ~ \del_i-\del_j, ~ \eps_i-\del_j, 
 ~ \del_j-\eps_i \} \\
&& \Delta_\evn = \{ \eps_i-\eps_j, ~ \del_i-\del_j \}, \quad 
\Delta_\odd = \{ \eps_i-\del_j, ~ \del_j-\eps_i \} \\
&& \overline{\Delta}_\evn = \Delta_\evn, \quad 
\overline{\Delta}_\odd = \Delta_\odd
\eno
where $1 \le i \ne j \le n$.
\\
$\dim\Delta_\evn = \dim\overline{\Delta}_\evn = 2n^2-2n$ and 
$\dim\Delta_\odd = \dim\overline{\Delta}_\odd = 2n^2$.
\\
Distinguished simple root system:
\beo
&& \alpha_1=\del_1-\del_2, ~ \dots, ~ \alpha_{n-1}=\del_{n-1}-\del_n,
~ \alpha_n=\del_n-\eps_1, \\
&& \alpha_{n+1}=\eps_1-\eps_2, ~ \dots,
~ \alpha_{2n-1}=\eps_{n-1}-\eps_{n}
\eno
Number of simple roots = $2n-1$ ($\ne$ rank); the simple roots are not
independent:
\[
\alpha_1 + 2\alpha_2 + \dots + n\alpha_n + (n-1) \alpha_{n+1} + \dots
+ 2\alpha_{2n-2} + \alpha_{2n-1} = 0 
\]
Distinguished positive roots ($1 \le i < j \le m$ and 
$1 \le k < l \le n$): 
\beo
\del_k-\del_l &=& \alpha_k + \dots + \alpha_{l-1} \\
\eps_i-\eps_j &=& \alpha_{n+i} + \dots + \alpha_{n+j-1} \\
\del_k-\eps_i &=& \alpha_k + \dots + \alpha_{n+i-1} 
\eno
Sums of even/odd distinguished positive roots:
\beo
&& 2\rho_0 = (n-1)\eps_1 + (n-3)\eps_2 + (n-5)\eps_3 + \dots 
- (n-3)\eps_{n-1} - (n-1)\eps_n \\
&& \hspace{12mm} + (n-1)\del_1 + (n-3)\del_2 + (n-5)\del_3 + \dots  
- (n-3)\del_{n-1} - (n-1)\del_n \\
&& 2\rho_1 = n(\del_1 + \dots + \del_n - \eps_1 - \dots - \eps_n)
\eno
Distinguished Dynkin diagram:
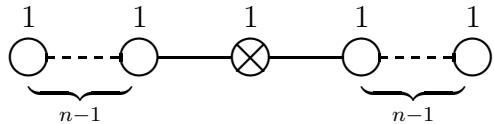
\begin{center}
\begin{picture}(160,20)
\thicklines
\multiput(0,0)(42,0){5}{\circle{14}}
\put(0,15){\makebox(0.4,0.6){1}}
\put(42,15){\makebox(0.4,0.6){1}}
\put(84,15){\makebox(0.4,0.6){1}}
\put(126,15){\makebox(0.4,0.6){1}}
\put(168,15){\makebox(0.4,0.6){1}}
\put(79,-5){\line(1,1){10}}\put(79,5){\line(1,-1){10}}
\put(7,0){\dashbox{3}(28,0)}
\put(49,0){\line(1,0){28}}
\put(91,0){\line(1,0){28}}
\put(133,0){\dashbox{3}(28,0)}
\put(0,-10){$\underbrace{~~~~~~~~~~}_{n-1}$}
\put(126,-10){$\underbrace{~~~~~~~~~~}_{n-1}$}
\end{picture}
\end{center}
\vspace{5mm}
Distinguished Cartan matrix:
\[
\left(\begin{array}{rrrrr|r|rrrrr}
2 & -1 & 0 & \cdots & 0 & \cdots & \cdots &&& \cdots & 0 \\
-1 & \ddots & \ddots & \ddots &&&&&&& \vdots \\
0 & \ddots && \ddots & 0 &&&&&& \\
\vdots & \ddots & \ddots & \ddots & -1 & \ddots &&&&& \\
0 && 0 & -1 & 2 & -1 & \ddots &&&& \vdots \\ \hline
\vdots &&& \ddots & -1 & 0 & 1 & \ddots &&& \vdots \\ \hline
\vdots &&&& \ddots & -1 & 2 & -1 & 0 && 0 \\
&&&&& \ddots & -1 & \ddots & \ddots & \ddots & \vdots \\
&&&&&& 0 & \ddots && \ddots & 0 \\
&&&&&&& \ddots & \ddots & \ddots & -1 \\
0 & \cdots &&& \cdots & \cdots & 0 & \cdots & 0 & -1 & 2 \\
\end{array}\right)
\]
Longest distinguished root:
\[
-\alpha_0 = \alpha_1 + \dots + \alpha_{2n-1} = \delta_1 - \eps_n
\]
Distinguished extended Dynkin diagram:
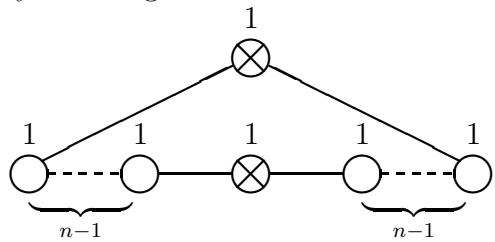
\vspace{5mm}
\begin{center}
\begin{picture}(180,40)
\thicklines
\multiput(0,0)(42,0){5}{\circle{14}}
\put(84,44){\circle{14}}
\put(0,15){\makebox(0.4,0.6){1}}
\put(42,15){\makebox(0.4,0.6){1}}
\put(84,15){\makebox(0.4,0.6){1}}
\put(126,15){\makebox(0.4,0.6){1}}
\put(168,15){\makebox(0.4,0.6){1}}
\put(84,59){\makebox(0.4,0.6){1}}
\put(79,-5){\line(1,1){10}}\put(79,5){\line(1,-1){10}}
\put(79,39){\line(1,1){10}}\put(79,49){\line(1,-1){10}}
\put(5,5){\line(2,1){72}}
\put(163,5){\line(-2,1){72}}
\put(7,0){\dashbox{3}(28,0)}
\put(49,0){\line(1,0){28}}
\put(91,0){\line(1,0){28}}
\put(133,0){\dashbox{3}(28,0)}
\put(0,-10){$\underbrace{~~~~~~~~~~}_{n-1}$}
\put(126,-10){$\underbrace{~~~~~~~~~~}_{n-1}$}
\end{picture}
\end{center}
\vspace{7mm}
Factor group $\autout(\cG) = \aut(\cG)/\autint(\cG)$: 
\beo
&& \autout(\cG) = \ZZ_2\times\ZZ_2 \smbox{for} A(n,n) \smbox{with} 
n \ne 1 \\
&& \autout(\cG) = \ZZ_2 \smbox{for} A(1,1)
\eno

\clearpage

\begin{table}[htbp]
\caption{The basic Lie superalgebra $B(m,n) = osp(2m+1|2n)$.
\label{table26}}
\end{table}

\noindent
Structure: $\cG_\evn = so(2m+1) \oplus sp(2n)$ and 
$\cG_\odd = (2m+1,2n)$, type II. 
\\
Rank: $m+n$, dimension: $2(m+n)^2+m+3n$.
\\
Root system:
\beo
&& \Delta = \{ \pm\eps_i\pm\eps_j, ~ \pm\eps_i, ~ \pm\del_k\pm\del_l,
 ~ \pm 2\del_k, ~ \pm\eps_i\pm\del_k, ~ \pm\del_k \} \\
&& \Delta_\evn = \{ \pm\eps_i\pm\eps_j, ~ \pm\eps_i, ~ \pm\del_k\pm\del_l,
 ~ \pm 2\del_k \}, \quad 
\Delta_\odd = \{ \pm\eps_i\pm\del_k, ~ \pm\del_k \} \\
&& \overline{\Delta}_\evn = \{ \pm\eps_i\pm\eps_j, ~ \pm\eps_i, 
~ \pm\del_k\pm\del_l \}, \quad 
\overline{\Delta}_\odd = \{ \pm\eps_i\pm\del_k \} 
\eno
where $1 \le i \ne j \le m$ and $1 \le k \ne l \le n$.
\\
$\dim\Delta_\evn = 2m^2+2n^2$, $\dim\Delta_\odd = 4mn+2n$,
$\dim\overline{\Delta}_\evn = 2m^2+2n^2-2n$, 
$\dim\overline{\Delta}_\odd = 4mn$.
\\
Distinguished simple root system:
\beo
&& \alpha_1=\del_1-\del_2, ~ \dots, ~ \alpha_{n-1}=\del_{n-1}-\del_n, 
~ \alpha_n=\del_n-\eps_1, \\ 
&& \alpha_{n+1}=\eps_1-\eps_2, ~ \dots, ~ \alpha_{n+m-1} =
\eps_{m-1}-\eps_m, ~ \alpha_{n+m}=\eps_m
\eno
Distinguished positive roots ($1 \le i < j \le m$ and 
$1 \le k < l \le n$):
\beo
\del_k-\del_l &=& \alpha_k + \dots + \alpha_{l-1} \\ 
\del_k+\del_l &=& \alpha_k + \dots + \alpha_{l-1} 
+ 2\alpha_l + \dots + 2\alpha_{n+m} \\
2\del_k &=& 2\alpha_k + \dots + 2\alpha_{n+m} \\
\eps_i-\eps_j &=& \alpha_{n+i} + \dots + \alpha_{n+j-1} \\ 
\eps_i+\eps_j &=& \alpha_{n+i} + \dots + \alpha_{n+j-1} 
+ 2\alpha_{n+j} + \dots + 2\alpha_{n+m} \\
\eps_i &=& \alpha_{n+i} + \dots + \alpha_{n+m} \\
\del_k-\eps_i &=& \alpha_k + \dots + \alpha_{n+i-1}  \\
\del_k+\eps_i &=& \alpha_k + \dots + \alpha_{n+i-1} 
+ 2\alpha_{n+i} + \dots + 2\alpha_{n+m}
\eno
Sums of even/odd distinguished positive roots:
\beo
&& 2\rho_0 = (2m-1)\eps_1 + (2m-3)\eps_2 + \dots + 3\eps_{m-1} 
+ \eps_m \\
&& \hspace{12mm} + 2n\del_1 + (2n-2)\del_2 + \dots + 4\del_{n-1} 
+ 2\del_n \\
&& 2\rho_1 = (2m+1)(\del_1 + \dots + \del_n)
\eno
Distinguished Dynkin diagram:
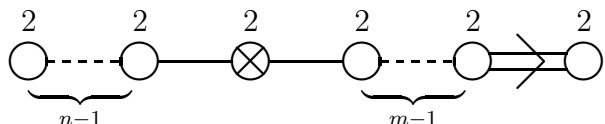
\begin{center}
\begin{picture}(200,20)
\thicklines
\multiput(0,0)(42,0){6}{\circle{14}}
\put(0,15){\makebox(0.4,0.6){2}}
\put(42,15){\makebox(0.4,0.6){2}}
\put(84,15){\makebox(0.4,0.6){2}}
\put(126,15){\makebox(0.4,0.6){2}}
\put(168,15){\makebox(0.4,0.6){2}}
\put(210,15){\makebox(0.4,0.6){2}}
\put(79,-5){\line(1,1){10}}\put(79,5){\line(1,-1){10}}
\put(7,0){\dashbox{3}(28,0)}
\put(49,0){\line(1,0){28}}
\put(91,0){\line(1,0){28}}
\put(133,0){\dashbox{3}(28,0)}
\put(174,-3){\line(1,0){30}}
\put(174,3){\line(1,0){30}}
\put(195,0){\line(-1,1){10}}\put(195,0){\line(-1,-1){10}}
\put(0,-10){$\underbrace{~~~~~~~~~~}_{n-1}$}
\put(126,-10){$\underbrace{~~~~~~~~~~}_{m-1}$}
\end{picture}
\end{center}
\vspace{5mm}
Distinguished Cartan matrix:
\[
\left(\begin{array}{rrrrr|r|rrrrr}
2 & -1 & 0 & \cdots & 0 & \cdots & \cdots &&& \cdots & 0 \\
-1 & \ddots & \ddots & \ddots &&&&&&& \vdots \\
0 & \ddots && \ddots & 0 &&&&&& \\
\vdots & \ddots & \ddots & \ddots & -1 & \ddots &&&&& \\
0 && 0 & -1 & 2 & -1 & \ddots &&&& \vdots \\ \hline
\vdots &&& \ddots & -1 & 0 & 1 & \ddots &&& \vdots \\ \hline
\vdots &&&& \ddots & -1 & 2 & -1 & 0 && 0 \\
&&&&& \ddots & -1 & \ddots & \ddots & \ddots & \vdots \\
&&&&&& 0 & \ddots &\ddots & -1 & 0 \\
&&&&&&& \ddots & -1 & 2 & -1 \\
0 & \cdots &&& \cdots & \cdots & 0 & \cdots & 0 & -2 & 2 \\
\end{array}\right)
\]
Longest distinguished root:
\[
-\alpha_0 = 2\alpha_1 + \dots + 2\alpha_{n+m} = 2\del_1
\]
Distinguished extended Dynkin diagram:
\begin{center}
\begin{picture}(260,20)
\thicklines
\multiput(0,0)(42,0){7}{\circle{14}}
\put(0,15){\makebox(0.4,0.6){1}}
\put(42,15){\makebox(0.4,0.6){2}}
\put(84,15){\makebox(0.4,0.6){2}}
\put(126,15){\makebox(0.4,0.6){2}}
\put(168,15){\makebox(0.4,0.6){2}}
\put(210,15){\makebox(0.4,0.6){2}}
\put(252,15){\makebox(0.4,0.6){2}}
\put(121,-5){\line(1,1){10}}\put(121,5){\line(1,-1){10}}
\put(6,-3){\line(1,0){30}}
\put(6,3){\line(1,0){30}}
\put(27,0){\line(-1,1){10}}\put(27,0){\line(-1,-1){10}}
\put(49,0){\dashbox{3}(28,0)}
\put(91,0){\line(1,0){28}}
\put(133,0){\line(1,0){28}}
\put(175,0){\dashbox{3}(28,0)}
\put(216,-3){\line(1,0){30}}
\put(216,3){\line(1,0){30}}
\put(237,0){\line(-1,1){10}}\put(237,0){\line(-1,-1){10}}
\put(42,-10){$\underbrace{~~~~~~~~~~}_{n-1}$}
\put(168,-10){$\underbrace{~~~~~~~~~~}_{m-1}$}
\end{picture}
\end{center}
\vspace{7mm}
Factor group $\autout(\cG) = \aut(\cG)/\autint(\cG) = \II$.

\clearpage

\begin{table}[htbp]
\caption{The basic Lie superalgebra $B(0,n) = osp(1|2n)$.
\label{table27}}
\end{table}

\noindent
Structure: $\cG_\evn = sp(2n)$ and $\cG_\odd = (2n)$, type II.
\\
Rank: $n$, dimension: $2n^2+3n$.
\\
Root system:
\beo
&& \Delta = \{ \pm\del_k\pm\del_l, ~ \pm 2\del_k, ~ \pm\del_k \} \\
&& \Delta_\evn = \{ \pm\del_k\pm\del_l, ~ \pm 2\del_k \}, \quad 
\Delta_\odd = \{ \pm\del_k \} \\
&& \overline{\Delta}_\evn = \{ \pm\del_k\pm\del_l \}, \quad 
\overline{\Delta}_\odd = \emptyset 
\eno
where $1 \le k \ne l \le n$.
\\
$\dim\Delta_\evn = 2n^2$, $\dim\Delta_\odd = 2n$, 
$\dim\overline{\Delta}_\evn = 2n^2-2n$, $\dim\overline{\Delta}_\odd = 0$.
\\
Simple root system:
\[
\alpha_1=\del_1-\del_2, ~ \dots, ~ \alpha_{n-1}=\del_{n-1}-\del_n, 
~ \alpha_n=\del_n
\]
Positive roots ($1 \le k < l \le n$):
\beo
\del_k-\del_l &=& \alpha_k + \dots + \alpha_{l-1} \\
\del_k+\del_l &=& \alpha_k + \dots + \alpha_{l-1} 
+ 2\alpha_l + \dots + 2\alpha_{n+m} \\
2\del_k &=& 2\alpha_k + \dots + 2\alpha_{n+m} \\
\del_k &=& \alpha_k + \dots + \alpha_{n+m}
\eno
Sums of even/odd positive roots:
\beo
&& 2\rho_0 =  2n\del_1 + (2n-2)\del_2 + \dots + 4\del_{n-1} 
+ 2\del_n \\
&& 2\rho_1 = \del_1 + \dots + \del_n
\eno
Dynkin diagram:
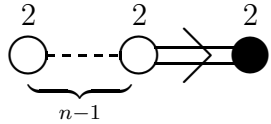
\begin{center}
\begin{picture}(80,20)
\thicklines
\put(0,0){\circle{14}}
\put(42,0){\circle{14}}
\put(84,0){\circle*{14}}
\put(0,15){\makebox(0.4,0.6){2}}
\put(42,15){\makebox(0.4,0.6){2}}
\put(84,15){\makebox(0.4,0.6){2}}
\put(7,0){\dashbox{3}(28,0)}
\put(48,-3){\line(1,0){30}}
\put(48,3){\line(1,0){30}}
\put(69,0){\line(-1,1){10}}\put(69,0){\line(-1,-1){10}}
\put(0,-10){$\underbrace{~~~~~~~~~~}_{n-1}$}
\end{picture}
\end{center}
\vspace{5mm}
Cartan matrix:
\[
\left(\begin{array}{rrrrrr}
2 & -1 & 0 & \cdots & \cdots & 0 \\
-1 & 2 & \ddots & \ddots & & \vdots \\
0 & \ddots & \ddots & \ddots & \ddots & \vdots \\
\vdots & \ddots & \ddots & 2 & -1 & 0 \\
\vdots & & \ddots & -1 & 2 & -1 \\
0 & \cdots & \cdots & 0 & -2 & 2 \\
\end{array}\right)
\]
Longest distinguished root:
\[
-\alpha_0 = 2\alpha_1 + \dots + 2\alpha_n = 2\del_1
\]
Extended Dynkin diagram:
\begin{center}
\begin{picture}(140,20)
\thicklines
\put(0,0){\circle{14}}
\put(42,0){\circle{14}}
\put(84,0){\circle{14}}
\put(126,0){\circle*{14}}
\put(0,15){\makebox(0.4,0.6){1}}
\put(42,15){\makebox(0.4,0.6){2}}
\put(84,15){\makebox(0.4,0.6){2}}
\put(126,15){\makebox(0.4,0.6){2}}
\put(6,-3){\line(1,0){30}}
\put(6,3){\line(1,0){30}}
\put(27,0){\line(-1,1){10}}\put(27,0){\line(-1,-1){10}}
\put(49,0){\dashbox{3}(28,0)}
\put(90,-3){\line(1,0){30}}
\put(90,3){\line(1,0){30}}
\put(111,0){\line(-1,1){10}}\put(111,0){\line(-1,-1){10}}
\put(42,-10){$\underbrace{~~~~~~~~~~}_{n-1}$}
\end{picture}
\end{center}
\vspace{7mm}
Factor group $\autout(\cG) = \aut(\cG)/\autint(\cG) = \II$.

\clearpage

\begin{table}[htbp]
\caption{The basic Lie superalgebra $C(n+1) = osp(2|2n)$.
\label{table28}}
\end{table}

\noindent
Structure: $\cG_\evn = so(2) \oplus sp(2n)$ and 
$\cG_\odd = (2n) \oplus (2n)$, type I. 
\\
Rank: $n+1$, dimension: $2n^2+5n+1$.
\\
Root system:
\beo
&& \Delta = \{ \pm\del_k\pm\del_l, ~ \pm 2\del_k, ~ \pm\eps\pm\del_k \} \\
&& \Delta_\evn = \{ \pm\del_k\pm\del_l, ~ \pm 2\del_k \}, \quad 
\Delta_\odd = \{ \pm\eps\pm\del_k \} \\
&& \overline{\Delta}_\evn = \Delta_\evn, \quad 
\overline{\Delta}_\odd = \Delta_\odd 
\eno
where $1 \le k \ne l \le n$.
\\
$\dim\Delta_\evn = \dim\overline{\Delta}_\evn = 2n^2$ and
$\dim\Delta_\odd = \dim\overline{\Delta}_\odd = 4n$.
\\
Distinguished simple root system:
\[
\alpha_1=\eps-\del_1, ~ \alpha_2=\del_1-\del_2, ~ \dots,
~ \alpha_n=\del_{n-1}-\del_n, ~ \alpha_{n+1}=2\del_n 
\]
Distinguished positive roots ($1 \le k < l \le n$):
\beo
\del_k-\del_l &=& \alpha_{k+1} + \dots + \alpha_l \\
\del_k+\del_l &=& \alpha_{k+1} + \dots + \alpha_l 
+ 2\alpha_{l+1} + \dots + 2\alpha_{n+1} \\
2\del_k &=& 2\alpha_{k+1} + \dots + 2\alpha_n + \alpha_{n+1} 
 ~ (k \ne n) \qquad 2\del_n = \alpha_{n+1} \\
\eps-\del_k &=& \alpha_1 + \dots + \alpha_k \\
\eps+\del_k &=& \alpha_1 + \dots + \alpha_k 
+ 2\alpha_{k+1} + \dots + 2\alpha_n + \alpha_{n+1} ~ (k < n) \\
\eps+\del_n &=& \alpha_1 + \dots + \alpha_{n+1}
\eno
Sums of even/odd distinguished positive roots:
\beo
&& 2\rho_0 = 2n\del_1 + (2n-2)\del_2 + \dots + 4\del_{n-1} 
+ 2\del_n \\
&& 2\rho_1 = 2n\eps
\eno
Distinguished Dynkin diagram:
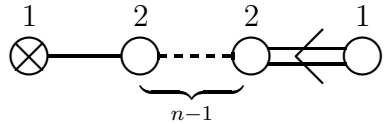
\begin{center}
\begin{picture}(120,20)
\thicklines
\multiput(0,0)(42,0){4}{\circle{14}}
\put(0,15){\makebox(0.4,0.6){1}}
\put(42,15){\makebox(0.4,0.6){2}}
\put(84,15){\makebox(0.4,0.6){2}}
\put(126,15){\makebox(0.4,0.6){1}}
\put(-5,-5){\line(1,1){10}}\put(-5,5){\line(1,-1){10}}
\put(7,0){\line(1,0){28}}
\put(49,0){\dashbox{3}(28,0)}
\put(101,0){\line(1,1){10}}\put(101,0){\line(1,-1){10}}
\put(90,-3){\line(1,0){30}}
\put(90,3){\line(1,0){30}}
\put(42,-10){$\underbrace{~~~~~~~~~~}_{n-1}$}
\end{picture}
\end{center}
\vspace{5mm}
Distinguished Cartan matrix:
\[
\left(\begin{array}{r|rrrrr}
0 & 1 & 0 & \cdots & \cdots & 0 \\ \hline
-1 & 2 & -1 & 0 && 0 \\
0 & -1 & \ddots & \ddots & \ddots & \vdots \\
0 & 0 & \ddots & \ddots & -1 & 0 \\
\vdots && \ddots & -1 &  2 & -2 \\
0 & \cdots & \cdots & 0 & -1 & 2 \\
\end{array}\right)
\]
Longest distinguished root:
\[
-\alpha_0 = \alpha_1 + 2\alpha_2 + \dots + 2\alpha_{n+1} + \alpha_n 
= \eps + \del_1
\]
Distinguished extended Dynkin diagram:
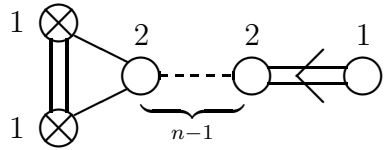
\vspace{3mm}
\begin{center}
\begin{picture}(140,20)
\thicklines
\put(11,20){\circle{14}}
\put(11,-20){\circle{14}}
\put(-5,20){\makebox(0.4,0.6){1}}
\put(-5,-20){\makebox(0.4,0.6){1}}
\put(42,0){\circle{14}}
\put(84,0){\circle{14}}
\put(126,0){\circle{14}}
\put(42,15){\makebox(0.4,0.6){2}}
\put(84,15){\makebox(0.4,0.6){2}}
\put(126,15){\makebox(0.4,0.6){1}}
\put(6,15){\line(1,1){10}}\put(6,25){\line(1,-1){10}}
\put(6,-25){\line(1,1){10}}\put(6,-15){\line(1,-1){10}}
\put(17,-15){\line(2,1){20}}
\put(17,15){\line(2,-1){20}}
\put(8,-14){\line(0,1){28}}
\put(14,-14){\line(0,1){28}}
\put(49,0){\dashbox{3}(28,0)}
\put(101,0){\line(1,1){10}}\put(101,0){\line(1,-1){10}}
\put(90,-3){\line(1,0){30}}
\put(90,3){\line(1,0){30}}
\put(42,-10){$\underbrace{~~~~~~~~~~}_{n-1}$}
\end{picture}
\end{center}
\vspace{7mm}
Factor group $\autout(\cG) = \aut(\cG)/\autint(\cG) = \ZZ_2$.

\clearpage

\begin{table}[htbp]
\caption{The basic Lie superalgebra $D(m,n) = osp(2m|2n)$.
\label{table29}}
\end{table}

\noindent
Structure: $\cG_\evn = so(2m) \oplus sp(2n) $ and 
$\cG_\odd = (2m,2n)$, type II.
\\
Rank: $m+n$, dimension: $2(m+n)^2-m+n$.
\\
Root system:
\beo
&& \Delta = \{ \pm\eps_i\pm\eps_j, ~ \pm\del_k\pm\del_l, ~ \pm 2\del_k, 
~ \pm\eps_i\pm\del_k \} \\ 
&& \Delta_\evn = \{ \pm\eps_i\pm\eps_j, ~ \pm\del_k\pm\del_l, 
~ \pm 2\del_k \}, \quad \Delta_\odd = \{ \pm\eps_i\pm\del_k \} \\
&& \overline{\Delta}_\evn = \Delta_\evn, \quad 
\overline{\Delta}_\odd = \Delta_\odd 
\eno
where $1 \le i \ne j \le m$ and $1 \le k \ne l \le n$.
\\
$\dim\Delta_\evn = \dim\overline{\Delta}_\evn = 2m^2+2n^2-2m$ and
$\dim\Delta_\odd = \dim\overline{\Delta}_\odd = 4mn$.
\\
Distinguished simple root system:
\beo
&& \alpha_1=\del_1-\del_2, ~ \dots, ~ \alpha_{n-1}=\del_{n-1}-\del_n,
~ \alpha_n=\del_n-\eps_1, \\
&& \alpha_{n+1}=\eps_1-\eps_2, ~ \dots, ~ \alpha_{n+m-1}=
\eps_{m-1}-\eps_m, ~ \alpha_{n+m}=\eps_{m-1}+\eps_m 
\eno
Distinguished positive roots ($1 \le i < j \le m$ and 
$1 \le k < l \le n$):
\beo
\del_k-\del_l &=& \alpha_k + \dots + \alpha_{l-1} \\
\del_k+\del_l &=& \alpha_k + \dots + \alpha_{l-1} + 2\alpha_l 
+ \dots + 2\alpha_{n+m-2} + \alpha_{n+m-1} + \alpha_{n+m} \\
2\del_k &=& 2\alpha_k + \dots + 2\alpha_{n+m-2} + \alpha_{n+m-1} 
+ \alpha_{n+m} \\
\eps_i-\eps_j &=& \alpha_{n+i} + \dots + \alpha_{n+j-1} \\
\eps_i+\eps_j &=& \alpha_{n+i} + \dots + \alpha_{n+j-1} + 2\alpha_{n+j} 
+ \dots + 2\alpha_{n+m-2} + \alpha_{n+m-1} + \alpha_{n+m} ~ (j < m-1) \\
\eps_i+\eps_{m-1} &=& \alpha_{n+i} + \dots + \alpha_{n+m-2} + 
+ \alpha_{n+m-1} + \alpha_{n+m} \\
\eps_i+\eps_m &=& \alpha_{n+i} + \dots + \alpha_{n+m-2} + \alpha_{n+m} \\
\del_k-\eps_i &=& \alpha_k + \dots + \alpha_{n+i-1} \\
\del_k+\eps_i &=& \alpha_k + \dots + \alpha_{n+i-1} + 2\alpha_{n+i} 
+ \dots + 2\alpha_{n+m-2} + \alpha_{n+m-1} + \alpha_{n+m} ~ (j < m-1) \\
\del_k+\eps_{m-1} &=& \alpha_k + \dots + \alpha_{n+m-2} + 
+ \alpha_{n+m-1} + \alpha_{n+m} \\
\del_k+\eps_m &=& \alpha_k + \dots + \alpha_{n+m-2} + \alpha_{n+m}
\eno
Sums of even/odd distinguished positive roots:
\beo
&& 2\rho_0 = (2m-2)\eps_1 + (2m-4)\eps_2 + \dots + 2\eps_{m-1}
+ 2n\del_1 + (2n-2)\del_2 + \dots + 4\del_{n-1} + 2\del_n \\
&& 2\rho_1 = 2m(\del_1 + \dots + \del_n)
\eno
\newpage
Distinguished Dynkin diagram:
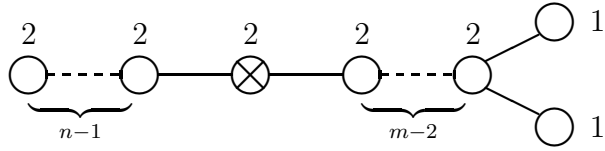
\begin{center}
\begin{picture}(200,20)
\thicklines
\multiput(0,0)(42,0){5}{\circle{14}}
\put(0,15){\makebox(0.4,0.6){2}}
\put(42,15){\makebox(0.4,0.6){2}}
\put(84,15){\makebox(0.4,0.6){2}}
\put(126,15){\makebox(0.4,0.6){2}}
\put(168,15){\makebox(0.4,0.6){2}}
\put(79,-5){\line(1,1){10}}\put(79,5){\line(1,-1){10}}
\put(7,0){\dashbox{3}(28,0)}
\put(49,0){\line(1,0){28}}
\put(91,0){\line(1,0){28}}
\put(133,0){\dashbox{3}(28,0)}
\put(173,5){\line(2,1){20}}\put(173,-5){\line(2,-1){20}}
\put(199,20){\circle{14}}
\put(215,20){\makebox(0.4,0.6){1}}
\put(199,-20){\circle{14}}
\put(215,-20){\makebox(0.4,0.6){1}}
\put(0,-10){$\underbrace{~~~~~~~~~~}_{n-1}$}
\put(126,-10){$\underbrace{~~~~~~~~~~}_{m-2}$}
\end{picture}
\end{center}
\vspace{5mm}
Distinguished Cartan matrix:
\[
\left(\begin{array}{rrrrr|r|rrrrr}
2 & -1 & 0 & \cdots & 0 & \cdots & \cdots &&& \cdots & 0 \\
-1 & \ddots & \ddots & \ddots &&&&&&& \vdots \\
0 & \ddots && \ddots & 0 &&&&&& \\
\vdots & \ddots & \ddots & \ddots & -1 & \ddots &&&&& \\
0 && 0 & -1 & 2 & -1 & \ddots &&&& \vdots \\ \hline
\vdots &&& \ddots & -1 & 0 & 1 & \ddots &&& \vdots \\ \hline
\vdots &&&& \ddots & -1 & 2 & -1 & 0 && 0 \\
&&&&& \ddots & -1 & \ddots & \ddots & \ddots & \vdots \\
&&&&&& 0 & \ddots & \ddots & -1 & -1 \\
&&&&&&& \ddots & -1 & 2 & 0 \\
0 & \cdots &&& \cdots & \cdots & 0 & \cdots & -1 & 0 & 2 \\
\end{array}\right)
\]
Longest distinguished root:
\[
-\alpha_0 = 2\alpha_1 + \dots + 2\alpha_{n+m-2} + \alpha_{n+m-1}
+ \alpha_{n+m} = 2\del_1
\]
Distinguished extended Dynkin diagram:
\begin{center}
\begin{picture}(260,20)
\thicklines
\multiput(0,0)(42,0){6}{\circle{14}}
\put(0,15){\makebox(0.4,0.6){1}}
\put(42,15){\makebox(0.4,0.6){2}}
\put(84,15){\makebox(0.4,0.6){2}}
\put(126,15){\makebox(0.4,0.6){2}}
\put(168,15){\makebox(0.4,0.6){2}}
\put(210,15){\makebox(0.4,0.6){2}}
\put(121,-5){\line(1,1){10}}\put(121,5){\line(1,-1){10}}
\put(6,-3){\line(1,0){30}}
\put(6,3){\line(1,0){30}}
\put(27,0){\line(-1,1){10}}\put(27,0){\line(-1,-1){10}}
\put(49,0){\dashbox{3}(28,0)}
\put(91,0){\line(1,0){28}}
\put(133,0){\line(1,0){28}}
\put(175,0){\dashbox{3}(28,0)}
\put(215,5){\line(2,1){20}}\put(215,-5){\line(2,-1){20}}
\put(241,20){\circle{14}}
\put(257,20){\makebox(0.4,0.6){1}}
\put(241,-20){\circle{14}}
\put(257,-20){\makebox(0.4,0.6){1}}
\put(42,-10){$\underbrace{~~~~~~~~~~}_{n-1}$}
\put(168,-10){$\underbrace{~~~~~~~~~~}_{m-2}$}
\end{picture}
\end{center}
\vspace{7mm}
Factor group $\autout(\cG) = \aut(\cG)/\autint(\cG) = \ZZ_2$.

\clearpage

\begin{table}[htbp]
\caption{The basic Lie superalgebra $F(4)$.\label{table30}}
\end{table}

\noindent
Structure: $\cG_\evn = sl(2) \oplus so(7)$ and $\cG_\odd = (2,8)$, 
type II. 
\\
Rank: $4$, dimension: $40$.
\\
Root system:
\beo
&& \Delta = \{ \pm\del, ~ \pm\eps_i\pm\eps_j, ~ \pm\eps_i, ~ 
{\half} (\pm\eps_1\pm\eps_2\pm\eps_3\pm\del) \} \\
&& \Delta_\evn = \{ \pm\del, ~ \pm\eps_i\pm\eps_j, ~ \pm\eps_i \}, \quad 
\Delta_\odd = \{ {\half} (\pm\eps_1\pm\eps_2\pm\eps_3\pm\del) \} \\
&& \overline{\Delta}_\evn = \Delta_\evn, \quad 
\overline{\Delta}_\odd = \Delta_\odd 
\eno
where $1 \le i \ne j \le 3$.
\\
$\dim\Delta_\evn = \dim\overline{\Delta}_\evn = 20$ and
$\dim\Delta_\odd = \dim\overline{\Delta}_\odd = 16$.
\\
Distinguished simple root system:
\[
\alpha_1 = {\half} (\del-\eps_1-\eps_2-\eps_3),
 ~ \alpha_2 = \eps_3, ~ \alpha_3 = \eps_2-\eps_3,
 ~ \alpha_4 = \eps_1-\eps_2 
\]
Distinguished positive roots ($1 \le i < j \le 3$):
\beo
\eps_i-\eps_j &=& \alpha_3, ~ \alpha_4, ~ \alpha_3+\alpha_4 \\
\eps_i+\eps_j &=& 2\alpha_2+\alpha_3, ~ 2\alpha_2+\alpha_3+\alpha_4,
 ~ 2\alpha_2+2\alpha_3+\alpha_4 \\
\eps_i &=& \alpha_2, ~ \alpha_2+\alpha_3, ~ \alpha_2+\alpha_3+\alpha_4 \\
\del &=& 2\alpha_1+3\alpha_2+2\alpha_3+\alpha_4 \\
\half(\del\pm\eps_1\pm\eps_2\pm\eps_3) &=& \alpha_1,
 ~ \alpha_1+\alpha_2, ~ \alpha_1+\alpha_2+\alpha_3,
 ~ \alpha_1+\alpha_2+\alpha_3+\alpha_4,
 ~ \alpha_1+2\alpha_2+\alpha_3, \\
&& \alpha_1+2\alpha_2+\alpha_3+\alpha_4, ~ 
\alpha_1+2\alpha_2+2\alpha_3+\alpha_4, ~ 
\alpha_1+3\alpha_2+2\alpha_3+\alpha_4
\eno
Sums of even/odd distinguished positive roots:
\beo
&& 2\rho_0 = 5\eps_1 + 3\eps_2 + \eps_3 + \del \\
&& 2\rho_1 = 4\del
\eno
Distinguished Dynkin diagram:
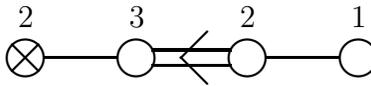
\begin{center}
\begin{picture}(120,20)
\thicklines
\put(0,0){\circle{14}}
\put(42,0){\circle{14}}
\put(84,0){\circle{14}}
\put(126,0){\circle{14}}
\put(0,15){\makebox(0.4,0.6){2}}
\put(42,15){\makebox(0.4,0.6){3}}
\put(84,15){\makebox(0.4,0.6){2}}
\put(126,15){\makebox(0.4,0.6){1}}
\put(-5,-5){\line(1,1){10}}\put(-5,5){\line(1,-1){10}}
\put(7,0){\line(1,0){28}}
\put(48,-3){\line(1,0){30}}
\put(48,3){\line(1,0){30}}
\put(59,0){\line(1,1){10}}\put(59,0){\line(1,-1){10}}
\put(91,0){\line(1,0){28}}
\end{picture}
\end{center}
Distinguished Cartan matrix:
\[
\left(\begin{array}{rrrr}
0 & 1 & 0 & 0 \\ -1 & 2 & -2 & 0 \\ 0 & -1 & 2 & -1 \\ 0 & 0 & -1 & 2 \\
\end{array}\right)
\]
Longest distinguished root:
\[
-\alpha_0 = 2\alpha_1 + 3\alpha_2 + 2\alpha_3 + \alpha_4 = \del
\]
Distinguished extended Dynkin diagram:
\begin{center}
\begin{picture}(180,20)
\thicklines
\multiput(0,0)(42,0){5}{\circle{14}}
\put(0,15){\makebox(0.4,0.6){1}}
\put(42,15){\makebox(0.4,0.6){2}}
\put(84,15){\makebox(0.4,0.6){3}}
\put(126,15){\makebox(0.4,0.6){2}}
\put(168,15){\makebox(0.4,0.6){1}}
\put(37,-5){\line(1,1){10}}\put(37,5){\line(1,-1){10}}
\put(6,-4){\line(1,0){30}}
\put(7,0){\line(1,0){28}}
\put(6,4){\line(1,0){30}}
\put(27,0){\line(-1,1){10}}\put(27,0){\line(-1,-1){10}}
\put(49,0){\line(1,0){28}}
\put(90,-3){\line(1,0){30}}
\put(90,3){\line(1,0){30}}
\put(101,0){\line(1,1){10}}\put(101,0){\line(1,-1){10}}
\put(133,0){\line(1,0){28}}
\end{picture}
\end{center}
Factor group $\autout(\cG) = \aut(\cG)/\autint(\cG) = \II$.

\clearpage

\begin{table}[htbp]
\caption{The basic Lie superalgebra $G(3)$.\label{table31}}
\end{table}

\noindent
Structure: $\cG_\evn = sl(2) \oplus G_2$ and $\cG_\odd = (2,7)$, 
type II.
\\
Rank: $3$, dimension: $31$.
\\
Root system:
\beo
&& \Delta = \{ \pm 2\del, ~ \pm\eps_i, ~ \eps_i-\eps_j, ~ 
\pm\del, ~ \pm\eps_i\pm\del  \} \\
&& \Delta_\evn = \{ \pm 2\del, ~ \pm\eps_i, ~ \eps_i-\eps_j \}, \quad 
\Delta_\odd = \{ \pm\del, ~ \pm\eps_i\pm\del \} \\
&& \overline{\Delta}_\evn = \{ \pm\eps_i, ~ \eps_i-\eps_j \}, \quad 
\overline{\Delta}_\odd = \{ \pm\eps_i\pm\del \} 
\eno
where $1 \le i \ne j \le 3$ and $\eps_1+\eps_2+\eps_3 = 0$.
\\
$\dim\Delta_\evn = 14$, $\dim\Delta_\odd = 14$, 
$\dim\overline{\Delta}_\evn = 12$, $\dim\overline{\Delta}_\odd = 12$.
\\
Distinguished simple root system:
\[
\alpha_1 = \del+\eps_3, ~ \alpha_2 = \eps_1, ~ \alpha_3 = \eps_2-\eps_1
\]
Distinguished positive roots:
\beo
&& \smbox{even roots:} \alpha_2, ~ \alpha_3, ~ \alpha_2+\alpha_3,
 ~ 2\alpha_2+\alpha_3, ~  3\alpha_2+\alpha_3, ~ 3\alpha_2+2\alpha_3,
 ~ 2\alpha_1+4\alpha_2+2\alpha_3 \\
&& \smbox{odd roots:} \alpha_1, ~ \alpha_1+\alpha_2,
 ~ \alpha_1+\alpha_2+\alpha_3, ~ \alpha_1+2\alpha_2+\alpha_3,
 ~ \alpha_1+3\alpha_2+\alpha_3, \\ 
&& \phantom{odd roots:} \alpha_1+3\alpha_2+2\alpha_3,
 ~ \alpha_1+4\alpha_2+2\alpha_3
\eno
Sums of even/odd distinguished positive roots:
\beo
&& 2\rho_0 = 2\eps_1 + 4\eps_2 - 2\eps_3 + 2\del \\
&& 2\rho_1 = 7\del
\eno
Distinguished Dynkin diagram:
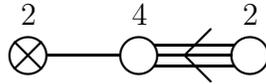
\begin{center}
\begin{picture}(80,20)
\thicklines
\put(0,0){\circle{14}}
\put(42,0){\circle{14}}
\put(84,0){\circle{14}}
\put(0,15){\makebox(0.4,0.6){2}}
\put(42,15){\makebox(0.4,0.6){4}}
\put(84,15){\makebox(0.4,0.6){2}}
\put(-5,-5){\line(1,1){10}}\put(-5,5){\line(1,-1){10}}
\put(7,0){\line(1,0){28}}
\put(48,-4){\line(1,0){30}}
\put(49,0){\line(1,0){28}}
\put(48,4){\line(1,0){30}}
\put(59,0){\line(1,1){10}}\put(59,0){\line(1,-1){10}}
\end{picture}
\end{center}
Distinguished Cartan matrix:
\[
\left(\begin{array}{rrr}
0 & 1 & 0 \\ -1 & 2 & -3 \\ 0 & -1 & 2 \\
\end{array}\right)
\]
Longest distinguished root:
\[
-\alpha_0 = 2\alpha_1 + 4\alpha_2 + 2\alpha_3 = 2\del
\]
Distinguished extended Dynkin diagram:
\begin{center}
\begin{picture}(140,20)
\thicklines
\multiput(0,0)(42,0){4}{\circle{14}}
\put(0,15){\makebox(0.4,0.6){1}}
\put(42,15){\makebox(0.4,0.6){2}}
\put(84,15){\makebox(0.4,0.6){4}}
\put(126,15){\makebox(0.4,0.6){2}}
\put(37,-5){\line(1,1){10}}\put(37,5){\line(1,-1){10}}
\put(7,-2){\line(1,0){29}}
\put(7,2){\line(1,0){29}}
\put(5,-5){\line(1,0){32}}
\put(5,5){\line(1,0){32}}
\put(27,0){\line(-1,1){10}}\put(27,0){\line(-1,-1){10}}
\put(49,0){\line(1,0){28}}
\put(90,-4){\line(1,0){30}}
\put(91,0){\line(1,0){28}}
\put(90,4){\line(1,0){30}}
\put(101,0){\line(1,1){10}}\put(101,0){\line(1,-1){10}}
\end{picture}
\end{center}
\vspace{7mm}
Factor group $\autout(\cG) = \aut(\cG)/\autint(\cG) = \II$.

\clearpage

\begin{table}[htbp]
\caption{The basic Lie superalgebra $D(2,1;\alpha)$.\label{table32}}
\end{table}

\noindent
Structure: $\cG_\evn = sl(2) \oplus sl(2) \oplus sl(2)$ and 
$\cG_\odd = (2,2,2)$, type II. 
\\
Rank: $3$, dimension: $17$.
\\
Root system:
\beo
&& \Delta = \{ \pm 2\eps_i, ~ \pm\eps_1\pm\eps_2\pm\eps_3 \} \\
&& \Delta_\evn = \{ \pm 2\eps_i \}, \quad 
\Delta_\odd = \{ \pm\eps_1\pm\eps_2\pm\eps_3 \} \\
&& \overline{\Delta}_\evn = \Delta_\evn, \quad 
\overline{\Delta}_\odd = \Delta_\odd 
\eno
where $1 \le i \le 3$.
\\
$\dim\Delta_\evn = \dim\overline{\Delta}_\evn = 6$ and
$\dim\Delta_\odd = \dim\overline{\Delta}_\odd = 8$.
\\
Distinguished simple root system:
\[
\alpha_1 = \eps_1-\eps_2-\eps_3, \alpha_2 = 2\eps_2, \alpha_3 = 2\eps_3
\]
Distinguished positive roots ($1 \le < j \le 3$):
\beo
&& \smbox{even roots:} \alpha_2, ~ \alpha_3, 
 ~ 2\alpha_1+\alpha_2+\alpha_3 \\ 
&& \smbox{odd roots:} \alpha_1, ~ \alpha_1+\alpha_2,
 ~ \alpha_1+\alpha_3, ~ \alpha_1+\alpha_2+\alpha_3 
\eno
Sums of even/odd distinguished positive roots:
\beo
&& 2\rho_0 = 2\eps_1 + 2\eps_2 + 2\eps_3 \\
&& 2\rho_1 = 4\eps_1
\eno
Distinguished Dynkin diagram:
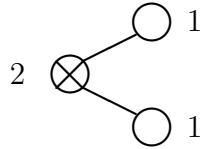
\begin{center}
\begin{picture}(60,20)
\thicklines
\put(10,0){\circle{14}}
\put(-10,0){\makebox(0.4,0.6){2}}
\put(5,-5){\line(1,1){10}}\put(5,5){\line(1,-1){10}}
\put(15,5){\line(2,1){20}}\put(15,-5){\line(2,-1){20}}
\put(41,20){\circle{14}}
\put(57,20){\makebox(0.4,0.6){1}}
\put(41,-20){\circle{14}}
\put(57,-20){\makebox(0.4,0.6){1}}
\end{picture}
\end{center}
\vspace{5mm}
Distinguished Cartan matrix:
\[
\left(\begin{array}{rrr}
0 & 1 & \alpha \\ -1 & 2 & 0 \\ -1 & 0 & 2 \\
\end{array}\right)
\]
Longest root (in the distinguished root system):
\[
-\alpha_0 = 2\alpha_1 + \alpha_2 + \alpha_3 = 2\eps_1
\]
\newpage
Distinguished extended Dynkin diagram:
\begin{center}
\begin{picture}(90,20)
\thicklines
\put(0,0){\circle{14}}
\put(42,0){\circle{14}}
\put(0,15){\makebox(0.4,0.6){1}}
\put(42,15){\makebox(0.4,0.6){2}}
\put(37,-5){\line(1,1){10}}\put(37,5){\line(1,-1){10}}
\put(7,0){\line(1,0){28}}
\put(47,5){\line(2,1){20}}\put(47,-5){\line(2,-1){20}}
\put(73,20){\circle{14}}
\put(89,20){\makebox(0.4,0.6){1}}
\put(73,-20){\circle{14}}
\put(89,-20){\makebox(0.4,0.6){1}}
\end{picture}
\end{center}
\vspace{7mm}
Factor group $\autout(\cG) = \aut(\cG)/\autint(\cG)$:
\beo
&& \autout(\cG) = \II \smbox{for generic} \alpha \\
&& \autout(\cG) = \ZZ_2 \smbox{for} \alpha = 1,-2,-1/2 \\
&& \autout(\cG) = \ZZ_3 \smbox{for} \alpha = e^{2i\pi/3}, e^{4i\pi/3}
\eno

\clearpage

\begin{table}[htbp]
\caption{Distinguished Dynkin diagrams of the basic Lie
superalgebras.\label{table5}}
\end{table}
\[
\begin{array}{llc}
A(m,n) &&
\begin{picture}(160,20)
\thicklines
\multiput(0,0)(42,0){5}{\circle{14}}
\put(0,15){\makebox(0.4,0.6){1}}
\put(42,15){\makebox(0.4,0.6){1}}
\put(84,15){\makebox(0.4,0.6){1}}
\put(126,15){\makebox(0.4,0.6){1}}
\put(168,15){\makebox(0.4,0.6){1}}
\put(79,-5){\line(1,1){10}}\put(79,5){\line(1,-1){10}}
\put(7,0){\dashbox{3}(28,0)}
\put(49,0){\line(1,0){28}}
\put(91,0){\line(1,0){28}}
\put(133,0){\dashbox{3}(28,0)}
\put(0,-10){$\underbrace{~~~~~~~~~~}_{n}$}
\put(126,-10){$\underbrace{~~~~~~~~~~}_{m}$}
\end{picture}
\cr &&\cr&&\cr
B(m,n) &&
\begin{picture}(200,20)
\thicklines
\multiput(0,0)(42,0){6}{\circle{14}}
\put(0,15){\makebox(0.4,0.6){2}}
\put(42,15){\makebox(0.4,0.6){2}}
\put(84,15){\makebox(0.4,0.6){2}}
\put(126,15){\makebox(0.4,0.6){2}}
\put(168,15){\makebox(0.4,0.6){2}}
\put(210,15){\makebox(0.4,0.6){2}}
\put(79,-5){\line(1,1){10}}\put(79,5){\line(1,-1){10}}
\put(7,0){\dashbox{3}(28,0)}
\put(49,0){\line(1,0){28}}
\put(91,0){\line(1,0){28}}
\put(133,0){\dashbox{3}(28,0)}
\put(174,-3){\line(1,0){30}}
\put(174,3){\line(1,0){30}}
\put(195,0){\line(-1,1){10}}\put(195,0){\line(-1,-1){10}}
\put(0,-10){$\underbrace{~~~~~~~~~~}_{n-1}$}
\put(126,-10){$\underbrace{~~~~~~~~~~}_{m-1}$}
\end{picture}
\cr &&\cr&&\cr
B(0,n) &&
\begin{picture}(80,20)
\thicklines
\put(0,0){\circle{14}}
\put(42,0){\circle{14}}
\put(84,0){\circle*{14}}
\put(0,15){\makebox(0.4,0.6){2}}
\put(42,15){\makebox(0.4,0.6){2}}
\put(84,15){\makebox(0.4,0.6){2}}
\put(7,0){\dashbox{3}(28,0)}
\put(48,-3){\line(1,0){30}}
\put(48,3){\line(1,0){30}}
\put(69,0){\line(-1,1){10}}\put(69,0){\line(-1,-1){10}}
\put(0,-10){$\underbrace{~~~~~~~~~~}_{n-1}$}
\end{picture}
\cr &&\cr&&\cr
C(n+1) &&
\begin{picture}(120,20)
\thicklines
\multiput(0,0)(42,0){4}{\circle{14}}
\put(0,15){\makebox(0.4,0.6){1}}
\put(42,15){\makebox(0.4,0.6){2}}
\put(84,15){\makebox(0.4,0.6){2}}
\put(126,15){\makebox(0.4,0.6){1}}
\put(-5,-5){\line(1,1){10}}\put(-5,5){\line(1,-1){10}}
\put(7,0){\line(1,0){28}}
\put(49,0){\dashbox{3}(28,0)}
\put(101,0){\line(1,1){10}}\put(101,0){\line(1,-1){10}}
\put(90,-3){\line(1,0){30}}
\put(90,3){\line(1,0){30}}
\put(42,-10){$\underbrace{~~~~~~~~~~}_{n-1}$}
\end{picture}
\cr &&\cr &&\cr
D(m,n) &&
\begin{picture}(200,20)
\thicklines
\multiput(0,0)(42,0){5}{\circle{14}}
\put(0,15){\makebox(0.4,0.6){2}}
\put(42,15){\makebox(0.4,0.6){2}}
\put(84,15){\makebox(0.4,0.6){2}}
\put(126,15){\makebox(0.4,0.6){2}}
\put(168,15){\makebox(0.4,0.6){2}}
\put(79,-5){\line(1,1){10}}\put(79,5){\line(1,-1){10}}
\put(7,0){\dashbox{3}(28,0)}
\put(49,0){\line(1,0){28}}
\put(91,0){\line(1,0){28}}
\put(133,0){\dashbox{3}(28,0)}
\put(173,5){\line(2,1){20}}\put(173,-5){\line(2,-1){20}}
\put(199,20){\circle{14}}
\put(215,20){\makebox(0.4,0.6){1}}
\put(199,-20){\circle{14}}
\put(215,-20){\makebox(0.4,0.6){1}}
\put(0,-10){$\underbrace{~~~~~~~~~~}_{n-1}$}
\put(126,-10){$\underbrace{~~~~~~~~~~}_{m-2}$}
\end{picture}
\cr &&\cr&&\cr
F(4) &&
\begin{picture}(120,20)
\thicklines
\put(0,0){\circle{14}}
\put(42,0){\circle{14}}
\put(84,0){\circle{14}}
\put(126,0){\circle{14}}
\put(0,15){\makebox(0.4,0.6){2}}
\put(42,15){\makebox(0.4,0.6){3}}
\put(84,15){\makebox(0.4,0.6){2}}
\put(126,15){\makebox(0.4,0.6){1}}
\put(-5,-5){\line(1,1){10}}\put(-5,5){\line(1,-1){10}}
\put(7,0){\line(1,0){28}}
\put(48,-3){\line(1,0){30}}
\put(48,3){\line(1,0){30}}
\put(59,0){\line(1,1){10}}\put(59,0){\line(1,-1){10}}
\put(91,0){\line(1,0){28}}
\end{picture}
\cr &&\cr&&\cr
G(3) &&
\begin{picture}(80,20)
\thicklines
\put(0,0){\circle{14}}
\put(42,0){\circle{14}}
\put(84,0){\circle{14}}
\put(0,15){\makebox(0.4,0.6){2}}
\put(42,15){\makebox(0.4,0.6){4}}
\put(84,15){\makebox(0.4,0.6){2}}
\put(-5,-5){\line(1,1){10}}\put(-5,5){\line(1,-1){10}}
\put(7,0){\line(1,0){28}}
\put(48,-4){\line(1,0){30}}
\put(49,0){\line(1,0){28}}
\put(48,4){\line(1,0){30}}
\put(59,0){\line(1,1){10}}\put(59,0){\line(1,-1){10}}
\end{picture}
\cr &&\cr&&\cr
D(2,1;\alpha) &&
\begin{picture}(60,20)
\thicklines
\put(10,0){\circle{14}}
\put(-10,0){\makebox(0.4,0.6){2}}
\put(5,-5){\line(1,1){10}}\put(5,5){\line(1,-1){10}}
\put(15,5){\line(2,1){20}}\put(15,-5){\line(2,-1){20}}
\put(41,20){\circle{14}}
\put(57,20){\makebox(0.4,0.6){1}}
\put(41,-20){\circle{14}}
\put(57,-20){\makebox(0.4,0.6){1}}
\end{picture}
\end{array}
\]

\clearpage

\begin{table}[htbp]
\caption{Dynkin diagrams of the basic Lie superalgebras of rank $\le$ 4
\label{table6}} 
\end{table}

\[
\begin{array}{ccccccc}
sl(1|1) &\hspace{10mm}&
\begin{picture}(20,20)
\thicklines
\put(0,0){\circle{14}}
\put(0,15){\makebox(0.4,0.6){1}}
\put(-5,-5){\line(1,1){10}}\put(-5,5){\line(1,-1){10}}
\end{picture}
&\hspace{20mm}&
osp(1|2) &\hspace{10mm}&
\begin{picture}(20,20)
\thicklines
\put(0,0){\circle*{14}}
\put(0,15){\makebox(0.4,0.6){1}}
\end{picture}
\cr
\end{array}
\]

\[
\begin{array}{ccccccccccc}
sl(1|2) &\hspace{5mm}&
\begin{picture}(50,20)
\thicklines
\put(0,0){\circle{14}}
\put(42,0){\circle{14}}
\put(0,15){\makebox(0.4,0.6){1}}
\put(42,15){\makebox(0.4,0.6){1}}
\put(7,0){\line(1,0){28}}
\put(-5,-5){\line(1,1){10}}\put(-5,5){\line(1,-1){10}}
\end{picture}
&&
\begin{picture}(50,20)
\thicklines
\put(0,0){\circle{14}}
\put(42,0){\circle{14}}
\put(0,15){\makebox(0.4,0.6){1}}
\put(42,15){\makebox(0.4,0.6){1}}
\put(7,0){\line(1,0){28}}
\put(-5,-5){\line(1,1){10}}\put(-5,5){\line(1,-1){10}}
\put(37,-5){\line(1,1){10}}\put(37,5){\line(1,-1){10}}
\end{picture}
&\hspace{10mm}&
osp(2|2) &\hspace{5mm}&
\begin{picture}(50,20)
\thicklines
\put(0,0){\circle{14}}
\put(42,0){\circle{14}}
\put(0,15){\makebox(0.4,0.6){2}}
\put(42,15){\makebox(0.4,0.6){2}}
\put(6,-3){\line(1,0){30}}
\put(6,3){\line(1,0){30}}
\put(-5,-5){\line(1,1){10}}\put(-5,5){\line(1,-1){10}}
\put(37,-5){\line(1,1){10}}\put(37,5){\line(1,-1){10}}
\end{picture}
&&
\begin{picture}(50,20)
\thicklines
\put(0,0){\circle{14}}
\put(42,0){\circle{14}}
\put(0,15){\makebox(0.4,0.6){2}}
\put(42,15){\makebox(0.4,0.6){2}}
\put(6,-3){\line(1,0){30}}
\put(6,3){\line(1,0){30}}
\put(17,0){\line(1,-1){10}}\put(17,0){\line(1,1){10}}
\put(-5,-5){\line(1,1){10}}\put(-5,5){\line(1,-1){10}}
\end{picture}
\cr \cr
osp(1|4) &\hspace{5mm}&
\begin{picture}(50,20)
\thicklines
\put(0,0){\circle{14}}
\put(42,0){\circle*{14}}
\put(0,15){\makebox(0.4,0.6){2}}
\put(42,15){\makebox(0.4,0.6){2}}
\put(6,-3){\line(1,0){30}}
\put(6,3){\line(1,0){30}}
\put(27,0){\line(-1,1){10}}\put(27,0){\line(-1,-1){10}}
\end{picture}
&&
&\hspace{10mm}&
osp(3|2) &\hspace{5mm}&
\begin{picture}(50,20)
\thicklines
\put(0,0){\circle{14}}
\put(42,0){\circle{14}}
\put(0,15){\makebox(0.4,0.6){2}}
\put(42,15){\makebox(0.4,0.6){2}}
\put(6,-3){\line(1,0){30}}
\put(6,3){\line(1,0){30}}
\put(27,0){\line(-1,1){10}}\put(27,0){\line(-1,-1){10}}
\put(-5,-5){\line(1,1){10}}\put(-5,5){\line(1,-1){10}}
\end{picture}
&&
\begin{picture}(50,20)
\thicklines
\put(0,0){\circle{14}}
\put(42,0){\circle*{14}}
\put(0,15){\makebox(0.4,0.6){1}}
\put(42,15){\makebox(0.4,0.6){2}}
\put(6,-3){\line(1,0){30}}
\put(6,3){\line(1,0){30}}
\put(27,0){\line(-1,1){10}}\put(27,0){\line(-1,-1){10}}
\put(-5,-5){\line(1,1){10}}\put(-5,5){\line(1,-1){10}}
\end{picture}
\cr
\end{array}
\]

\[
\begin{array}{ccccccc}
sl(1|3) &\hspace{10mm}&
\begin{picture}(100,20)
\thicklines
\multiput(0,0)(42,0){3}{\circle{14}}
\put(0,15){\makebox(0.4,0.6){1}}
\put(42,15){\makebox(0.4,0.6){1}}
\put(84,15){\makebox(0.4,0.6){1}}
\put(7,0){\line(1,0){28}}
\put(49,0){\line(1,0){28}}
\put(-5,-5){\line(1,1){10}}\put(-5,5){\line(1,-1){10}}
\end{picture}
&&
\begin{picture}(100,20)
\thicklines
\multiput(0,0)(42,0){3}{\circle{14}}
\put(0,15){\makebox(0.4,0.6){1}}
\put(42,15){\makebox(0.4,0.6){1}}
\put(84,15){\makebox(0.4,0.6){1}}
\put(7,0){\line(1,0){28}}
\put(49,0){\line(1,0){28}}
\put(-5,-5){\line(1,1){10}}\put(-5,5){\line(1,-1){10}}
\put(37,-5){\line(1,1){10}}\put(37,5){\line(1,-1){10}}
\end{picture}
&&
\cr \cr
sl(2|2) &\hspace{10mm}&
\begin{picture}(100,20)
\thicklines
\multiput(0,0)(42,0){3}{\circle{14}}
\put(0,15){\makebox(0.4,0.6){1}}
\put(42,15){\makebox(0.4,0.6){1}}
\put(84,15){\makebox(0.4,0.6){1}}
\put(7,0){\line(1,0){28}}
\put(49,0){\line(1,0){28}}
\put(37,-5){\line(1,1){10}}\put(37,5){\line(1,-1){10}}
\end{picture}
&&
\begin{picture}(100,20)
\thicklines
\multiput(0,0)(42,0){3}{\circle{14}}
\put(0,15){\makebox(0.4,0.6){1}}
\put(42,15){\makebox(0.4,0.6){1}}
\put(84,15){\makebox(0.4,0.6){1}}
\put(7,0){\line(1,0){28}}
\put(49,0){\line(1,0){28}}
\put(-5,-5){\line(1,1){10}}\put(-5,5){\line(1,-1){10}}
\put(37,-5){\line(1,1){10}}\put(37,5){\line(1,-1){10}}
\put(79,-5){\line(1,1){10}}\put(79,5){\line(1,-1){10}}
\end{picture}
&&
\begin{picture}(100,20)
\thicklines
\multiput(0,0)(42,0){3}{\circle{14}}
\put(0,15){\makebox(0.4,0.6){1}}
\put(42,15){\makebox(0.4,0.6){1}}
\put(84,15){\makebox(0.4,0.6){1}}
\put(7,0){\line(1,0){28}}
\put(49,0){\line(1,0){28}}
\put(-5,-5){\line(1,1){10}}\put(-5,5){\line(1,-1){10}}
\put(79,-5){\line(1,1){10}}\put(79,5){\line(1,-1){10}}
\end{picture}
\cr \cr
osp(1|6) &\hspace{10mm}&
\begin{picture}(100,20)
\thicklines
\multiput(0,0)(42,0){2}{\circle{14}}
\put(84,0){\circle*{14}}
\put(0,15){\makebox(0.4,0.6){2}}
\put(42,15){\makebox(0.4,0.6){2}}
\put(84,15){\makebox(0.4,0.6){2}}
\put(7,0){\line(1,0){28}}
\put(48,-3){\line(1,0){30}}
\put(48,3){\line(1,0){30}}
\put(69,0){\line(-1,1){10}}\put(69,0){\line(-1,-1){10}}
\end{picture}
&&&& 
\cr \cr
osp(2|4) &\hspace{10mm}&
\begin{picture}(100,20)
\thicklines
\multiput(0,0)(42,0){3}{\circle{14}}
\put(0,15){\makebox(0.4,0.6){1}}
\put(42,15){\makebox(0.4,0.6){2}}
\put(84,15){\makebox(0.4,0.6){1}}
\put(7,0){\line(1,0){28}}
\put(48,-3){\line(1,0){30}}
\put(48,3){\line(1,0){30}}
\put(59,0){\line(1,-1){10}}\put(59,0){\line(1,1){10}}
\put(-5,-5){\line(1,1){10}}\put(-5,5){\line(1,-1){10}}
\end{picture}
&&
\begin{picture}(100,20)
\thicklines
\multiput(0,0)(42,0){3}{\circle{14}}
\put(0,15){\makebox(0.4,0.6){2}}
\put(42,15){\makebox(0.4,0.6){2}}
\put(84,15){\makebox(0.4,0.6){1}}
\put(7,0){\line(1,0){28}}
\put(48,-3){\line(1,0){30}}
\put(48,3){\line(1,0){30}}
\put(59,0){\line(1,-1){10}}\put(59,0){\line(1,1){10}}
\put(-5,-5){\line(1,1){10}}\put(-5,5){\line(1,-1){10}}
\put(37,-5){\line(1,1){10}}\put(37,5){\line(1,-1){10}}
\end{picture}
&& 
\begin{picture}(50,20)
\thicklines
\put(0,0){\circle{14}}
\put(0,15){\makebox(0.4,0.6){2}}
\put(5,5){\line(2,1){20}}\put(5,-5){\line(2,-1){20}}
\put(31,20){\circle{14}}
\put(47,20){\makebox(0.4,0.6){1}}
\put(31,-20){\circle{14}}
\put(47,-20){\makebox(0.4,0.6){1}}
\put(26,15){\line(1,1){10}}\put(26,25){\line(1,-1){10}}
\put(26,-25){\line(1,1){10}}\put(26,-15){\line(1,-1){10}}
\put(28,-14){\line(0,1){28}}\put(34,-14){\line(0,1){28}}
\end{picture}
\cr \cr \cr
osp(3|4) &\hspace{10mm}&
\begin{picture}(100,20)
\thicklines
\multiput(0,0)(42,0){3}{\circle{14}}
\put(0,15){\makebox(0.4,0.6){2}}
\put(42,15){\makebox(0.4,0.6){2}}
\put(84,15){\makebox(0.4,0.6){2}}
\put(7,0){\line(1,0){28}}
\put(48,-3){\line(1,0){30}}
\put(48,3){\line(1,0){30}}
\put(69,0){\line(-1,1){10}}\put(69,0){\line(-1,-1){10}}
\put(37,-5){\line(1,1){10}}\put(37,5){\line(1,-1){10}}
\end{picture}
&&
\begin{picture}(100,20)
\thicklines
\multiput(0,0)(42,0){2}{\circle{14}}
\put(84,0){\circle*{14}}
\put(0,15){\makebox(0.4,0.6){1}}
\put(42,15){\makebox(0.4,0.6){2}}
\put(84,15){\makebox(0.4,0.6){2}}
\put(7,0){\line(1,0){28}}
\put(48,-3){\line(1,0){30}}
\put(48,3){\line(1,0){30}}
\put(69,0){\line(-1,1){10}}\put(69,0){\line(-1,-1){10}}
\put(-5,-5){\line(1,1){10}}\put(-5,5){\line(1,-1){10}}
\end{picture}
&& 
\begin{picture}(100,20)
\thicklines
\multiput(0,0)(42,0){2}{\circle{14}}
\put(84,0){\circle*{14}}
\put(0,15){\makebox(0.4,0.6){2}}
\put(42,15){\makebox(0.4,0.6){2}}
\put(84,15){\makebox(0.4,0.6){2}}
\put(7,0){\line(1,0){28}}
\put(48,-3){\line(1,0){30}}
\put(48,3){\line(1,0){30}}
\put(69,0){\line(-1,1){10}}\put(69,0){\line(-1,-1){10}}
\put(-5,-5){\line(1,1){10}}\put(-5,5){\line(1,-1){10}}
\put(37,-5){\line(1,1){10}}\put(37,5){\line(1,-1){10}}
\end{picture}
\cr \cr \cr
osp(4|2) &\hspace{10mm}&
\begin{picture}(100,20)
\thicklines
\multiput(0,0)(42,0){3}{\circle{14}}
\put(0,15){\makebox(0.4,0.6){1}}
\put(42,15){\makebox(0.4,0.6){2}}
\put(84,15){\makebox(0.4,0.6){1}}
\put(7,0){\line(1,0){28}}
\put(48,-3){\line(1,0){30}}
\put(48,3){\line(1,0){30}}
\put(59,0){\line(1,-1){10}}\put(59,0){\line(1,1){10}}
\put(37,-5){\line(1,1){10}}\put(37,5){\line(1,-1){10}}
\end{picture}
&&
\begin{picture}(50,20)
\thicklines
\put(0,0){\circle{14}}
\put(0,15){\makebox(0.4,0.6){1}}
\put(5,5){\line(2,1){20}}\put(5,-5){\line(2,-1){20}}
\put(31,20){\circle{14}}
\put(47,20){\makebox(0.4,0.6){1}}
\put(31,-20){\circle{14}}
\put(47,-20){\makebox(0.4,0.6){1}}
\put(-5,-5){\line(1,1){10}}\put(-5,5){\line(1,-1){10}}
\end{picture}
&& 
\begin{picture}(50,20)
\thicklines
\put(0,0){\circle{14}}
\put(0,15){\makebox(0.4,0.6){1}}
\put(5,5){\line(2,1){20}}\put(5,-5){\line(2,-1){20}}
\put(31,20){\circle{14}}
\put(47,20){\makebox(0.4,0.6){1}}
\put(31,-20){\circle{14}}
\put(47,-20){\makebox(0.4,0.6){1}}
\put(-5,-5){\line(1,1){10}}\put(-5,5){\line(1,-1){10}}
\put(26,15){\line(1,1){10}}\put(26,25){\line(1,-1){10}}
\put(26,-25){\line(1,1){10}}\put(26,-15){\line(1,-1){10}}
\put(28,-14){\line(0,1){28}}\put(34,-14){\line(0,1){28}}
\end{picture}
\cr \cr \cr
osp(5|2) &\hspace{10mm}&
\begin{picture}(100,20)
\thicklines
\multiput(0,0)(42,0){3}{\circle{14}}
\put(0,15){\makebox(0.4,0.6){2}}
\put(42,15){\makebox(0.4,0.6){2}}
\put(84,15){\makebox(0.4,0.6){2}}
\put(7,0){\line(1,0){28}}
\put(48,-3){\line(1,0){30}}
\put(48,3){\line(1,0){30}}
\put(69,0){\line(-1,1){10}}\put(69,0){\line(-1,-1){10}}
\put(-5,-5){\line(1,1){10}}\put(-5,5){\line(1,-1){10}}
\end{picture}
&&
\begin{picture}(100,20)
\thicklines
\multiput(0,0)(42,0){3}{\circle{14}}
\put(0,15){\makebox(0.4,0.6){1}}
\put(42,15){\makebox(0.4,0.6){2}}
\put(84,15){\makebox(0.4,0.6){2}}
\put(7,0){\line(1,0){28}}
\put(48,-3){\line(1,0){30}}
\put(48,3){\line(1,0){30}}
\put(69,0){\line(-1,1){10}}\put(69,0){\line(-1,-1){10}}
\put(-5,-5){\line(1,1){10}}\put(-5,5){\line(1,-1){10}}
\put(37,-5){\line(1,1){10}}\put(37,5){\line(1,-1){10}}
\end{picture}
&& 
\begin{picture}(100,20)
\thicklines
\multiput(0,0)(42,0){2}{\circle{14}}
\put(84,0){\circle*{14}}
\put(0,15){\makebox(0.4,0.6){1}}
\put(42,15){\makebox(0.4,0.6){2}}
\put(84,15){\makebox(0.4,0.6){2}}
\put(7,0){\line(1,0){28}}
\put(48,-3){\line(1,0){30}}
\put(48,3){\line(1,0){30}}
\put(69,0){\line(-1,1){10}}\put(69,0){\line(-1,-1){10}}
\put(37,-5){\line(1,1){10}}\put(37,5){\line(1,-1){10}}
\end{picture}
\cr \cr
\end{array}
\]

\[
\begin{array}{ccccc}
sl(1|4) &\hspace{10mm}&
\begin{picture}(140,20)
\thicklines
\multiput(0,0)(42,0){4}{\circle{14}}
\put(0,15){\makebox(0.4,0.6){1}}
\put(42,15){\makebox(0.4,0.6){1}}
\put(84,15){\makebox(0.4,0.6){1}}
\put(126,15){\makebox(0.4,0.6){1}}
\put(7,0){\line(1,0){28}}
\put(49,0){\line(1,0){28}}
\put(91,0){\line(1,0){28}}
\put(-5,-5){\line(1,1){10}}\put(-5,5){\line(1,-1){10}}
\end{picture}
&&
\begin{picture}(140,20)
\thicklines
\multiput(0,0)(42,0){4}{\circle{14}}
\put(0,15){\makebox(0.4,0.6){1}}
\put(42,15){\makebox(0.4,0.6){1}}
\put(84,15){\makebox(0.4,0.6){1}}
\put(126,15){\makebox(0.4,0.6){1}}
\put(7,0){\line(1,0){28}}
\put(49,0){\line(1,0){28}}
\put(91,0){\line(1,0){28}}
\put(-5,-5){\line(1,1){10}}\put(-5,5){\line(1,-1){10}}
\put(37,-5){\line(1,1){10}}\put(37,5){\line(1,-1){10}}
\end{picture}
\cr \cr
&&
\begin{picture}(140,20)
\thicklines
\multiput(0,0)(42,0){4}{\circle{14}}
\put(0,15){\makebox(0.4,0.6){1}}
\put(42,15){\makebox(0.4,0.6){1}}
\put(84,15){\makebox(0.4,0.6){1}}
\put(126,15){\makebox(0.4,0.6){1}}
\put(7,0){\line(1,0){28}}
\put(49,0){\line(1,0){28}}
\put(91,0){\line(1,0){28}}
\put(37,-5){\line(1,1){10}}\put(37,5){\line(1,-1){10}}
\put(79,-5){\line(1,1){10}}\put(79,5){\line(1,-1){10}}
\end{picture}
&&
\cr \cr
sl(2|3) &\hspace{10mm}&
\begin{picture}(140,20)
\thicklines
\multiput(0,0)(42,0){4}{\circle{14}}
\put(0,15){\makebox(0.4,0.6){1}}
\put(42,15){\makebox(0.4,0.6){1}}
\put(84,15){\makebox(0.4,0.6){1}}
\put(126,15){\makebox(0.4,0.6){1}}
\put(7,0){\line(1,0){28}}
\put(49,0){\line(1,0){28}}
\put(91,0){\line(1,0){28}}
\put(79,-5){\line(1,1){10}}\put(79,5){\line(1,-1){10}}
\end{picture}
&&
\begin{picture}(140,20)
\thicklines
\multiput(0,0)(42,0){4}{\circle{14}}
\put(0,15){\makebox(0.4,0.6){1}}
\put(42,15){\makebox(0.4,0.6){1}}
\put(84,15){\makebox(0.4,0.6){1}}
\put(126,15){\makebox(0.4,0.6){1}}
\put(7,0){\line(1,0){28}}
\put(49,0){\line(1,0){28}}
\put(91,0){\line(1,0){28}}
\put(37,-5){\line(1,1){10}}\put(37,5){\line(1,-1){10}}
\put(79,-5){\line(1,1){10}}\put(79,5){\line(1,-1){10}}
\put(121,-5){\line(1,1){10}}\put(121,5){\line(1,-1){10}}
\end{picture}
\cr \cr
&&
\begin{picture}(140,20)
\thicklines
\multiput(0,0)(42,0){4}{\circle{14}}
\put(0,15){\makebox(0.4,0.6){1}}
\put(42,15){\makebox(0.4,0.6){1}}
\put(84,15){\makebox(0.4,0.6){1}}
\put(126,15){\makebox(0.4,0.6){1}}
\put(7,0){\line(1,0){28}}
\put(49,0){\line(1,0){28}}
\put(91,0){\line(1,0){28}}
\put(-5,-5){\line(1,1){10}}\put(-5,5){\line(1,-1){10}}
\put(37,-5){\line(1,1){10}}\put(37,5){\line(1,-1){10}}
\put(121,-5){\line(1,1){10}}\put(121,5){\line(1,-1){10}}
\end{picture}
&&
\begin{picture}(140,20)
\thicklines
\multiput(0,0)(42,0){4}{\circle{14}}
\put(0,15){\makebox(0.4,0.6){1}}
\put(42,15){\makebox(0.4,0.6){1}}
\put(84,15){\makebox(0.4,0.6){1}}
\put(126,15){\makebox(0.4,0.6){1}}
\put(7,0){\line(1,0){28}}
\put(49,0){\line(1,0){28}}
\put(91,0){\line(1,0){28}}
\put(-5,-5){\line(1,1){10}}\put(-5,5){\line(1,-1){10}}
\put(37,-5){\line(1,1){10}}\put(37,5){\line(1,-1){10}}
\put(79,-5){\line(1,1){10}}\put(79,5){\line(1,-1){10}}
\put(121,-5){\line(1,1){10}}\put(121,5){\line(1,-1){10}}
\end{picture}
\cr \cr
&&
\begin{picture}(140,20)
\thicklines
\multiput(0,0)(42,0){4}{\circle{14}}
\put(0,15){\makebox(0.4,0.6){1}}
\put(42,15){\makebox(0.4,0.6){1}}
\put(84,15){\makebox(0.4,0.6){1}}
\put(126,15){\makebox(0.4,0.6){1}}
\put(7,0){\line(1,0){28}}
\put(49,0){\line(1,0){28}}
\put(91,0){\line(1,0){28}}
\put(37,-5){\line(1,1){10}}\put(37,5){\line(1,-1){10}}
\put(121,-5){\line(1,1){10}}\put(121,5){\line(1,-1){10}}
\end{picture}
&&
\begin{picture}(140,20)
\thicklines
\multiput(0,0)(42,0){4}{\circle{14}}
\put(0,15){\makebox(0.4,0.6){1}}
\put(42,15){\makebox(0.4,0.6){1}}
\put(84,15){\makebox(0.4,0.6){1}}
\put(126,15){\makebox(0.4,0.6){1}}
\put(7,0){\line(1,0){28}}
\put(49,0){\line(1,0){28}}
\put(91,0){\line(1,0){28}}
\put(-5,-5){\line(1,1){10}}\put(-5,5){\line(1,-1){10}}
\put(121,-5){\line(1,1){10}}\put(121,5){\line(1,-1){10}}
\end{picture}
\cr \cr
osp(1|8) &\hspace{10mm}&
\begin{picture}(140,20)
\thicklines
\multiput(0,0)(42,0){3}{\circle{14}}
\put(126,0){\circle*{14}}
\put(0,15){\makebox(0.4,0.6){2}}
\put(42,15){\makebox(0.4,0.6){2}}
\put(84,15){\makebox(0.4,0.6){2}}
\put(126,15){\makebox(0.4,0.6){2}}
\put(7,0){\line(1,0){28}}
\put(49,0){\line(1,0){28}}
\put(90,-3){\line(1,0){30}}
\put(90,3){\line(1,0){30}}
\put(111,0){\line(-1,1){10}}\put(111,0){\line(-1,-1){10}}
\end{picture}
&&
\cr \cr
osp(2|6) &\hspace{10mm}&
\begin{picture}(140,20)
\thicklines
\multiput(0,0)(42,0){4}{\circle{14}}
\put(0,15){\makebox(0.4,0.6){1}}
\put(42,15){\makebox(0.4,0.6){2}}
\put(84,15){\makebox(0.4,0.6){2}}
\put(126,15){\makebox(0.4,0.6){2}}
\put(7,0){\line(1,0){28}}
\put(49,0){\line(1,0){28}}
\put(90,-3){\line(1,0){30}}
\put(90,3){\line(1,0){30}}
\put(101,0){\line(1,-1){10}}\put(101,0){\line(1,1){10}}
\put(-5,-5){\line(1,1){10}}\put(-5,5){\line(1,-1){10}}
\end{picture}
&&
\begin{picture}(140,20)
\thicklines
\multiput(0,0)(42,0){4}{\circle{14}}
\put(0,15){\makebox(0.4,0.6){2}}
\put(42,15){\makebox(0.4,0.6){2}}
\put(84,15){\makebox(0.4,0.6){2}}
\put(126,15){\makebox(0.4,0.6){2}}
\put(7,0){\line(1,0){28}}
\put(49,0){\line(1,0){28}}
\put(90,-3){\line(1,0){30}}
\put(90,3){\line(1,0){30}}
\put(101,0){\line(1,-1){10}}\put(101,0){\line(1,1){10}}
\put(-5,-5){\line(1,1){10}}\put(-5,5){\line(1,-1){10}}
\put(37,-5){\line(1,1){10}}\put(37,5){\line(1,-1){10}}
\end{picture}
\cr \cr \cr
&&
\begin{picture}(140,20)
\thicklines
\multiput(0,0)(42,0){4}{\circle{14}}
\put(0,15){\makebox(0.4,0.6){2}}
\put(42,15){\makebox(0.4,0.6){2}}
\put(84,15){\makebox(0.4,0.6){2}}
\put(126,15){\makebox(0.4,0.6){2}}
\put(7,0){\line(1,0){28}}
\put(49,0){\line(1,0){28}}
\put(90,-3){\line(1,0){30}}
\put(90,3){\line(1,0){30}}
\put(101,0){\line(1,-1){10}}\put(101,0){\line(1,1){10}}
\put(37,-5){\line(1,1){10}}\put(37,5){\line(1,-1){10}}
\put(79,-5){\line(1,1){10}}\put(79,5){\line(1,-1){10}}
\end{picture}
&&
\begin{picture}(90,20)
\thicklines
\put(0,0){\circle{14}}
\put(42,0){\circle{14}}
\put(0,15){\makebox(0.4,0.6){2}}
\put(42,15){\makebox(0.4,0.6){2}}
\put(7,0){\line(1,0){28}}
\put(47,5){\line(2,1){20}}\put(47,-5){\line(2,-1){20}}
\put(73,20){\circle{14}}
\put(89,20){\makebox(0.4,0.6){1}}
\put(73,-20){\circle{14}}
\put(89,-20){\makebox(0.4,0.6){1}}
\put(68,15){\line(1,1){10}}\put(68,25){\line(1,-1){10}}
\put(68,-25){\line(1,1){10}}\put(68,-15){\line(1,-1){10}}
\put(70,-14){\line(0,1){28}}\put(76,-14){\line(0,1){28}}
\end{picture}
\cr \cr \cr
osp(3|6) &\hspace{10mm}&
\begin{picture}(140,20)
\thicklines
\multiput(0,0)(42,0){3}{\circle{14}}
\put(126,0){\circle*{14}}
\put(0,15){\makebox(0.4,0.6){1}}
\put(42,15){\makebox(0.4,0.6){2}}
\put(84,15){\makebox(0.4,0.6){2}}
\put(126,15){\makebox(0.4,0.6){2}}
\put(7,0){\line(1,0){28}}
\put(49,0){\line(1,0){28}}
\put(90,-3){\line(1,0){30}}
\put(90,3){\line(1,0){30}}
\put(111,0){\line(-1,1){10}}\put(111,0){\line(-1,-1){10}}
\put(-5,-5){\line(1,1){10}}\put(-5,5){\line(1,-1){10}}
\end{picture}
&&
\begin{picture}(140,20)
\thicklines
\multiput(0,0)(42,0){3}{\circle{14}}
\put(126,0){\circle*{14}}
\put(0,15){\makebox(0.4,0.6){2}}
\put(42,15){\makebox(0.4,0.6){2}}
\put(84,15){\makebox(0.4,0.6){2}}
\put(126,15){\makebox(0.4,0.6){2}}
\put(7,0){\line(1,0){28}}
\put(49,0){\line(1,0){28}}
\put(90,-3){\line(1,0){30}}
\put(90,3){\line(1,0){30}}
\put(111,0){\line(-1,1){10}}\put(111,0){\line(-1,-1){10}}
\put(-5,-5){\line(1,1){10}}\put(-5,5){\line(1,-1){10}}
\put(37,-5){\line(1,1){10}}\put(37,5){\line(1,-1){10}}
\end{picture}
\cr \cr
&&
\begin{picture}(140,20)
\thicklines
\multiput(0,0)(42,0){3}{\circle{14}}
\put(126,0){\circle*{14}}
\put(0,15){\makebox(0.4,0.6){2}}
\put(42,15){\makebox(0.4,0.6){2}}
\put(84,15){\makebox(0.4,0.6){2}}
\put(126,15){\makebox(0.4,0.6){2}}
\put(7,0){\line(1,0){28}}
\put(49,0){\line(1,0){28}}
\put(90,-3){\line(1,0){30}}
\put(90,3){\line(1,0){30}}
\put(111,0){\line(-1,1){10}}\put(111,0){\line(-1,-1){10}}
\put(37,-5){\line(1,1){10}}\put(37,5){\line(1,-1){10}}
\put(79,-5){\line(1,1){10}}\put(79,5){\line(1,-1){10}}
\end{picture}
&&
\begin{picture}(140,20)
\thicklines
\multiput(0,0)(42,0){4}{\circle{14}}
\put(0,15){\makebox(0.4,0.6){2}}
\put(42,15){\makebox(0.4,0.6){2}}
\put(84,15){\makebox(0.4,0.6){2}}
\put(126,15){\makebox(0.4,0.6){2}}
\put(7,0){\line(1,0){28}}
\put(49,0){\line(1,0){28}}
\put(90,-3){\line(1,0){30}}
\put(90,3){\line(1,0){30}}
\put(111,0){\line(-1,1){10}}\put(111,0){\line(-1,-1){10}}
\put(79,-5){\line(1,1){10}}\put(79,5){\line(1,-1){10}}
\end{picture}
\cr \cr
osp(4|4) &\hspace{10mm}&
\begin{picture}(140,20)
\thicklines
\multiput(0,0)(42,0){4}{\circle{14}}
\put(0,15){\makebox(0.4,0.6){1}}
\put(42,15){\makebox(0.4,0.6){2}}
\put(84,15){\makebox(0.4,0.6){2}}
\put(126,15){\makebox(0.4,0.6){1}}
\put(7,0){\line(1,0){28}}
\put(49,0){\line(1,0){28}}
\put(90,-3){\line(1,0){30}}
\put(90,3){\line(1,0){30}}
\put(101,0){\line(1,-1){10}}\put(101,0){\line(1,1){10}}
\put(-5,-5){\line(1,1){10}}\put(-5,5){\line(1,-1){10}}
\put(37,-5){\line(1,1){10}}\put(37,5){\line(1,-1){10}}
\put(79,-5){\line(1,1){10}}\put(79,5){\line(1,-1){10}}
\end{picture}
&&
\begin{picture}(140,20)
\thicklines
\multiput(0,0)(42,0){4}{\circle{14}}
\put(0,15){\makebox(0.4,0.6){1}}
\put(42,15){\makebox(0.4,0.6){2}}
\put(84,15){\makebox(0.4,0.6){2}}
\put(126,15){\makebox(0.4,0.6){1}}
\put(7,0){\line(1,0){28}}
\put(49,0){\line(1,0){28}}
\put(90,-3){\line(1,0){30}}
\put(90,3){\line(1,0){30}}
\put(101,0){\line(1,-1){10}}\put(101,0){\line(1,1){10}}
\put(37,-5){\line(1,1){10}}\put(37,5){\line(1,-1){10}}
\end{picture}
\cr \cr \cr
&&
\begin{picture}(140,20)
\thicklines
\multiput(0,0)(42,0){4}{\circle{14}}
\put(0,15){\makebox(0.4,0.6){2}}
\put(42,15){\makebox(0.4,0.6){2}}
\put(84,15){\makebox(0.4,0.6){2}}
\put(126,15){\makebox(0.4,0.6){1}}
\put(7,0){\line(1,0){28}}
\put(49,0){\line(1,0){28}}
\put(90,-3){\line(1,0){30}}
\put(90,3){\line(1,0){30}}
\put(101,0){\line(1,-1){10}}\put(101,0){\line(1,1){10}}
\put(-5,-5){\line(1,1){10}}\put(-5,5){\line(1,-1){10}}
\put(79,-5){\line(1,1){10}}\put(79,5){\line(1,-1){10}}
\end{picture}
&&
\begin{picture}(90,20)
\thicklines
\put(0,0){\circle{14}}
\put(42,0){\circle{14}}
\put(0,15){\makebox(0.4,0.6){2}}
\put(42,15){\makebox(0.4,0.6){2}}
\put(37,-5){\line(1,1){10}}\put(37,5){\line(1,-1){10}}
\put(7,0){\line(1,0){28}}
\put(47,5){\line(2,1){20}}\put(47,-5){\line(2,-1){20}}
\put(73,20){\circle{14}}
\put(89,20){\makebox(0.4,0.6){1}}
\put(73,-20){\circle{14}}
\put(89,-20){\makebox(0.4,0.6){1}}
\end{picture}
\cr \cr \cr \cr
&&
\begin{picture}(90,20)
\thicklines
\put(0,0){\circle{14}}
\put(42,0){\circle{14}}
\put(0,15){\makebox(0.4,0.6){2}}
\put(42,15){\makebox(0.4,0.6){2}}
\put(-5,-5){\line(1,1){10}}\put(-5,5){\line(1,-1){10}}
\put(37,-5){\line(1,1){10}}\put(37,5){\line(1,-1){10}}
\put(7,0){\line(1,0){28}}
\put(47,5){\line(2,1){20}}\put(47,-5){\line(2,-1){20}}
\put(73,20){\circle{14}}
\put(89,20){\makebox(0.4,0.6){1}}
\put(73,-20){\circle{14}}
\put(89,-20){\makebox(0.4,0.6){1}}
\put(68,15){\line(1,1){10}}\put(68,25){\line(1,-1){10}}
\put(68,-25){\line(1,1){10}}\put(68,-15){\line(1,-1){10}}
\put(70,-14){\line(0,1){28}}\put(76,-14){\line(0,1){28}}
\end{picture}
&&
\begin{picture}(90,20)
\thicklines
\put(0,0){\circle{14}}
\put(42,0){\circle{14}}
\put(0,15){\makebox(0.4,0.6){1}}
\put(42,15){\makebox(0.4,0.6){2}}
\put(-5,-5){\line(1,1){10}}\put(-5,5){\line(1,-1){10}}
\put(7,0){\line(1,0){28}}
\put(47,5){\line(2,1){20}}\put(47,-5){\line(2,-1){20}}
\put(73,20){\circle{14}}
\put(89,20){\makebox(0.4,0.6){1}}
\put(73,-20){\circle{14}}
\put(89,-20){\makebox(0.4,0.6){1}}
\put(68,15){\line(1,1){10}}\put(68,25){\line(1,-1){10}}
\put(68,-25){\line(1,1){10}}\put(68,-15){\line(1,-1){10}}
\put(70,-14){\line(0,1){28}}\put(76,-14){\line(0,1){28}}
\end{picture}
\cr \cr \cr
\end{array}
\]

\[
\begin{array}{ccccc}
osp(5|4) &\hspace{10mm}&
\begin{picture}(140,20)
\thicklines
\multiput(0,0)(42,0){4}{\circle{14}}
\put(0,15){\makebox(0.4,0.6){2}}
\put(42,15){\makebox(0.4,0.6){2}}
\put(84,15){\makebox(0.4,0.6){2}}
\put(126,15){\makebox(0.4,0.6){2}}
\put(7,0){\line(1,0){28}}
\put(49,0){\line(1,0){28}}
\put(90,-3){\line(1,0){30}}
\put(90,3){\line(1,0){30}}
\put(111,0){\line(-1,1){10}}\put(111,0){\line(-1,-1){10}}
\put(37,-5){\line(1,1){10}}\put(37,5){\line(1,-1){10}}
\end{picture}
&&
\begin{picture}(140,20)
\thicklines
\multiput(0,0)(42,0){4}{\circle{14}}
\put(0,15){\makebox(0.4,0.6){2}}
\put(42,15){\makebox(0.4,0.6){2}}
\put(84,15){\makebox(0.4,0.6){2}}
\put(126,15){\makebox(0.4,0.6){2}}
\put(7,0){\line(1,0){28}}
\put(49,0){\line(1,0){28}}
\put(90,-3){\line(1,0){30}}
\put(90,3){\line(1,0){30}}
\put(111,0){\line(-1,1){10}}\put(111,0){\line(-1,-1){10}}
\put(-5,-5){\line(1,1){10}}\put(-5,5){\line(1,-1){10}}
\put(37,-5){\line(1,1){10}}\put(37,5){\line(1,-1){10}}
\put(79,-5){\line(1,1){10}}\put(79,5){\line(1,-1){10}}
\end{picture}
\cr \cr
&&
\begin{picture}(140,20)
\thicklines
\multiput(0,0)(42,0){4}{\circle{14}}
\put(0,15){\makebox(0.4,0.6){1}}
\put(42,15){\makebox(0.4,0.6){2}}
\put(84,15){\makebox(0.4,0.6){2}}
\put(126,15){\makebox(0.4,0.6){2}}
\put(7,0){\line(1,0){28}}
\put(49,0){\line(1,0){28}}
\put(90,-3){\line(1,0){30}}
\put(90,3){\line(1,0){30}}
\put(111,0){\line(-1,1){10}}\put(111,0){\line(-1,-1){10}}
\put(-5,-5){\line(1,1){10}}\put(-5,5){\line(1,-1){10}}
\put(79,-5){\line(1,1){10}}\put(79,5){\line(1,-1){10}}
\end{picture}
&&
\begin{picture}(140,20)
\thicklines
\multiput(0,0)(42,0){3}{\circle{14}}
\put(126,0){\circle*{14}}
\put(0,15){\makebox(0.4,0.6){1}}
\put(42,15){\makebox(0.4,0.6){2}}
\put(84,15){\makebox(0.4,0.6){2}}
\put(126,15){\makebox(0.4,0.6){2}}
\put(7,0){\line(1,0){28}}
\put(49,0){\line(1,0){28}}
\put(90,-3){\line(1,0){30}}
\put(90,3){\line(1,0){30}}
\put(111,0){\line(-1,1){10}}\put(111,0){\line(-1,-1){10}}
\put(-5,-5){\line(1,1){10}}\put(-5,5){\line(1,-1){10}}
\put(37,-5){\line(1,1){10}}\put(37,5){\line(1,-1){10}}
\put(79,-5){\line(1,1){10}}\put(79,5){\line(1,-1){10}}
\end{picture}
\cr \cr
&&
\begin{picture}(140,20)
\thicklines
\multiput(0,0)(42,0){3}{\circle{14}}
\put(126,0){\circle*{14}}
\put(0,15){\makebox(0.4,0.6){1}}
\put(42,15){\makebox(0.4,0.6){2}}
\put(84,15){\makebox(0.4,0.6){2}}
\put(126,15){\makebox(0.4,0.6){2}}
\put(7,0){\line(1,0){28}}
\put(49,0){\line(1,0){28}}
\put(90,-3){\line(1,0){30}}
\put(90,3){\line(1,0){30}}
\put(111,0){\line(-1,1){10}}\put(111,0){\line(-1,-1){10}}
\put(37,-5){\line(1,1){10}}\put(37,5){\line(1,-1){10}}
\end{picture}
&&
\begin{picture}(140,20)
\thicklines
\multiput(0,0)(42,0){3}{\circle{14}}
\put(126,0){\circle*{14}}
\put(0,15){\makebox(0.4,0.6){2}}
\put(42,15){\makebox(0.4,0.6){2}}
\put(84,15){\makebox(0.4,0.6){2}}
\put(126,15){\makebox(0.4,0.6){2}}
\put(7,0){\line(1,0){28}}
\put(49,0){\line(1,0){28}}
\put(90,-3){\line(1,0){30}}
\put(90,3){\line(1,0){30}}
\put(111,0){\line(-1,1){10}}\put(111,0){\line(-1,-1){10}}
\put(-5,-5){\line(1,1){10}}\put(-5,5){\line(1,-1){10}}
\put(79,-5){\line(1,1){10}}\put(79,5){\line(1,-1){10}}
\end{picture}
\cr \cr \cr
osp(6|2) &\hspace{10mm}&
\begin{picture}(140,20)
\thicklines
\multiput(0,0)(42,0){4}{\circle{14}}
\put(0,15){\makebox(0.4,0.6){1}}
\put(42,15){\makebox(0.4,0.6){2}}
\put(84,15){\makebox(0.4,0.6){2}}
\put(126,15){\makebox(0.4,0.6){1}}
\put(7,0){\line(1,0){28}}
\put(49,0){\line(1,0){28}}
\put(90,-3){\line(1,0){30}}
\put(90,3){\line(1,0){30}}
\put(101,0){\line(1,-1){10}}\put(101,0){\line(1,1){10}}
\put(79,-5){\line(1,1){10}}\put(79,5){\line(1,-1){10}}
\end{picture}
&&
\begin{picture}(90,20)
\thicklines
\put(0,0){\circle{14}}
\put(42,0){\circle{14}}
\put(0,15){\makebox(0.4,0.6){1}}
\put(42,15){\makebox(0.4,0.6){2}}
\put(37,-5){\line(1,1){10}}\put(37,5){\line(1,-1){10}}
\put(7,0){\line(1,0){28}}
\put(47,5){\line(2,1){20}}\put(47,-5){\line(2,-1){20}}
\put(73,20){\circle{14}}
\put(89,20){\makebox(0.4,0.6){1}}
\put(73,-20){\circle{14}}
\put(89,-20){\makebox(0.4,0.6){1}}
\put(68,15){\line(1,1){10}}\put(68,25){\line(1,-1){10}}
\put(68,-25){\line(1,1){10}}\put(68,-15){\line(1,-1){10}}
\put(70,-14){\line(0,1){28}}\put(76,-14){\line(0,1){28}}
\end{picture}
\cr \cr \cr \cr
&&
\begin{picture}(90,20)
\thicklines
\put(0,0){\circle{14}}
\put(42,0){\circle{14}}
\put(0,15){\makebox(0.4,0.6){1}}
\put(42,15){\makebox(0.4,0.6){2}}
\put(-5,-5){\line(1,1){10}}\put(-5,5){\line(1,-1){10}}
\put(37,-5){\line(1,1){10}}\put(37,5){\line(1,-1){10}}
\put(7,0){\line(1,0){28}}
\put(47,5){\line(2,1){20}}\put(47,-5){\line(2,-1){20}}
\put(73,20){\circle{14}}
\put(89,20){\makebox(0.4,0.6){1}}
\put(73,-20){\circle{14}}
\put(89,-20){\makebox(0.4,0.6){1}}
\end{picture}
&&
\begin{picture}(90,20)
\thicklines
\put(0,0){\circle{14}}
\put(42,0){\circle{14}}
\put(0,15){\makebox(0.4,0.6){2}}
\put(42,15){\makebox(0.4,0.6){2}}
\put(-5,-5){\line(1,1){10}}\put(-5,5){\line(1,-1){10}}
\put(7,0){\line(1,0){28}}
\put(47,5){\line(2,1){20}}\put(47,-5){\line(2,-1){20}}
\put(73,20){\circle{14}}
\put(89,20){\makebox(0.4,0.6){1}}
\put(73,-20){\circle{14}}
\put(89,-20){\makebox(0.4,0.6){1}}
\end{picture}
\cr \cr \cr
osp(7|2) &\hspace{10mm}&
\begin{picture}(140,20)
\thicklines
\multiput(0,0)(42,0){3}{\circle{14}}
\put(126,0){\circle*{14}}
\put(0,15){\makebox(0.4,0.6){1}}
\put(42,15){\makebox(0.4,0.6){2}}
\put(84,15){\makebox(0.4,0.6){2}}
\put(126,15){\makebox(0.4,0.6){2}}
\put(7,0){\line(1,0){28}}
\put(49,0){\line(1,0){28}}
\put(90,-3){\line(1,0){30}}
\put(90,3){\line(1,0){30}}
\put(111,0){\line(-1,1){10}}\put(111,0){\line(-1,-1){10}}
\put(79,-5){\line(1,1){10}}\put(79,5){\line(1,-1){10}}
\end{picture}
&&
\begin{picture}(140,20)
\thicklines
\multiput(0,0)(42,0){4}{\circle{14}}
\put(0,15){\makebox(0.4,0.6){1}}
\put(42,15){\makebox(0.4,0.6){2}}
\put(84,15){\makebox(0.4,0.6){2}}
\put(126,15){\makebox(0.4,0.6){2}}
\put(7,0){\line(1,0){28}}
\put(49,0){\line(1,0){28}}
\put(90,-3){\line(1,0){30}}
\put(90,3){\line(1,0){30}}
\put(111,0){\line(-1,1){10}}\put(111,0){\line(-1,-1){10}}
\put(37,-5){\line(1,1){10}}\put(37,5){\line(1,-1){10}}
\put(79,-5){\line(1,1){10}}\put(79,5){\line(1,-1){10}}
\end{picture}
\cr \cr
&&
\begin{picture}(140,20)
\thicklines
\multiput(0,0)(42,0){4}{\circle{14}}
\put(0,15){\makebox(0.4,0.6){1}}
\put(42,15){\makebox(0.4,0.6){2}}
\put(84,15){\makebox(0.4,0.6){2}}
\put(126,15){\makebox(0.4,0.6){2}}
\put(7,0){\line(1,0){28}}
\put(49,0){\line(1,0){28}}
\put(90,-3){\line(1,0){30}}
\put(90,3){\line(1,0){30}}
\put(111,0){\line(-1,1){10}}\put(111,0){\line(-1,-1){10}}
\put(-5,-5){\line(1,1){10}}\put(-5,5){\line(1,-1){10}}
\put(37,-5){\line(1,1){10}}\put(37,5){\line(1,-1){10}}
\end{picture}
&&
\begin{picture}(140,20)
\thicklines
\multiput(0,0)(42,0){4}{\circle{14}}
\put(0,15){\makebox(0.4,0.6){2}}
\put(42,15){\makebox(0.4,0.6){2}}
\put(84,15){\makebox(0.4,0.6){2}}
\put(126,15){\makebox(0.4,0.6){2}}
\put(7,0){\line(1,0){28}}
\put(49,0){\line(1,0){28}}
\put(90,-3){\line(1,0){30}}
\put(90,3){\line(1,0){30}}
\put(111,0){\line(-1,1){10}}\put(111,0){\line(-1,-1){10}}
\put(-5,-5){\line(1,1){10}}\put(-5,5){\line(1,-1){10}}
\end{picture}
\cr \cr
\end{array}
\]

\clearpage

\begin{table}[htbp]
\caption{Distinguished extended Dynkin diagrams of the basic Lie
superalgebras.\label{table16}}
\end{table}
\[
\begin{array}{llc}
A(m,n)^{(1)} &&
\begin{picture}(180,40)
\thicklines
\multiput(0,0)(42,0){5}{\circle{14}}
\put(84,44){\circle{14}}
\put(0,15){\makebox(0.4,0.6){1}}
\put(42,15){\makebox(0.4,0.6){1}}
\put(84,15){\makebox(0.4,0.6){1}}
\put(126,15){\makebox(0.4,0.6){1}}
\put(168,15){\makebox(0.4,0.6){1}}
\put(84,59){\makebox(0.4,0.6){1}}
\put(79,-5){\line(1,1){10}}\put(79,5){\line(1,-1){10}}
\put(79,39){\line(1,1){10}}\put(79,49){\line(1,-1){10}}
\put(5,5){\line(2,1){72}}
\put(163,5){\line(-2,1){72}}
\put(7,0){\dashbox{3}(28,0)}
\put(49,0){\line(1,0){28}}
\put(91,0){\line(1,0){28}}
\put(133,0){\dashbox{3}(28,0)}
\put(0,-10){$\underbrace{~~~~~~~~~~}_{m}$}
\put(126,-10){$\underbrace{~~~~~~~~~~}_{n}$}
\end{picture}
\cr &&\cr &&\cr
B(m,n)^{(1)} &&
\begin{picture}(260,20)
\thicklines
\multiput(0,0)(42,0){7}{\circle{14}}
\put(0,15){\makebox(0.4,0.6){1}}
\put(42,15){\makebox(0.4,0.6){2}}
\put(84,15){\makebox(0.4,0.6){2}}
\put(126,15){\makebox(0.4,0.6){2}}
\put(168,15){\makebox(0.4,0.6){2}}
\put(210,15){\makebox(0.4,0.6){2}}
\put(252,15){\makebox(0.4,0.6){2}}
\put(121,-5){\line(1,1){10}}\put(121,5){\line(1,-1){10}}
\put(6,-3){\line(1,0){30}}
\put(6,3){\line(1,0){30}}
\put(27,0){\line(-1,1){10}}\put(27,0){\line(-1,-1){10}}
\put(49,0){\dashbox{3}(28,0)}
\put(91,0){\line(1,0){28}}
\put(133,0){\line(1,0){28}}
\put(175,0){\dashbox{3}(28,0)}
\put(216,-3){\line(1,0){30}}
\put(216,3){\line(1,0){30}}
\put(237,0){\line(-1,1){10}}\put(237,0){\line(-1,-1){10}}
\put(42,-10){$\underbrace{~~~~~~~~~~}_{n-1}$}
\put(168,-10){$\underbrace{~~~~~~~~~~}_{m-1}$}
\end{picture}
\cr &&\cr &&\cr
B(0,n)^{(1)} &&
\begin{picture}(140,20)
\thicklines
\put(0,0){\circle{14}}
\put(42,0){\circle{14}}
\put(84,0){\circle{14}}
\put(126,0){\circle*{14}}
\put(0,15){\makebox(0.4,0.6){1}}
\put(42,15){\makebox(0.4,0.6){2}}
\put(84,15){\makebox(0.4,0.6){2}}
\put(126,15){\makebox(0.4,0.6){2}}
\put(6,-3){\line(1,0){30}}
\put(6,3){\line(1,0){30}}
\put(27,0){\line(-1,1){10}}\put(27,0){\line(-1,-1){10}}
\put(49,0){\dashbox{3}(28,0)}
\put(90,-3){\line(1,0){30}}
\put(90,3){\line(1,0){30}}
\put(111,0){\line(-1,1){10}}\put(111,0){\line(-1,-1){10}}
\put(42,-10){$\underbrace{~~~~~~~~~~}_{n-1}$}
\end{picture}
\cr &&\cr &&\cr &&\cr
C(n+1)^{(1)} &&
\begin{picture}(160,20)
\thicklines
\put(11,20){\circle{14}}
\put(11,-20){\circle{14}}
\put(-5,20){\makebox(0.4,0.6){1}}
\put(-5,-20){\makebox(0.4,0.6){1}}
\put(42,0){\circle{14}}
\put(84,0){\circle{14}}
\put(126,0){\circle{14}}
\put(42,15){\makebox(0.4,0.6){2}}
\put(84,15){\makebox(0.4,0.6){2}}
\put(126,15){\makebox(0.4,0.6){1}}
\put(6,15){\line(1,1){10}}\put(6,25){\line(1,-1){10}}
\put(6,-25){\line(1,1){10}}\put(6,-15){\line(1,-1){10}}
\put(17,-15){\line(2,1){20}}
\put(17,15){\line(2,-1){20}}
\put(8,-14){\line(0,1){28}}
\put(14,-14){\line(0,1){28}}
\put(49,0){\dashbox{3}(28,0)}
\put(101,0){\line(1,1){10}}\put(101,0){\line(1,-1){10}}
\put(90,-3){\line(1,0){30}}
\put(90,3){\line(1,0){30}}
\put(42,-10){$\underbrace{~~~~~~~~~~}_{n-1}$}
\end{picture}
\cr &&\cr &&\cr &&\cr
D(m,n)^{(1)} &&
\begin{picture}(260,20)
\thicklines
\multiput(0,0)(42,0){6}{\circle{14}}
\put(0,15){\makebox(0.4,0.6){1}}
\put(42,15){\makebox(0.4,0.6){2}}
\put(84,15){\makebox(0.4,0.6){2}}
\put(126,15){\makebox(0.4,0.6){2}}
\put(168,15){\makebox(0.4,0.6){2}}
\put(210,15){\makebox(0.4,0.6){2}}
\put(121,-5){\line(1,1){10}}\put(121,5){\line(1,-1){10}}
\put(6,-3){\line(1,0){30}}
\put(6,3){\line(1,0){30}}
\put(27,0){\line(-1,1){10}}\put(27,0){\line(-1,-1){10}}
\put(49,0){\dashbox{3}(28,0)}
\put(91,0){\line(1,0){28}}
\put(133,0){\line(1,0){28}}
\put(175,0){\dashbox{3}(28,0)}
\put(215,5){\line(2,1){20}}\put(215,-5){\line(2,-1){20}}
\put(241,20){\circle{14}}
\put(257,20){\makebox(0.4,0.6){1}}
\put(241,-20){\circle{14}}
\put(257,-20){\makebox(0.4,0.6){1}}
\put(42,-10){$\underbrace{~~~~~~~~~~}_{n-1}$}
\put(168,-10){$\underbrace{~~~~~~~~~~}_{m-2}$}
\end{picture}
\cr &&\cr &&\cr
F(4)^{(1)} &&
\begin{picture}(180,20)
\thicklines
\multiput(0,0)(42,0){5}{\circle{14}}
\put(0,15){\makebox(0.4,0.6){1}}
\put(42,15){\makebox(0.4,0.6){2}}
\put(84,15){\makebox(0.4,0.6){3}}
\put(126,15){\makebox(0.4,0.6){2}}
\put(168,15){\makebox(0.4,0.6){1}}
\put(37,-5){\line(1,1){10}}\put(37,5){\line(1,-1){10}}
\put(6,-4){\line(1,0){30}}
\put(7,0){\line(1,0){28}}
\put(6,4){\line(1,0){30}}
\put(27,0){\line(-1,1){10}}\put(27,0){\line(-1,-1){10}}
\put(49,0){\line(1,0){28}}
\put(90,-3){\line(1,0){30}}
\put(90,3){\line(1,0){30}}
\put(101,0){\line(1,1){10}}\put(101,0){\line(1,-1){10}}
\put(133,0){\line(1,0){28}}
\end{picture}
\cr &&\cr &&\cr
G(3)^{(1)} &&
\begin{picture}(140,20)
\thicklines
\multiput(0,0)(42,0){4}{\circle{14}}
\put(0,15){\makebox(0.4,0.6){1}}
\put(42,15){\makebox(0.4,0.6){2}}
\put(84,15){\makebox(0.4,0.6){4}}
\put(126,15){\makebox(0.4,0.6){2}}
\put(37,-5){\line(1,1){10}}\put(37,5){\line(1,-1){10}}
\put(7,-2){\line(1,0){29}}
\put(7,2){\line(1,0){29}}
\put(5,-5){\line(1,0){32}}
\put(5,5){\line(1,0){32}}
\put(27,0){\line(-1,1){10}}\put(27,0){\line(-1,-1){10}}
\put(49,0){\line(1,0){28}}
\put(90,-4){\line(1,0){30}}
\put(91,0){\line(1,0){28}}
\put(90,4){\line(1,0){30}}
\put(101,0){\line(1,1){10}}\put(101,0){\line(1,-1){10}}
\end{picture}
\cr &&\cr &&\cr
D(2,1;\alpha)^{(1)} &&
\begin{picture}(90,20)
\thicklines
\put(0,0){\circle{14}}
\put(42,0){\circle{14}}
\put(0,15){\makebox(0.4,0.6){1}}
\put(42,15){\makebox(0.4,0.6){2}}
\put(37,-5){\line(1,1){10}}\put(37,5){\line(1,-1){10}}
\put(7,0){\line(1,0){28}}
\put(47,5){\line(2,1){20}}\put(47,-5){\line(2,-1){20}}
\put(73,20){\circle{14}}
\put(89,20){\makebox(0.4,0.6){1}}
\put(73,-20){\circle{14}}
\put(89,-20){\makebox(0.4,0.6){1}}
\end{picture}
\end{array}
\]

\clearpage

\begin{table}[htbp]
\centering
\begin{tabular}{|l|l|l|l|} \hline
~~~~~ $\cG$ & ~~~~~~~~~~ $\cG_\evn$ & ~~~~~~~~~~~ $\cG^\phi$ 
& ~~~~~~~~~~~~~~~ $\cG_\evn^\phi$ \\
\hline
$A(m,n)$ & $sl(m) \oplus sl(n) \oplus U(1)$ 
&$sl(m|n;\RR)$ &$sl(m,\RR) \oplus sl(n,\RR) \oplus \RR$ \\ 
&&$sl(m|n;\HH)$ &$su^*(m) \oplus su^*(n) \oplus \RR$ \\
&&$su(p,m-p|q,n-q)$ &$su(p,m-p) \oplus su(q,n-q) \oplus i\RR$ \\
\hline
$A(n,n)$ & $sl(n) \oplus sl(n)$ 
&$sl(n|n;\RR)$ &$sl(n,\RR) \oplus sl(n,\RR)$ \\ 
&&$sl(n|n;\HH)$ &$su^*(n) \oplus su^*(n)$ \\
&&$su(p,n-p|q,n-q)$ &$su(p,n-p) \oplus su(q,n-q)$ \\
\hline
$B(m,n)$ & $so(2m+1) \oplus sp(2n)$ 
&$osp(p,2m+1-p|2n;\RR)$ &$so(p,2m+1-p) \oplus sp(2n,\RR)$ \\
$B(0,n)$ & $sp(2n)$ 
&$osp(1|2n;\RR)$ &$sp(2n,\RR)$ \\
\hline
$C(n+1)$ & $so(2) \oplus sp(2n)$ 
&$osp(2|2n;\RR)$ &$so(2) \oplus sp(2n,\RR)$ \\
&&$osp(2|2q,2n-2q;\HH)$ &$so^*(2) \oplus sp(2q,2n-2q)$ \\
\hline
$D(m,n)$ & $so(2m) \oplus sp(2n)$ 
&$osp(p,2m-p|2n;\RR)$ &$so(p,2m-p) \oplus sp(2n,\RR)$ \\
&&$osp(2m|2q,2n-2q;\HH)$ &$so^*(2m) \oplus sp(2q,2n-2q)$ \\
\hline
$F(4)$ & $sl(2) \oplus so(7)$ 
&$F(4;0)$ &$sl(2,\RR) \oplus so(7)$ \\ 
&&$F(4;3)$ &$sl(2,\RR) \oplus so(1,6)$ \\
&&$F(4;2)$ &$sl(2,\RR) \oplus so(2,5)$ \\
&&$F(4;1)$ &$sl(2,\RR) \oplus so(3,4)$ \\
\hline
$G(3)$ & $sl(2) \oplus G_2$ 
&$G(3;0)$ &$sl(2,\RR) \oplus G_{2,0}$ \\
&&$G(3;1)$ &$sl(2,\RR) \oplus G_{2,2}$ \\
\hline
$D(2,1;\alpha)$ & $sl(2) \oplus sl(2) \oplus sl(2)$ 
&$D(2,1;\alpha;0)$ &$sl(2,\RR) \oplus sl(2,\RR) \oplus sl(2,\RR)$ \\
&&$D(2,1;\alpha;1)$ &$su(2) \oplus su(2) \oplus sl(2,\RR)$ \\
&&$D(2,1;\alpha;2)$ &$sl(2,\CC) \oplus sl(2,\RR)$ \\
\hline
$P(n)$ & $sl(n)$ 
&$P(n;\RR)$ &$sl(n,\RR)$ \\
\hline
$Q(n)$ & $sl(n)$ 
&$Q(n;\RR)$ &$sl(n,\RR)$ \\
&&$HQ(n)$ &$su^*(n)$ \\
&&$UQ(p,n-p)$ &$su(p,n-p)$ \\
\hline
\end{tabular}
\caption{Real forms of the classical Lie superalgebras.
\label{table12}}
\end{table}

\clearpage

\begin{table}[htbp]
\centering
\begin{tabular}{|c|c|c|}
\hline
$\cG$ & $\cK$ & $\fund\cG ~ / ~ \cK$ \\ 
\hline && \\
$sl(m|n)$
 &$sl(p+1|p)$ &$\cR_{p/2} \oplus (m-p-1)\cR_0 \oplus (n-p) 
 \cR''_0$ \\
 && \\
 &$sl(p|p+1)$ &$\cR''_{p/2} \oplus (m-p)\cR_0 \oplus (n-p-1) 
 \cR''_0$ \\
 && \\
 \hline && \\
$osp(2m|2n)$
 &$osp(2k|2k)$ 
 &$\cR''_{k-1/2} \oplus (2n-2k)\cR''_0 \oplus (2m-2k+1)\cR_0$ \\
 && \\
 &$osp(2k+2|2k)$ 
 &$\cR_{k} \oplus(2m-2k-1)\cR_0 \oplus (2n-2k)\cR''_0$ \\
 && \\
 &$sl(p+1|p)$
 &$2\cR_{p/2} \oplus 2(m-p-1)\cR_0 \oplus 2(n-p)\cR''_0$ \\
 && \\
 &$sl(p|p+1)$
 &$2\cR''_{p/2} \oplus 2(n-p-1)\cR''_0 \oplus 2(m-p)\cR_0$ \\
 && \\
 \hline && \\
$osp(2m+1|2n)$
 &$\left.\begin{array}{c} osp(2k|2k) \\ osp(2k-1|2k) \end{array} 
 \right\}$
 &$\cR''_{k-1/2} \oplus (2n-2k)\cR''_0 \oplus (2m-2k+2)\cR_0$ \\ 
 && \\
 &$\left.\begin{array}{c} osp(2k+2|2k) \\ osp(2k+1|2k) \end{array} 
 \right\}$
 &$\cR_{k} \oplus (2m-2k)\cR_0 \oplus (2n-2k)\cR''_0$ \\
 && \\
 &$sl(p+1|p)$
 &$2\cR_{p/2} \oplus 2(m-p-1)\cR_0 \oplus \cR_0 \oplus 
 2(n-p)\cR''_0$ \\
 && \\
 &$sl(p|p+1)$
 &$2\cR''_{p/2} \oplus 2(n-p-1)\cR''_0 \oplus \cR_0 \oplus 
 2(m-p)\cR_0$ \\
 && \\
\hline && \\
$osp(2|2n)$
 &$osp(2|2)$
 &$\cR''_{1/2} \oplus \cR_0 \oplus (2n-2)\cR''_0$ \\ && \\ 
 &$sl(1|2)$
 &$2\cR''_{1/2} \oplus (2n-4)\cR''_0$ \\ && \\
\hline
\end{tabular}
\caption{$osp(1|2)$ decompositions of the fundamental representations 
of the basic Lie superalgebras (regular cases).\label{table7}} 
\end{table}

$\see$ $osp(1|2)$ decompositions

\clearpage

\begin{table}[htbp]
\centering
\begin{tabular}{|c|c|c|}
\hline
$\cG$ & $\cK$ & $\fund\cG ~ / ~ \cK$ \\ 
\hline && \\
$osp(2n+2|2n)$
 &$\begin{array}{c}osp(2k+1|2k) \oplus \\ osp(2n-2k+1|2n-2k) 
 \end{array}$ 
 &$\cR_{k} \oplus \cR_{n-k}$ \\ && \\
 \hline && \\
$osp(2n-2|2n)$
 &$\begin{array}{c}osp(2k-1|2k) \oplus \\ osp(2n-2k-1|2n-2k) 
 \end{array}$
 &$\cR''_{k-1/2} \oplus \cR''_{n-k-1/2}$ \\ && \\
 \hline && \\
$osp(2n|2n)$
 &$\begin{array}{c}osp(2k+1|2k) \oplus \\ osp(2n-2k-1|2n-2k) 
 \end{array}$
 &$\cR_{k} \oplus \cR''_{n-k-1/2}$ \\ && \\
 &$\begin{array}{c}osp(2k-1|2k) \oplus \\ osp(2n-2k+1|2n-2k) 
 \end{array}$
 &$\cR_{n-k} \oplus \cR''_{k-1/2}$ \\
\hline
\end{tabular}
\caption{$osp(1|2)$ decompositions of the fundamental representations 
of the basic Lie superalgebras (singular cases).\label{table8}} 
\end{table}

$\see$ $osp(1|2)$ decompositions

\clearpage

\begin{table}[htbp]
\caption{$osp(1|2)$ decompositions of the adjoint representations of
the basic Lie superalgebras (regular cases).\label{table9}} 
\end{table}
\beo
&&\frac{\ad osp(2m|2n)}{osp(2k|2k)} = \\
&& ~~~~~ \cR_{2k-1} ~ \oplus ~ \cR_{2k-5/2} ~ \oplus ~ \cR_{2k-3} ~ 
\oplus ~ \cR_{2k-9/2} ~ \oplus ~ \dots ~ \oplus ~ \cR_{3/2} ~ 
\oplus ~ \cR_1 \\
&& ~~~~~ \oplus ~ (2m-2k+1) \cR_{k-1/2} ~ \oplus ~ 2(n-k) 
\cR'_{k-1/2} ~ \oplus ~ 2(2m-2k+1)(n-k) \cR'_0 \\
&& ~~~~~ \oplus ~ [(2m-2k+1)(m-k) + (2n-2k+1)(n-k)] \cR_0 \\ 
&&\\
&&\frac{\ad osp(2m|2n)}{osp(2k+2|2k)} = \\
&& ~~~~~ \cR_{2k-1/2} ~ \oplus ~ \cR_{2k-1} ~ \oplus ~ \cR_{2k-5/2} ~ 
\oplus ~ \cR_{2k-3} ~ \oplus ~ \dots ~ \oplus ~ \cR_{3/2} ~ \oplus ~ 
\cR_1 \\
&& ~~~~~ \oplus ~ (2m-2k-1) \cR_k ~ \oplus ~ 2(n-k) \cR'_k ~ 
\oplus ~ 2(2m-2k-1)(n-k) \cR'_0 \\
&& ~~~~~ \oplus ~ [(2m-2k-1)(m-k-1) + (2n-2k+1)(n-k)] \cR_0 \\ 
&&\\
&&\frac{\ad osp(2m|2n)}{sl(2k+1|2k)} = \\
&& ~~~~~ \cR_{2k} ~ \oplus ~ 3\cR_{2k-1} ~ \oplus ~ \cR_{2k-2} ~ 
\oplus ~ \dots ~ \oplus ~ \cR_2 ~ \oplus ~ 3\cR_1 ~ \oplus ~ \cR_0 \\
&& ~~~~~ \oplus ~ 3\cR_{2k-1/2} ~ \oplus ~ \cR_{2k-3/2} ~ \oplus ~ 
3\cR_{2k-5/2} ~ \oplus ~ \dots ~ \oplus ~ 3\cR_{3/2} ~ \oplus ~ 
\cR_{1/2} \\
&& ~~~~~ \oplus ~ 4(m-2k-1) \cR_{k} ~ \oplus ~ 4(n-2k)\cR'_{k} ~ 
\oplus ~ 4(m-2k-1)(n-2k) \cR'_0 \\
&& ~~~~~ \oplus ~ [(2m-4k-3)(m-2k-1) + (2n-4k+1)(n-2k)] \cR_0 \\
&&\\
&&\frac{\ad osp(2m|2n)}{sl(2k-1|2k)} = \\
&& ~~~~~ 3\cR_{2k-1} ~ \oplus ~ \cR_{2k-2} ~ \oplus ~ 3\cR_{2k-3} ~ 
\oplus ~ \dots ~ \oplus ~ \cR_2 ~ \oplus ~ 3\cR_1 ~ \oplus ~ \cR_0 \\
&& ~~~~~ \oplus ~ \cR_{2k-3/2} ~ \oplus ~ 3\cR_{2k-5/2} ~ \oplus ~ 
\cR_{2k-7/2} ~ \oplus ~ \dots ~ \oplus ~ 3\cR_{3/2} ~ \oplus ~ 
\cR_{1/2} \\
&& ~~~~~ \oplus ~ 4(m-2k+1) \cR_{k-1/2} ~ \oplus ~ 4(n-2k)\cR'_{k-1/2} 
 ~ \oplus ~ 4(m-2k+1)(n-2k) \cR'_0 \\
&& ~~~~~ \oplus ~ [(2m-4k+1)(m-2k+1) + (2n-4k+1)(n-2k)] \cR_0 \\
&&\\
&&\frac{\ad osp(2m|2n)}{sl(2k|2k+1)} = \\
&& ~~~~~ 3\cR_{2k} ~ \oplus ~ \cR_{2k-1} ~ \oplus ~ 3\cR_{2k-2} ~ 
\oplus ~ \dots ~ \oplus ~ 3\cR_2 ~ \oplus ~ \cR_1 ~ \oplus ~ 3\cR_0 \\
&& ~~~~~ \oplus ~ \cR_{2k-1/2} ~ \oplus ~ 3\cR_{2k-3/2} ~ \oplus ~ 
\cR_{2k-5/2} ~ \oplus ~ \dots ~ \oplus ~ \cR_{3/2} ~ \oplus ~ 
3\cR_{1/2} \\
&& ~~~~~ \oplus ~ 4(m-2k) \cR'_{k} ~ \oplus ~ 4(n-2k-1)\cR_{k} ~ 
\oplus ~ 4(m-2k)(n-2k-1) \cR'_0 \\
&& ~~~~~ \oplus ~ [(2m-4k-1)(m-2k) + (2n-4k-1)(n-2k-1)] \cR_0 \\
&&\\
&&\frac{\ad osp(2m|2n)}{sl(2k|2k-1)} = \\
&& ~~~~~ \cR_{2k-1} ~ \oplus ~ 3\cR_{2k-2} ~ \oplus ~ \cR_{2k-3} ~ 
\oplus ~ \dots ~ \oplus ~ 3\cR_2 ~ \oplus ~ \cR_1 ~ \oplus ~ 3\cR_0 \\
&& ~~~~~ \oplus ~ 3\cR_{2k-3/2} ~ \oplus ~ \cR_{2k-5/2} ~ \oplus ~ 
3\cR_{2k-7/2} ~ \oplus ~ \dots ~ \oplus ~ \cR_{3/2} ~ \oplus ~ 
3\cR_{1/2} \\
&& ~~~~~ \oplus ~ 4(m-2k) \cR'_{k-1/2} ~ \oplus ~ 4(n-2k+1)\cR_{k-1/2} 
 ~ \oplus ~ 4(m-2k)(n-2k+1) \cR'_0 \\
&& ~~~~~ \oplus ~ [(2m-4k-1)(m-2k) + (2n-4k+3)(n-2k+1)] \cR_0 \\
&&\\
&&\frac{\ad osp(2m+1|2n)}{osp(2k|2k)} = 
\frac{\ad osp(2m+1|2n)}{osp(2k-1|2k)} = \\ 
&& ~~~~~ \cR_{2k-1} ~ \oplus ~ \cR_{2k-5/2} ~ \oplus ~ \cR_{2k-3} ~ 
\oplus ~ \cR_{2k-9/2} ~ \oplus ~ \dots ~ \oplus ~ \cR_{3/2} ~ 
\oplus ~ \cR_1 \\
&& ~~~~~ \oplus ~ 2(m-k+1) \cR_{k-1/2} ~ \oplus ~ 2(n-k) 
\cR'_{k-1/2} ~ \oplus ~ 4(m-k+1)(n-k) \cR'_0 \\
&& ~~~~~ \oplus ~ [(2m-2k+1)(m-k+1) + (2n-2k+1)(n-k)] \cR_0 \\
&&\\
&&\frac{\ad osp(2m+1|2n)}{osp(2k+2|2k)} = 
\frac{\ad osp(2m+1|2n)}{osp(2k+1|2k)} = \\ 
&& ~~~~~ \cR_{2k-1/2} ~ \oplus ~ \cR_{2k-1} ~ \oplus ~ \cR_{2k-5/2} ~ 
\oplus ~ \cR_{2k-3} ~ \oplus ~ \dots ~ \oplus ~ \cR_{3/2} ~ \oplus ~ 
\cR_1 \\
&& ~~~~~ \oplus ~ 2(n-k) \cR'_k ~ \oplus ~ 2(m-k) \cR_k ~ \oplus ~ 
4(m-k)(n-k) \cR'_0 \\
&& ~~~~~ \oplus ~ [(2m-2k-1)(m-k) + (2n-2k+1)(n-k)] \cR_0 \\
&&\\
&&\frac{\ad osp(2|2n)}{osp(2|2)} = 
\cR_1 ~ \oplus ~ \cR_{1/2} ~ \oplus ~ (2n-2) \cR'_{1/2} ~ \oplus ~ 
(2n^2-3n+1) \cR_0 ~ \oplus ~ (2n-2) \cR'_0 \\
&&\\
&&\frac{\ad osp(2|2n)}{sl(1|2)} = 
3\cR_1 ~ \oplus ~ \cR_{1/2} ~ \oplus ~ (4n-8)\cR'_{1/2} ~ \oplus ~ 
(2n^2-7n+7) \cR_0
\eno

$\see$ $osp(1|2)$ decompositions

\clearpage

\begin{table}[htbp]
\caption{$osp(1|2)$ decompositions of the adjoint representations of
the basic Lie superalgebras (singular cases).\label{table10}} 
\end{table}
\beo
&&\frac{\ad osp(2n+2|2n)}{osp(2k+1|2k) ~ \oplus ~ osp(2n-2k+1|2n-2k)} = \\
&& ~~~~~ \cR_{2n-2k-1} ~ \oplus ~ \cR_{2n-2k-3} ~ \oplus ~ \dots ~ 
\oplus ~ \cR_1 ~ \oplus ~ \cR_{2n-2k-1/2} ~ \oplus ~ \cR_{2n-2k-3/2}
 ~ \oplus ~ \dots \\
&& ~~~~~ \oplus ~ \cR_{3/2} ~ \oplus ~ \cR_{2k-1} ~ \oplus ~ 
\cR_{2k-3} ~ \oplus ~ \dots ~ \oplus ~ \cR_1 ~ \oplus ~ \cR_{2k-1/2}
 ~ \oplus ~ \cR_{2k-3/2} \\
&& ~~~~~ \oplus ~ \dots ~ \oplus ~ \cR_{3/2} ~ \oplus ~ \cR_n ~ 
\oplus ~ \cR_{n-1} ~ \oplus ~ \dots ~ \oplus ~ \cR_{n-2k} \\
&& ~~~~~ \oplus ~ \cR_{n-1/2} ~ \oplus ~ \cR_{n-3/2} ~ \oplus ~ 
\dots ~ \oplus ~ \cR_{n-2k+1/2} \\
&&\\
&&\frac{\ad [osp(2n-2|2n)]}{osp(2k-1|2k) ~ \oplus ~ osp(2n-2k-1|2n-2k)} = \\
&& ~~~~~ \cR_{2n-2k-1} ~ \oplus ~ \cR_{2n-2k-3} ~ \oplus ~ \dots ~ 
\oplus ~ \cR_1 ~ \oplus ~ \cR_{2n-2k-5/2} ~ \oplus ~ \cR_{2n-2k-7/2}
 ~ \oplus ~ \dots \\
&& ~~~~~ \oplus ~ \cR_{3/2} ~ \oplus ~ \cR_{2k-1} ~ \oplus ~ 
\cR_{2k-3} ~ \oplus ~ \dots ~ \oplus ~ \cR_1 ~ \oplus ~ \cR_{2k-5/2}
 ~ \oplus ~ \cR_{2k-7/2} \\ 
&& ~~~~~ \oplus ~ \dots ~ \oplus ~ \cR_{3/2} ~ \oplus ~ \cR_{n-1} ~ 
\oplus ~ \cR_{n-2} ~ \oplus ~ \dots ~ \oplus ~ \cR_{n-2k} \\
&& ~~~~~ \oplus ~ \cR_{n-3/2} ~ \oplus ~ \cR_{n-5/2} ~ \oplus ~ 
\dots ~ \oplus ~ \cR_{n-2k+1/2} \\
&&\\
&&\frac{\ad [osp(2n|2n)]}{osp(2k+1|2k) ~ \oplus ~ osp(2n-2k-1|2n-2k)} = \\
&& ~~~~~ \cR_{2n-2k-1} ~ \oplus ~ \cR_{2n-2k-3} ~ \oplus ~ \dots ~ 
\oplus ~ \cR_1 ~ \oplus ~ \cR_{2n-2k-5/2} ~ \oplus ~ \cR_{2n-2k-7/2}
 ~ \oplus ~ \dots \\
&& ~~~~~ \oplus ~ \cR_{3/2} ~ \oplus ~ \cR_{2k-1} ~ \oplus ~ 
\cR_{2k-3} ~ \oplus ~ \dots ~ \oplus ~ \cR_1 ~ \oplus ~ \cR_{2k-1/2}
 ~ \oplus ~ \cR_{2k-3/2} \\
&& ~~~~~ \oplus ~ \dots ~ \oplus ~ \cR_{3/2} ~ \oplus ~ \cR_{n-1} ~ 
\oplus ~ \cR_{n-2} ~ \oplus ~ \dots ~ \oplus ~ \cR_{n-2k} \\
&& ~~~~~ \oplus ~ \cR_{n-1/2} ~ \oplus ~ \cR_{n-3/2} ~ \oplus ~ 
\dots ~ \oplus ~ \cR_{n-2k-1/2} \\
&&\\
&&\frac{\ad [osp(2n|2n)]}{osp(2k-1|2k) ~ \oplus ~ osp(2n-2k+1|2n-2k)} = \\ 
&& ~~~~~ \cR_{2n-2k-1} ~ \oplus ~ \cR_{2n-2k-3} ~ \oplus ~ \dots ~ 
\oplus ~ \cR_1 ~ \oplus ~ \cR_{2n-2k-1/2} ~ \oplus ~ \cR_{2n-2k-3/2}
 ~ \oplus ~ \dots \\ 
&& ~~~~~ \oplus ~ \cR_{3/2} ~ \oplus ~ \cR_{2k-1} ~ \oplus ~ 
\cR_{2k-3} ~ \oplus ~ \dots ~ \oplus ~ \cR_1 ~ \oplus ~ \cR_{2k-5/2}
 ~ \oplus ~ \cR_{2k-7/2} \\ 
&& ~~~~~ \oplus ~ \dots ~ \oplus ~ \cR_{3/2} ~ \oplus ~ \cR_{n-1} ~ 
\oplus ~ \cR_{n-2} ~ \oplus ~ \dots ~ \oplus ~ \cR_{n-2k+1} \\ 
&& ~~~~~ \oplus ~ \cR_{n-1/2} ~ \oplus ~ \cR_{n-3/2} ~ \oplus ~ 
\dots ~ \oplus ~ \cR_{n-2k+1/2} 
\eno

$\see$ $osp(1|2)$ decompositions

\clearpage

\begin{table}[t]
\centering
\begin{tabular}{|c|c|c|c|} \hline
$\cG$ &SSA &Decomposition of the &Decomposition of the \\
 &in $\cG$ &fundamental of $\cG$ &adjoint of $\cG$ \\ \hline
 &&& \\
$A(0,1)$
 &$A(0,1)$ &$\cR''_{1/2}$ &$\cR_1 \oplus \cR_{1/2}$ \\ 
 &&& \\ 
$A(0,2)$
 &$A(0,1)$ &$\cR''_{1/2} \oplus \cR''_0$
 &$\cR_1 \oplus \cR_{1/2} \oplus 2\cR'_{1/2} \oplus \cR_0$\\
 &&& \\ 
$A(1,1)$
 &$A(0,1)$ &$\cR''_{1/2} \oplus \cR_0$ 
 &$\cR_1 \oplus 3\cR_{1/2}$ \\
 &&& \\ 
$A(0,3)$
 &$A(0,1)$ &$\cR''_{1/2} \oplus 2\cR''_0$ 
 &$\cR_1 \oplus \cR_{1/2} \oplus 4\cR'_{1/2} \oplus 4\cR_0$ \\ 
 &&& \\ 
$A(1,2)$
 &$A(1,2)$ &$\cR''_{1}$ 
 &$\cR_2 \oplus \cR_{3/2} \oplus \cR_1 \oplus \cR_{1/2}$ \\
 &$A(0,1)$ &$\cR''_{1/2} \oplus \cR_0 \oplus \cR''_0$
 &$\cR_1 \oplus 3\cR_{1/2} \oplus 2\cR'_{1/2} \oplus 2\cR_0 \oplus
 2\cR'_0$ \\ 
 &$A(1,0)$ &$\cR_{1/2} \oplus 2\cR''_0$ 
 &$\cR_1 \oplus 5\cR_{1/2} \oplus 4\cR_0$ \\ &&& \\ \hline 
\end{tabular}
\caption{$osp(1|2)$ decompositions of the $A(m,n)$ superalgebras up to
rank 4. \label{table17}}
\end{table}

\begin{table}[p]
\centering
\begin{tabular}{|c|c|c|c|} \hline
$\cG$ &SSA &Decomposition of the &Decomposition of the \\
 &in $\cG$ &fundamental of $\cG$ &adjoint of $\cG$ \\ \hline
 &&& \\
$B(0,2)$
 &$B(0,1)$ &$\cR''_{1/2} \oplus 2\cR''_0$ 
 &$\cR_1 \oplus 2\cR'_{1/2} \oplus 3\cR_0$ \\
 &&& \\ 
$B(1,1)$
 &$B(1,1)$ &$\cR_1$ &$\cR_{3/2} \oplus \cR_1$ \\
 &$C(2), B(0,1)$ &$\cR''_{1/2} \oplus 2\cR_0$
 &$\cR_1 \oplus 2\cR_{1/2} \oplus \cR_0$ \\
 &&& \\ 
$B(0,3)$
 &$B(0,1)$ &$\cR''_{1/2} \oplus 4\cR''_0$ 
 &$\cR_1 \oplus 4\cR'_{1/2} \oplus 10\cR_0$ \\
 &&& \\ 
$B(1,2)$ 
 &$B(1,2)$ &$\cR''_{3/2}$ &$\cR_3 \oplus \cR_{3/2} \oplus \cR_1$ \\
 &$B(1,1)$ &$\cR_1 \oplus 2\cR''_0$ 
 &$\cR_{3/2} \oplus \cR_1 \oplus 2\cR'_1 \oplus 3\cR_0$ \\ 
 &$C(2), B(0,1)$ &$\cR''_{1/2} \oplus 2\cR_0 \oplus 2\cR''_0$ 
 &$\cR_1 \oplus 2\cR_{1/2} \oplus 2\cR'_{1/2} \oplus 4\cR_0
 \oplus 4\cR'_0$ \\ 
 &$C(2) \oplus B(0,1), A(0,1)$ &$2\cR''_{1/2} \oplus \cR_0$ 
 &$3\cR_1 \oplus 3\cR_{1/2} \oplus \cR_0$ \\ &&& \\ 
$B(2,1)$ 
 &$D(2,1), B(1,1)$ &$\cR_1 \oplus 2\cR_0$
 &$\cR_{3/2} \oplus 3\cR_1 \oplus \cR_0$ \\
 &$C(2), B(0,1)$ &$\cR''_{1/2} \oplus 4\cR_0$ 
 &$\cR_1 \oplus 4\cR_{1/2} \oplus 6\cR_0$ \\ 
 &$A(1,0)$ &$2\cR_{1/2} \oplus \cR_0$ 
 &$\cR_1 \oplus 3\cR_{1/2} \oplus 2\cR'_{1/2} \oplus 3\cR_0$ \\ 
 &&& \\ \hline  
\end{tabular}
\caption{$osp(1|2)$ decompositions of the $B(m,n)$ superalgebras of rank
2 and 3. \label{table18}}
\end{table}

\clearpage

\begin{table}[h]
\centering
\begin{tabular}{|c|c|c|c|} \hline
$\cG$ &SSA &Decomposition of the &Decomposition of the \\
 &in $\cG$ &fundamental of $\cG$ &adjoint of $\cG$ \\ \hline
 &&& \\
$B(0,4)$
 &$B(0,1)$ &$\cR''_{1/2} \oplus 6\cR''_0$ 
 &$\cR_1 \oplus 6 \cR'_{1/2} \oplus 21 \cR_0$ \\
 &&& \\ 
$B(1,3)$
 &$B(1,2)$ &$\cR''_{3/2} \oplus 2\cR''_0$ 
 &$\cR_3 \oplus \cR_{3/2} \oplus 2\cR'_{3/2} \oplus \cR_1
 \oplus 3\cR_0$ \\ 
 &$B(1,1)$ &$\cR_1 \oplus 4\cR''_0$ 
 &$\cR_{3/2} \oplus \cR_1 \oplus 4\cR'_1 \oplus 10\cR_0$ \\ 
 &$C(2) \oplus B(0,1), A(0,1)$ &$2\cR''_{1/2} \oplus \cR_0 \oplus 2\cR''_0$
 &$3\cR_1 \oplus 3\cR_{1/2} \oplus 4\cR'_{1/2} \oplus 4\cR_0 \oplus 
 2\cR'_0$ \\ 
 &$C(2), B(0,1)$ &$\cR''_{1/2} \oplus 4\cR''_0 \oplus 2\cR_0$
 &$\cR_1 \oplus 2\cR_{1/2} \oplus 4\cR'_{1/2} \oplus 11\cR_0 \oplus 
 9\cR'_0$ \\ &&& \\ 
$B(2,2)$
 &$B(2,2)$ &$\cR_2$ 
 &$\cR_{7/2} \oplus \cR_3 \oplus \cR_{3/2} \oplus \cR_1$ \\ 
 &$D(2,2), B(1,2)$ &$\cR''_{3/2} \oplus 2\cR_0$
 &$\cR_3 \oplus 3\cR_{3/2} \oplus \cR_1 \oplus \cR_0$ \\
 &$D(2,1), B(1,1)$ &$\cR_1 \oplus 2\cR''_0 \oplus 2\cR_0$
 &$\cR_{3/2} \oplus 3\cR_1 \oplus 2\cR'_1 \oplus 4\cR_0
 \oplus 4\cR'_0$ \\ 
 &$\left.\begin{array}{c} D(2,1) \oplus B(0,1) \\ B(1,1) \oplus C(2)
 \end{array} \right\}$
 &$\cR_1 \oplus \cR''_{1/2} \oplus \cR_0$ 
 &$2\cR_{3/2} \oplus 4\cR_1 \oplus 2\cR_{1/2}$ \\ 
 &$C(2) \oplus C(2), A(0,1)$ &$2\cR''_{1/2} \oplus 3\cR_0$
 &$3\cR_1 \oplus 7\cR_{1/2} \oplus 4\cR_0$ \\ 
 &$C(2), B(0,1)$ &$\cR''_{1/2} \oplus 4\cR_0 \oplus 2\cR''_0$ 
 &$\cR_1 \oplus 4\cR_{1/2} \oplus 2\cR'_{1/2} \oplus 
 9\cR_0 \oplus 8\cR'_0$ \\ 
 &$A(1,0)$ &$2\cR_{1/2} \oplus \cR_0 \oplus 2\cR''_0$ 
 &$\cR_1 \oplus 7\cR_{1/2} \oplus 2\cR'_{1/2} \oplus 6\cR_0 \oplus 2
 \cR'_0$ \\ &&& \\ 
$B(3,1)$ 
 &$D(2,1), B(1,1)$ &$\cR_1 \oplus 4\cR_0$ 
 &$\cR_{3/2} \oplus 5\cR_1 \oplus 6\cR_0$ \\ 
 &$C(2), B(0,1)$ &$\cR''_{1/2} \oplus 6\cR_0$ 
 &$\cR_1 \oplus 6\cR_{1/2} \oplus 15\cR_0$ \\ 
 &$A(1,0)$ &$2\cR_{1/2} \oplus 3\cR_0$
 &$\cR_1 \oplus 3\cR_{1/2} \oplus 6\cR'_{1/2} \oplus 6\cR_0$ \\ 
 &&& \\ \hline 
\end{tabular}
\caption{$osp(1|2)$ decompositions of the $B(m,n)$ superalgebras of 
rank 4. \label{table19}}
\end{table}

\clearpage

\begin{table}[htbp]
\centering
\begin{tabular}{|c|c|c|c|} \hline
$\cG$ &SSA &Decomposition of the &Decomposition of the \\
 &in $\cG$ &fundamental of $\cG$ &adjoint of $\cG$ \\ \hline
 &&& \\
$C(3)$ 
 &$A(0,1)$ &$2\cR''_{1/2}$
 &$3\cR_1 \oplus\cR_{1/2} \oplus\cR_0$ \\
 &$C(2)$ &$\cR''_{1/2} \oplus \cR_0 \oplus 2\cR''_0$ 
 &$\cR_1 \oplus \cR_{1/2} \oplus 2\cR'_{1/2} \oplus 3\cR_0 \oplus
 2\cR'_0$ \\ &&& \\ 
$C(4)$
 &$A(0,1)$ &$2\cR''_{1/2} \oplus 2\cR''_0$ 
 &$3\cR_1 \oplus \cR_{1/2} \oplus 4\cR'_{1/2} \oplus 4\cR_0$ \\
 &$C(2)$ &$\cR''_{1/2} \oplus \cR_0 \oplus 4\cR''_0$ 
 &$\cR_1 \oplus \cR_{1/2} \oplus 4\cR'_{1/2} \oplus 10\cR_0 \oplus
 4\cR'_0$ \\ 
 &&& \\ \hline
\end{tabular}
\caption{$osp(1|2)$ decompositions of the $C(n+1)$ superalgebras up to
rank 4. \label{table20}}
\end{table}

\vspace{20mm}

\begin{table}[hbp]
\centering
\begin{tabular}{|c|c|c|c|} \hline
$\cG$ &SSA &Decomposition of the &Decomposition of the \\
 &in $\cG$ &fundamental of $\cG$ &adjoint of $\cG$ \\ \hline
 &&& \\
$D(2,1)$
 &$D(2,1)$ &$\cR_1 \oplus \cR_0$ 
 &$\cR_{3/2} \oplus 2\cR_1$ \\ 
 &$C(2)$ &$\cR''_{1/2} \oplus 3\cR_0$ 
 &$\cR_1 \oplus 3\cR_{1/2} \oplus 3\cR_0$ \\ 
 &$A(1,0)$ &$2\cR_{1/2}$ 
 &$\cR_1 \oplus 3\cR_{1/2} \oplus 3\cR_0$ \\ 
 &&& \\ 
$D(2,2)$
 &$D(2,2)$ &$\cR''_{3/2} \oplus \cR_0$ 
 &$\cR_3 \oplus 2\cR_{3/2} \oplus \cR_1$ \\
 &$D(2,1)$ &$\cR_1 \oplus \cR_0 \oplus 2\cR''_0$ 
 &$\cR_{3/2} \oplus 2\cR_1 \oplus 2\cR'_1 \oplus 5\cR_0$ \\ 
 &$C(2)$ &$\cR''_{1/2} \oplus 3\cR_0 \oplus 2\cR''_0$ 
 &$\cR_1 \oplus 3\cR_{1/2} \oplus 2\cR'_{1/2} \oplus 6\cR_0 
 \oplus 6\cR'_0$ \\ 
 &$C(2) \oplus C(2), A(0,1)$ &$\cR''_{1/2} \oplus 2\cR_0$
 &$3\cR_1 \oplus 5\cR_{1/2} \oplus 2\cR_0$ \\
 &$B(1,1) \oplus B(0,1)$ &$\cR_1 \oplus \cR''_{1/2}$ 
 &$2\cR_{3/2} \oplus 3\cR_1 \oplus \cR_{1/2}$ \\ 
 &$A(1,0)$ &$2\cR_{1/2} \oplus 2\cR''_0$ 
 &$\cR_1 \oplus 7\cR_{1/2} \oplus 6\cR_0$ \\ &&& \\ 
$D(3,1)$
 &$D(2,1)$ &$\cR_1 \oplus 3\cR_0$ 
 &$\cR_{3/2} \oplus 4\cR_1 \oplus 3\cR_0$ \\ 
 &$C(2)$ &$\cR''_{1/2} \oplus 5\cR_0$ 
 &$\cR_1 \oplus 5\cR_{1/2} \oplus 10\cR_0$ \\
 &$A(1,0)$ &$2\cR_{1/2} \oplus 2\cR_0$
 &$\cR_1 \oplus 3\cR_{1/2} \oplus 4\cR'_{1/2} \oplus 4\cR_0$ \\
 &&& \\ \hline
\end{tabular}
\caption{$osp(1|2)$ decompositions of the $D(m,n)$ superalgebras up to
rank 4. \label{table21}}
\end{table}

\clearpage

\begin{table}[p]
\centering
\begin{tabular}{|c|c|} \hline
SSA &Decomposition of the \\ in $\cG$ &adjoint of $\cG$ \\ \hline & \\
$A(1,0)$ &$\cR_1 \oplus 7\cR_{1/2} \oplus 14\cR_0$ \\ 
$A(0,1)$ &$\cR_1 \oplus 3\cR_{1/2} \oplus 6\cR'_{1/2} \oplus 6\cR_0
 \oplus 2\cR'_0$ \\ 
$C(2)$ &$5\cR_1 \oplus 3\cR_{1/2} \oplus 6\cR_0$ \\ 
$D(2,1;2)$ &$\cR_{3/2} \oplus 2\cR'_{3/2} \oplus 2\cR_1 \oplus 
 2\cR'_{1/2} \oplus 3\cR_0$ \\ & \\ \hline
\end{tabular}
\caption{$osp(1|2)$ decompositions of the superalgebra $F(4)$.
\label{table23}}
\end{table}

\begin{table}[htbp]
\centering
\begin{tabular}{|c|c|} \hline
SSA &Decomposition of the \\ in $\cG$ &adjoint of $\cG$ \\ \hline & \\
$A(1,0)$ &$\cR_1 \oplus 3\cR_{1/2} \oplus 4\cR'_{1/2} \oplus 3\cR_0
 \oplus 2\cR'_0$ \\ 
$A(1,0)'$ &$2\cR'_{3/2} \oplus \cR_1 \oplus 3\cR_{1/2} \oplus 3\cR_0$ \\ 
 & \\ 
$B(0,1)$ &$\cR_1 \oplus 6\cR_{1/2} \oplus 8\cR_0$ \\ & \\ 
$B(1,1)$ &$\cR_{3/2} \oplus 2\cR'_{3/2} \oplus \cR_1 \oplus 3\cR_0 \oplus
 2\cR'_0$ \\ 
 & \\ 
$D(2,1;3)$ &$\cR_2 \oplus \cR_{3/2} \oplus 3\cR_1$ \\ & \\ \hline
\end{tabular}
\caption{$osp(1|2)$ decompositions of the superalgebra $G(3)$.
\label{table22}}
\end{table} 

\begin{table}[b]
\centering
\begin{tabular}{|c|c|c|} \hline
SSA &Decomposition of the &Decomposition of the \\
 in $\cG$ &fundamental of $\cG$ &adjoint of $\cG$ \\ \hline
 && \\
$D(2,1)$ &$\cR_1 \oplus \cR_0$ 
 &$\cR_{3/2} \oplus 2\cR_1$ \\ 
$C(2)$ &$\cR''_{1/2} \oplus 3\cR_0$ 
 &$\cR_1 \oplus 3\cR_{1/2} \oplus 3 \cR_0$ \\ 
$A(1,0)$ &$2\cR_{1/2}$ 
 &$\cR_1 \oplus 3\cR_{1/2} \oplus 3\cR_0$ \\
 && \\ \hline
\end{tabular}
\caption{$osp(1|2)$ decompositions of the superalgebra $D(2,1;\alpha)$.
\label{table24}}
\end{table}

\clearpage

\begin{table}[t]
\centering
\begin{tabular}{|c|c|c|c|} \hline
$\cG$ &SSA &Decomposition of the &Decomposition of the \\
 &in $\cG$ &fundamental of $\cG$ &adjoint of $\cG$ \\ \hline
&&& \\
$A(0,1)$
 &$A(0,1)$ &$\pi''(\half,\half)$ & $\pi(0,1)$ \\ 
&&& \\ 
$A(0,2)$
 &$A(0,1)$ &$\pi''(\half,\half) \oplus \pi''(0,0)$ & $\pi(0,1) \oplus
 \pi'(\half,\half)  \oplus\pi'(-\half,\half) \oplus \pi(0,0)$ \\ 
&&& \\ 
$A(1,1)$
 &$A(0,1)$ &$\pi''(\half,\half) \oplus \pi(0,0)$ & $\pi(0,1) \oplus
 \pi(\half,\half) \oplus \pi(-\half,\half)$ \\
&&& \\ 
$A(0,3)$
 &$A(0,1)$ &$\pi''(\half,\half) \oplus 2\pi''(0,0)$ & $\pi(0,1) \oplus
 2\pi'(\half,\half) \oplus 2\pi'(-\half,\half) \oplus 4\pi(0,0)$ \\ 
&&& \\ 
$A(1,2)$
 &$A(1,2)$ &$\pi''(1,1)$ & $\pi(0,2) \oplus \pi(0,1)$ \\ &&& \\
 &$A(0,1)$ &$\pi''(\half,\half) \oplus \pi(0,0) \oplus \pi''(0,0)$ 
 & $\begin{array}{c}
 \pi(0,1) \oplus \pi(\half,\half) \oplus \pi(-\half,\half) \\
 \oplus \pi'(\half,\half) \oplus \pi'(-\half,\half) \oplus 2\pi(0,0) 
 \oplus 2\pi'(0,0) \end{array}$ \\ &&& \\
 &$A(1,0)$ &$\pi(\half,\half) \oplus 2\pi''(0,0)$ & $\pi(0,1) \oplus 
 2\pi(\half,\half) \oplus 2\pi(-\half,\half) \oplus 4\pi(0,0)$ \\
&&& \\ \hline
\end{tabular}
\caption{$sl(1|2)$ decompositions of the $A(m,n)$ superalgebras up to
rank 4. \label{table33}}
\end{table}

\begin{table}[p]
\centering
\begin{tabular}{|c|c|c|c|} \hline
$\cG$ &SSA &Decomposition of the &Decomposition of the \\
 &in $\cG$ &fundamental of $\cG$ &adjoint of $\cG$ \\ \hline
&&& \\
$B(1,1)$
 & $C(2)$ &$\pi''(0,\half) \oplus \pi(0,0)$ & $\pi(0,1) \oplus
 \pi(0,\half)$ \\ &&& \\
$B(1,2)$
 & $C(2)$ & $\pi''(0,\half) \oplus \pi(0,0) \oplus 2\pi''(0,0) $ 
 & $\begin{array}{c} \pi(0,1) \oplus \pi(0,\half) \oplus 2\pi'(0,\half) \\ 
 \oplus 3\pi(0,0) \oplus 2\pi'(0,0)
 \end{array}$ \\ &&& \\
 & $A(0,1)$ &$\pi''(\half,\half) \oplus \pi''(-\half,\half) \oplus \pi(0,0)$ 
 & $\begin{array}{c} \pi(0,1) \oplus \pi(1,1) \oplus \pi(-1,1) \\ 
 \oplus \pi(\half,\half) \oplus \pi(-\half,\half) \oplus \pi(0,0) 
 \end{array}$ \\
&&& \\ 
$B(2,1)$
 &$C(2)$
 &$\pi''(0,\half) \oplus 3\pi(0,0)$ & $\pi(0,1) \oplus 3\pi(0,\half) \oplus 
 3\pi(0,0)$ \\ &&& \\
 &$A(1,0)$ &$\pi(\half,\half)\oplus\pi(-\half,\half) \oplus \pi(0,0)$ &
 $\begin{array}{c} \pi(0,1)\oplus \pi(\frac{3}{2},\half) \oplus
 \pi(-\frac{3}{2},\half) \\
 \oplus \pi'(\half,\half) \oplus \pi'(-\half,\half) \oplus \pi(0,0) 
 \end{array}$ \\
&&& \\ \hline
\end{tabular}
\caption{$sl(1|2)$ decompositions of the $B(m,n)$ superalgebras of rank
2 and 3. \label{table34}}
\end{table}

\clearpage

\begin{table}[htbp]
\centering
\begin{tabular}{|c|c|c|c|} \hline
$\cG$ &SSA &Decomposition of the &Decomposition of the \\
 &in $\cG$ &fundamental of $\cG$ &adjoint of $\cG$ \\ \hline
&&& \\
$B(1,3)$
 &$A(0,1)$
 &$\begin{array}{c} \pi''(\half,\half) \oplus \pi''(-\half,\half) \\ 
 \oplus \pi(0,0) \oplus 2\pi''(0,0) \end{array}$
 & $\begin{array}{c} \pi(0,1) \oplus \pi(1,1) \oplus \pi(-1,1) \\
 \oplus \pi(\half,\half) \oplus \pi(-\half,\half) \oplus 
 2\pi'(\half,\half) \\ 
 \oplus 2\pi'(-\half,\half) \oplus 4\pi(0,0) \oplus 2\pi'(0,0) 
 \end{array}$ \\ &&& \\
 &$C(2)$ 
 &$\pi''(0,\half) \oplus 4\pi''(0,0) \oplus \pi(0,0)$ 
 &$\begin{array}{c}
 \pi(0,1) \oplus \pi(0,\half) \oplus 4\pi'(0,\half) \\ 
 \oplus 10\pi(0,0) \oplus 4\pi'(0,0) \end{array}$ \\
&&& \\ 
$B(2,2)$
 &$2 ~ C(2)$ 
 & $2\pi''(0,\half) \oplus \pi(0,0)$ 
 &$3\pi(0,1)\oplus 2\pi(0,\half) \oplus \pi(0;-\half,\half;0)$ \\ &&& \\
 &$A(0,1)$ 
 & $\pi''(\half,\half) \oplus \pi''(-\half,\half) \oplus 3\pi(0,0)$
 &$\begin{array}{c} \pi(0,1)\oplus \pi(1,1) \oplus \pi(-1,1) \\
 \oplus 3\pi'(\half,\half) \oplus 3\pi'(-\half,\half) \oplus 4\pi(0,0) 
 \end{array}$ \\ &&& \\
 & $C(2)$ &$\pi''(0,\half) \oplus 3\pi(0,0) \oplus 2\pi''(0,0)$ 
 & $\begin{array}{c}
 \pi(0,1) \oplus 3\pi(0,\half) \oplus 2\pi'(0,\half)\\ 
 \oplus 6\pi(0,0) \oplus 6\pi'(0,0) \end{array}$ \\ &&& \\
 &$A(1,0)$ 
 &$\begin{array}{c} \pi(\half,\half) \oplus \pi(-\half,\half) \\
 \oplus \pi(0,0) \oplus 2\pi''(0,0) \end{array}$
 & $\begin{array}{c}
 \pi(0,1) \oplus \pi(\frac{3}{2},\half) \oplus \pi(-\frac{3}{2},\half) \\ 
 \oplus 2\pi(\half,\half) \oplus 2\pi(-\half,\half) \oplus 
 \pi'(\half,\half) \\ 
 \oplus \pi'(-\half,\half) \oplus 4\pi(0,0) \oplus 2\pi'(0,0) 
 \end{array}$ \\
&&& \\ 
$B(3,1)$
 &$C(2)$ 
 &$\pi''(0,\half) \oplus 5\pi(0,0)$ & $\pi(0,1)\oplus 5\pi(0,\half)
 \oplus 10\pi(0,0)$ \\ &&& \\
 &$A(1,0)$ 
 &$\pi(\half,\half) \oplus \pi(-\half,\half) \oplus 3\pi(0,0)$ 
 & $\begin{array}{c} \pi(0,1) \oplus \pi(\frac{3}{2},\half) \oplus 
 \pi(-\frac{3}{2},\half) \\ 
 \oplus 3\pi'(\half,\half) \oplus 3\pi'(-\half,\half) \oplus 4\pi(0,0) 
 \end{array}$ \\
&&& \\ \hline
\end{tabular}
\caption{$sl(1|2)$ decompositions of the $B(m,n)$ superalgebras of rank 4. 
\label{table35}} 
\end{table}

\clearpage

\begin{table}[t]
\centering
\begin{tabular}{|c|c|c|c|} \hline
$\cG$ &SSA &Decomposition of the &Decomposition of the \\
 &in $\cG$ &fundamental of $\cG$ &adjoint of $\cG$ \\ \hline
&&& \\
$C(3)$
 &$A(0,1)$ 
 &$\pi''(\half,\half) \oplus \pi''(-\half,\half)$ & $\pi(0,1) \oplus \pi(1,1) 
 \oplus \pi(-1,1) \oplus \pi(0,0)$ \\ 
 &$C(2)$ 
 &$\pi''(0,\half) \oplus 2\pi''(0,0)$ & $\pi(0,1) \oplus 2\pi'(0,\half) 
 \oplus 3\pi(0,0)$ \\ 
&&& \\
$C(4)$
 &$A(0,1)$
 &$\pi''(\half,\half) \oplus \pi''(-\half,\half) \oplus 2\pi''(0,0)$
 &$\begin{array}{c} \pi(0,1) \oplus \pi(1,1) \oplus \pi(-1,1) \\ 
 \oplus 2\pi'(\half,\half) \oplus 2\pi'(-\half,\half) \oplus 4\pi(0,0) 
 \end{array}$ \\ &&& \\
 &$C(2)$ 
 &$\pi''(0,\half) \oplus 4\pi''(0,0)$ & $\pi(0,1) \oplus 4\pi'(0,\half) 
 \oplus 10\pi(0,0)$ \\
&&& \\ \hline
\end{tabular}
\caption{$sl(1|2)$ decompositions of the $C(n+1)$ superalgebras up to
rank 4. \label{table36}}
\end{table}

\begin{table}[p]
\centering
\begin{tabular}{|c|c|c|c|} \hline
$\cG$ &SSA &Decomposition of the &Decomposition of the \\
 &in $\cG$ &fundamental of $\cG$ &adjoint of $\cG$ \\ \hline
&&& \\
$D(2,1)$
 &$C(2)$ 
 &$\pi''(0,\half) \oplus 2\pi(0,0)$ &$\pi(0,1) \oplus
 2\pi(0,\half) \oplus \pi(0,0)$ \\ 
 &$A(1,0)$ &$\pi(\half,\half) \oplus \pi(-\half,\half)$ 
 & $\pi(0,1)\oplus \pi(\frac{3}{2},\half) \oplus 
 \pi(-\frac{3}{2},\half) \oplus \pi(0,0)$ \\
&&& \\ 
$D(2,2)$
 &$C(2)$ 
 &$\pi''(0,\half) \oplus 2\pi(0,0) \oplus 2\pi''(0,0)$ 
 & $\begin{array}{c} \pi(0,1)\oplus 2\pi(0,\half) \oplus 2\pi'(0,\half) \\ 
 \oplus 4\pi(0,0) \oplus 4\pi'(0,0) \end{array}$ \\ &&& \\
 &$ 2 ~ C(2)$
 &$2\pi''(0,\half)$ & $3\pi(0,1)\oplus \pi(0;-\half,\half;0)$ \\ &&& \\
 & $A(0,1) $ 
 & $\pi''(\half,\half) \oplus \pi''(-\half,\half) \oplus 2\pi(0,0)$ 
 &$\begin{array}{c} \pi(0,1)\oplus \pi(1,1) \oplus \pi(-1,1) \\
 \oplus 2\pi(\half,\half) \oplus 2\pi(-\half,\half) \oplus 2\pi(0,0) 
 \end{array}$\\ &&& \\
 &$A(1,0)$ & $\pi(\half,\half) \oplus \pi(-\half,\half) \oplus 2\pi''(0,0)$ 
 & $\begin{array}{c} \pi(0,1) \oplus \pi(\frac{3}{2},\half) \oplus 
 \pi(-\frac{3}{2},\half) \\ 
 \oplus 2\pi(\half,\half) \oplus 2\pi(-\half,\half) \oplus4 \pi(0,0) 
 \end{array}$ \\
&&& \\ 
$D(3,1)$
 &$C(2)$ &$\pi''(0,\half) \oplus 4\pi(0,0)$ 
 &$\pi(0,1) \oplus 4\pi(0,\half) \oplus 6\pi(0,0)$ \\ &&& \\
 &$A(1,0)$ &$\pi(\half,\half) \oplus \pi(-\half,\half) \oplus 2\pi(0,0)$
 &$\begin{array}{c} \pi(0,1) \oplus \pi(\frac{3}{2},\half) \oplus 
 \pi(-\frac{3}{2},\half) \\ 
 \oplus 2\pi'(\half,\half) \oplus 2\pi'(-\half,\half) \oplus 2\pi(0,0) 
 \end{array}$ \\
&&& \\ \hline
\end{tabular}
\caption{$sl(1|2)$ decompositions of the $D(m,n)$ superalgebras up to
rank 4. \label{table37}}
\end{table}

\clearpage

\begin{table}[htbp]
\centering
\begin{tabular}{|c|c|c|} \hline
$\cG$ & SSA &Decomposition of the \\
 & in $\cG$ &adjoint of $\cG$ \\ \hline && \\
$F(4)$ 
 &$A(1,0)$ 
 &$(0,1) \oplus 3 \pi(\frac{1}{6},\half) \oplus
 3\pi(-\frac{1}{6},\half) \oplus 8\pi(0,0)$ \\ && \\
 &$A(0,1)$ 
 &$\begin{array}{c} \pi(0,1) \oplus \pi(1,\half) \oplus \pi(-1,\half) 
 \oplus 4 \pi(0,0) \\
 \oplus 2\pi'(\half,\half) \oplus 2\pi'(-\half,\half) \oplus 
 2\pi'(0,\half) \end{array}$ \\ && \\
 &$C(2)$ 
 &$\begin{array}{c} \pi(0,1) \oplus 2\pi(1,1) \oplus 2\pi(-1,1) \\
 \oplus \pi(\frac{5}{2},\half) \oplus \pi(-\frac{5}{2},\half) \oplus 
 4\pi(0,0) \end{array}$ \\ && \\ 
 \hline && \\
$G(3)$ 
 & $A(1,0)$ 
 &$\begin{array}{c} \pi(0,1) \oplus \pi(\frac{5}{6},\half) \oplus 
 \pi(-\frac{5}{6},\half) \oplus \pi'(\frac{1}{6},\half) \\
 \oplus \pi'(-\frac{1}{6},\half) \oplus \pi'(\frac{1}{2},\half) \oplus 
 \pi'(-\frac{1}{2},\half) \oplus \pi(0,0) \end{array}$ \\ && \\ 
 &$A(1,0)'$ 
  &$\begin{array}{c} \pi(0,1) \oplus \pi(\frac{7}{2},\half) \oplus 
  \pi(-\frac{7}{2},\half) \\
 \oplus \pi'(\frac{3}{2},\frac{3}{2}) \oplus
 \pi'(-\frac{3}{2},\frac{3}{2}) \oplus \pi(0,0) \end{array}$ \\ && \\
 &$C(2)$ 
 &$\pi(0,1) \oplus 2\pi(\frac{1}{4},\half) \oplus
 2\pi(-\frac{1}{4},\half) \oplus \pi(0,\frac{1}{2})$ \\ && \\ 
 \hline && \\
$D(2,1;\alpha)$
 & $A(1,0)$
 & $\pi(0,1) \oplus \pi(\half(2\alpha+1),\half) \oplus 
 \pi(-\half(2\alpha+1),\half) \oplus \pi(0,0)$ \\ && \\ 
 & $A(1,0)'$
 & $\pi(0,1) \oplus \pi(\half\left(\frac{1-\alpha}{1+\alpha}\right)
 ,\half) \oplus \pi(-\half\left(\frac{1-\alpha}{1+\alpha}\right),\half) 
 \oplus \pi(0,0)$ \\ && \\
 & $C(2)$
 & $\pi(0,1) \oplus \pi(\half\left(\frac{2+\alpha}{\alpha}\right)
 ,\half) \oplus \pi(-\half\left(\frac{2+\alpha}{\alpha}\right),\half) 
 \oplus \pi(0,0)$ \\ && \\
 \hline
\end{tabular}
\caption{$sl(1|2)$ decompositions of the exceptional superalgebras. 
\label{table38}}
\end{table}

\vspace{20mm}

Let us remark that for $D(2,1;\alpha)$ from any $sl(1|2)$ decomposition
one gets the two others by replacing $\alpha$ by one of the values 
$\alpha^{-1}$, $-1-\alpha$, $\displaystyle{\frac{-\alpha}{1+\alpha}}$. 
This corresponds to isomorphic versions of the exceptional superalgebra
$D(2,1;\alpha)$ ($\see$). One can check this triality-like property,
which certainly deserves some developments, by the studying the
completely odd Dynkin diagram of $D(2,1;\alpha)$.

\end{document}